\documentclass{aa}
\usepackage{natbib}
\usepackage{amsmath}
\usepackage[utf8]{inputenc}
\usepackage{arydshln}

\usepackage{graphicx}
\usepackage{txfonts}
\usepackage{hyperref}
\usepackage{pdflscape}

\newcommand\kms{km$\,$s$^{-1}$}
\newcommand\Msol{M$_{\odot}$}

\newcommand{\hi}{H\,{\sc i}}

\title{Evolution of compact groups from intermediate to final stages:}
\subtitle{A case study of the \hi \ content of HCG 16}
\titlerunning{\hi \ study of HCG 16}
\author{M.~G.~Jones\inst{\ref{inst1}}\thanks{E-mail: mjones@iaa.es} \and L.~Verdes-Montenegro\inst{\ref{inst1}}\thanks{E-mail: lourdes@iaa.es} \and 
A.~Damas-Segovia\inst{\ref{inst1}} \and 
S.~Borthakur\inst{\ref{inst2}} \and 
M.~Yun\inst{\ref{inst3}} \and 
A.~del~Olmo\inst{\ref{inst1}} \and J.~Perea\inst{\ref{inst1}} \and 
J. Rom\'{a}n\inst{\ref{inst1}} \and 
S.~Luna\inst{\ref{inst1}} \and 
D.~Lopez~Gutierrez\inst{\ref{inst4}} \and
B.~Williams\inst{\ref{inst6}} \and
F.~P.~A.~Vogt\inst{\ref{inst5},\ref{inst7}} \and 
J.~Garrido\inst{\ref{inst1}} \and S.~Sanchez\inst{\ref{inst1}} \and
J.~Cannon\inst{\ref{inst4}} \and
P.~Ram\'{i}rez-Moreta\inst{\ref{inst1}}}
\authorrunning{Jones et al.}

\date{\today}

\institute{
Instituto de Astrof\'{i}sica de Andaluc\'{i}a (CSIC), Glorieta de la Astronom\'{i}a, 18008 Granada, Spain\label{inst1}
\and
School of Earth and Space Exploration, Arizona State University, 781 Terrace Mall, Tempe, AZ 85287, USA\label{inst2}
\and
Astronomy Department, University of Massachusetts, Amherst, MA 01003, USA\label{inst3}
\and
Department of Physics \& Astronomy, Macalester College, 1600 Grand Avenue, Saint Paul, MN 55105, USA\label{inst4}
\and
Department of Physics and Astronomy, University of Delaware, Newark, DE 19716\label{inst6}
\and
European Southern Observatory, Av. Alonso de Cordova 3107, 763 0355 Vitacura, Santiago, Chile\label{inst5}
\and
ESO Fellow\label{inst7}
}

\abstract{
    Hickson Compact Group (HCG) 16 is a prototypical compact group of galaxies in an intermediate stage of the evolutionary sequence proposed by \citet{Verdes-Montenegro+2001}, where its galaxies are losing gas to the intra-group medium (IGrM). The group hosts galaxies that are \hi-normal, \hi-poor, centrally active with both AGN and starbursts, as well as a likely new member and a $\sim$160 kpc long \hi \ tidal feature. Despite being a well-studied group at all wavelengths, no previous study of HCG 16 has focused on its extraordinary \hi \ component.
}{
    The characteristics of HCG 16 make it an ideal case study for exploring which processes are likely to dominate the late stages of evolution in compact groups, and ultimately determine their end states. In order to build a coherent picture of the evolution of this group we make use of the multi-wavelength data available, but focus particularly on \hi \ as a tracer of interactions and evolutionary phase.
}{
    We reprocess archival VLA L-band observations of HCG 16 using the multi-scale CLEAN algorithm to accurately recover diffuse features. Tidal features and galaxies are separated in 3 dimensions using the \texttt{SlicerAstro} package. The \hi \ deficiency of the separated galaxies is assessed against the benchmark of recent scaling relations of isolated galaxies. This work has been performed with particular attention to reproducibility and is accompanied by a complete workflow to reproduce all the final data products, figures, and results.
}{
    Despite the clear disruption of the \hi \ component of HCG 16 we find that it is not globally \hi \ deficient, even though HCG 16a and b have lost the majority of their \hi \ and almost 50\% of the group's \hi \ is in the IGrM. The \hi \ content of HCG 16d shows highly disturbed kinematics, with only a marginal velocity gradient that is almost perpendicular to its optical major axis. The $\sim$160 kpc long tail extending towards the South-East appears to be part of an even larger structure which spatially and kinematically connects NGC 848 to the North-West corner of the group.
}{
    This study indicates that in the recent past ($\sim$1 Gyr) galaxies HCG 16a and b likely underwent major interactions that unbound gas without triggering significant star formation. This gas was then swept away by a high speed, close encounter with NGC 848. The starburst events HCG 16c and d, likely initiated by their mutual interaction, have triggered galactic winds which, in the case of HCG 16d, appears to have disrupted its \hi \ reservoir.
    The tidal features still connected to all these galaxies indicate that more \hi \ will soon be lost to the IGrM, while that which remains in the discs will likely be consumed by star formation episodes triggered by their on-going interaction. This is expected to result in a collection of gas-poor galaxies embedded in a diffuse \hi \ structure, which will gradually (over several Gyr) be evaporated by the UV background, resembling the final stage of the evolutionary model of compact groups.
}

\begin{document}

\maketitle

\section{Introduction}

Hickson compact groups \citep[HCGs,][]{Hickson1982} of galaxies are systems characterised by a high local density while being located in low density environments when viewed at larger scales. This high density combined with low velocity dispersions \citep{Hickson+1992} in many cases leads them to exhibit multiple physical processes associated with galaxy--galaxy interactions: tidal tails and bridges visible optically, in atomic gas (\hi), or both \citep[e.g.][]{Verdes-Montenegro+1997,Sulentic+2001,Verdes-Montenegro+2005,Serra+2013,Konstantopoulos+2013}; intragroup diffuse X-ray emission \citep{Belsole+2003,Desjardins+2013,OSullivan+2014b}; shock excitation from starburst winds or galaxy--tidal debris collisions \citep{Rich+2010,Vogt+2013,Cluver+2013}; anomalous star formation (SF) activity, molecular gas content and morphological transformations \citep{Tzanavaris+2010,Plauchu-Frayn+2012,Alatalo+2015,Eigenthaler+2015,Zucker+2016,Lisenfeld+2017}, among others.

Single dish \hi \ studies of HCGs \citep{Williams+Rood1987,Huchtmeier1997} revealed that most are deficient in \hi. \citet{Verdes-Montenegro+2001} expanded on this discovery by performing a comprehensive study of the total \hi \ contents of 72 HCGs observed with single dish telescopes, together with an analysis of the spatial distribution and kinematics of the \hi \ gas within a subset of 16 HCGs observed with the VLA (Very Large Array). As a result of the analysis the authors proposed an evolutionary sequence in which compact group galaxies become increasingly \hi \ deficient as the group evolves. In phase 1 of the sequence the \hi \ is relatively unperturbed and found mostly in the discs of the galaxies, with the remaining gas found in incipient tidal tails. In phase 2 30--60\% of the total \hi \ mass forms tidal features. In Phase 3a most, if not all, of the \hi \ has been stripped from the discs of the galaxies and is either found in tails, or is not detected. The least common phase, 3b, involves groups where the \hi \ gas seems to form a large cloud with a single velocity gradient that contains all the galaxies. However, of the four groups proposed to be in this phase, HCGs 18 and 54 are now thought to be false groups \citep{Verdes-Montenegro+2001,Verdes-Montenegro+2002} and the \hi \ distribution in HCG 26 probably does not fulfil the necessary criteria (Damas-Segovia et al. in prep), leaving only HCG 49 and raising the question whether phase 3b is a genuine phase of CG evolution. A slightly different evolutionary sequence was proposed by \citet{Konstantopoulos+2010}, where the evolution follows a similar sequence, but all groups are split into two categories: a) those where the gas is mostly consumed by SF in the galactic discs before major interactions can strip it, leading to late-time dry mergers, and b) those where the gas is removed from the galaxies through tidal stripping early on in the group's evolution, leading to a hot, diffuse intra-group medium (IGrM) at late times.

\citet{Borthakur+2010,Borthakur+2015} compared single dish \hi \ spectra of HCGs obtained with the GBT (Green Bank Telescope) and VLA \hi \ maps to demonstrate that some HCGs have a diffuse \hi \ component that was not detected by the VLA and can extend to up to 1000 \kms \ in velocity width. The fraction of \hi \ in this component seemed to be greater for groups with larger \hi \ deficiencies, and thus makes up some, but not all, of the ``missing'' \hi \ reported by \citet{Verdes-Montenegro+2001}. The connection between the \hi \ content and distribution, SF activity and X-ray emission has been the subject of numerous studies \citep[e.g.][]{Ponman+1996,Rasmussen+2008,Bitsakis+2011,Martinez-Badenes+2012,Desjardins+2013,OSullivan+2014a,OSullivan+2014b}, however, how the observed \hi \ depletion occurs and, more generally, how the groups might evolve from phase 2 to 3, remains far from understood.

Assuming the proposed evolutionary scenario is correct, detailed studies of phase 2 groups are of special relevance for addressing the unknowns above, because in these groups the processes that drive the transformation to phase 3 HCGs should be at work. HCG 16 is a prototypical example of this intermediate phase of evolution. Its \hi \ gas is in the process of leaving the discs of the galaxies and filling the intragroup medium with significant amounts of \hi \ in tidal tails, but the group has not yet become globally \hi \ deficient. HCG 16 also hosts an array of other ongoing processes that will likely shape its future evolution: active galactic nuclei (AGN), a new member, starburst events and the accompanying winds and shocks. Many of these have been studied in detail in an extensive set of papers focusing on the group or a small sample of groups including HCG 16 \citep{Ribeiro+1996,Mendes+1998,Rich+2010,Vogt+2013,Konstantopoulos+2013,OSullivan+2014a,OSullivan+2014b}, however, to date there has been no study specifically of the group's extraordinary \hi \ component, which will be the focus of this work.

The aim of this paper is then to shed light on how the final stages of evolution in HCGs are reached by performing a census of the on-going physical processes in HCG 16, identifying those that could be influencing the fate of the \hi \ in the group and its evolution towards a phase 3 morphology.

In the following section we give a brief overview of HCG 16, in \S\ref{sec:data} we describe the observations and standard data reduction, \S\ref{sec:HIsep} covers the separation of \hi \ into galaxies and tidal features. In \S\ref{sec:results} we present our results for the group as a whole and the individual galaxies, and in \S\ref{sec:discuss} we discuss their interpretation and attempt to construct a coherent picture of the evolution of the group. Throughout this paper we assume a distance of $55.2 \pm 3.3$ Mpc for HCG 16 and all its constituent galaxies.

\section{Overview of HCG 16}

HCG 16 was first identified in the Atlas of Peculiar Galaxies \citep{Arp1966}, Arp 318, and later classified as a compact group by \citet{Hickson1982}. Since then it has been referenced in approximately 100 journal articles and is thus an extremely well-studied group with a large amount of multi-wavelength data. However, this work represents the first focused investigation of its \hi \ component.

The core group contains 4 disc galaxies, each with stellar mass of the order $10^{10}$--$10^{11}$ \Msol, that all fall within a projected separation of just 7\arcmin \ (120 kpc). There is a fifth, similar sized, member of the group (NGC 848) to the South-East that was identified as being associated with the core group through optical spectroscopy \citep{deCarvalho+1997}, and was later shown to be connected in \hi \ also \citep{Verdes-Montenegro+2001}. \citet{deCarvalho+1997} also identified two other dwarf galaxy members of HCG 16, PGC 8210 to the South-West and 2MASS J02083670-0956140 to the North-West. The latter is not considered in this work as it falls outside of the primary beam of the \hi \ observations. The basic optical properties of the others are summarised in Table \ref{tab:optprops}.

In \S\ref{sec:results} we discuss the details of each galaxy individually, here we provide a brief overview of their properties for readers unfamiliar with this group. Figure \ref{fig:optim} shows an optical $grz$ image (from the Dark Energy Camera Legacy Survey) of the group. From North-West to South-East the galaxies are HCG 16b, a, c, d, and NGC 848. PGC 8210 is to the South-West of the core group. The 4 galaxies in the core group form 2 interacting pairs, HCG 16a and b, and HCG 16c and d. In the first pair both galaxies host an AGN, but have limited SF activity, while the second pair does not host AGN and both galaxies are currently undergoing nuclear starburst events. NGC 848 is physically connected to the core group by an enormous \hi \ tail, while PGC 8210 appears quite separate and shows no evidence in \hi \ for a past interaction with the core group.

\begin{table*}
\centering
\caption{HCG 16 galaxies}
\label{tab:optprops}
\begin{tabular}{cccccccc}
\hline \hline
HCG ID & Other name & RA            & Dec          & Type & $v_\mathrm{opt}/\mathrm{km\,s^{-1}}$ & $D_{25}$/\arcsec & $L_\mathrm{B}/\mathrm{L_\odot}$    \\ \hline
HCG 16a    & NGC 835    & 2h 09m 24.6s & -10$^{\circ}$ 08\arcmin \ 09\arcsec & Sab  & $4073$\textsuperscript{a} & 76  & $10.27 \pm 0.05$\\
HCG 16b    & NGC 833    & 2h 09m 20.8s & -10$^{\circ}$ 07\arcmin \ 59\arcsec & SABa   & $3864$\textsuperscript{a} & 89  & $10.14 \pm 0.05$ \\
HCG 16c    & NGC 838    & 2h 09m 38.5s & -10$^{\circ}$ 08\arcmin \ 48\arcsec & S0a  & $3849$\textsuperscript{b} & 69  & $10.11 \pm 0.02$ \\
HCG 16d    & NGC 839    & 2h 09m 42.9s & -10$^{\circ}$ 11\arcmin \ 03\arcsec & S0a  & $3874$\textsuperscript{b} & 87  & $9.97 \pm 0.02$ \\
           & NGC 848    & 2h 10m 17.6s & -10$^{\circ}$ 19\arcmin \ 17\arcsec & SBab & $4045$\textsuperscript{b} & 89  & $10.09 \pm 0.04$ \\
           & PGC 8210   & 2h 09m 06.0s & -10$^{\circ}$ 19\arcmin \ 13\arcsec & Sc   & $3972$\textsuperscript{a} & 72\textsuperscript{c}  & $9.37 \pm 0.18$  \\ \hline
\end{tabular}
\tablefoot{Columns: (1) name in HCG catalogue, (2) other name, (3) right ascension (J2000), (4) declination (J2000), (5) morphological type, from HyperLeda (\url{http://leda.univ-lyon1.fr/}), (6) heliocentric velocity from optical spectra (references below), (7) optical isophotal diameter at 25 mag arcsec$^2$ in B-band \citep[from RC3][unless indicated otherwise]{RC3}, (8) logarithm of B-band luminosity calculated following \citet{Fernandez-Lorenzo+2012} with values from HyperLeda and the morphologies and velocities given in this table, the quoted errors ignore distance uncertainty ($\pm 0.05$ dex).\\
\textsuperscript{a} \citet{Ribeiro+1996},
\textsuperscript{b} \citet{Diaz-Gimenez+2012},
\textsuperscript{c} \citet{Paturel+2000}.
}
\end{table*}

\section{Observations and data reduction}
\label{sec:data}

\subsection{\hi \ data}
\label{sec:HIdata}

HCG 16 was mapped with the VLA in C and D configurations in 1999 and 1989 respectively. These data were reduced using AIPS (Astronomical Image Processing System) and presented in \citet{Verdes-Montenegro+2001} and \citet{Borthakur+2010}. In this work we have re-reduced the raw data using \texttt{CASA} \citep[Common Astronomy Software Applications,][]{CASA}\footnote{\url{https://casa.nrao.edu/}} and re-imaged them using multi-scale CLEAN in the \texttt{CASA} task \texttt{tclean}. The \hi \ line emission was imaged over the velocity range 3246 \kms \ to 4557 \kms \ with a resolution of 21 \kms. The dataset was imaged twice to generate two cubes using Brigg's robust weighting parameters of 2 and 0. While almost all of the following analysis relies on the robust=2 cube, the robust=0 is useful to see some parts of the highest column density gas with finer angular resolution. The multi-scale CLEAN angular scales used in these two cubes are 0, 8, 16, 24, and 40 pixels in the robust=2 cube, and 0, 4, 8, 16, 24, and 40 pixels in the robust=0 cube, where each pixel was 4\arcsec \ across in both cases. The resulting beam sizes of these two cubes were 37.2\arcsec \ $\times$ 30.3\arcsec \ and 19.4\arcsec \ $\times$ 14.8\arcsec \ respectively. At the assumed distance of 55.2 Mpc, 20\arcsec corresponds to a projected distance of 5.4 kpc. The robust=2 and robust=0 cubes have rms noises of 0.36 mJy/beam and 0.40 mJy/beam respectively, which correspond to 3$\sigma$ \hi \ column density sensitivities of 2.2 $\times$ 10$^{19}$ cm$^{-2}$ and 9.6 $\times$ 10$^{19}$ cm$^{-2}$ at a velocity resolution of 21 \kms. An interactive, 3-dimensional figure displaying the robust=2 cube created using the X3D pathway introduced in \citet{Vogt+2016} is available at \url{http://amiga.iaa.es/FCKeditor/UserFiles/X3D/HCG16/HCG16.html}.

To create a source mask within which the 0th moment and total integrated flux could be calculated we made use of the \texttt{SoFiA} package \citep[Source Finding in Astronomy,][]{SoFiA,Serra+2015}, using smoothing kernels over spatial scales approximately equal to one and two times the (robust=2) beam size, over one and three channels, and clipping at 3.5$\sigma_{\mathrm{rms}}$ (shown with contours in Figure \ref{fig:mom0}, right panel). A reliability threshold of 100\% was set to remove spurious noise spikes, the sources were merged and the final mask dilated using \texttt{SoFiA}'s mask optimisation tools in order to include all flux associated with the group. An equivalent procedure with a threshold of 5$\sigma_{\mathrm{rms}}$ and without dilating the mask was used to produce the 1st moment map of the group \hi \ emission (Figure \ref{fig:mom1}). In addition we made a more traditional source mask based on a 3$\sigma_{\mathrm{rms}}$ clipping in each channel (of the robust=2 cube) using \texttt{CASA}. A comparison of the moments generated by these two masks can be seen in Figure \ref{fig:mom0}.

The spatial and spectral smoothing performed by \texttt{SoFiA} results in a more extended mask (even though the threshold is 3.5$\sigma_{\mathrm{rms}}$ rather than 3$\sigma_{\mathrm{rms}}$), which should include more of the low column density emission. The standard 3$\sigma_{\mathrm{rms}}$ mask is only used for visual representation as it more clearly separates the higher column density features (precisely because it excludes the fainter emission). In all the following sections and analysis we use the robust=2 cube and the \texttt{SoFiA} source mask, unless explicitly stated otherwise. 

\subsection{Optical images}
\label{sec:opt_data}

Throughout this paper we compare \hi \ features with optical images from the Dark Energy Camera Legacy Survey (DECaLS, \url{http://www.legacysurvey.org/decamls}). The three DECaLS bands ($g$,$r$, and $z$) have surface brightness limits in the field of interest of 28.5, 28.7 and 28.0 mag arcsec$^{-2}$ (3$\sigma$ in 10\arcsec$\times$10\arcsec \ boxes), respectively. 
This is the deepest image that covers the entire field of which we are aware, thus we focus on this image to look for faint optical features which may be associated with extended \hi \ features. However, Hubble Space Telescope and Spitzer images of the group have been published by \citet{Konstantopoulos+2013}.

We used the DECaLS images as published, which were processed with the automated Dark Energy Camera Community Pipeline. This processing slightly over-subtracts the sky in the vicinity of large galaxies, negatively impacting the sensitivity for large scale faint features near large galaxies (like those in HCG 16). As this work is focused on the \hi \ component of the group, we note this issue, but do not reprocess the images.

\section{Separation of \hi \ features}
\label{sec:HIsep}

The \hi \ content of HCG 16 is enormously complicated, with multiple blended galaxies and tidal features. Therefore, in order to study the properties of each galaxy and tidal feature, the \hi \ emission must be separated into these individual objects wherever possible. By doing this we can assess whether each galaxy has a high, normal, or low \hi \ content (irrespective of the global group \hi \ content), if they have well ordered rotation and morphology, or have been disrupted by interactions, and where the gas in tidal features likely originated. The answers to these questions will in turn help to build a consistent picture of how the group has evolved to date and how it might proceed in the future.

To perform this separation we used the \texttt{SlicerAstro} package \citep{SlicerAstro,Punzo+2017}. The \texttt{SoFiA} mask was imported into \texttt{SlicerAstro} and divided into sub-regions corresponding to galaxies or specific tidal features. By importing the same mask we ensure that all of the flux included in the integrated measurement is assigned to a particular galaxy or feature. This is especially important for obtaining the \hi \ masses of individual galaxies. Although the higher column density regions are more straightforward to separate into distinct sources, this would lead to the \hi \ flux being underestimated as low column density emission would be systematically missing, while the scaling relations we will use to define the deficiency were calculated from single dish observations that include all the \hi \ flux of target (isolated) galaxies. The final separation was made through an iterative comparison of the channel maps, optical images, 3-dimensional visualisations, and moment maps of the individual galaxies. This process was unavoidably subjective to some degree, particularly in the region of HCG 16c and HCG 16d where emission smoothly changes from gas associated with the galaxies to multiple high column density tidal features. However, despite the resulting large uncertainties in the galaxy parameters this is still an instructive tool for assessing the likely history of the group, as discussed in the following sections. 
More specifics of this process are described in \S\ref{sec:indv_gals}, along with the results for individual galaxies once they have been separated from surrounding features.

\section{Results}
\label{sec:results}

In this section we present the results of our \hi \ analysis, first for the group as a whole and then for each galaxy.

\begin{figure*}
    \centering
    \includegraphics[width=0.8\textwidth]{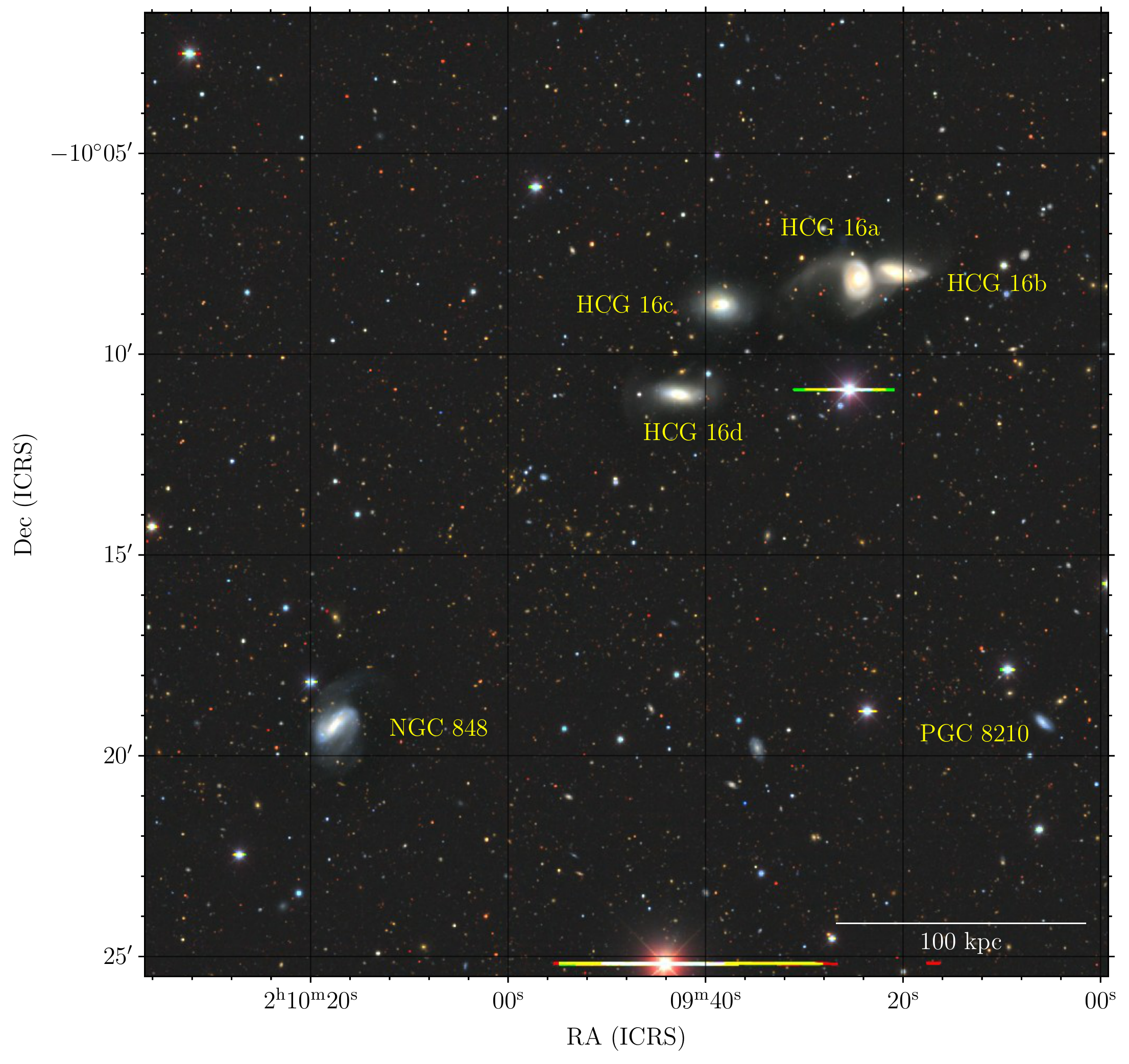}
    \caption{DECaLS $grz$ colour image of HCG 16 with the member galaxies labelled.}
    \label{fig:optim}
\end{figure*}
\begin{table*}
\centering
\caption{\hi \ properties of galaxies and tidal features in HCG 16}
\label{tab:HIprops}
\begin{tabular}{lcccccc}
\hline\hline
Object        & $S_{\mathrm{HI}}/\mathrm{Jy\,km\,s^{-1}}$ & $\bar{v}/\mathrm{km\,s^{-1}}$ & $\sigma_{v}/\mathrm{km\,s^{-1}}$ & $\log{M_{\mathrm{HI}}/\mathrm{M}_{\odot}}$ & \hi \ deficiency           \\ \hline
HCG 16a       & 1.97    & 4026      & 120        & 9.15 & $0.69 \pm 0.21$  \\
HCG 16b       & 0.69    & 3882      & 61         & 8.70 & $1.01 \pm 0.21$  \\
HCG 16c       & 5.14    & 3819      & 89         & 9.57 & $0.12 \pm 0.21$  \\
HCG 16d       & 7.48    & 3901      & 56         & 9.73 & $-0.18 \pm 0.21$ \\
NGC 848       & 8.53    & 3987      & 69         & 9.79 & $-0.12 \pm 0.21$ \\
PGC 8210      & 1.14    & 3978      & 48         & 8.91 & $0.07 \pm 0.26$ \\
NW tail       & 5.50    & 3788      & 106         & 9.60 & \\
NE tail       & 2.98    & 3888      & 51        & 9.33 & \\
E clump       & 0.72    & 3910      & 23         & 8.72 & \\
S clump       & 1.25    & 4071      & 20        & 8.95 & \\
SE tail       & 4.74    & 3997      & 45         & 9.53 & \\
cd bridge     & 5.97    & 3944      & 148         & 9.63 & \\
NGC 848S tail & 1.08    & 4030      & 43         & 8.89 & \\ \hdashline
NGC 848S loop & 1.75    & 4034      & 24         & 9.10 & \\ \hline
\end{tabular}
\tablefoot{Columns: (1) object name, (2) \hi \ integrated flux, (3) flux-weighted mean velocity, (4) flux-weighted velocity dispersion, (5) logarithm of \hi \ mass, (6) \hi \ deficiency (calculated in \S\ref{sec:indv_gals}). The final component is separated from the rest as it was not included without the \texttt{SoFiA} mask as it is low significance. It also does not contribute to the global flux measurement.}
\end{table*}

\subsection{Global \hi \ morphology}

Figure \ref{fig:mom0} (left) shows the 0th moment map of the \hi \ emission created using a standard 3$\sigma_{\mathrm{rms}}$ clipping in each channel. This map excludes much of the lowest column density emission, making many features easier to distinguish by eye. The galaxies HCG 16a, b, c, d, NGC 848, and PGC 8210 are all detected along with tidal features across the whole group that appear to connect HCG 16a and b to HCG 16c, HCG 16c to d, and HCG 16d to NGC 848. The most striking tidal feature is the South-East tail, which stretches over a projected distance of $\sim$160 kpc, connecting NGC 848 to the core group. There are no indications of an interaction between PGC 8210 and the rest of the group.

Figure \ref{fig:mom0} (right) shows the 0th moment again, but this time made using the \texttt{SoFiA}-generated mask (\S\ref{sec:HIdata}) and displayed as contours overlaid on the DECaLS $grz$ image. 
This mask includes more of the low column density emission than the previous mask and thus provides a more complete measurement of the \hi \ content, but makes most individual features more difficult to identify by eye. This masking is used throughout the following analysis as we want to include low column density emission because a large fraction of the group's \hi \ is in tidal features.

In Figure \ref{fig:mom1} we show the 1st moment of the entire group (using a \texttt{SoFiA} mask with a 5$\sigma_{\mathrm{rms}}$ threshold). 
Due to the high spatial density of the group, the galaxies and tidal features are superposed and confused in this image, complicating its interpretation. Without separating out individual objects and features, clear signs of ordered rotation are only visible in NGC 848 and PGC 8210, although in the latter cases it is barely larger than the beam. The velocity field of each galaxy is presented in \S\ref{sec:indv_gals} after separating them from the rest of the emission in the group.

\subsection{\hi \ flux and mass}

\begin{landscape}
\begin{figure}
    \includegraphics[width=\columnwidth]{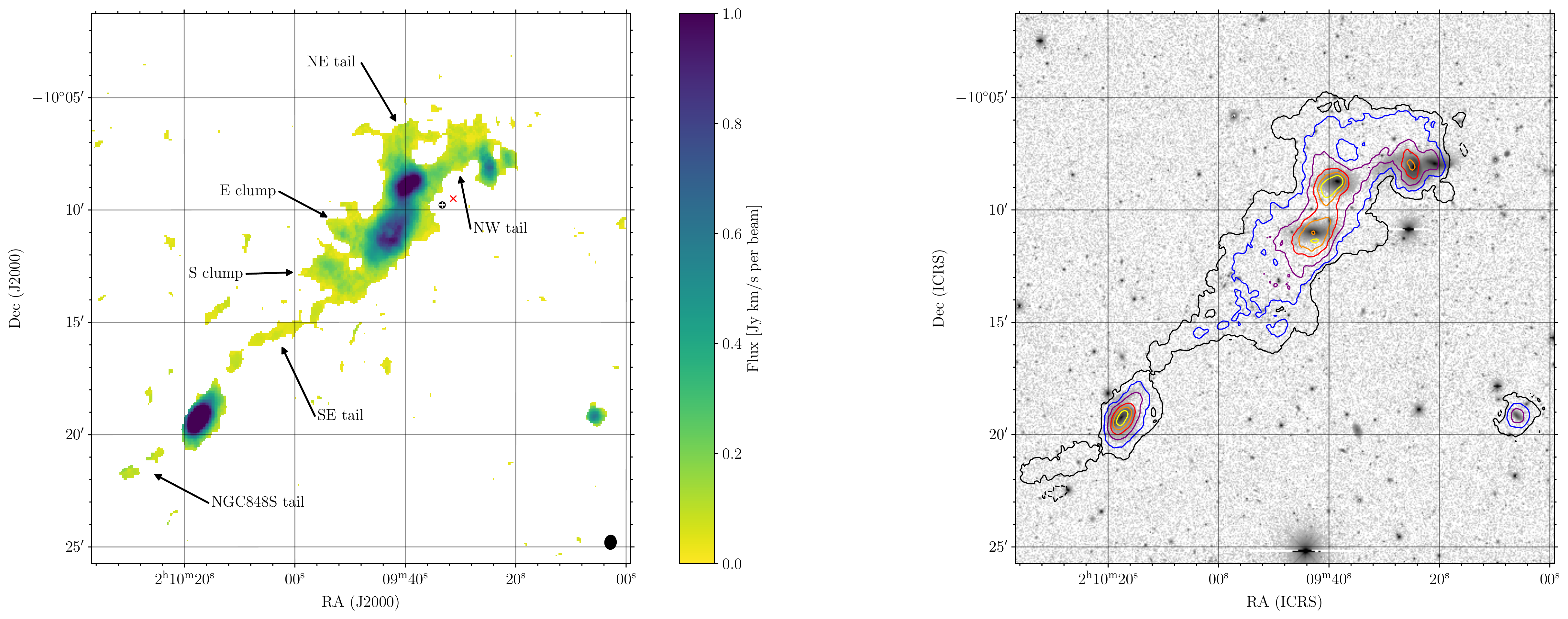}
    \caption{\textit{Left}: Moment 0 map (primary beam corrected) of the \hi \ emission of HCG 16 calculated using a 3$\sigma_{\mathrm{rms}}$ mask in each channel (this mask is intended only for visual purposes, all the analysis uses the \texttt{SoFiA}-generated mask, shown in the right panel).
    The black ellipse in the lower right corner indicates the beam size (37.2\arcsec \ $\times$ 30.3\arcsec \ or 10.0 kpc $\times$ 8.1 kpc) and the small red x shows the centre of the GBT pointing from \cite{Borthakur+2010}. The extended features that we separated from the galaxies are indicated by arrows, with the exception of the emission joining galaxies HCG 16c and d. The filled black circle and the white cross indicate the pointing centres of the VLA D and C array data respectively. \textit{Right}: Moment 0 contours (uncorrected for primary beam) overlaid on a DECaLS $r$-band image. In this case the map was generated using the \texttt{SoFiA}-generated mask described in \S\ref{sec:HIdata}, which includes more extended emission. The galaxies in the main band of \hi \ emission going from top-right (NW) to bottom-left (SE) are: HCG 16b, a, c, d, NGC 848, and the single galaxy to the lower-right of the group is PGC 8210. Contour levels: -2.45, 2.45, 9.80, 24.4, 49.0, 73.5, and 98.0 $\times 10^{19} \; \mathrm{cm}^{-2}$, where $2.2 \times 10^{19} \; \mathrm{cm}^{-2}$ corresponds to the 3$\sigma$ sensitivity in one channel. In order of increasing flux the contours are coloured: black (dashed), black, blue, purple, red, orange, and yellow.}
    \label{fig:mom0}
\end{figure}
\end{landscape}

\begin{figure*}
    \centering
    \includegraphics[width=0.8\textwidth]{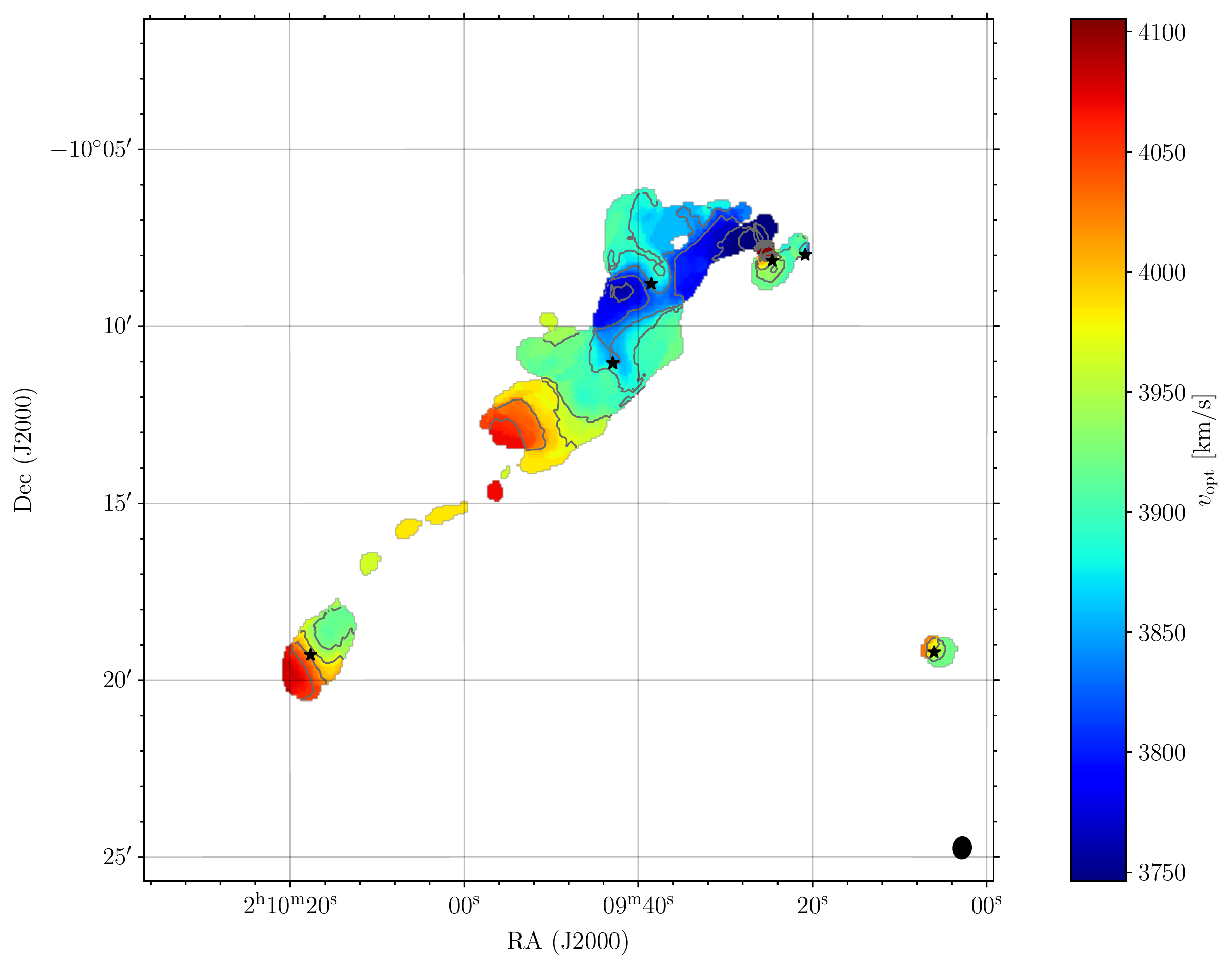}
    \caption{Moment 1 map of the \hi \ emission calculated using the \texttt{SoFiA}-generated mask with a threshold of 5$\sigma_{\mathrm{rms}}$ and without dilation. The rainbow colour scale indicates the recessional (optical) velocity in \kms. The black ellipse in the lower right indicates the beam size, the grey lines are isovelocity contours separated by 40 \kms, and the small black star symbols indicate the locations of the optical centres of the galaxies in the group.}
    \label{fig:mom1}
\end{figure*}

\begin{figure}
    \centering
    \includegraphics[width=\columnwidth]{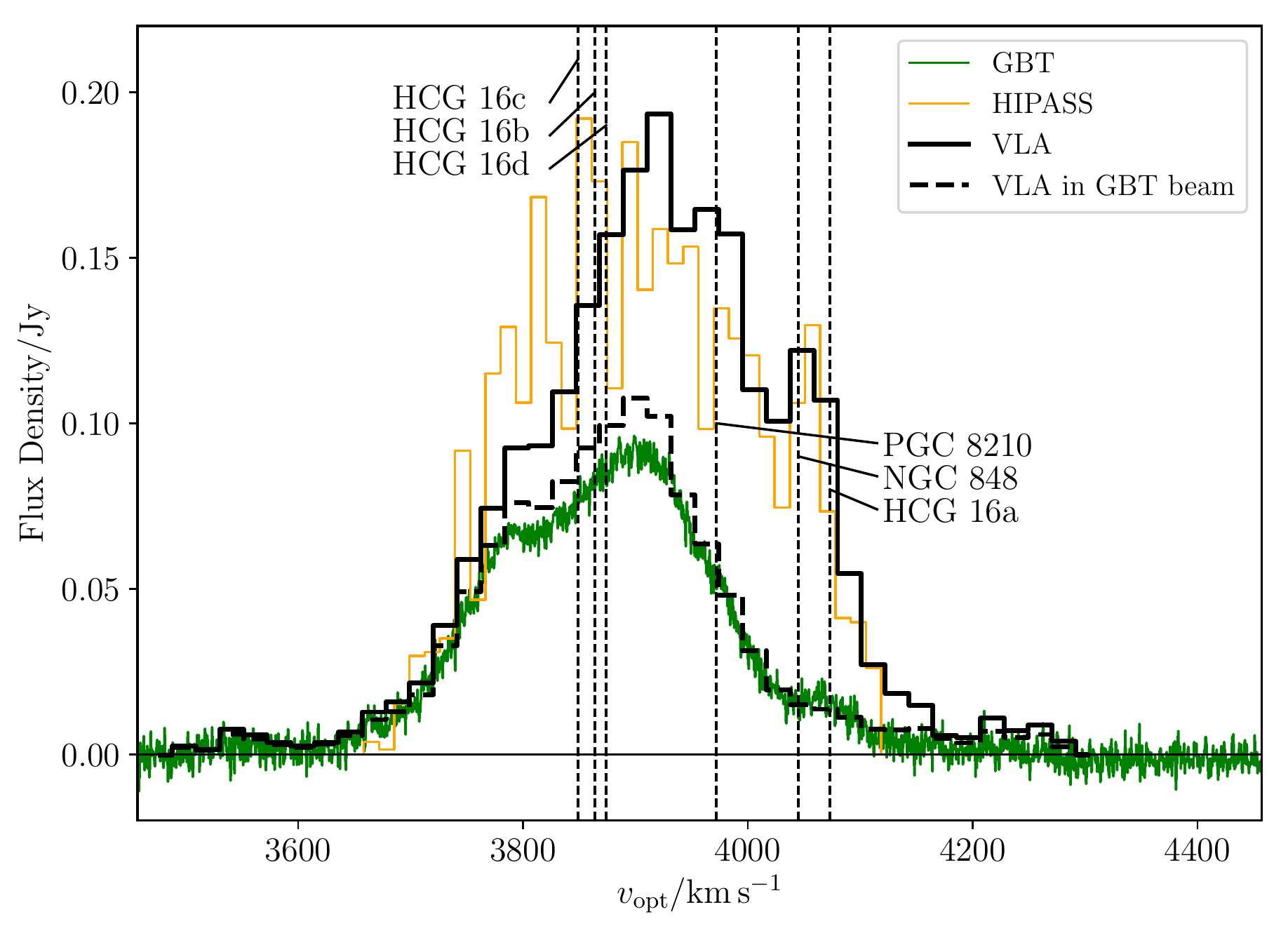}
    \caption{The integrated VLA spectrum of the entire group (thick black line) compared to the GBT spectrum of the core group \citep{Borthakur+2010} (high spectral resolution green line), the VLA spectrum within the GBT beam area (thick black dashed line), and the spectrum extracted from the HIPASS cube (orange line). Vertical dashed lines indicate the optical redshifts of the member galaxies.}
    \label{fig:group_spec}
\end{figure}

Integrating the entire \hi \ emission shown in Figure \ref{fig:mom0} (right) gives the total \hi \ flux of HCG 16 as 47.2 Jy \kms \ and its mass as $\log{M_\mathrm{HI}/M_\odot} = 10.53 \pm 0.05$ (using the standard formula, $M_\mathrm{HI}/\mathrm{M_{\odot}} = 235600 \times [D/\mathrm{Mpc}]^2 \times [S_\mathrm{HI}/\mathrm{Jy \, km \, s^{-1}}]$). This is considerably higher than the values of \citet{Borthakur+2010}, 21.5 Jy \kms \ and $10.19 \pm 0.05 \times 10^{10}$ \Msol, based on the single dish spectrum taken with GBT. However, as pointed out in that work, this difference arises because the GBT HPBW is 9.1\arcmin \ and the pointing was centred on the group core (red x in the left panel of Figure \ref{fig:mom0}). This means that the majority of the flux from the tail extending to the SE, NGC 848 and PGC 8210 is missing from the GBT spectrum. The moment 0 map was weighted with a 2D Gaussian window (FWHM of 9.1\arcmin) centred on the GBT pointing centre to estimate the fraction of the total emission that the GBT observations would have been able to detect. This gives an \hi \ mass of $\log{M_\mathrm{HI}/M_\odot} = 10.24 \pm 0.05$ \Msol \ and a flux of 24.2 Jy \kms, which is about 10\% higher than the GBT measurement (Figure \ref{fig:group_spec}). This slight difference could arise from calibration or baseline uncertainties, or simply because a Gaussian is not a completely accurate representation of the beam response. 

We compare our total flux measurement (47.2 Jy \kms) to that in HIPASS. As HCG 16 is extended over many arcmin it is a marginally resolved source even for the Parkes telescope. Therefore, we cannot use the HIPASS catalogue values, which assume it is a point source. Using the HIPASS cube in this region of the sky we perform a source extraction with \texttt{SoFiA}, applying a threshold limit of 4.5$\sigma$ over smoothing kernels of 3 and 6 spatial pixels and 3 and 7 velocity pixels (each pixel is 4\arcmin \ $\times$ 4\arcmin \ $\times$ 13 \kms). The resulting spectrum is shown in Figure \ref{fig:group_spec}. The integrated flux in HIPASS is 43.8 Jy \kms, which is within 10\% of our measurement with the VLA. Given the difficulty in absolute calibration \citep[e.g.][]{vanZee+1997}, this is about the level of agreement that is to be expected. However, there are some additional discrepancies with this comparison which we discuss further in Appendix \ref{sec:flux_disc}.

\subsection{Individual galaxies}
\label{sec:indv_gals}

\begin{figure*}
    \centering
    \includegraphics[width=\textwidth]{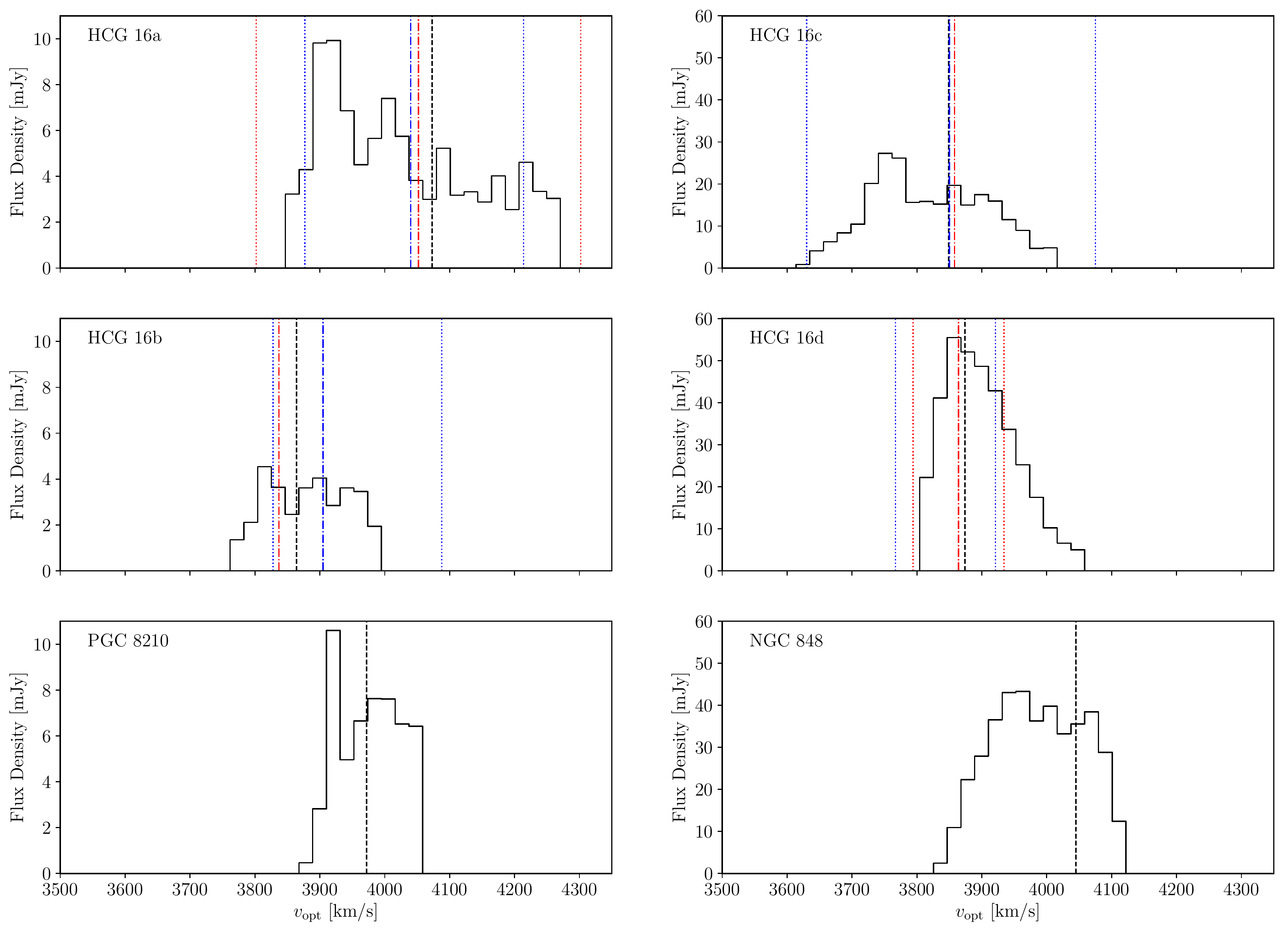}
    \caption{Spectral profiles of each of the 6 galaxies detected in \hi \ near the core group, calculated from the separation of tidal features and galaxies performed in \S\ref{sec:HIdata}. The black vertical dashed lines show the optical redshifts of each galaxy as given in Table \ref{tab:optprops}. The dot-dashed and dotted vertical lines show the central and extreme velocities from the rotation curve measurements of \citet{Rubin+1991} in red and \citet{Mendes+1998} in blue. These measurements are only available for the four core galaxies, also \citet{Rubin+1991} does not specify $V_\mathrm{max}$ values for HCG 16b or c due to the peculiar shape of their rotation curves. The left panels have vertical scales that go to 11 mJy, whereas the scales on the right are a factor of approximately 6 higher (60 mJy).}
    \label{fig:spectra}
\end{figure*}

\begin{figure*}
    \centering
    \includegraphics[width=\textwidth]{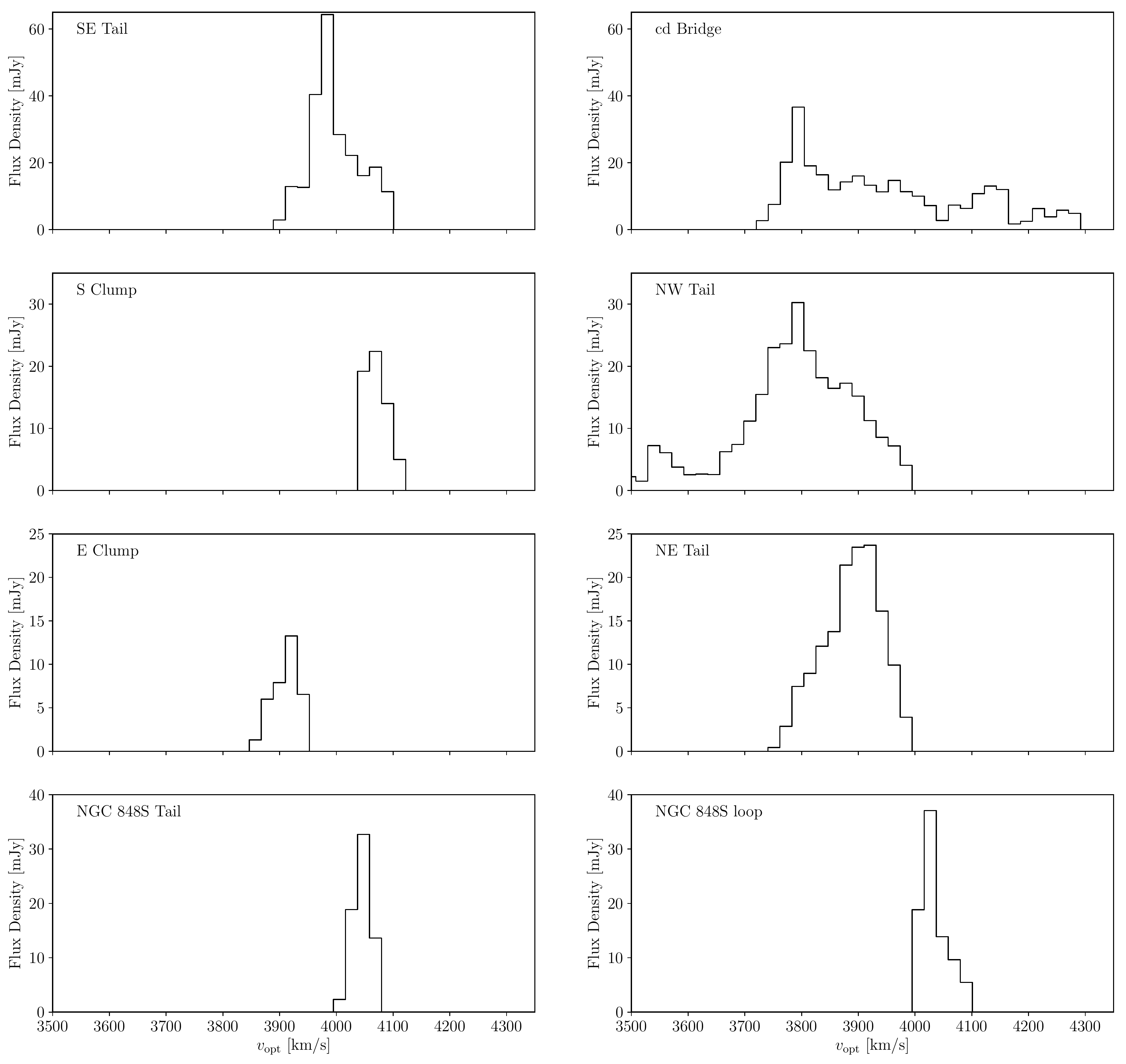}
    \caption{\hi \ spectral profiles of each of the 8 separate tidal features in the group, calculated from the separation of tidal features and galaxies performed in \S\ref{sec:HIdata}. Each row has a different vertical scale.}
    \label{fig:tidal_spectra}
\end{figure*}

Table \ref{tab:HIprops} shows the basic \hi \ properties of the individual galaxies and tidal features based on the separation performed in \S\ref{sec:HIsep}. We have chosen to use the flux-weighted mean velocity and flux-weighted velocity dispersion to quantify the centre and width of each emission profile, rather than more common measures e.g. $W_{50}$, because some of the profiles, particularly those of the strongly disturbed galaxies or tidal features, do not follow typical profile shapes of \hi \ sources (Figures \ref{fig:spectra} \& \ref{fig:tidal_spectra}). The measurements of the B-band luminosity, $L_{\mathrm{B}}$, of each galaxy (Table \ref{tab:optprops}) were used as inputs to the \hi \ scaling relation of \citet{Jones+2018} in order to estimate their expected \hi \ masses if they were in isolation, and thus their \hi \ deficiencies. We note that although it has been fairly common in past works to consider the morphology of a galaxy when calculating \hi \ deficiency, here we choose to ignore it. There are two main reasons for this:
\begin{itemize}
\item \citet{Jones+2018} suggest that morphology should be ignored unless the sample has a large fraction of ellipticals, because their piece-wise scaling relations (split by morphology) are quite uncertain due to the small number of galaxies that are not Sb-Sc in the AMIGA \citep[Analysis of the interstellar Medium of Isolated GAlaxies,][]{Verdes-Montenegro+2005b} reference sample. Hence, using these piece-wise relations, in the case of HCG 16, would trade a small bias for a large uncertainty that is difficult to accurately quantify.
\item Galaxies are expected to undergo morphological transformation in compact groups as they are stripped of their gas and perturbed by interactions. Therefore, it is perhaps misguided to use their present day morphologies in these scaling relations at all.
\end{itemize}

Our analysis indicates that only galaxies HCG 16a and b are missing neutral gas compared to their expected \hi \ content if they were in isolation, all the remaining galaxies are consistent with having a normal quantity of \hi. The group as a whole (galaxies

\begin{figure*}
    \centering
    \includegraphics[width=\textwidth]{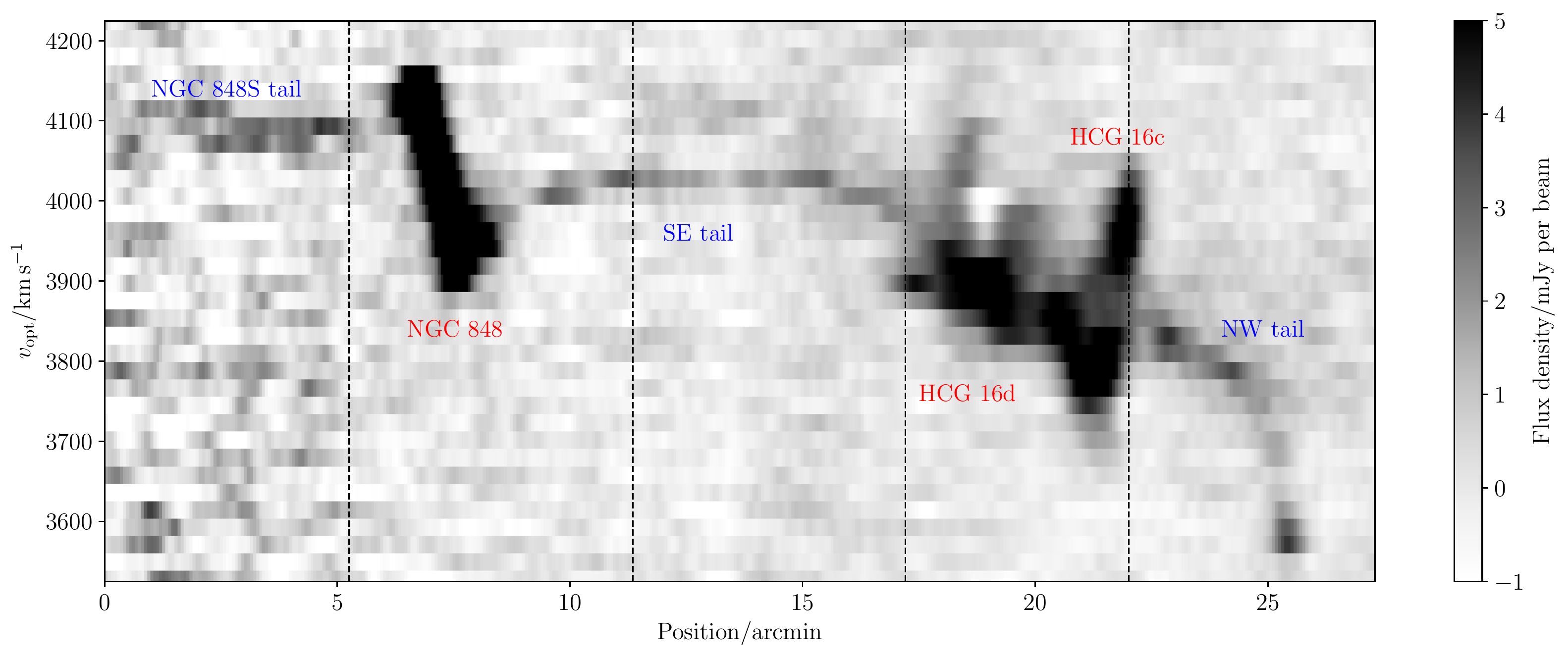}
    \includegraphics[width=\textwidth]{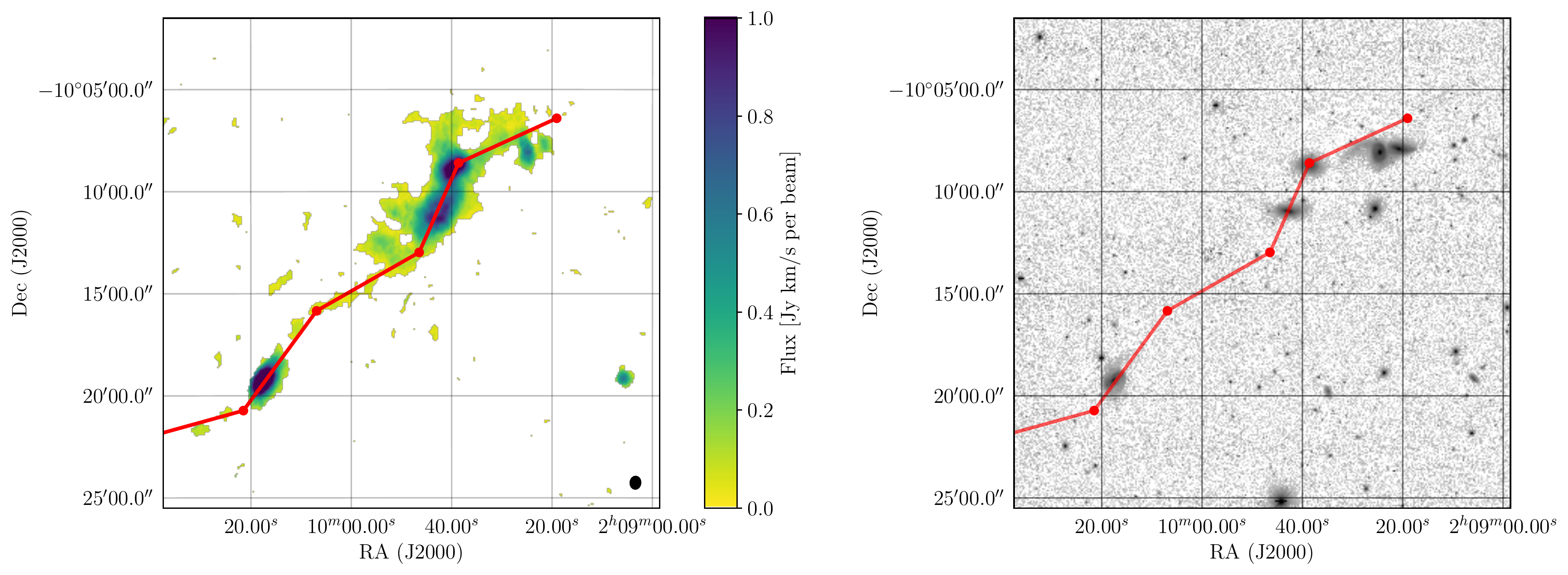}
    \caption{\textit{Top}: A segmented position--velocity diagram showing the spatially and kinematically continuous \hi \ emission spanning HCG 16. The vertical dashed lines show the locations of the nodes making up the segmented slice through the data cube. Galaxies are labelled in red and tidal features in blue. Note that the noise in this plot increases significantly near the left edge as this region is approaching the primary beam edge. \textit{Bottom}: The left panel shows the position--velocity slice as a segmented red line on top of the moment 0 map (using the standard 3$\sigma$ threshold mask as in the left panel of Figure \ref{fig:mom0}) and the right panel shows the same line but overlaid on the DECaLS $r$-band image.}
    \label{fig:pv_plot}
\end{figure*}

\begin{figure*}
    \centering
    \includegraphics[height=75mm]{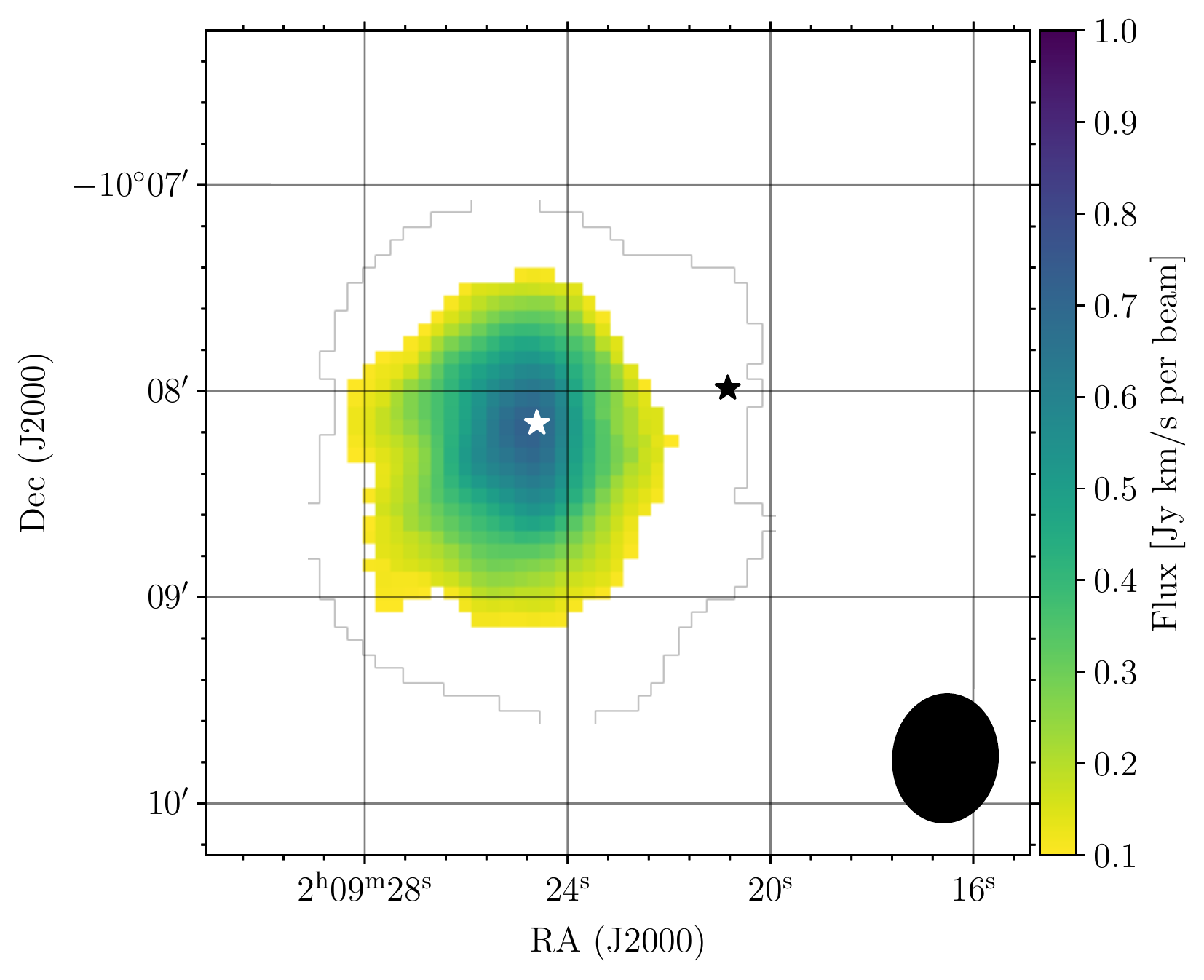}
    \includegraphics[height=75mm]{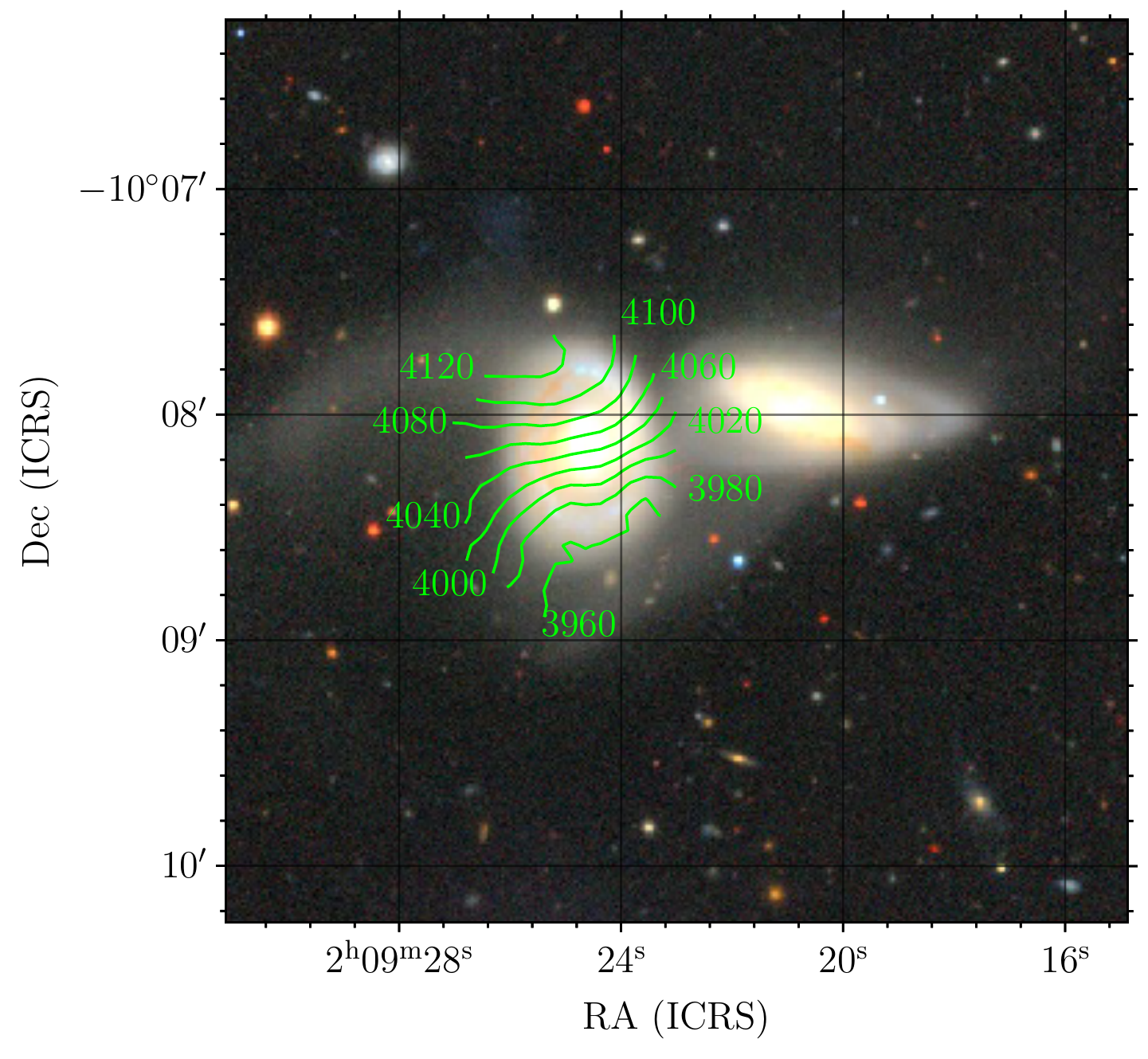}
    \caption{\textit{Left}: The moment 0 map of HCG 16a. The white and black stars indicate the optical centres of HCG 16a and b respectively, and the black ellipse shows the beam. \textit{Right:} Moment 1 contours for HCG 16a overlaid on the DECaLS $grz$ image of HCG 16a and b. The green lines are separated by 20 \kms.}
    \label{fig:HCG16a_moms}
\end{figure*}

\noindent
and extended features) has an \hi \ mass of $(3.39 \pm 0.29) \times 10^{10}$ \Msol \ (accounting only for distance uncertainty) and a global \hi \ deficiency of $-0.12 \pm 0.09$.\footnote{If we were to assume that the HIPASS flux is correct and that ours is an overestimate then the \hi \ deficiency would be $-0.08 \pm 0.09$.} Thus the group as a whole does not appear to be missing \hi \ gas (it is marginally \hi-rich), but the galaxy pair HCG 16a and b has likely lost the vast majority of its original \hi \ content. The probable fate of this lost gas is discussed in the following sections.

Here we note that had we used the \citet{Haynes+Giovanelli1984} scaling relation instead of the updated \citet{Jones+2018} relation, then we would have found the global \hi \ deficiency to be 0.05, which is approximately 2$\sigma$ higher (more deficient). The differences between these relations are discussed in detail in \citet{Jones+2018}, but in this case the most important point is that the uncertainties in the measurements of $L_{\mathrm{B}}$ for the reference sample used to calibrate the relation, results in the \citet{Haynes+Giovanelli1984} $M_{\mathrm{HI}}$--$L_{\mathrm{B}}$ scaling relation overestimating the \hi \ deficiency of galaxies.

The uncertainty estimates for the \hi \ deficiencies in Table \ref{tab:HIprops} are dominated by the scatter in the $M_{\mathrm{HI}}$--$L_{\mathrm{B}}$ scaling relation and also have small contributions due to the uncertainty in $L_{\mathrm{B}}$ and the group distance, $55.2 \pm 3.3$ Mpc, which was estimated using the \citet{Mould+2000} local flow model as described in \citet{Jones+2018}, which has corrections for flow towards the Virgo cluster, the Great Attractor, and the Shapley supercluster. The uncertainties in the masses of each component are challenging to estimate because these values strongly depend on where subjective boundaries are drawn between tidal gas and gas in a galactic disc. As \hi \ deficiency is the quantity of interest for the galaxies and the scatter in the scaling relation (0.2 dex) results in an uncertainty of 40-60\%, which is expected to dominate the error budget, we do not attempt to estimate the \hi \ mass uncertainty due to the subjective measurements. The same issue applies to the tidal features, but is more severe as they are mostly made up of lower column density gas. Thus their individual mass measurements should be treated with caution.

\subsubsection{HCG 16a}
\label{sec:HCG16a}

HCG 16a is an Sab galaxy that has been classified as an active star forming galaxy in IR \citep{Zucker+2016} and hosts an AGN (Seyfert 2) that has been confirmed with both optical line ratios \citep{Veron-Cetty+Veron2010,Martinez+2010} and X-rays \citep{Turner+2001,Oda+2018}. The galaxy also hosts a ring of SF (star formation) that is clearly visible in the IR and UV \citep{Tzanavaris+2010,Bitsakis+2014}. In terms of its \hi, HCG 16a is the second most deficient galaxy in the group, apparently missing approximately $5.5 \times 10^{9}$ \Msol \ of \hi.

The discs of HCG 16a and b overlap in the plane of the sky which complicates the separation of their \hi \ emission. However, in velocity the gas that is co-spatial with HCG 16a forms a clear, continuous, and steep gradient, as would be expected from an inclined disc galaxy. This is most readily seen in the \href{http://amiga.iaa.es/FCKeditor/UserFiles/X3D/HCG16/HCG16.html}{X3D interactive figure} which accompanies this paper. The emission associated with HCG 16b forms a small, offset clump at approximately the same velocity as the lower velocity "horn" of the HCG 16a profile (Figure \ref{fig:spectra}). The two galaxies are connected in \hi \ by a faint bridge of emission, visible in the channel maps (available in the electronic version) between velocities 3858 and 4006 \kms. Given the size of the beam and the fact that the optical discs of HCG 16a and b overlap, it is not possible to reliably assign the emission to either galaxy. Therefore we simply split it approximately half-way between the two sources. The uncertainty in where this bridge should be split introduces a negligible error to the \hi \ properties of HCG 16a as the entirety of the emission that we attribute to HCG 16b would only increase the \hi \ mass of HCG 16a by 0.13 dex.

When viewed projected in the plane of the sky there is an \hi \ tail which connects HCG 16a to HCG 16c (the NW tail, see Figure \ref{fig:mom0}, left panel). There is also an accompanying stellar tail \citep[identified by][]{Rubin+1991} that extends in the same direction away from HCG 16a (clearly visible in the Figures \ref{fig:optim} \& \ref{fig:pv_plot}). However, when the \hi \ data cube is studied in 3D it is clear that the \hi \ tail does not form a kinematic connection with HCG 16a and the apparent connection is the result of a projection effect. In Figure \ref{fig:pv_plot} it can be seen that as the NW tail is traced away from HCG 16c its velocity decreases from $\sim$3800 \kms \ to $\sim$3500 \kms, whereas HCG 16a's \hi \ profile covers the approximate range 3850-4300 \kms \ (Figure \ref{fig:spectra} \& \ref{fig:HCG16a_moms} right panel). This tail then ends in a dense clump, which may have formed a tidal dwarf galaxy (TDG), discussed further in \S\ref{sec:TDGs}.

Figure \ref{fig:HCG16a_moms} shows the 0th and 1st moments of the \hi \ emission. From these two maps it can be seen that the high column density gas in the immediate vicinity of the optical disc of HCG 16a is relatively undisturbed, despite that large amount of missing \hi. The moment 0 map has a regular oval shape and is centrally peaked (almost coincident with the optical centre). While the velocity field appears quite regular in the centre, the line of nodes forms an `S' shape in the outer regions, likely indicating the presence of a warp in the disc \citep[e.g.][]{Bosma1978}. Given the stellar tails in the optical and the clear on-going interaction with HCG 16b such an asymmetry is not unexpected.

\citet{Mendes+1998} also found that the rotation in the inner region of HCG 16a is very regular, rising quickly and becoming flat well within one of our beam widths. The velocity extent of our \hi \ spectrum (Figure \ref{fig:spectra}) approximately agrees with that implied by their H$\alpha$ rotation curve. However, \citet{Rubin+1991} found the conflicting result that the H$\alpha$ rotation has an anomalous structure, declining on one side of the galaxy at large radii. As our \hi \ map extends to much larger radii we would expect to see a continuation of this in our data, but we do not. Therefore, the \hi \ velocity field appears more consistent with the rotation curve of \citet{Mendes+1998}. Having said this, the major axis position angle given in that work is almost exactly aligned N-S, whereas in \hi \ it is clearly rotated (counter-clockwise). We expect this is due to a combination of beam smearing and disturbances in the outer regions of the disc.

\subsubsection{HCG 16b}
\label{sec:HCG16b}

\begin{figure*}
    \centering
    \includegraphics[height=75mm]{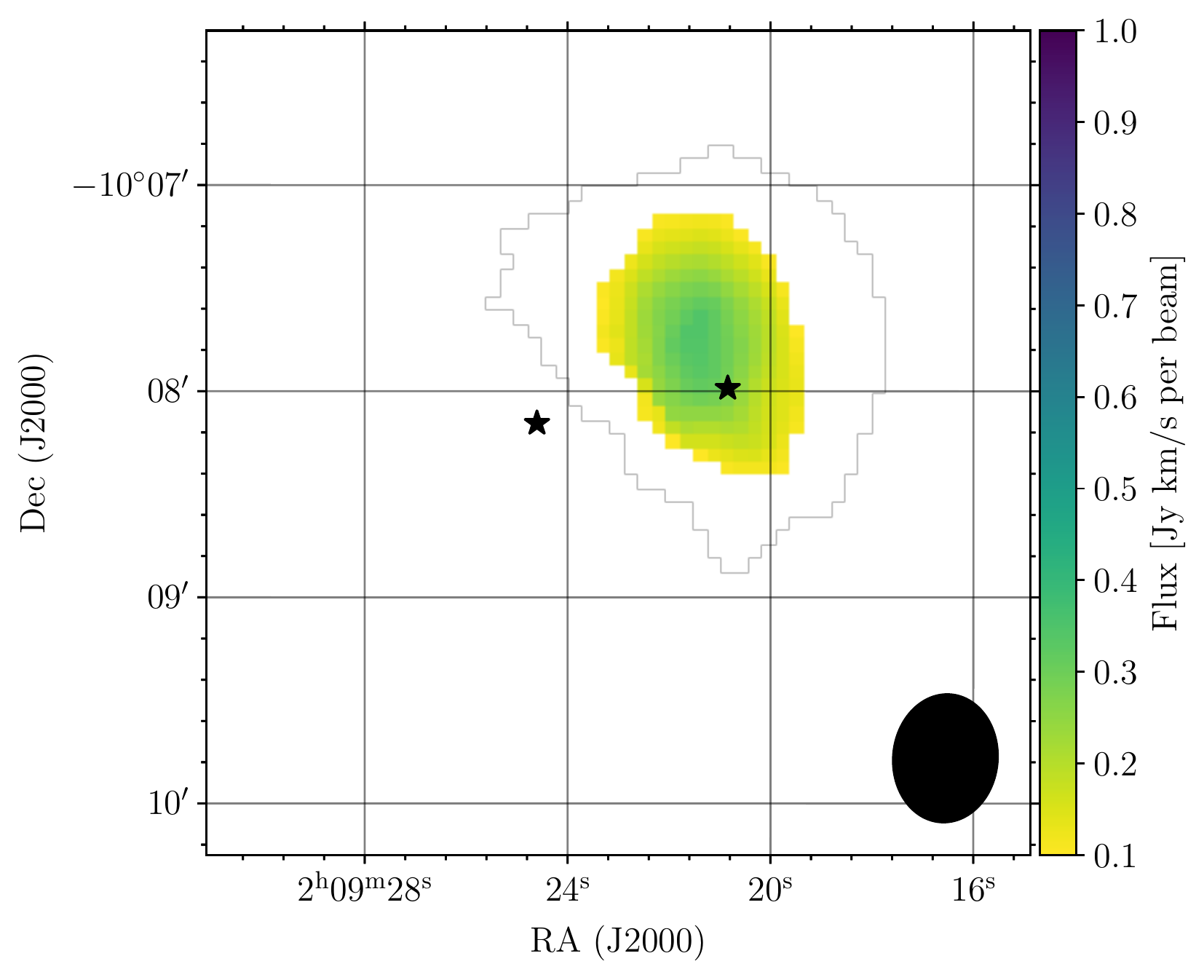}
    \includegraphics[height=75mm]{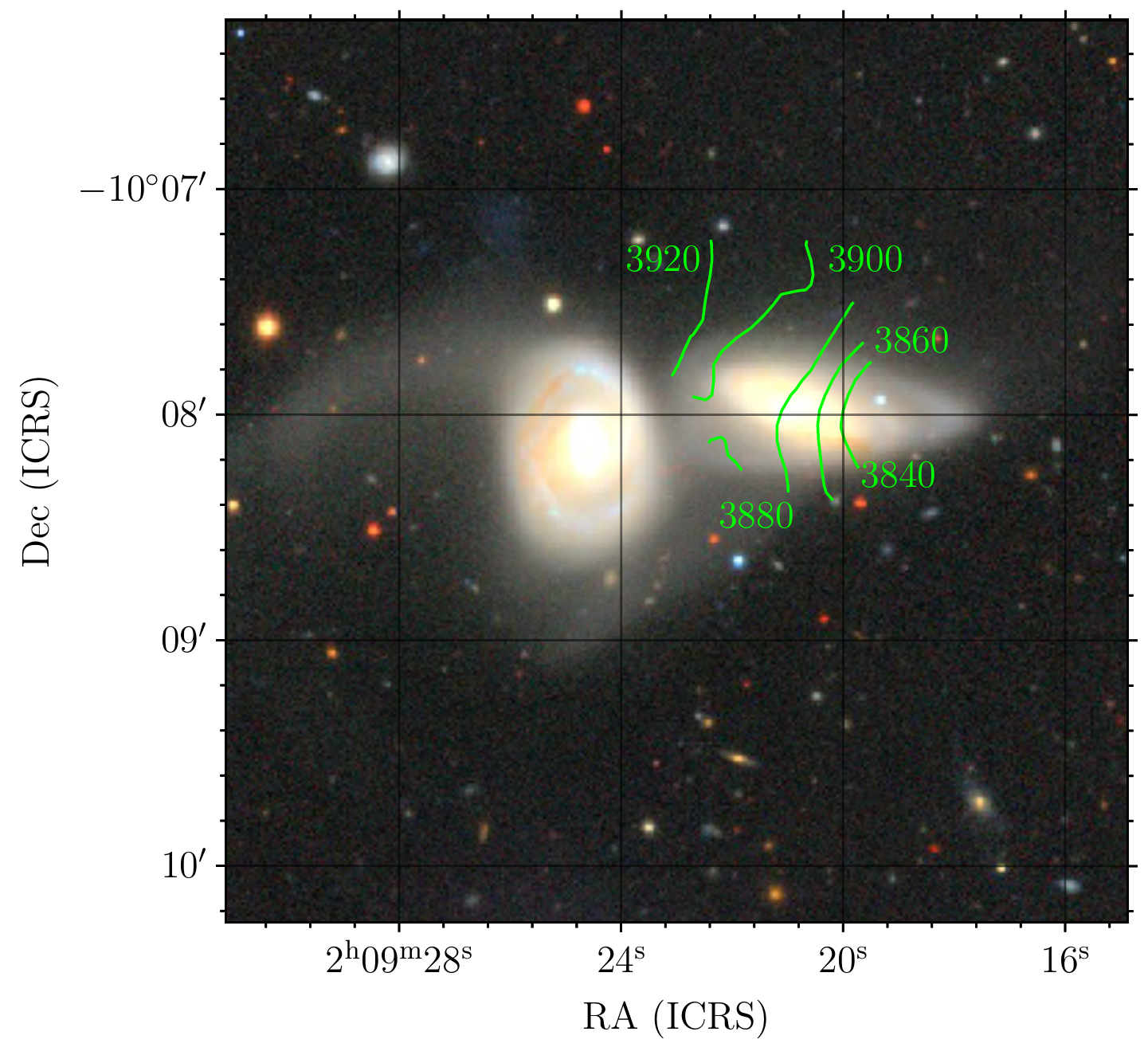}
    \caption{\textit{Left}: The moment 0 map of HCG 16b. The black stars indicate the optical centres of HCG 16a and b, and the black ellipse shows the beam. \textit{Right:} Moment 1 contours for HCG 16b overlaid on the DECaLS $grz$ image of HCG 16a and b. The green lines are separated by 20 \kms. The signature of a small, consistent velocity gradient across the optical disc is visible, but the \hi \ distribution is strongly off centre and disturbed.}
    \label{fig:HCG16b_moms}
\end{figure*}

HCG 16b is an Sa galaxy and the most \hi \ deficient galaxy in the group, probably having lost about $4.6 \times 10^{9}$ \Msol \ of \hi.\footnote{It is correct that this galaxy has lost less \hi \ than HCG 16a, but is more \hi \ deficient. This is because \hi \ deficiency is defined as the logarithmic decrement, not the additive decrement.} The galaxy has a central source identified as an AGN with X-rays and line ratios \citep{Turner+2001,Martinez+2010,Oda+2018}, it has also been classified as a LINER \citep{Veron-Cetty+Veron2010}. This is the only galaxy in the group that was classified as quiescent in terms of its WISE IR colours \citep{Zucker+2016}.

As described above, HCG 16b has an \hi \ bridge connecting to the much more \hi \ massive HCG 16a. In the absence of more information the faint bridge is simply split approximately half-way between the centres of emission of each object in each channel. One may argue that this bridge should be classified as a separate tidal feature, however, given the fact that the outer regions of the optical discs of the two galaxies blend together and the resolution of the \hi \ cube, this is not a practical suggestion. While it is difficult to quantify the resulting uncertainty in the \hi \ mass of HCG 16b due to the simplistic separation, it is likely that the procedure assigns too much emission to HCG 16b rather than HCG 16a, because the latter is both more optically luminous and has a higher \hi \ mass, thus more of the bridge is probably gravitationally bound to it. Therefore, the key result that HCG 16b has lost around 90\% or more of its expected \hi \ content would be unchanged.

Figure \ref{fig:HCG16b_moms} shows that the little remaining \hi \ in HCG 16b is very off centre compared to its optical disc. The velocity field shows a small, but clear velocity gradient aligned with the major axis of the optical disc, suggesting that at least some of the remaining \hi \ is likely still rotating with the optical disc. However, this gradient is disrupted on the eastern side of the galaxy where there is a gas extension to the North-East that connects to emission around the NW tail and that around HCG 16a.

For HCG 16b there is dramatic disagreement between the rotation curves of \citet{Rubin+1991} and \citet{Mendes+1998}, with them even disagreeing on the direction of rotation. The direction that we see in \hi \ agrees with that of \citet{Rubin+1991}, with redshifted emission occurring to the NE of the galaxy centre and blueshifted to the SW. If we take the \citet{Rubin+1991} systemic velocity then it appears that we are mostly detecting gas on the redshifted side of the disc, with little contribution from the blueshifted side. This would also imply that the entire velocity gradient seen in \citet{Mendes+1998} is on the redshifted side of the galaxy.

\subsubsection{HCG 16c}
\label{sec:HCG16c}

\begin{figure*}
    \centering
    \includegraphics[height=75mm]{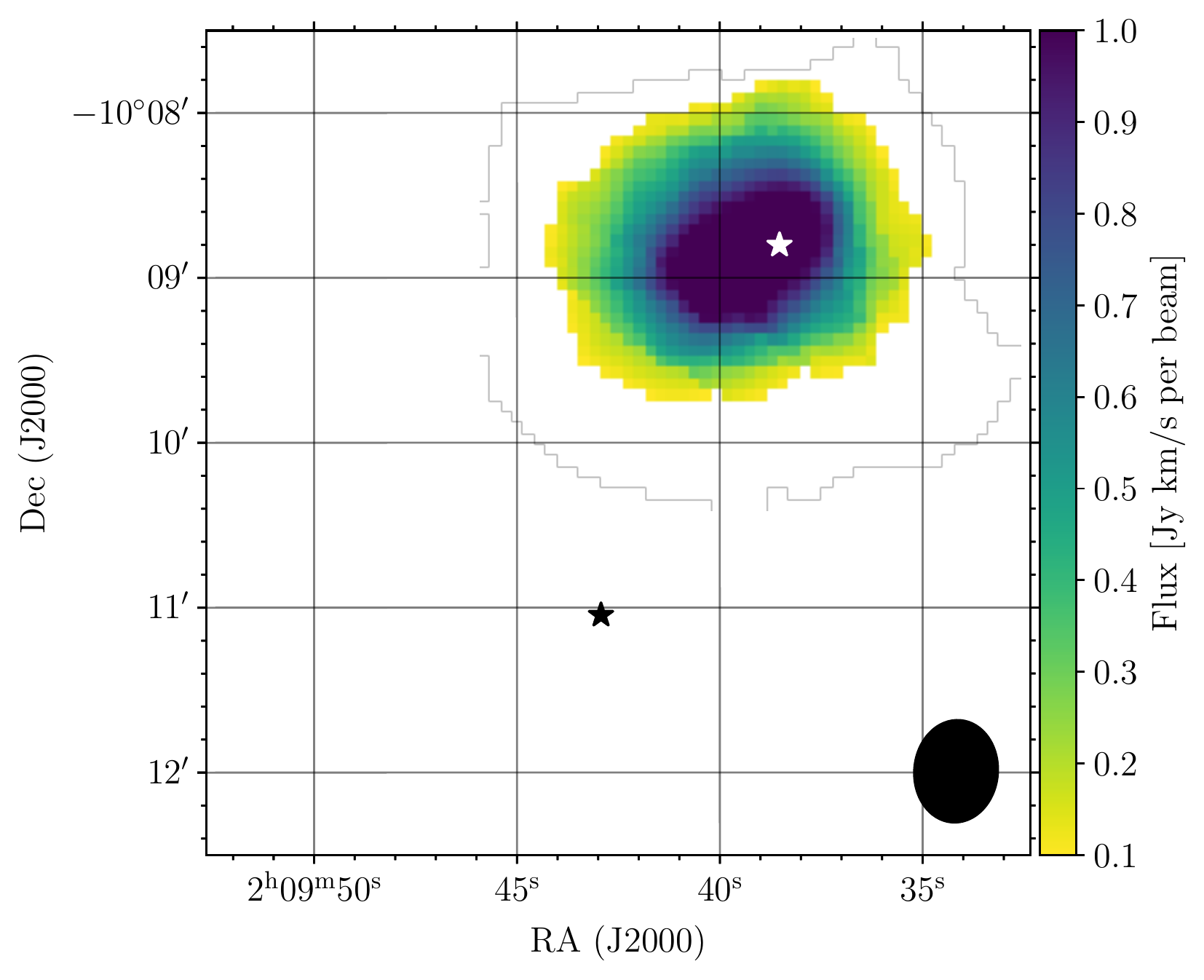}
    \includegraphics[height=75mm]{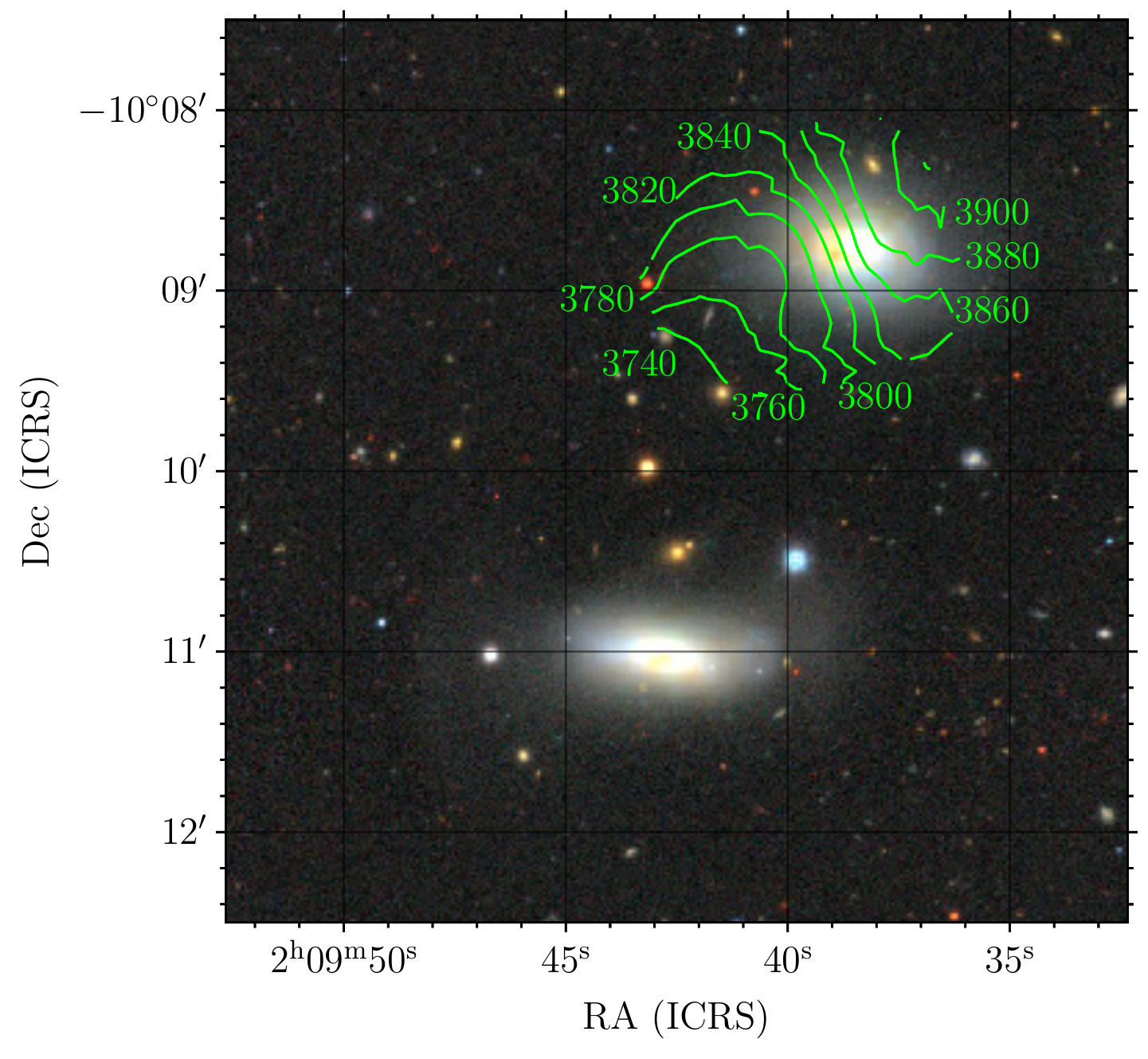}
    \caption{\textit{Left}: The moment 0 map of HCG 16c. The white and black stars indicate the optical centres of HCG 16c and d respectively, and the black ellipse shows the beam. \textit{Right:} Moment 1 contours for HCG 16c overlaid on the DECaLS $grz$ image of HCG 16c and d. The green lines are separated by 20 \kms.}
    \label{fig:HCG16c_moms}
\end{figure*}

HCG 16c is classified as an S0-a galaxy and a LIRG (Luminous Infrared Galaxy). It is also the most actively star-forming galaxy in the group. It has been classified as both a pure starburst based on X-ray observations \citep{Turner+2001} and a composite object, with both AGN and SF activity, using optical line ratios \citep{Martinez+2010}. The overall morphology of the galaxy is very reminiscent of M82, with the nuclear starburst driving a bipolar galactic wind that was studied in detail by \citet{Vogt+2013}.

HCG 16c is apparently involved in a number of ongoing interactions: the NW tail extends between HCG 16c towards HCG 16a, the NE tail extends from HCG 16c away from the rest of the group, and there is an \hi \ bridge connecting HCG 16c and d, all of which can be seen in Figure \ref{fig:mom0}, the \href{http://amiga.iaa.es/FCKeditor/UserFiles/X3D/HCG16/HCG16.html}{X3D plot}, and in the channel maps (available in the electronic version). All three of these features contain more than $10^{9}$ \Msol \ of \hi, that is, they are comparable in \hi \ mass to the galaxy itself, yet only considering the \hi \ emission in the region we identified as the \hi \ disc of HCG 16c the galaxy is not found to be \hi \ deficient. This strongly suggests that the majority of the neutral gas surrounding HCG 16c likely originated elsewhere (discussed further in \S\ref{sec:discuss}). 

We trace the NW tail from the central velocity and position of HCG 16c extending towards HCG 16a in the plane of the sky, but away from it in velocity space, ending when separated from HCG 16a by over 300 \kms. The NE tail appears to originate from the receding edge of HCG 16c and extends out to the NE, but then curves back towards HCG 16a, overlapping both spatially and in velocity with the NW tail. In this region there is a great deal of extended emission of uncertain origin and it is impossible to reliably separate the two features anywhere except at their bases where the column density is high. Therefore, we assign emission at different velocities that is mostly co-spatial (on the plane of the sky) with the bulk of each feature. This means the fluxes of these two tails should be treated with great caution, however the measurement of the overall flux of the extended emission is not impacted by this issue. Finally, there is the cd bridge. This is predominantly made up of the high column density \hi \ emission that forms a bridge between galaxies HCG 16c and d, however, all the remaining emission in the vicinity of galaxies HCG 16c and d that had not been assigned to any of the galaxies or other features listed above, was also assigned to the cd bridge, thus it is more poorly defined than the other extended features we describe.

Despite the evidence for interactions listed above, the centre of the \hi \ disc of HCG 16c still has a consistent velocity gradient across it, indicating that it is rotating and has not been completely disrupted (Figure \ref{fig:HCG16c_moms}). However, the \hi \ distribution is clearly extended (asymmetrically) in the direction of HCG 16d and the outer regions of the velocity field also trail off in that direction, indicating that there are major disturbances in the outer parts of the disc.

The \citet{Rubin+1991} and \citet{Mendes+1998} rotation curves are again quite different for this source, but this is because the spectral slit that \citet{Rubin+1991} used was aligned with the R-band photometric major axis, which is rotated about 40$^\circ$ with respect to the kinematic axis identified by \citet{Mendes+1998}. The kinematic major axis of the \hi \ data (Figure \ref{fig:HCG16c_moms}, right) appears approximately consistent with this position angle (120$^\circ$) obtained from the H$\alpha$ line. The \citet{Mendes+1998} velocity field is more or less regular, but the range of velocities seen is considerably larger than we see in \hi. This may indicate that the rotation curve declines beyond the central region that they measure, or that their rotation curve was contaminated by H$\alpha$ associated with the central outflow, as suggested by \citet{Vogt+2013}.

Figure \ref{fig:pv_plot} shows the peculiar kinematics of the NW tail. Rather than connecting to the outer edge of the disc of HCG 16c, as is typical for tidal tails (in fact this is visible on the opposite side of the same plot, around NGC 848), the NW tail appears to intersect HCG 16c at its central velocity. This behaviour leads us to consider 3 competing hypotheses which we will discuss in the following section: 1) the gas is being accreted directly to the central regions of the galaxy and providing fuel for the nuclear starburst, 2) the gas is being ejected from the central regions of the galaxy by the nuclear starburst, or 3) the entire feature is superposed with HCG 16c and that the coinciding velocities do not correspond to a spatial connection.

\subsubsection{HCG 16d}
\label{sec:HCG16d}

\begin{figure*}
    \centering
    \includegraphics[height=75mm]{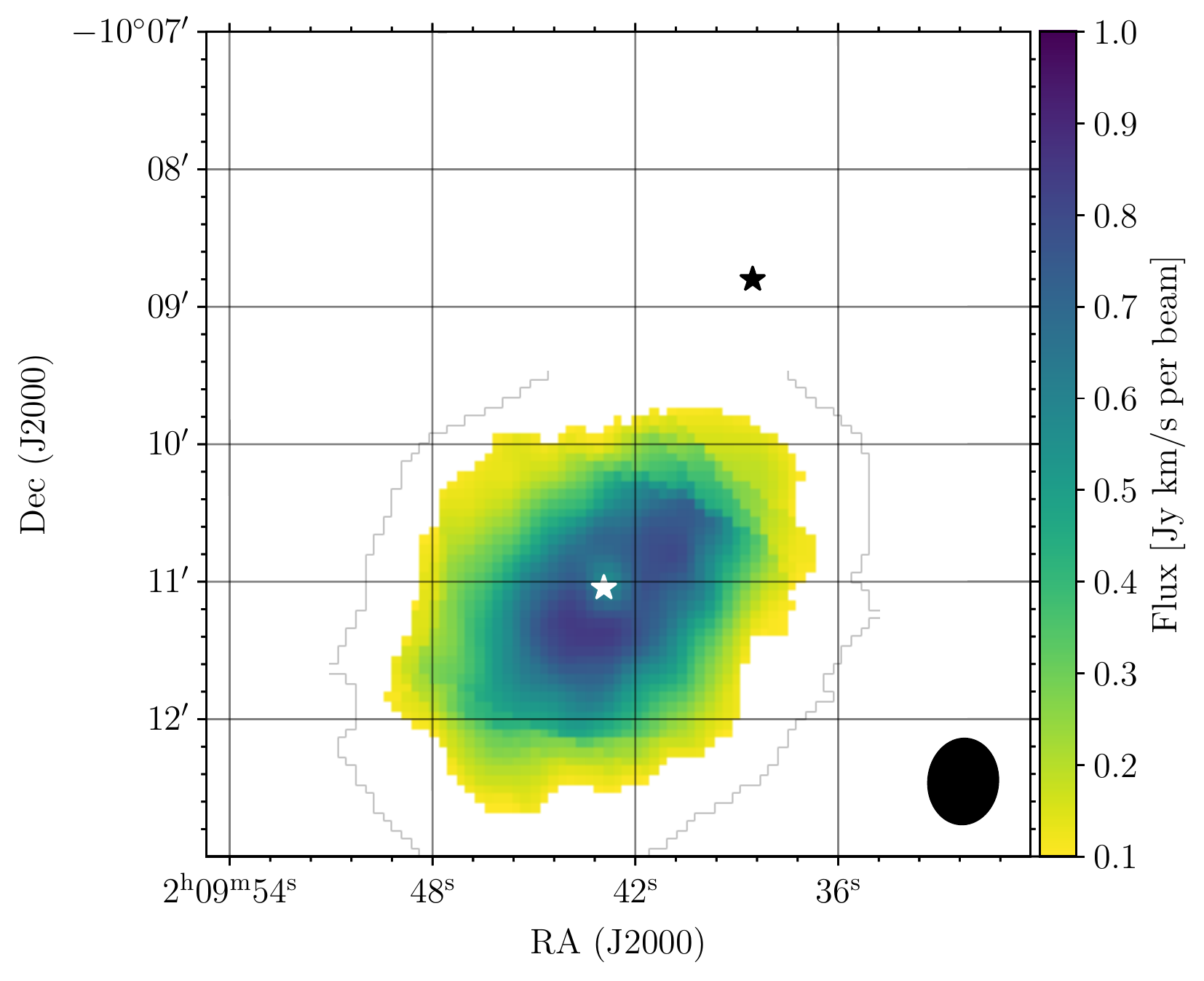}
    \includegraphics[height=75mm]{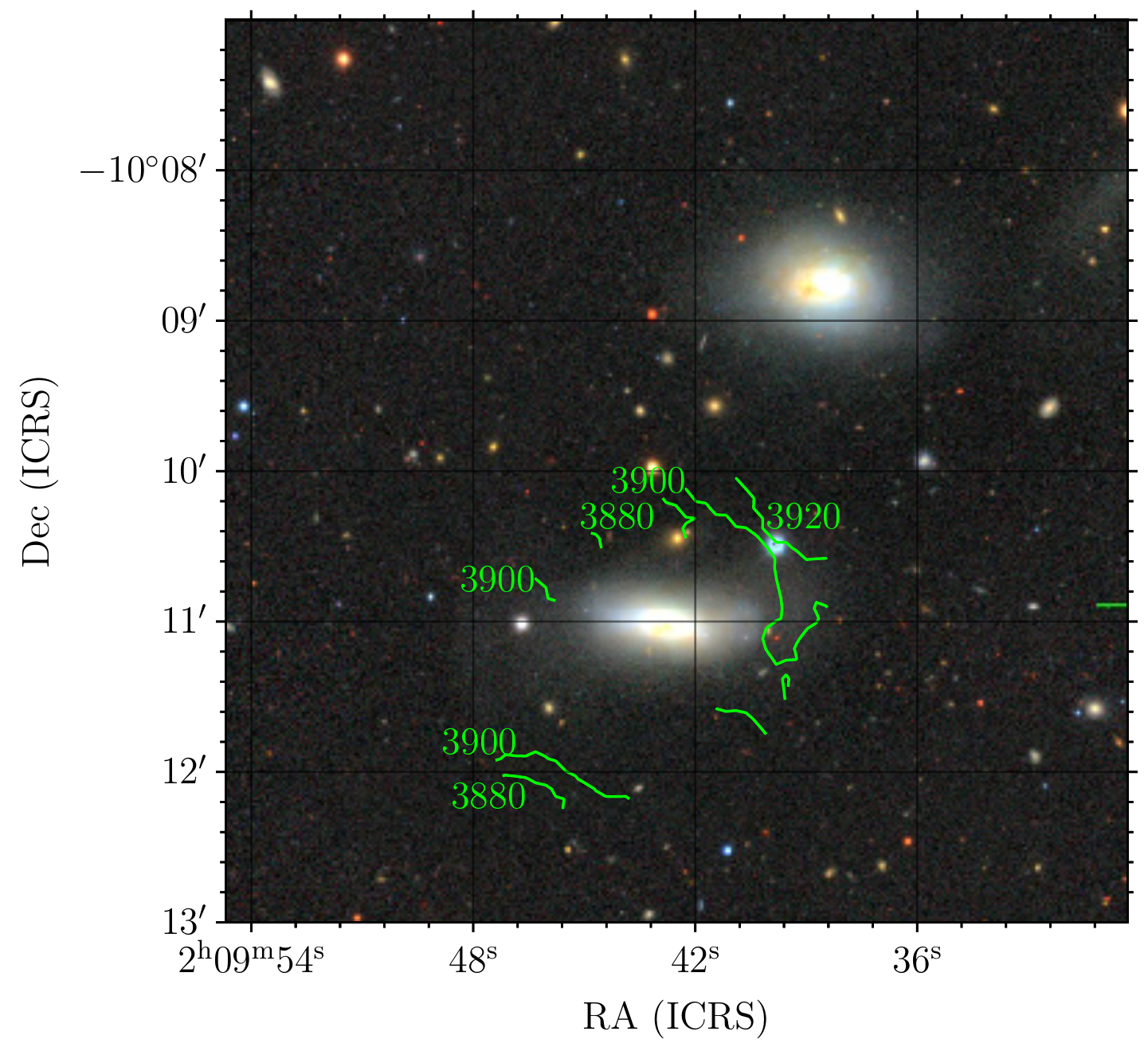}
    \caption{\textit{Left}: The moment 0 map of HCG 16d. The white and black stars indicate the optical centres of HCG 16d and c respectively, and the black ellipse shows the beam. A central depression is clearly visible around the white star marking the centre of HCG 16d. \textit{Right:} Moment 1 contours for HCG 16d overlaid on the DECaLS $grz$ image of HCG 16d and c. The central region of the galaxy is removed as the \hi \ absorption feature would contaminate the moment 1 map there. The green lines are separated by 20 \kms. While the velocity field is highly irregular there is a slight gradient roughly aligned with the \textit{minor} axis of the optical disc.}
    \label{fig:HCG16d_moms}
\end{figure*}

\begin{figure}
    \includegraphics[width=\columnwidth]{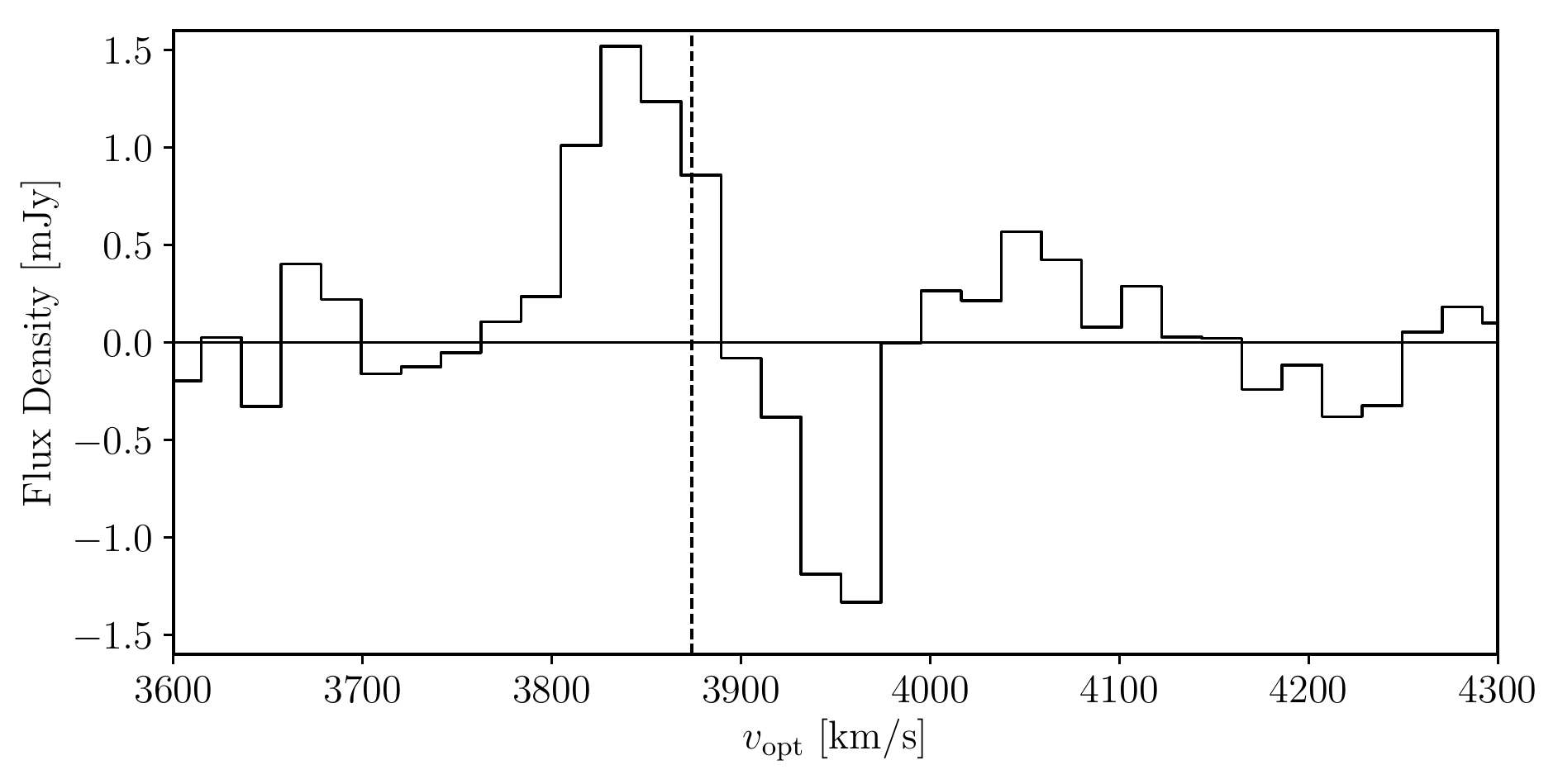}
    \caption{The \hi \ spectral profile of the centre of HCG 16d. The dashed vertical line shows the systemic velocity as determined from optical lines. The profile was extracted from the robust=0 cube (higher resolution) by placing a double Gaussian window, with the shape and orientation of the synthesized beam, at the position of the optical centre of the galaxy.}
    \label{fig:absorp_spec}
\end{figure}

Similarly to HCG 16c, HCG 16d is a LIRG, classified as S0-a, and is morphologically similar to M82. It contains a central source classified as both a LINER and as a LINER/Seyfert 2 double nucleus \citep{Ribeiro+1996,deCarvalho+1999} with optical line ratios, or as an AGN in X-rays observations with XMM \citep{Turner+2001}. However, HST observations show that the suggested double nucleus is instead a group of star clusters \citep{Konstantopoulos+2013} and Chandra X-ray observations indicate that the hard X-rays might originate solely from X-ray binaries, not an AGN \citep{OSullivan+2014a,Oda+2018}, while IFU observations suggest that the LINER line ratios in the optical may arise from shock excitation driven by the on-going starburst \citep{Rich+2010}.

Upon first inspection the integrated \hi \ emission (Figure \ref{fig:HCG16d_moms}) appears almost like a face-on galaxy with a central \hi \ hole, however, this is quite misleading and does not agree with the optical image, which shows a highly inclined disc. The apparent central hole is instead \hi \ absorption in front of a central continuum source. The \hi \ depression is the same shape as the beam in both the robust=2 and robust=0 cubes (i.e. at two different resolutions) and the spectral profile at that position switches from a flux density of approximately 1.5 mJy to -1.5 mJy at the central velocity of the galaxy (Figure \ref{fig:absorp_spec}). 

This absorption feature is redshifted relative to the central velocity (from optical spectra) by about 100 \kms. A regular rotating disc would form a symmetric absorption feature about the central velocity \citep{Morganti+2018}, so the fact this feature is redshifted suggests it could be gas falling towards the centre and fuelling the starburst event. However, a 100 \kms \ velocity shift is comfortably within what might be expected from orbiting gas in a galaxy of this size. Given the resolution of the data it is not possible to distinguish between these two possibilities, or indeed that the absorption may be due to an intervening clump of stripped gas in the group.

The absorption also means that the integrated flux for HCG 16d will be underestimated. To estimate how much the integrated flux is reduced we made a crude linear interpolation across the absorption feature between 3879 and 4006 \kms. This gives the missing flux as 0.1 Jy \kms, which is barely more than 1\% of the total \hi \ emission flux of HCG 16d. Indeed, the absorption feature is not apparent in its integrated spectrum (Figure \ref{fig:spectra}). Therefore, we ignored this feature for our analysis of the \hi \ deficiency.

When considering the kinematics of the gas in HCG 16d there is an apparent contradiction between \hi \ and H$\alpha$. In \hi \ there is only a weak velocity gradient in approximately the North-South direction, almost perpendicular to the optical major axis (Figure \ref{fig:HCG16d_moms}). In contrast, the existing H$\alpha$ rotation curve of the inner regions of the galaxy \citep{Rubin+1991,Mendes+1998,Rich+2010} shows significant rotation ($V_{\mathrm{rot}} \approx 100$ \kms) and has its major axis almost aligned with the optical disc, that is, approximately perpendicular to the slight gradient in the \hi \ emission. The full extent of the H$\alpha$ velocity field in \citet{Mendes+1998} is about 15\arcsec, less than the beam size of the \hi \ data, thus it is possible that the gas could undergo a major kinematic warp (or other distortion) beyond this region, meaning the \hi \ and H$\alpha$ results are not necessarily in conflict. However, as the H$\alpha$ kinematic major axis is aligned with the optical disc of the galaxy (which extends to much larger radii) this seems unlikely.

The \hi \ data thus demonstrates two key points: a) that there is a continuum source at the centre of HCG 16d, although the \hi \ data offers no information on its nature, and b) the \hi \ gas appears to be completely kinematically disconnected from the optical disc of the galaxy. These points raise the question "is the \hi \ truly associated with the galaxy or just superposed on it?", which we will discuss in the following section.

There are also a number of tidal features in the vicinity of HCG 16d, most notable is the SE tail that extends for about 10\arcmin \ (160 kpc) towards NGC 848. Although this feature connects HCG 16d and NGC 848 in both velocity and the plane of the sky (Figure \ref{fig:pv_plot}) there is no evidence for an accompanying optical tail emanating from HCG 16d. Most of the emission from the tail is contained within only 4 channels (a range of 84 \kms), although there are some clumps of emission which we have assigned to this feature near HCG 16d that extend further in velocity space such that the entire profile of the SE tail covers 10 channels (Figure \ref{fig:tidal_spectra}). The small velocity range along the length of the tail likely indicates that it is probably almost aligned with the plane of the sky. HCG 16d is also accompanied by 2 small, dense clumps on the East and South, and at approximately the same velocity as HCG 16d. Each is several times $10^8$ \Msol \ in \hi \ mass. These clumps are discussed further in \S\ref{sec:TDGs}. There is also a high \hi \ column density bridge connecting HCG 16d and c. The complexity of this region of the group makes separation of these features highly subjective and thus their \hi \ properties should be treated with caution.

\subsubsection{NGC 848}
\label{sec:NGC848}

\begin{figure*}
    \centering
    \includegraphics[height=75mm]{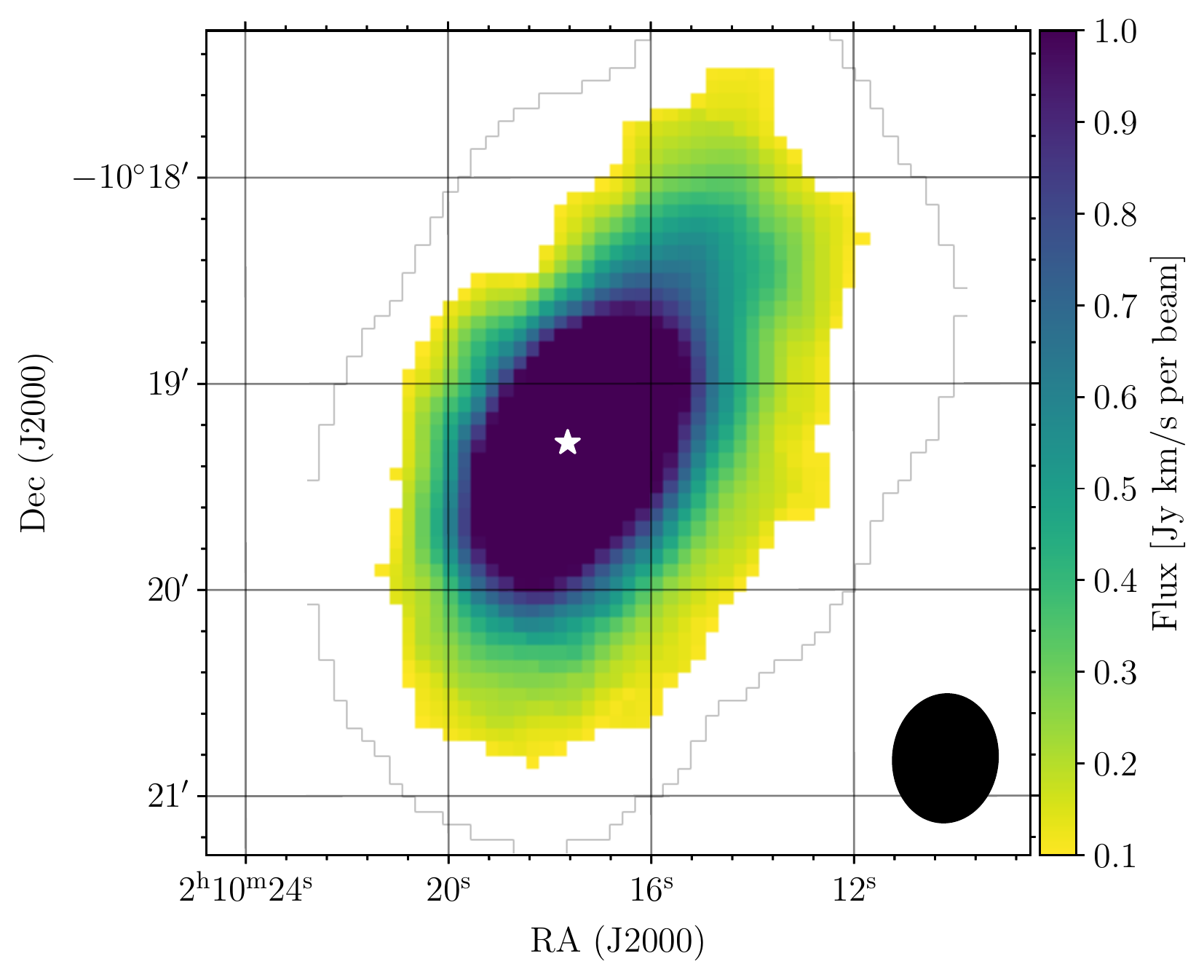}
    \includegraphics[height=75mm]{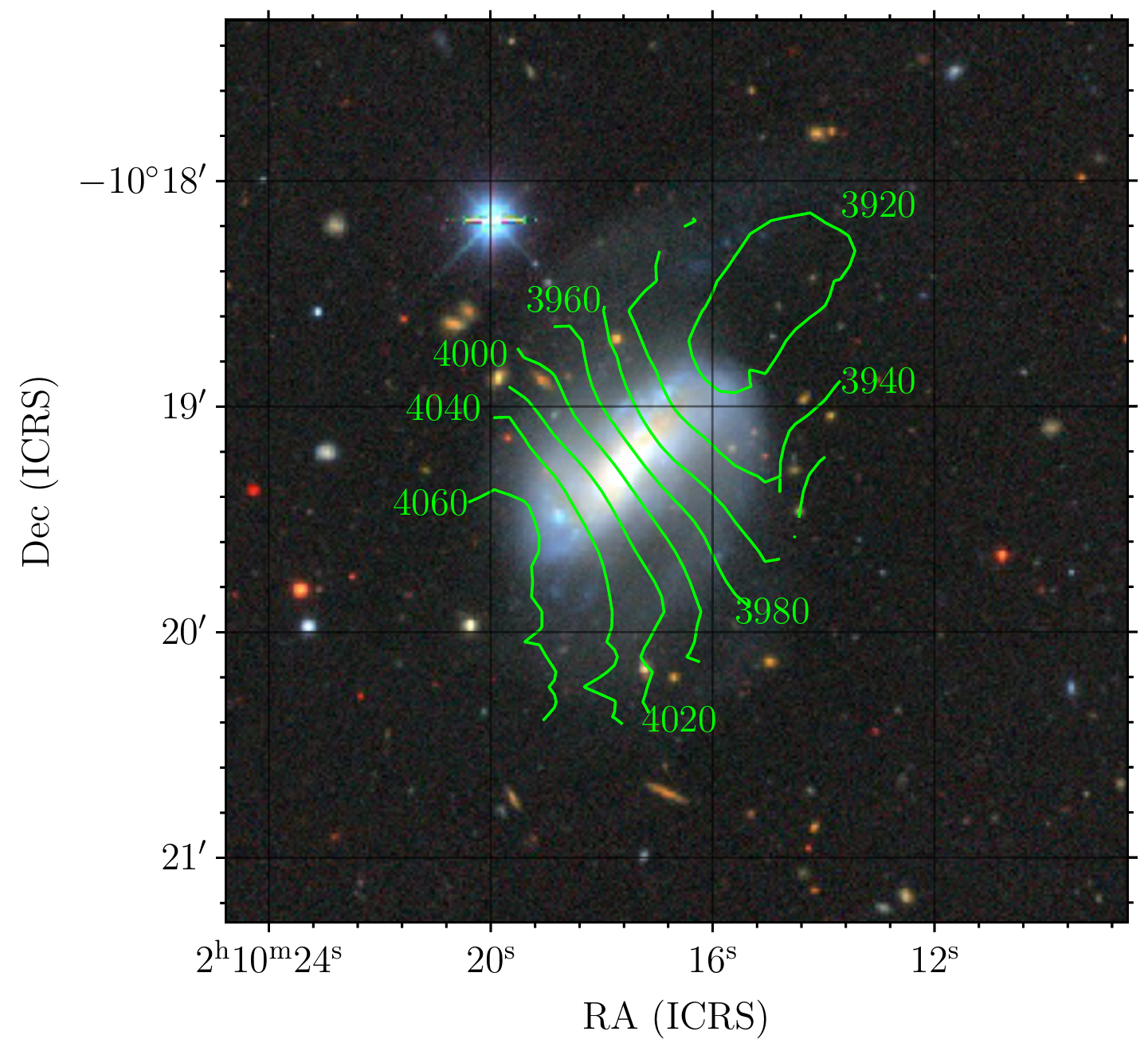}
    \caption{\textit{Left}: The moment 0 map of NGC 848. The white star indicates the optical centre, and the black ellipse shows the beam. \textit{Right:} Moment 1 contours for NGC 848 overlaid on the DECaLS $grz$ image. The green lines are separated by 20 \kms.}
    \label{fig:NGC848_moms}
\end{figure*}

NGC 848 is a barred spiral galaxy (SBab) that is separated from the rest of the group by about 10\arcmin, or around 160 kpc, and is another galaxy in the group undergoing a starburst \citep{Ribeiro+1996,deCarvalho+1999}. Although this galaxy was not included in the original Hickson catalogue, once spectra were obtained of the galaxies it was immediately noticed that it was at the same redshift as the core group (Table \ref{tab:optprops}) and therefore likely associated \citep{Ribeiro+1996}. However, it was not until there were \hi \ interferometric observations that the physical connection, in the form of a 160 kpc tidal tail, was discovered \citep{Verdes-Montenegro+2001}. 

Despite its connection to this enormous tidal feature (visible in Figure \ref{fig:mom0}) the galaxy itself has an entirely normal global \hi \ content (Table \ref{tab:HIprops}). This alone indicates that the gas in the tail probably did not (for the most part) originate from NGC 848, but from somewhere else in the group, especially as the total \hi \ mass in the tail is approximately the same as the \hi \ mass of NGC 848. There is another tail connected to NGC 848, which appears to originate from its receding side and extends to the South-East, that we refer to as the NGC 848S tail (Figure \ref{fig:mom0}). At least part of this feature is included in the \texttt{SoFiA} mask, however it may extend much further, looping back around towards the core group, but this is at low signal-to-noise and (if real) the emission in adjacent channels also shifts in position, such that smoothing spatially and in velocity does little to improve its signal-to-noise. Some parts of this faint feature are visible in the left panel of Figure \ref{fig:mom0} (as well as the \href{http://amiga.iaa.es/FCKeditor/UserFiles/X3D/HCG16/HCG16.html}{X3D plot}) slightly to the North of NGC 848 and the SE tail. We include this extended loop in Table \ref{tab:HIprops} and Figure \ref{fig:tidal_spectra}, but it is not included in the total integrated emission of the group as it does not have high enough significance to be included in the \texttt{SoFiA} mask.

The \hi \ distribution in the galaxy itself (Figure \ref{fig:NGC848_moms}) appears mostly regular, with a centroid coincident with the optical centre and a mostly uniform velocity field. However, the kinematic major and minor axes are not quite perpendicular, which is probably due to the presence of the bar \citep[e.g.][]{Bosma1981}. The \hi \ gas is also slightly extended towards the NW and SE, i.e. in the direction of the tidal tail, while in the optical image the spiral arms appear very loosely wrapped (Figure \ref{fig:NGC848_moms}), probably indicating that the outer disc is beginning to be unbound.

\subsubsection{PGC 8210}
\label{sec:PGC8210}

\begin{figure*}
    \centering
    \includegraphics[height=75mm]{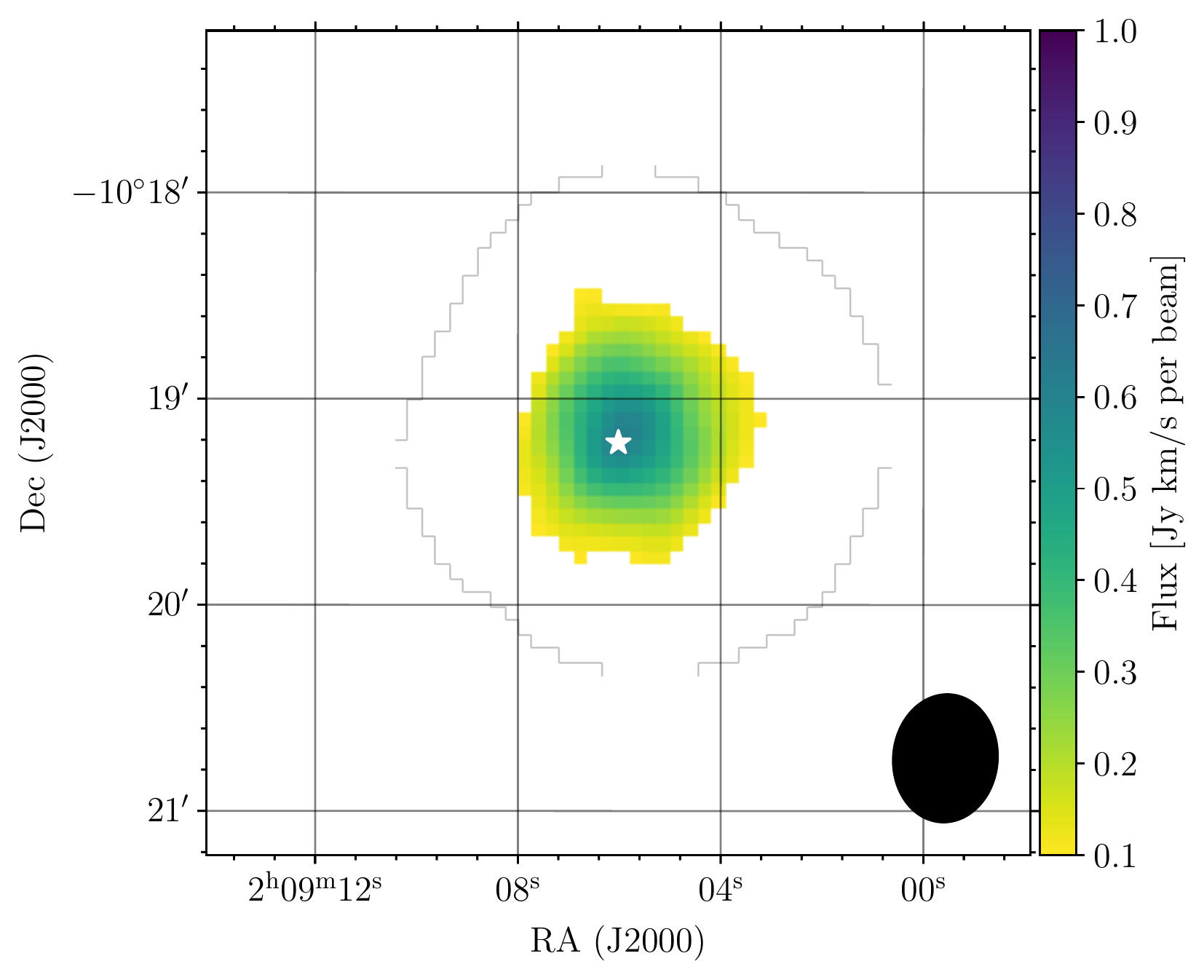}
    \includegraphics[height=75mm]{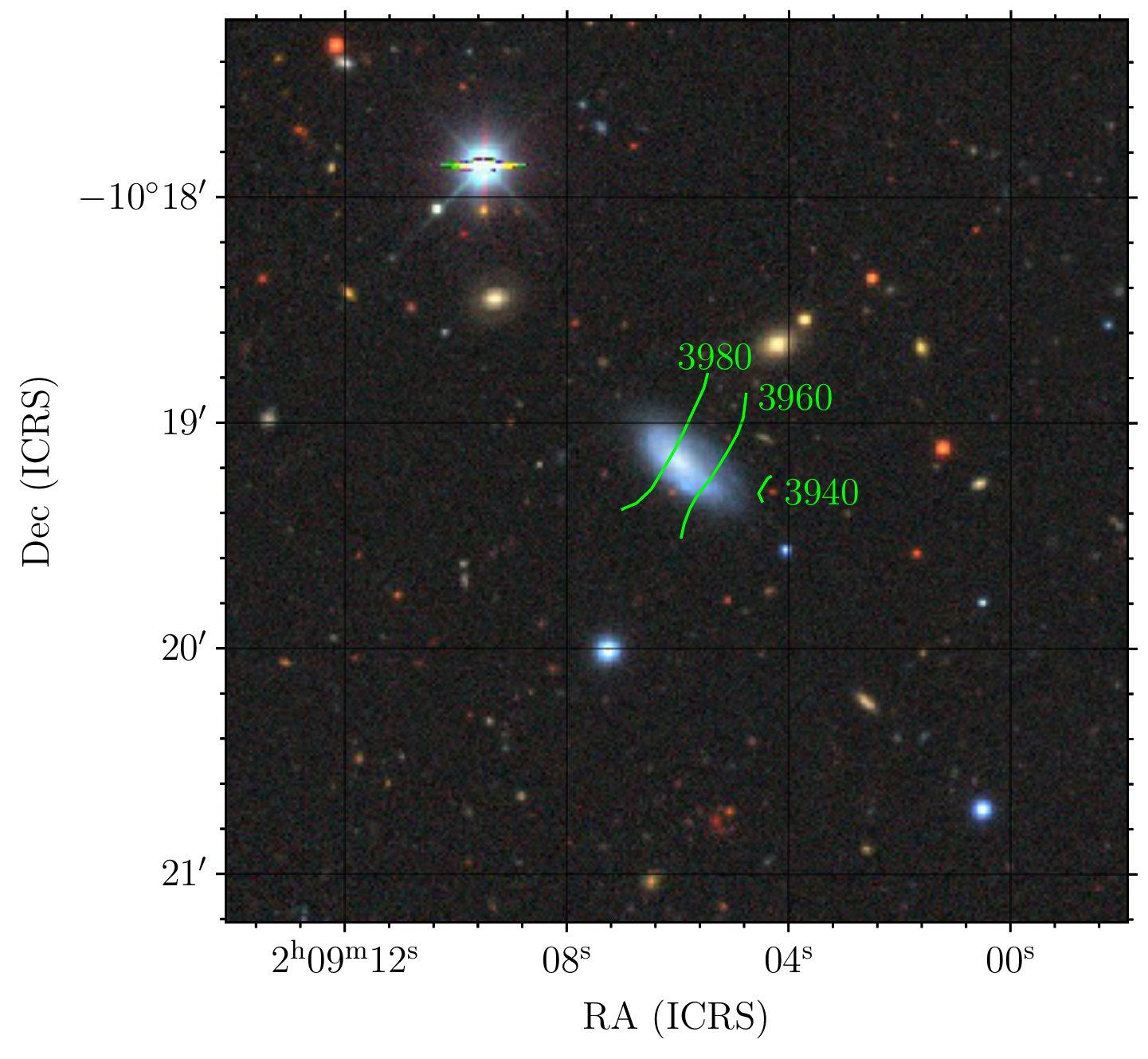}
    \caption{\textit{Left}: The moment 0 map of PGC 8210. The white star indicates the optical centre, and the black ellipse shows the beam. \textit{Right:} Moment 1 contours for PGC 8210 overlaid on the DECaLS $grz$ image. The green lines are separated by 20 \kms.}
    \label{fig:PGC8210_moms}
\end{figure*}

The final galaxy detected within the primary beam of the VLA observations is PGC 8210. This is another spiral galaxy but it is considerably smaller and less massive than the core members of the group. Its B-band luminosity is almost 1 dex lower than the galaxies in the core group and its \hi \ mass is considerably less than all but HCG 16b, which is extremely \hi \ deficient. Because the velocity of PGC 8210 is coincident with that of the group it is assumed to be part of the same structure, or at least about to join it. However, as shown in Figure \ref{fig:PGC8210_moms} the \hi \ distribution appears undisturbed (with the caveat that it is hardly larger than the beam), as does its optical disc, and its total \hi \ content is entirely normal. Therefore, it seems very unlikely that this galaxy has had any meaningful interaction with the group to date.

\subsection{Tidal dwarf galaxy candidates}
\label{sec:TDGs}

\begin{figure*}
    \centering
    \includegraphics[width=0.66\columnwidth]{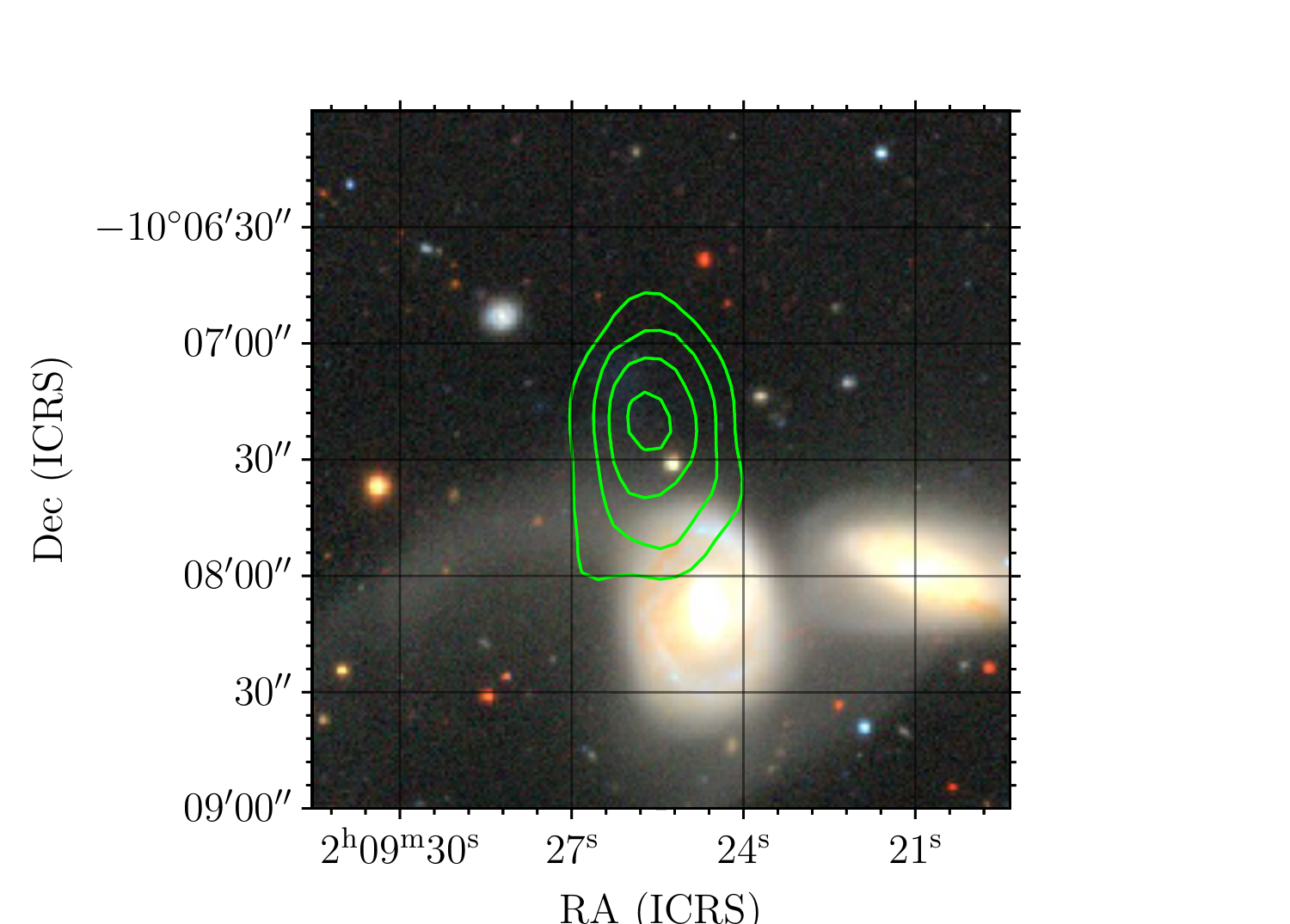}
    \includegraphics[width=0.66\columnwidth]{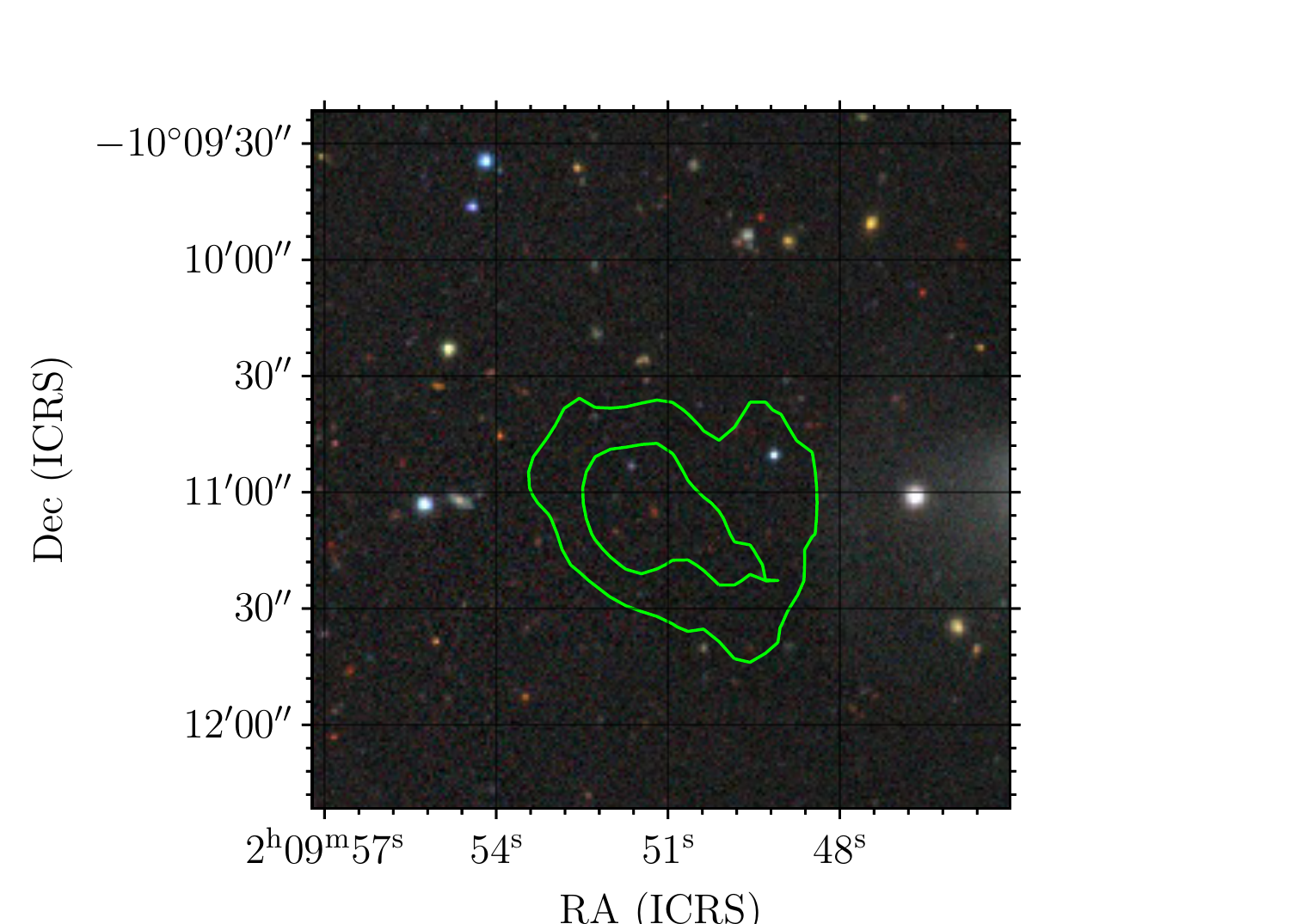}
    \includegraphics[width=0.66\columnwidth]{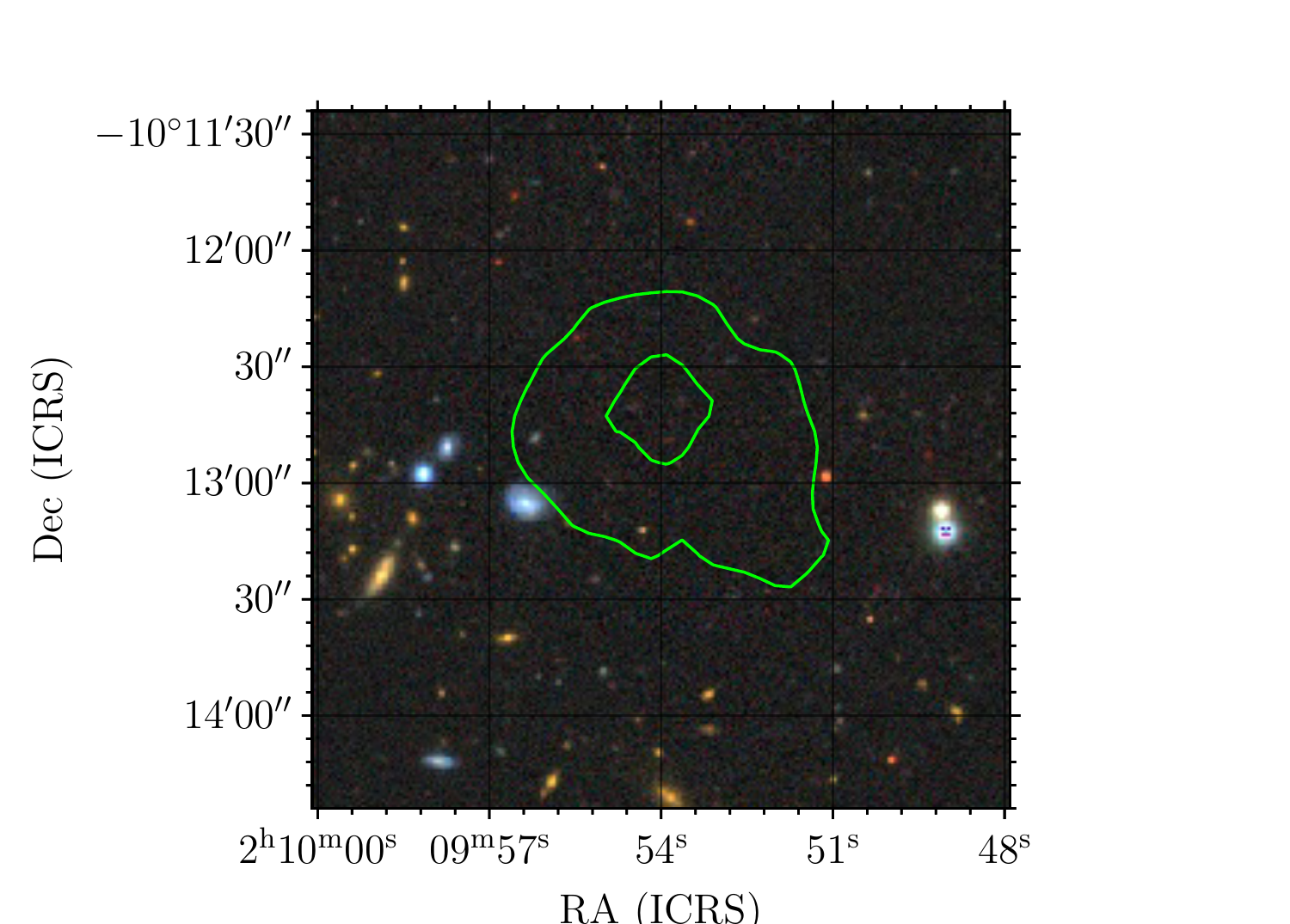}
    \includegraphics[width=0.66\columnwidth]{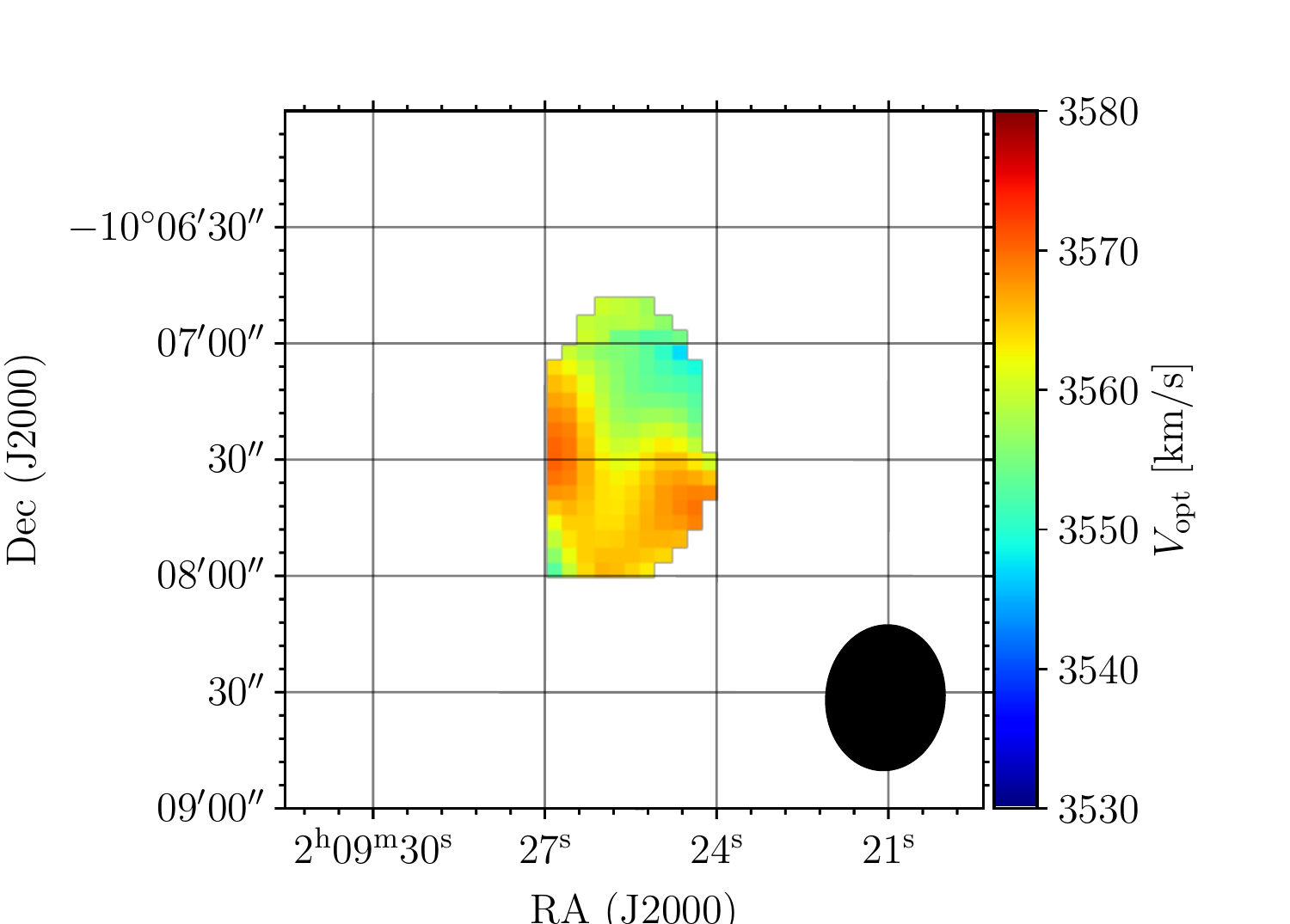}
    \includegraphics[width=0.66\columnwidth]{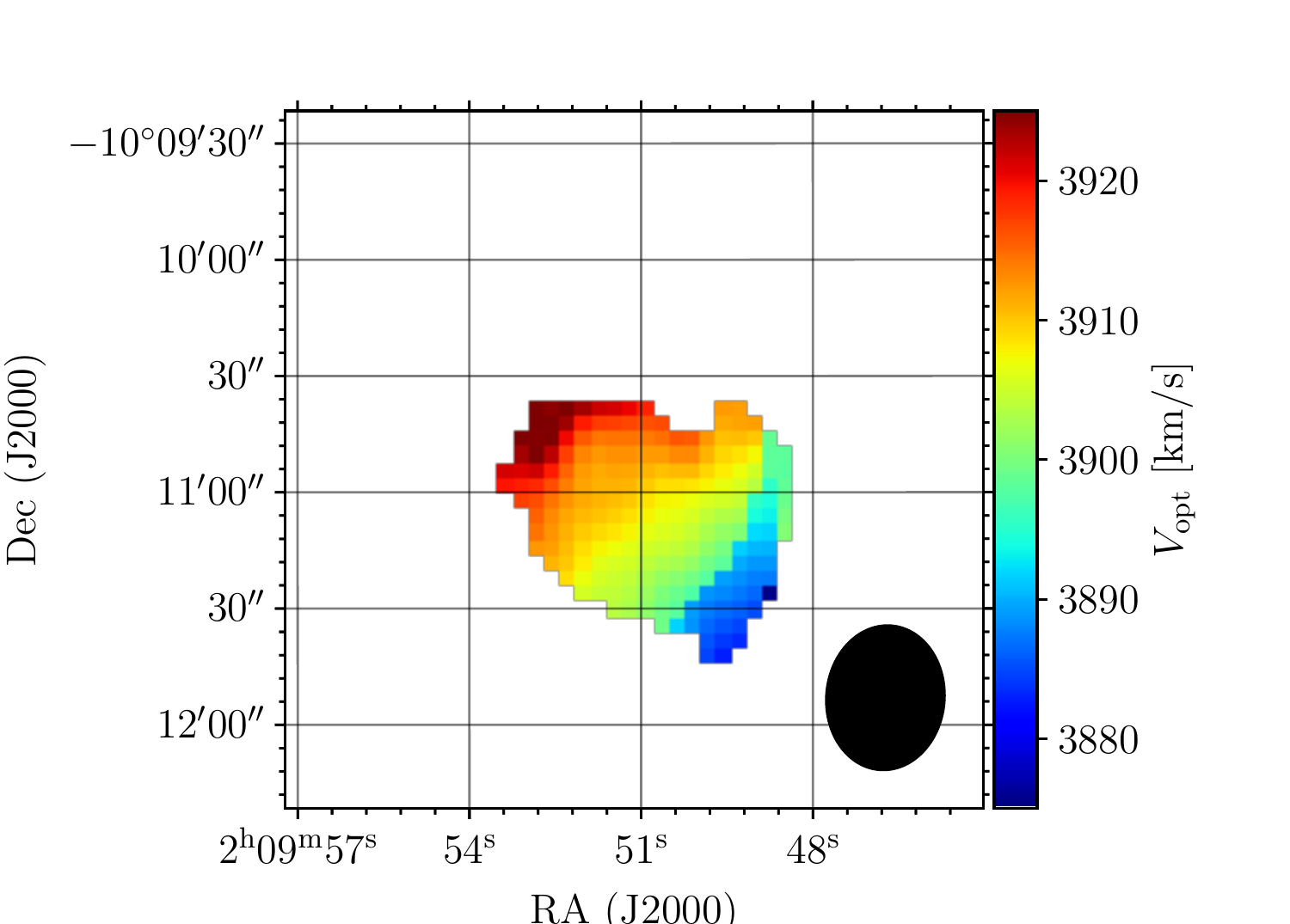}
    \includegraphics[width=0.66\columnwidth]{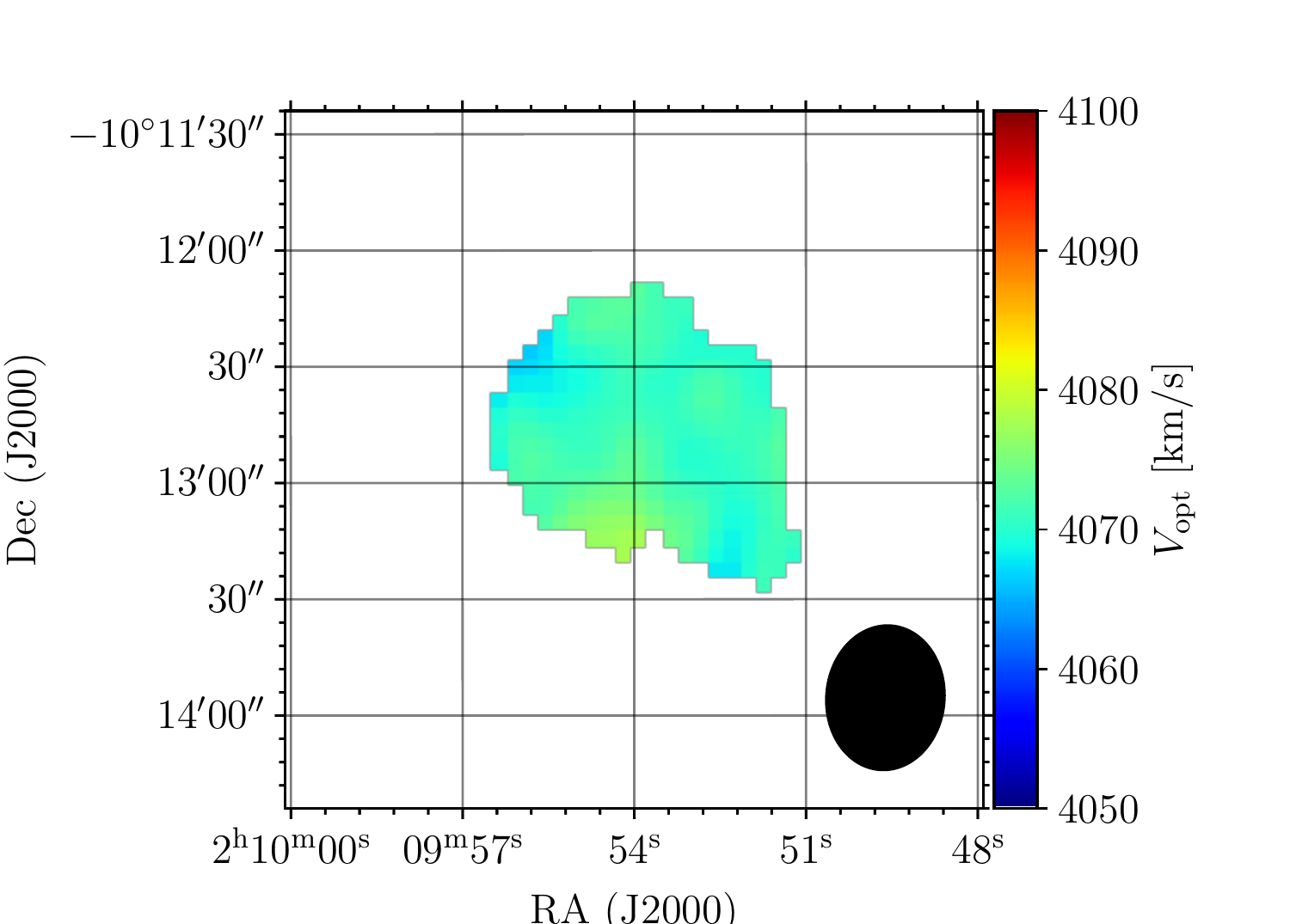}
    \caption{\textit{Top}: The moment 0 maps of the NW, E and S clumps (left to right) overlaid on the DECaLS $grz$ image. Contour levels: 0.98, 1.47, 1.96, 2.45 $\times 10^{20} \; \mathrm{cm}^{-2}$. \textit{Bottom:} Moment 1 maps of the same features with the beam shown as a black ellipse in the corner.}
    \label{fig:clump_moms}
\end{figure*}

We have identified three dense clumps of \hi \ emission in the group that are not associated with any of the main galaxies. The first two are listed in Table \ref{tab:HIprops}, the E and S clumps in the vicinity of HCG 16d, and the third is a dense clump at the north-western end of the NW tail, which is clearly seen as an excess at low velocities in the spectrum of the NW tail (Figure \ref{fig:tidal_spectra}). These could be candidate LSB (low surface brightness) dwarf galaxies or TDGs, both of which can occur in galaxy groups and can be rich in \hi \ \citep{Lee-Waddell+2012,Lee-Waddell+2016,Roman+2017,Spekkens+2018}, or transient clumps of stripped gas. Their moment 0 and moment 1 maps are shown in Figure \ref{fig:clump_moms} (NW, S and E clumps, left to right).

All three \hi \ clumps have masses well above $10^{8}$ \Msol, the minimum mass thought to be needed to form a long-lived TDG \citep{Bournaud+2006}. Their masses are 3.9, 5.2, and 9.0 $\times 10^{8}$ \Msol, respectively (going left to right in Figure \ref{fig:clump_moms}). While the S clump is the most massive (in \hi) it is also the most spatially extended, has the lowest peak column density, shows little evidence for rotation in its moment 1 map, and has no apparent optical counterpart. Therefore, despite its large mass it seems unlikely that the S clump is a long-lived TDG or a LSB galaxy, and instead is probably a transient feature associated with the SE tail. 

The E clump is in between the other two clumps in terms of \hi \ mass and column density, however, its moment 1 map shows a clear velocity gradient across it indicating that it may be rotating. Taking very approximately the \hi \ diameter as 1.5\arcmin \ and the rotation velocity as 25 \kms, the dynamical mass (assuming equilibrium) would be $1.7 \times 10^{9}$ \Msol, which is about 3 times its measured \hi \ mass. This simple estimate implies that there may need to be some dark matter component to explain the velocity gradient, if it is due to rotation. However, there are also a number of reasons why this estimate might be incorrect. For example, the assumption of dynamical equilibrium is unlikely to be correct and even if it were, the source is only 3 beams across, which means both its velocity field and spatial extent are likely to be heavily affected by beam smearing.

The NW clump has the smallest \hi \ mass of the three clumps, but the highest column density, peaking at over $2.45 \times 10^{20} \; \mathrm{cm}^{-2}$. The moment 1 map does not show clear evidence for rotation, however, it is barely 2 beams across and any potential signature of rotation may be confused with gas in the NW tail. TDGs are expected to be found at the tip of tidal tails as this clump is \citep[e.g.][]{Bournaud+2004} and unlike the other two clumps there is a clear, but very LSB optical counterpart visible in the DECaLS image at $\mathrm{RA = 2h\;9m\;26s}$, $\mathrm{Dec = -10^\circ}$ 7\arcmin \ 9\arcsec \ (top left panel of Figure \ref{fig:clump_moms} and right panel of Figure \ref{fig:HCG16a_moms}) and it is also just visible in GALEX \citep[Galaxy Evolution Explorer,][]{Martin+2005}. The blue colour of this optical counterpart suggests that it is the result of in situ SF, not old stars which might have been stripped along with the gas. This strengthens the case for this candidate TDG, but there remains the possibility that this is a gas-rich LSB dwarf which is having its \hi \ stripped, rather than a new galaxy that has formed as a result of the tidal interactions in the group. While it is not possible, with the current information, to rule out either of these hypotheses, we favour the TDG interpretation because this dwarf appears at the end of an enormous tidal feature and it seems implausible that either such a small galaxy stripped this gas from the other (much more massive) group members, or that it could have originally been sufficiently gas-rich to be the origin of the observed tidal gas.

\section{Discussion}
\label{sec:discuss}

\subsection{Gas consumption timescales}

\begin{table*}
\centering
\caption{Gas consumption times of HCG 16 core galaxies}
\label{tab:gas_time}
\begin{tabular}{lccccc}
\hline\hline
Object  & $\log{M_{\ast}/\mathrm{M_\odot}}$ & $\log{M_{\mathrm{HI}}/\mathrm{M_\odot}}$ & $\log{M_{\mathrm{H_2}}/\mathrm{M_\odot}}$  & SFR/$\mathrm{M_\odot}\,\mathrm{yr}^{-1}$ & $T_\mathrm{{con}}$/Gyr   \\ \hline
HCG 16a & 11.05  & 9.15 & 10.05 & 4.6 & 3.5 \\
HCG 16b & 10.84  & 8.70 & 9.17  & 0.46 & 5.6 \\
HCG 16c & 10.86  & 9.57 & 9.78  & 14.0  & 0.9 \\
HCG 16d & 10.61  & 9.73 & 10.01 & 16.7  & 1.2 \\ \hline
\end{tabular}
\tablefoot{Columns: (1) object name, (2) logarithm of stellar mass estimated from IR photometry \citep{Lenkic+2016}, (3) logarithm of \hi \ mass, (4) logarithm of the molecular hydrogen mass estimated from the CO observations compiled in \citet{Martinez-Badenes+2012}, (5) star formation rate estimated from the combination of UV and IR fluxes \citep{Lenkic+2016}, and (6) the gas consumption timescale taken to be $1.3(M_{\mathrm{HI}} + M_{\mathrm{H_2}})/\mathrm{SFR}$.}
\end{table*}

One way that a galaxy can become gas deficient is by converting its gas into stars without replenishment. Ideally the detailed SFH (star formation history) of a galaxy would be compared to its present day gas content, but in the absence of a SFH the gas consumption time given by the current SFR (star formation rate) can be used instead. If the galaxy is in the middle of converting its gas reservoir into stars then this should be apparent.

To estimate the gas consumption timescales of the cold gas in the galaxies we have taken stellar masses and SFRs from \citet{Lenkic+2016}, and molecular gas mass measurements from \citet{Martinez-Badenes+2012}. \citet{Lenkic+2016} use Spitzer IRAC (Infrared Array Camera) 3.6 and 4.5 $\mu$m photometry to estimate stellar masses based on the prescription of \citep{Eskew+2012}. To estimate SFRs these authors use the UV (2030 \AA, from Swift) and IR (24 $\mu$m) luminosities as proxies for the unobscured and obscured (re-radiated) emission due to SF, and combine the two to estimate the total SFR, which in theory allows them to avoid correcting for internal extinction. \citet{Martinez-Badenes+2012} use CO data from \citet{Boselli+1996,Leon+1998,Verdes-Montenegro+1998} to estimate molecular gas masses. They use a standard constant value of the CO-to-$\mathrm{H_{2}}$ conversion factor, $X = N_\mathrm{H_{2}}/I_\mathrm{CO} = 2 \times 10^{20} \; \mathrm{cm^{-2} \, K^{-1}\,km^{-1}\,s}$. As all the CO observations are single pointings they also extrapolate for emission beyond the primary beam by assuming a pure exponential disc with a scale length of $0.2 r_{25}$, where $r_{25}$ is the major optical 25 mag arcsec$^{-2}$ isophotal radius.

Table \ref{tab:gas_time} shows the gas consumption timescales of the four core galaxies of HCG 16, estimated by combining our measurements of the \hi \ masses with measurements of the H$_2$ masses \citep{Martinez-Badenes+2012} and the SFRs \citep{Lenkic+2016}. The total hydrogen mass is multiplied by 1.3 to account for all other elements, giving $T_{\mathrm{con}} = 1.3(M_{\mathrm{HI}} + M_{\mathrm{H_2}})/\mathrm{SFR}$. The galaxies are roughly split into two categories in terms of their gas consumption time, those that at their current SFR will exhaust their existing gas reservoirs in about a Gyr (HCG 16c and d) and those which will take several Gyr to do the same (HCG 16a and b).

HCG 16c and d are both starbursting LIRGs, so it is unsurprising that their gas consumption timescales are short. It is tempting to think that HCG 16a and b probably looked much the same approximately a Gyr in the past and now have slowed SFRs and have become gas deficient. However, as the group is not globally \hi \ deficient it seems unlikely that their atomic gas was converted to H$_2$ and consumed in SF, unless the group was gas-rich to begin with. But even if we allow for that possibility this scenario does not seem to agree with their gas and stellar content. To become depleted in \hi, but not in H$_2$, via SF would have required them to have undergone a SF episode, which would consume much of the molecular gas, and then for the remaining \hi \ to condense into H$_2$. The stellar populations (see \S\ref{sec:stellar_pop}) do not support this scenario and tidal stripping is a more natural explanation as it preferentially removes the most loosely bound gas, which is typically \hi, not H$_2$. However, this would imply that the encounter(s) responsible for stripping the \hi \ gas did not trigger a major SF event in these galaxies, despite the presence of a considerable amount of molecular gas.

\subsection{Hot gas in the IGrM}

\citet{Belsole+2003} and \citet{OSullivan+2014b} measured the hot diffuse gas component of the IGrM of HCG 16 with the XMM Newton and Chandra satellites respectively. The fact that this hot diffuse medium is detected at all is already unusual for a spiral-rich HCG, but in addition it is also quite massive. \citet{Belsole+2003} estimate the total hot gas component of the IGrM as $4.5 \times 10^{10}$ \Msol \ and \citet{OSullivan+2014b} estimate it as 5.0-9.0 $\times 10^{10}$ \Msol \ (after adjusting to a distance of 55.2 Mpc). As thoroughly discussed in \citet{OSullivan+2014b} the origin of such a large amount of hot gas is difficult to fully explain. The group is not virialised, so it is unlikely that the gas has a cosmic origin and has simply fallen into the group halo and been heated. The group is also not globally deficient in \hi, meaning that stripped gas cannot be the main source either. \citet{OSullivan+2014b} conclude that the most probable origin of the majority of the hot gas is the galactic winds of HCG 16c and d.

The hot gas in the vicinity of the group core is of course co-spatial with a large amount of \hi \ in tidal structures, demonstrating that the IGrM is multi-phase. This hot gas is unlikely to act as a source of additional cool gas due to its long cooling timescale \citep[7-10 Gyr,][]{OSullivan+2014b}, but it could negatively impact the lifetime of the \hi, which we discuss in section \S\ref{sec:HI_lifetime}.

\subsection{Stellar populations, star formation rates, and outflows}
\label{sec:stellar_pop}

\citet{OSullivan+2014a} used the STARLIGHT code and spectra from SDSS to model stellar populations in HCG 16b and c, the only two of the galaxies with spectra in SDSS. They concluded that HCG 16b is entirely dominated by an old stellar population with a characteristic age of $\sim$10 Gyr. They also find some evidence of a very minor SF event occurring at some point in the past few hundred Myr. This event was likely triggered by the on-going interactions with HCG 16a, but it represents a negligible fraction of the total stellar population. For HCG 16c the results were heavily dependent on the choice of stellar population models, but the general finding was that a significant minority of the stellar mass of HCG 16c was formed in a starburst event during the past few hundred Myrs, with the rest of the population being made up of old stars (5-10 Gyr).

As mentioned in the previous section \citet{Lenkic+2016} estimated the SFRs of all the core galaxies. Although there is no stellar population estimate for HCG 16a its central region appears similar in colour to HCG 16b, suggesting it is made up of an evolved stellar population. However, it is surrounded by a ring of SF that is bright in GALEX UV bands and greatly elevates the estimated SFR. In the case of HCG 16d the estimated SFR is very high ($\sim17 \; \mathrm{M_{\odot}\,yr^{-1}}$) as is expected for a LIRG.

Using WiFeS (Wide Field Spectrograph on the ANU 2.3 m telescope) \citet{Rich+2010} studied the biconical outflow from HCG 16d. This outflow, driven by the ongoing nuclear starburst, contains ionised and neutral gas, traced by H$\alpha$ and Na D lines. They also find A-type stellar absorption features throughout the stellar disc, and suggest that starbursting galaxies like HCG 16d might be progenitors for E+A galaxies. This A-type stellar population in the stellar disc indicates that the galaxy has undergone a global star formation event less than a Gyr ago, in addition to the current nuclear starburst (although they may represent different phases of the same sustained event). Despite this evidence of significant recent star formation, HCG 16d still has sizeable reservoirs of both molecular and atomic gas (Table \ref{tab:gas_time}), however, it is not clear whether the \hi \ component of this gas is truly associated with the galaxy, or is a chance superposition.

One reason to believe that the \hi \ gas might not be associated with the galaxy is because of its peculiar velocity structure, which is very disorderly and any potential gradient appears to be almost perpendicular to the stellar disc (Figure \ref{fig:HCG16d_moms}). Here we draw a comparison with the \hi \ distribution around M82, one of the best studied starbursting galaxies. The \hi \ distribution around M82 looks somewhat similar to HCG 16d, with the velocity gradient in \hi \ aligned with the outflow rather than with the stellar disc \citep[e.g.][]{Martini+2018}. In the case of M82 the \hi \ distribution is interpreted as gas that is entrained in the hot wind, although the details of the exact mechanism are uncertain. This lends support to the interpretation that this \hi \ gas observed in the vicinity of HCG 16d really is associated with it and that its anomalous velocity structure is a result of the current wind. However, it is also possible, and even likely, that both proposed scenarios are somewhat true. There is a great deal of extended \hi \ in the IGrM so it is very plausible that we have mistakenly attributed some of this to HCG 16d.

HCG 16c is another LIRG in the group, and like HCG 16d has a high SFR. \citet{Vogt+2013} studied the M82-like wind that also exists in this galaxy, also with the WiFeS instrument. In striking similarity with HCG 16d the wind also appears to be a nuclear starburst driven phenomenon and the galactic disc shows signs of an A-type stellar population throughout. This indicates that the recent past has been very similar for these two galaxies, with each experiencing a global SF event within the last Gyr and both currently undergoing a nuclear starburst that is powering a galactic wind. The simplest explanation for this synchronised evolution is that it is driven by their tidal interaction with each other. However, there is another plausible explanation, that the passage of NGC 848 through the group triggered these events in both galaxies at approximately the same time (we consider this time scale in the following subsection). \citet{Vogt+2013} argue that although NGC 848 could have triggered the event responsible for the A-type population, the timescales are not compatible for the ongoing starbursts, however, it is possible that the events were initially global and have since been funnelled to the nuclear regions.

Despite the apparent similarity in their recent SFHs and the presence of winds, the \hi \ properties of HCG 16c and d are quite disparate. While HCG 16d shows no signs of rotation and a possible velocity gradient along the minor axis, HCG 16c has a mostly regular \hi \ velocity field (Figure \ref{fig:HCG16c_moms}) except in its outskirts. \citet{Vogt+2013} discussed the different nature of the galactic winds in the two galaxies. The wind emanating from HCG 16c is still (mostly) confined to two bubbles above and below the disc within the \hi \ surrounding envelope, indicating that it is young (only a few Myr old). On the other hand the HCG 16d wind is biconical and apparently free streaming. These authors also argue that the primary driving mechanisms of the winds differ, with one being shock-excited (HGC 16d) and the other photoionised (HCG 16c). Given that these two galaxies are in the same environment and appear to have had similar recent interactions, the differences in these winds are probably due to pre-existing differences in the host galaxies or simply the different phases we are currently observing them in. We refer the reader to \citet{Rich+2010}, \citet{Vogt+2013}, and references therein, for further discussion on this topic.

\subsection{Tail age estimates}

The SE tail, which links the core group to NGC 848 was most likely formed by NGC 848 passing very close to the core group, unbinding (or attracting already loosely bound) \hi \ gas and stretching it out to form the $\sim$160 kpc long tail. Across most of its length the SE tail is visible in just 4 velocity channels (3942--4006 \kms), which suggests the feature is approximately aligned with the plane of the sky and that we can consider its projected length as almost equal to its physical length. Given the large separation between NGC 848 and the core group it is also reasonable to assume that when NGC 848 passed through the group it was travelling at approximately the escape velocity. Summing the stellar masses on the four core galaxies gives the total stellar mass of the core group as $4.42 \times 10^{10}$ \Msol \ (Table \ref{tab:gas_time}). Combining this with the stellar mass--halo mass relation from \citet[][their equation 3 and Table 2]{Matthee+2017} calculated from the EAGLE (Evolution and Assembly of GaLaxies and their Environments) simulations, we estimate the dark matter halo mass of the core group as $2.8 \times 10^{12}$ \Msol. This corresponds to an escape velocity of $\sim$400 \kms \ at the present separation. Assuming this velocity NGC 848 would have passed by the group approximately 400 Myr ago. It should be noted that this value is quite uncertain as the simplistic argument above hides many complexities, however, it is still useful as an order of magnitude guide.

The only disturbance to NGC 848 that is visible in the optical image is that its spiral arms appear quite loose and are extended to the North-West and the South-East. However, over the majority of the extent of the \hi \ tail there is no detectable optical counterpart. \citet{Konstantopoulos+2010} estimate that optical tidal features in groups will be dispersed within about 500 Myr. Given our age estimate of the \hi \ feature, this may explain why no optical counterpart is seen, however, it is also possible that the tail is formed of \hi \ gas that was already loosely bound and did not host any significant stellar population. In this case there may never have been an optical counterpart. We discard the possibility of in situ SF within the SE tail because at no point along its length does the column density rise to $10^{21} \; \mathrm{cm^{-2}}$ (the peak value is $1.3 \times 10^{20} \; \mathrm{cm^{-2}}$).

Following the arguments in \S3.5 of \citet{Borthakur+2015} we can also make an estimate of how long the \hi \ content of the SE tail can persist into the future. The peak column density along the spine of the SE tail is about $10^{20} \; \mathrm{cm^{-2}}$ and it is about 1\arcmin \ (16 kpc) in width. If we assume that the tail is a cylinder of gas then the average density is about 0.016 $\mathrm{cm^{-3}}$. \citet{Borthakur+2015} find that \hi \ becomes susceptible to ionisation from background radiation below densities of about $10^{-3} \; \mathrm{cm^{-3}}$ at column densities of about $10^{19} \; \mathrm{cm^{-2}}$ (their Figure 6). If we assume that the \hi \ clouds making up the SE tail are expanding with a fiducial velocity of 20 \kms \ then it will reach this threshold column density after about 1 Gyr and the threshold density after about 1.7 Gyr. Thus we expect this feature to survive for at least another Gyr.

\citet{Konstantopoulos+2013} use the colours (from SDSS images) of the optical tail extending eastward of HCG 16a to estimate an age of between 100 Myr and 1 Gyr. While in projection this tail looks like it is associated with the NW tail (Figure \ref{fig:mom0}), as discussed previously this \hi \ feature does not connect kinematically to HCG 16a, instead if this optical tail has an \hi \ counterpart it is probably the much smaller \hi \ feature visible on the eastern side of HCG 16a at 4027 \kms \ (see the channel maps in the electronic version). Given the loose constraint on the age of this tail and the density of the core group it is difficult to say what interaction is responsible for this tail. It could simply be the on going interaction between HCG 16a and b, or an interaction with HCG 16c and d, or perhaps even due to the recent passage of NGC 848.

\subsection{The NW tail: accreting gas, tidal tail, or outflow?}
\label{sec:nw_tail}

The NW tail intersects HCG 16c at its centre on the plane of the sky and at its central velocity in the spectral direction (Figure \ref{fig:pv_plot}). As HCG 16c is currently undergoing a SB event it is worthwhile considering if this could be a sign of low angular momentum, cool gas accreting onto the centre of HCG 16c and fuelling its starburst. However, before asserting such an exceptional hypothesis we should first examine and attempt to eliminate other more mundane scenarios. As mentioned previously, two other competing hypotheses are that this feature could be the result of an outflow or the chance superposition of a tidal tail. 

First, consider the outflow hypothesis. We have already argued that \hi \ gas in HCG 16d is being disrupted by its galactic wind and that there is also a wind emanating from HCG 16c. However, the NW tail is a well-collimated feature, albeit with a pronounced curve, whereas the gas in the vicinity of HCG 16d is disordered and even the \hi \ velocity field of M82 \citep{Martini+2018} does not show collimated features like this. Without invoking a mechanism to collimate the ouflowing neutral gas over many 10s of kpc it would not be possible to form such a feature with a galactic wind, therefore, we discard the possibility of the NW tail being an outflow.

Next, we discuss the possibility of a chance superposition. Given that the feature intersects the centre of HCG 16c in both velocity and position a chance superposition seems, at first, unlikely. However, we know already that NGC 848 must have passed extremely close to HCG 16d (and therefore HCG 16c) in order to form the SE tail. Also the SE tail and the NW tail appear as though they may be part of one continuous feature which passes through the core group (Figure \ref{fig:pv_plot}, top panel). This feature may trace the path of NGC 848 when it passed through the core group, with the NW tail being the leading tail formed from the gas surrounding HCG 16c and d as NGC 848 approached the group, and the SE tail being the trailing tail which became very extended as NGC 848 exited the opposite side of the group. In this scenario a chance superposition of the centre of HCG 16c and the NW tail is not nearly as contrived as it might otherwise be.

In summary, while it remains a possibility that the NW tail might be accreting onto HCG 16c's centre, given the other information about the likely recent past of the group it seems that a chance superposition is the most likely explanation, that is, that the NW tail is a tidal tail with the same redshift as the centre of HCG 16c, but not the same line-of-sight distance (owing to peculiar motions).

\subsection{Survival of \hi \ in the IGrM}
\label{sec:HI_lifetime}

We observe a considerable amount of \hi \ gas in HCG 16 that is not associated with any individual galaxy. Here we discuss the stability of this gas considering the on-going processes within the group. As noted by \citet{Verdes-Montenegro+2001}, HCGs seem to have two final stage morphologies\footnote{Here we ignored the common envelope phase as at least some of the few known examples were misidentified \citep[][and Damas-Segovia et al. in prep]{Verdes-Montenegro+2002} and we are no longer convinced this is a genuine phase.}; those with almost no remaining \hi, and those with \hi \ found only in extended features, not associated with individual galaxies. Which of these will be that fate of HCG 16?

\citet{Borthakur+2015} discussed the distribution and fate of diffuse \hi \ gas in 4 compact groups. These authors formulated an analytic approximation for the minimum distance an \hi \ cloud of a given column density can be from a starburst event forming stars at a given rate (their equation 3). The lowest column density contours in Figure \ref{fig:mom0} are 2.45 and $9.80 \times 10^{19} \; \mathrm{cm}^{-2}$, and these enclose much of the area around the galaxies in the core group. Using the fiducial column density of $5 \times 10^{19} \; \mathrm{cm}^{-2}$ and the expressions from \citet{Borthakur+2015} we estimate that such \hi \ clouds should not be stable within $\sim$100 kpc of either HCG 16c or d, yet this would rule out most of the core region of the group, where we already know there is \hi \ gas.

\citet{Borthakur+2015} found a similar apparent contradiction in HCG 31, but reasoned that the diffuse \hi \ probably followed a similar distribution to the higher column density \hi \ and was thus likely shielded from ionising photons. 
In HCG 16 it may be that the apparently low column density features are really made up of dense clumps, smaller than the resolution of our images ($\sim$30 \arcsec or about 8 kpc), which have their emission smeared out by the beam. Also the energy output from the starbursts in HCG 16c and d will be highly non-isotropic, which may mean that a small faction of the core group is very hostile to \hi \ clouds, but that the majority is not.

Using deep Chandra observations \citet{OSullivan+2014b} estimated the temperature of the hot, diffuse IGrM in HCG 16 as 0.3 keV ($3.5 \times 10^{6}$ K) and its number density as around $1 \times 10^{-3} \; \mathrm{cm}^{-3}$. Following \citet{Borthakur+2010} we estimate that the critical radius of \hi \ clouds to prevent evaporation due to conductive heating is about 2 kpc. Given the spatial resolution of our VLA data we cannot verify this directly, other than to say that the persistence of \hi \ in the IGrM of HCG 16 implies that the \hi \ clouds are larger than this limit. We also note that lack of correlation between the \hi \ properties and hot IGrM properties found in other HCGs \citep{Rasmussen+2008} disfavours conductive heating as being a key \hi \ removal mechanism in HCGs. 

The final source of energy which could have the potential to remove the \hi \ gas from the IGrM is the starburst-driven galactic winds emanating from HCG 16c and d. \citet{Rich+2010} and \citet{Vogt+2013} describe these winds that are driving out neutral gas from the discs of these two galaxies. Despite the dramatic nature of these winds, \citet{Borthakur+2013} argue that the energy release rate of an M82-like wind is similar to the cooling rate of the surrounding medium, so while a starburst can maintain a galactic scale wind while it is active, it is unlikely to have a major long-term effect on the temperature of the IGrM.

To summarise, we are not aware of any source that is likely to destroy a considerable fraction of the \hi \ currently observed in the IGrM on short timescales ($<$1 Gyr). Therefore, we expect the dominant effect to be gradual evaporation due to background UV radiation.

\subsection{Correspondence of \hi \ and optical extended features}

\begin{figure}
    \centering
    \includegraphics[width=\columnwidth]{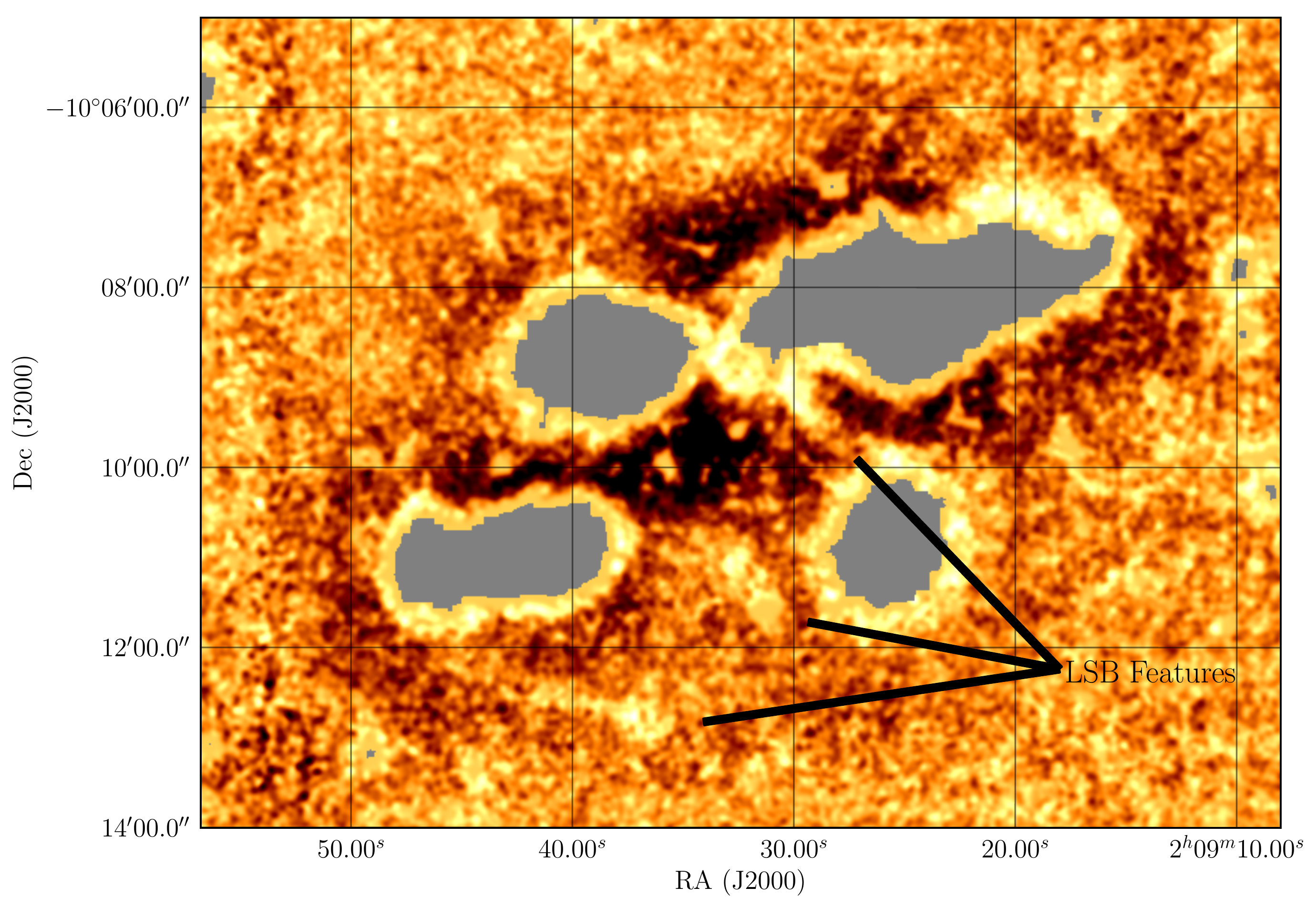}
    \includegraphics[width=\columnwidth]{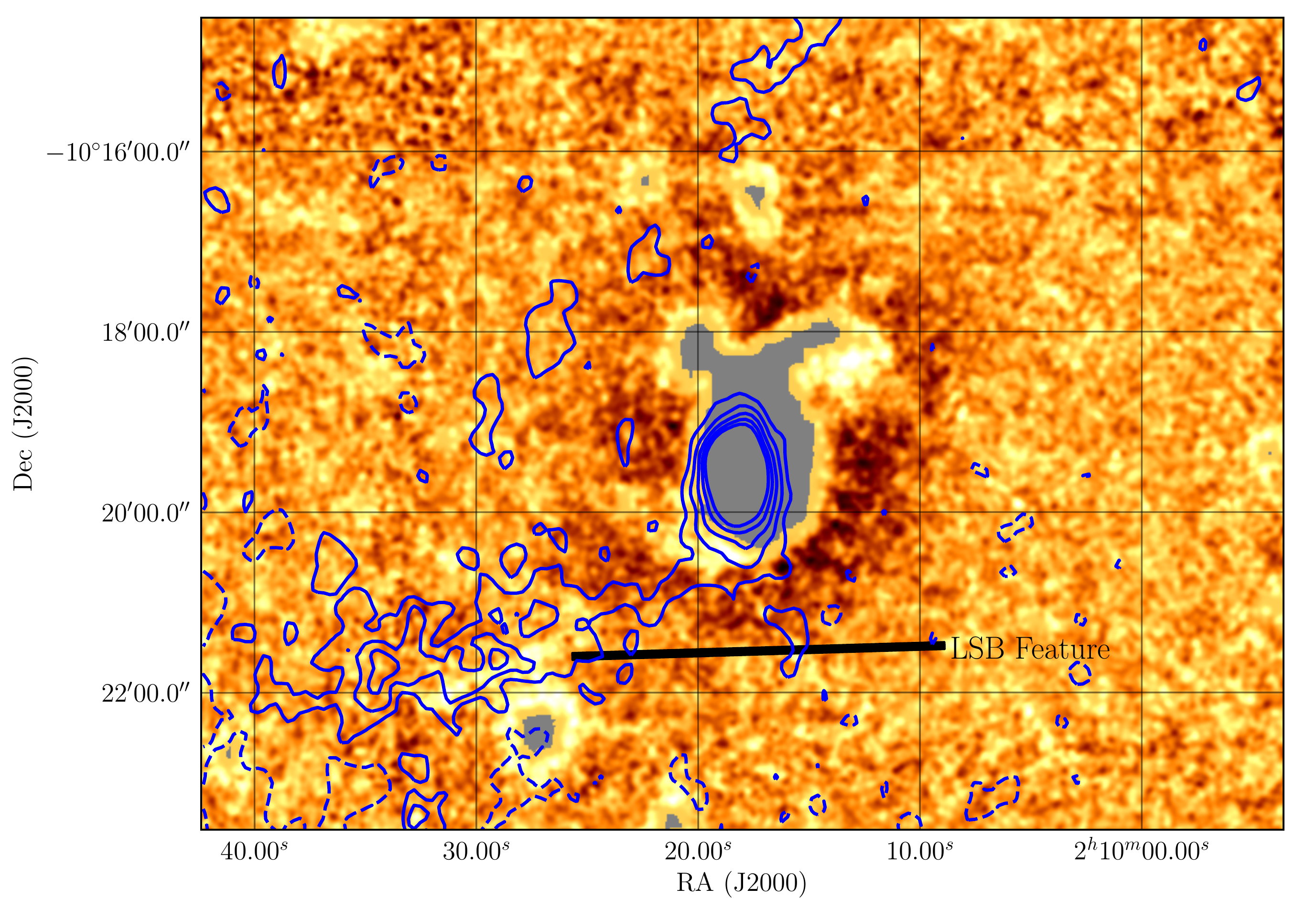}
    \caption{\textit{Top}: Enhanced DECaLS $gr$ image of the core group with LSB features highlighted. The black regions around the galaxies are due to over-subtraction of the sky that occurs in the DECaLS pipeline. The grey regions are the masked high surface brightness emission. \textit{Bottom}: As above but for the area surrounding NGC 848. This image also includes \hi \ contours from the combination of the 3 channels where the NGC 848S tail is most pronounced. Levels: -1, 1, 2, 3, 4, 5 $\times 0.05$ Jy \kms \ per beam, $\times 4.9 \times 10^{19} \; \mathrm{cm^{-2}}$, or $\times 0.39 \; \mathrm{M_{\odot}\,pc^{-2}}$}
    \label{fig:deep_opt}
\end{figure}

The DECaLS images were masked using the \texttt{Noisechisel} software \citep{Akhlaghi+2015}. Then we rebinned the images to a pixel size of 1\arcsec \ followed by smoothing with a kernel of 2 pixels. This enhances the diffuse emission of the data, allowing the lowest surface brightness features of the data, not detectable at the original pixel scale of $\approx0.27$\arcsec, to be identified. By comparing images in the $g$, $r$ and $z$ bands we identified 4 new LSB features (Figure \ref{fig:deep_opt}) which appear to be associated with HCG 16 (there are other LSB features which were attributed to background galaxies or clusters). The over-subtraction and fluctuation in the reference background level, especially close to extended galaxies, makes the surface brightness of the features unreliable, but we estimate the features to be between 27 and 28 mag arcsec$^{-2}$ in $r$-band.

In the core group we were unable to confidently identify any \hi \ structures which correspond to these LSB features. As stellar tails are generally thought to be shorter lived than \hi \ tails this likely indicates that these LSB features have a gas-poor origin. For example, it is possible that the most western feature originates from either HCG 16a or b, which are both \hi \ deficient. An alternative source for these features could also be disrupted dwarf members of the group that have been accreted by the larger members, leaving only faint stellar tails.

Near NGC 848 there is a very linear LSB feature, which was almost obscured by a star (Figure \ref{fig:deep_opt}, bottom). This feature is co-spatial with the \hi \ emission of the NGC 848S tail and is most likely its optical counterpart. It has a $g-r$ colour of $0.4 \pm 0.1$. The contours in the lower panel of Figure \ref{fig:deep_opt} are from the combination of 3 channels (4027-4069 \kms) where the NGC 848S tail is most prominent. The peak column density contour within the tail is $1.5 \times 10^{20} \; \mathrm{cm^{-2}}$, so in situ star formation is unlikely and we are probably detecting the old stellar population in this tail. As optical tidal tails are expected to be observed for only $\sim$500 Myr after formation, it is consistent with our age estimate of 400 Myr for the SE tail and for the interaction of NGC 848 with the rest of the group. If this tidal feature formed at approximately the same time then today it would be expected to be barely visible in the optical.

Even in these enhanced images there is still no indication of a stellar counterpart for the SE tail. The deepest of the three bands is $r$ with a 3$\sigma$ surface brightness sensitivity of 28.7 mag arcsec$^{-2}$ in a 10\arcsec$\times$10\arcsec \ box. This null detection supports the hypothesis that the SE tail was formed from loosely bound \hi \ gas that had no associated stellar counterpart.

\subsection{Morphological transformation}
\label{sec:morph_trans}

HCGs are known to host an excess of early-type galaxies, particularly S0s, and have a corresponding shortage of late-type galaxies relative to the field \citep{Hickson+1988}. This shortage of spirals and excess of S0s has led various authors to suggest that as spiral in CGs are stripped of gas they may evolve into S0s \citep[e.g.][]{Verdes-Montenegro+2001,Sulentic+2001}. In addition, CGs are found to have a pronounced gap in IR colours \citep{Johnson+2007}, referred to as the ``canyon'' or the IRTZ (Infra-Red Transition Zone), where galaxies appear to be in the late stages of shifting their morphology after passing the optical green valley \citep{Alatalo+2014} and beginning to lose their molecular gas \citep{Lisenfeld+2017}. Furthermore, in HCGs the IR canyon is dominated by lenticulars and early-types, again suggesting that spirals may be transitioning to later types in CGs.

HCG 16 contains two S0a galaxies, HCG 16c and d, which are both undergoing starbursts, as would be expected in the model proposed by \citet{Bekki+2011}, in which spirals transforms into S0s through repeated close interactions and the associated SF events. This raises the question, are HCG 16c and d currently undergoing morphological transformation from spirals to lenticulars?

In the DECaLS image (Figure \ref{fig:optim}) faint stellar shells are visible around both HCG 16c and d. Shells in massive galaxies have been widely regarded as the result of interactions with low mass companions on approximately radial orbits \citep[e.g.][]{Quinn1984,Dupraz+1986,Hernquist+1988,Hernquist+1989} and generally have been found around early-type, rather than late-type, galaxies \citep[e.g.][]{Malin+1983,Atkinson+2013}. However, more recently \citet{Pop+2018} performed a study of the occurrence of shells based on the Illustris cosmological simulations, finding that they can be caused by more equal mass interactions as well. In addition, they find that although for 1:10 mass ratio interactions (the minimum found to produce shells) the orbit must be almost purely radial, the more equal the mass ratio of the progenitors the wider the range of impact parameters that can produce shells. Thus, the simplest and most likely explanation for both HCG16c and d displaying shells is that they are the respective causes of each other's shells. However, this does not rule out the possibility that they both independently formed shells in mergers with now unseen companions, prior to their present interaction with each other.

The presence of shells in HCG 16c and d also hints that they are likely to have earlier morphologies in the near future, as most shells are found around early-type galaxies. However, the formation mechanism of S0s in groups and the field is still a topic of debate, with evidence that supports either a secular \citep{Guerou+2016,Rizzo+2018} or violent \citep{Laurikainen+2010,Querejeta+2015,Tapia+2017,Eliche-Moral+2018} evolutionary pathway from spirals to lenticulars, or perhaps both \citep{Fraser-McKelvie+2018}. It has even been suggested that spirals are not the progenitors at all and that lenticulars represent an old population distinct from both ellipitcal and spiral galaxies \citep{Gao+2018}.

It should also be noted that HCG 16c and d are not in the IRTZ/canyon because they are actively forming stars, which, almost by definition, IRTZ/canyon galaxies are not. However, they are already classified as S0a, indicating the lenticular-like appearance arose before any potential future transition. In fact the only galaxy in the group that has apparently traversed the canyon already is HCG 16b \citep{Zucker+2016}, which has clear spiral features. Of course, just because crossing the canyon was not accompanied by a morphological change for this galaxy does not mean it cannot be for others, but it does indicate that caution is needed when trying to interpret small samples.

Given this lack of consensus regarding the origin of S0s and what relation they may have to the IRTZ, it is not possible at present to say whether HCG 16c and d are likely undergoing a transformation from spiral to lenticular morphology, or whether they were lenticular already. However, this is a possibility that deserves further consideration both in HCG 16 and other CGs.

\section{Summary and outlook}
\label{sec:summary}

In this final section we first review our findings of the present state of the group as evidenced by the \hi \ data and the rich multi-wavelength data set described in the literature, then to conclude we propose a scenario which explains these findings and discuss the probable end state of the group.

\subsection{Summary}

Overall HCG 16 is not deficient in \hi, it has the expected quantity of \hi \ gas given the B-band luminosity of its members. However, while the total amount may be equivalent to that found in isolated galaxies, the gas is unevenly distributed throughout the group and its members. The north-western pair, HCG 16a and b, have both lost the vast majority of their expected \hi \ content, while the other members all have normal \hi \ masses. The remaining gas is spread out through the IGrM in a tangle of tidal tails, bridges and dense clumps, some of which may be TDGs. 

Despite being spread out across the entire group, the most plausible origin for the majority of the extended \hi \ gas is the HCG 16a and b pair. That pair is undergoing a strong interaction resulting in multiple optical and \hi \ tidal features between and around them. However, despite this interaction (and the presence of significant molecular gas reservoirs) they do not appear to have highly elevated SFRs (although HCG 16a does have a ring of SF in its outer disc) nor do their optical colours or stellar population models point to bursts of SF in the recent past. Given that these galaxies are both \hi \ deficient and have not recently converted large amounts of gas into stars, we conclude that the \hi \ gas must have been tidally stripped by interactions without triggering starburst events.

The pair at the centre of the group, HCG 16c and d, are also undergoing a strong interaction, but with quite the opposite outcome. Both galaxies are currently starbursting and have galactic winds powered by these events. They are also embedded in a tangled web of extended \hi \ features which form a high column density bridge between the two galaxies, tidal tails to the NE and NW, dense knots, and an enormous tail to the SE. The \hi \ kinematics of HCG 16c are mostly regular in its inner region, although the outer disc is somewhat disturbed. The NW tail connects in position and velocity to the centre of HCG 16c, presenting the possibility of cool gas with low angular momentum accreting directly to its centre and fuelling the starburst. However, given the interactions we traced through the group we favour an interpretation of this as a chance superposition in velocity space that does not correspond to the two objects being truly co-spatial in 3 dimensions. In HCG 16d the \hi \ kinematics appear to be completely disrupted. There is only a faint indication of a gradient in its velocity field and this is approximately perpendicular to the major axis of the disc. We interpret this as the starburst event disrupting the \hi \ and it becoming entrailed in the galactic wind. The rapid SFRs of these two galaxies mean that they will exhaust their gas reservoirs within about a Gyr, leaving them gas deficient without external replenishment.

The final large galaxy in the group, NGC 848, is separated from the other galaxies in optical emission by approximately 160 kpc, but this distance is traversed by the enormous SE tail, which forms an \hi \ connection between NGC 848 and the core group. Other than the unwinding edges of NGC 848's disc, this tidal tail has no apparent optical counterpart along its entire length, indicating that it is either too old for any accompanying stellar component to visibly persist to the present day, or that it was formed from loosely bound gas that did not have an associated stellar component to begin with. A simplistic estimate of the age of the tail, based on the assumption that NGC 848 is travelling at approximately the group's escape velocity, gives 400 Myr, on the upper end of how long a counterpart is expected to survive. Given the approximate nature of this estimate it is hardly conclusive either way, however, due to the abundance of neutral gas in the core group that is not associated with any galaxy we favour the latter interpretation.

Therefore, the dominant processes modifying the \hi \ content of HCG 16 are tidal stripping and star formation. Tidal interactions removed the majority of the \hi \ content of HCG 16a and b, and spread it out across the entire group, while SF in HCG 16c and d is rapidly consuming molecular gas (which will presumably be replenished from the available \hi \ reservoirs) and disrupting the \hi \ disc of HCG 16d. If this interpretation is correct then it would contradict the \citet{Konstantopoulos+2010} modified evolutionary sequence for HCGs because the HCG 16a and b pair would have been a gas-rich, strong interaction which did not result in gas being consumed by SF, whereas HCG 16c and d are another gas-rich pair which clearly has resulted in elevated SF, demonstrating not only that these two different results are possible, but that they are even possible in the same group. Thus the group cannot be classified as either a case where gas is mostly consumed by SF before major interactions occur, or as a case where tidal interactions remove the gas before it can be consumed by SF (the two distinct pathways in that scheme).

After reviewing the potential mechanisms for ionising the \hi \ currently in the IGrM we find that the hot component of the IGrM is not energetic enough to evaporate large ($>$2 kpc) \hi \ structures, that the on-going SF in the group does not appear to be strongly effecting the existing \hi \ in the IGrM, and that while the galactic wind of HCG 16d may currently be ejecting gas and energy into the IGrM, the effects of this are unlikely to persist once the starburst event has ended. Thus, evaporation by the UV background will likely be the principal mechanism for removing \hi \ from the IGrM on a long timescale ($>$1 Gyr).

\subsection{Global picture and future outlook}

With all of the above results in mind we have attempted to construct a coherent picture of the past evolution of the group that fits with all the available evidence. 

Strong tidal interactions involving HCG 16a and b likely unbound much of their \hi \ gas without triggering a major SF event. This unbound gas was then dragged through the centre of the group by the passage of NGC 848 about 0.5 Gyr ago. This close passage also started SF episodes in HCG 16c and d, generating the E+A-like spectra their discs have today. This passage of NGC 848 is traced by the SE tail and the NW tail, which together form a continuous structure spanning the entire group, and the latter of which may have formed a TDG at its tip. At present the SFRs of HCG 16c and d are highly elevated, now driven by their interaction with each other, while the SFRs in HCG 16a and b remain more modest, aside from a ring of SF activity in HCG 16a.

Over the next Gyr HCG 16c and d will likely convert, consume or expel much of their gas supply through SF leaving themselves \hi \ deficient like HCG 16a and b, though by different means. Meanwhile, the \hi \ in all the galaxies will continue to be stripped by tidal interactions. It is unclear whether NGC 848 is travelling fast enough to escape the group, or whether it will fall back to be the sole \hi-normal large galaxy in the group. The extended \hi \ features in the group are expected to persist for several Gyr as they are gradually evaporated by the UV background. This will result in HCG 16 resembling a phase 3a group where there is little or no \hi \ remaining in the galaxies (with the possible exception of NGC 848), but extended \hi \ features are still visible in the IGrM.

\begin{acknowledgements}
MGJ is supported by a Juan de la Cierva formaci\'{o}n fellowship. We also acknowledge support from the grants AYA2015-65973-C3-1-R and RTI2018-096228-B-C31 (MINECO/FEDER, UE). This work has been supported by the Spanish Science Ministry ``Centro de Excelencia Severo Ochoa” program under grant SEV-2017-0709. 
MGJ wishes to thank B. Koribalski, K. Lee-Waddell, and S. Cazzoli for helpful discussions. We also thank the referee for his thorough comments which helped to improve this paper.
This project used archival data from the VLA. The National Radio Astronomy Observatory is a facility of the National Science Foundation operated under cooperative agreement by Associated Universities, Inc.
This project used data obtained with the Dark Energy Camera (DECam), which was constructed by the Dark Energy Survey (DES) collaboration (full acknowledgement at \url{legacysurvey.org/acknowledgment}).
This research has made use of the NASA/IPAC Extragalactic Database (NED), which is operated by the Jet Propulsion Laboratory, California Institute of Technology, under contract with the National Aeronautics and Space Administration. We also acknowledge the use of the HyperLeda database \citep{HyperLeda}. This research made use of \texttt{APLpy}, an open-source plotting package for \texttt{Python} \citep{aplpy2012,aplpy2019}, \texttt{astropy} \citep{astropy1,astropy2}, \texttt{Aladin} \citep{aladin}, \texttt{mayavi} \citep{mayavi}, and \texttt{SAOImageDS9} \citep{ds9}.
\end{acknowledgements}

\bibliographystyle{aa}
\bibliography{refs}

\begin{thebibliography}{105}
\expandafter\ifx\csname natexlab\endcsname\relax\def\natexlab#1{#1}\fi

\bibitem[{{Akhlaghi} \& {Ichikawa}(2015)}]{Akhlaghi+2015}
{Akhlaghi}, M. \& {Ichikawa}, T. 2015, \apjs, 220, 1

\bibitem[{{Alatalo} {et~al.}(2015){Alatalo}, {Appleton}, {Lisenfeld},
  {Bitsakis}, {Lanz}, {Lacy}, {Charmandaris}, {Cluver}, {Dopita}, {Guillard},
  {Jarrett}, {Kewley}, {Nyland}, {Ogle}, {Rasmussen}, {Rich},
  {Verdes-Montenegro}, {Xu}, \& {Yun}}]{Alatalo+2015}
{Alatalo}, K., {Appleton}, P.~N., {Lisenfeld}, U., {et~al.} 2015, \apj, 812,
  117

\bibitem[{{Alatalo} {et~al.}(2014){Alatalo}, {Cales}, {Appleton}, {Kewley},
  {Lacy}, {Lisenfeld}, {Nyland}, \& {Rich}}]{Alatalo+2014}
{Alatalo}, K., {Cales}, S.~L., {Appleton}, P.~N., {et~al.} 2014, \apjl, 794,
  L13

\bibitem[{{Arp}(1966)}]{Arp1966}
{Arp}, H. 1966, The Astrophysical Journal Supplement Series, 14, 1

\bibitem[{{Astropy Collaboration} {et~al.}(2018){Astropy Collaboration},
  {Price-Whelan}, {Sip{\H o}cz}, {G{\"u}nther}, {Lim}, {Crawford}, {Conseil},
  {Shupe}, {Craig}, {Dencheva}, {Ginsburg}, {VanderPlas}, {Bradley},
  {P{\'e}rez-Su{\'a}rez}, {de Val-Borro}, {Aldcroft}, {Cruz}, {Robitaille},
  {Tollerud}, {Ardelean}, {Babej}, {Bach}, {Bachetti}, {Bakanov}, {Bamford},
  {Barentsen}, {Barmby}, {Baumbach}, {Berry}, {Biscani}, {Boquien}, {Bostroem},
  {Bouma}, {Brammer}, {Bray}, {Breytenbach}, {Buddelmeijer}, {Burke},
  {Calderone}, {Cano Rodr{\'{\i}}guez}, {Cara}, {Cardoso}, {Cheedella},
  {Copin}, {Corrales}, {Crichton}, {D'Avella}, {Deil}, {Depagne}, {Dietrich},
  {Donath}, {Droettboom}, {Earl}, {Erben}, {Fabbro}, {Ferreira}, {Finethy},
  {Fox}, {Garrison}, {Gibbons}, {Goldstein}, {Gommers}, {Greco}, {Greenfield},
  {Groener}, {Grollier}, {Hagen}, {Hirst}, {Homeier}, {Horton}, {Hosseinzadeh},
  {Hu}, {Hunkeler}, {Ivezi{\'c}}, {Jain}, {Jenness}, {Kanarek}, {Kendrew},
  {Kern}, {Kerzendorf}, {Khvalko}, {King}, {Kirkby}, {Kulkarni}, {Kumar},
  {Lee}, {Lenz}, {Littlefair}, {Ma}, {Macleod}, {Mastropietro}, {McCully},
  {Montagnac}, {Morris}, {Mueller}, {Mumford}, {Muna}, {Murphy}, {Nelson},
  {Nguyen}, {Ninan}, {N{\"o}the}, {Ogaz}, {Oh}, {Parejko}, {Parley}, {Pascual},
  {Patil}, {Patil}, {Plunkett}, {Prochaska}, {Rastogi}, {Reddy Janga},
  {Sabater}, {Sakurikar}, {Seifert}, {Sherbert}, {Sherwood-Taylor}, {Shih},
  {Sick}, {Silbiger}, {Singanamalla}, {Singer}, {Sladen}, {Sooley},
  {Sornarajah}, {Streicher}, {Teuben}, {Thomas}, {Tremblay}, {Turner},
  {Terr{\'o}n}, {van Kerkwijk}, {de la Vega}, {Watkins}, {Weaver}, {Whitmore},
  {Woillez}, {Zabalza}, \& {Astropy Contributors}}]{astropy2}
{Astropy Collaboration}, {Price-Whelan}, A.~M., {Sip{\H o}cz}, B.~M., {et~al.}
  2018, \aj, 156, 123

\bibitem[{{Astropy Collaboration} {et~al.}(2013){Astropy Collaboration},
  {Robitaille}, {Tollerud}, {Greenfield}, {Droettboom}, {Bray}, {Aldcroft},
  {Davis}, {Ginsburg}, {Price-Whelan}, {Kerzendorf}, {Conley}, {Crighton},
  {Barbary}, {Muna}, {Ferguson}, {Grollier}, {Parikh}, {Nair}, {Unther},
  {Deil}, {Woillez}, {Conseil}, {Kramer}, {Turner}, {Singer}, {Fox}, {Weaver},
  {Zabalza}, {Edwards}, {Azalee Bostroem}, {Burke}, {Casey}, {Crawford},
  {Dencheva}, {Ely}, {Jenness}, {Labrie}, {Lim}, {Pierfederici}, {Pontzen},
  {Ptak}, {Refsdal}, {Servillat}, \& {Streicher}}]{astropy1}
{Astropy Collaboration}, {Robitaille}, T.~P., {Tollerud}, E.~J., {et~al.} 2013,
  \aap, 558, A33

\bibitem[{{Atkinson} {et~al.}(2013){Atkinson}, {Abraham}, \&
  {Ferguson}}]{Atkinson+2013}
{Atkinson}, A.~M., {Abraham}, R.~G., \& {Ferguson}, A.~M.~N. 2013, \apj, 765,
  28

\bibitem[{{Bekki} \& {Couch}(2011)}]{Bekki+2011}
{Bekki}, K. \& {Couch}, W.~J. 2011, \mnras, 415, 1783

\bibitem[{{Belsole} {et~al.}(2003){Belsole}, {Sauvageot}, {Ponman}, \&
  {Bourdin}}]{Belsole+2003}
{Belsole}, E., {Sauvageot}, J.-L., {Ponman}, T.~J., \& {Bourdin}, H. 2003,
  \aap, 398, 1

\bibitem[{{Bitsakis} {et~al.}(2014){Bitsakis}, {Charmandaris}, {Appleton},
  {D{\'{\i}}az-Santos}, {Le Floc'h}, {da Cunha}, {Alatalo}, \&
  {Cluver}}]{Bitsakis+2014}
{Bitsakis}, T., {Charmandaris}, V., {Appleton}, P.~N., {et~al.} 2014, \aap,
  565, A25

\bibitem[{{Bitsakis} {et~al.}(2011){Bitsakis}, {Charmandaris}, {da Cunha},
  {D{\'{\i}}az-Santos}, {Le Floc'h}, \& {Magdis}}]{Bitsakis+2011}
{Bitsakis}, T., {Charmandaris}, V., {da Cunha}, E., {et~al.} 2011, \aap, 533,
  A142

\bibitem[{{Bonnarel} {et~al.}(2000){Bonnarel}, {Fernique}, {Bienaym{\'e}},
  {Egret}, {Genova}, {Louys}, {Ochsenbein}, {Wenger}, \& {Bartlett}}]{aladin}
{Bonnarel}, F., {Fernique}, P., {Bienaym{\'e}}, O., {et~al.} 2000, \aaps, 143,
  33

\bibitem[{{Borthakur} {et~al.}(2013){Borthakur}, {Heckman}, {Strickland},
  {Wild}, \& {Schiminovich}}]{Borthakur+2013}
{Borthakur}, S., {Heckman}, T., {Strickland}, D., {Wild}, V., \&
  {Schiminovich}, D. 2013, \apj, 768, 18

\bibitem[{{Borthakur} {et~al.}(2010){Borthakur}, {Yun}, \&
  {Verdes-Montenegro}}]{Borthakur+2010}
{Borthakur}, S., {Yun}, M.~S., \& {Verdes-Montenegro}, L. 2010, \apj, 710, 385

\bibitem[{{Borthakur} {et~al.}(2015){Borthakur}, {Yun}, {Verdes-Montenegro},
  {Heckman}, {Zhu}, \& {Braatz}}]{Borthakur+2015}
{Borthakur}, S., {Yun}, M.~S., {Verdes-Montenegro}, L., {et~al.} 2015, \apj,
  812, 78

\bibitem[{{Boselli} {et~al.}(1996){Boselli}, {Mendes de Oliveira}, {Balkowski},
  {Cayatte}, \& {Casoli}}]{Boselli+1996}
{Boselli}, A., {Mendes de Oliveira}, C., {Balkowski}, C., {Cayatte}, V., \&
  {Casoli}, F. 1996, \aap, 314, 738

\bibitem[{{Bosma}(1978)}]{Bosma1978}
{Bosma}, A. 1978, PhD thesis, PhD Thesis, Groningen Univ., (1978)

\bibitem[{{Bosma}(1981)}]{Bosma1981}
{Bosma}, A. 1981, \aj, 86, 1825

\bibitem[{{Bournaud} \& {Duc}(2006)}]{Bournaud+2006}
{Bournaud}, F. \& {Duc}, P.-A. 2006, \aap, 456, 481

\bibitem[{{Bournaud} {et~al.}(2004){Bournaud}, {Duc}, {Amram}, {Combes}, \&
  {Gach}}]{Bournaud+2004}
{Bournaud}, F., {Duc}, P.-A., {Amram}, P., {Combes}, F., \& {Gach}, J.-L. 2004,
  \aap, 425, 813

\bibitem[{{Cluver} {et~al.}(2013){Cluver}, {Appleton}, {Ogle}, {Jarrett},
  {Rasmussen}, {Lisenfeld}, {Guillard}, {Verdes-Montenegro}, {Antonucci},
  {Bitsakis}, {Charmandaris}, {Boulanger}, {Egami}, {Xu}, \&
  {Yun}}]{Cluver+2013}
{Cluver}, M.~E., {Appleton}, P.~N., {Ogle}, P., {et~al.} 2013, \apj, 765, 93

\bibitem[{{de Carvalho} \& {Coziol}(1999)}]{deCarvalho+1999}
{de Carvalho}, R.~R. \& {Coziol}, R. 1999, \aj, 117, 1657

\bibitem[{{de Carvalho} {et~al.}(1997){de Carvalho}, {Ribeiro}, {Capelato}, \&
  {Zepf}}]{deCarvalho+1997}
{de Carvalho}, R.~R., {Ribeiro}, A. L.~B., {Capelato}, H.~V., \& {Zepf}, S.~E.
  1997, The Astrophysical Journal Supplement Series, 110, 1

\bibitem[{{de Vaucouleurs} {et~al.}(1991){de Vaucouleurs}, {de Vaucouleurs},
  {Corwin}, {Buta}, {Paturel}, \& {Fouqu{\'e}}}]{RC3}
{de Vaucouleurs}, G., {de Vaucouleurs}, A., {Corwin}, Jr., H.~G., {et~al.}
  1991, {Third Reference Catalogue of Bright Galaxies. Volume I: Explanations
  and references. Volume II: Data for galaxies between 0$^{h}$ and 12$^{h}$.
  Volume III: Data for galaxies between 12$^{h}$ and 24$^{h}$.}

\bibitem[{{Desjardins} {et~al.}(2013){Desjardins}, {Gallagher}, {Tzanavaris},
  {Mulchaey}, {Brandt}, {Charlton}, {Garmire}, {Gronwall}, {Hornschemeier},
  {Johnson}, {Konstantopoulos}, \& {Zabludoff}}]{Desjardins+2013}
{Desjardins}, T.~D., {Gallagher}, S.~C., {Tzanavaris}, P., {et~al.} 2013, \apj,
  763, 121

\bibitem[{{D{\'\i}az-Gim{\'e}nez} {et~al.}(2012){D{\'\i}az-Gim{\'e}nez},
  {Mamon}, {Pacheco}, {Mendes de Oliveira}, \& {Alonso}}]{Diaz-Gimenez+2012}
{D{\'\i}az-Gim{\'e}nez}, E., {Mamon}, G.~A., {Pacheco}, M., {Mendes de
  Oliveira}, C., \& {Alonso}, M.~V. 2012, \mnras, 426, 296

\bibitem[{{Dupraz} \& {Combes}(1986)}]{Dupraz+1986}
{Dupraz}, C. \& {Combes}, F. 1986, \aap, 166, 53

\bibitem[{{Eigenthaler} {et~al.}(2015){Eigenthaler}, {Ploeckinger}, {Verdugo},
  \& {Ziegler}}]{Eigenthaler+2015}
{Eigenthaler}, P., {Ploeckinger}, S., {Verdugo}, M., \& {Ziegler}, B. 2015,
  \mnras, 451, 2793

\bibitem[{{Eliche-Moral} {et~al.}(2018){Eliche-Moral},
  {Rodr{\'{\i}}guez-P{\'e}rez}, {Borlaff}, {Querejeta}, \&
  {Tapia}}]{Eliche-Moral+2018}
{Eliche-Moral}, M.~C., {Rodr{\'{\i}}guez-P{\'e}rez}, C., {Borlaff}, A.,
  {Querejeta}, M., \& {Tapia}, T. 2018, \aap, 617, A113

\bibitem[{{Eskew} {et~al.}(2012){Eskew}, {Zaritsky}, \& {Meidt}}]{Eskew+2012}
{Eskew}, M., {Zaritsky}, D., \& {Meidt}, S. 2012, \aj, 143, 139

\bibitem[{{Fern{\'a}ndez Lorenzo} {et~al.}(2012){Fern{\'a}ndez Lorenzo},
  {Sulentic}, {Verdes-Montenegro}, {Ruiz}, {Sabater}, \&
  {S{\'a}nchez}}]{Fernandez-Lorenzo+2012}
{Fern{\'a}ndez Lorenzo}, M., {Sulentic}, J., {Verdes-Montenegro}, L., {et~al.}
  2012, \aap, 540, A47

\bibitem[{{Fraser-McKelvie} {et~al.}(2018){Fraser-McKelvie},
  {Arag{\'o}n-Salamanca}, {Merrifield}, {Tabor}, {Bernardi}, {Drory}, {Parikh},
  \& {Argudo-Fern{\'a}ndez}}]{Fraser-McKelvie+2018}
{Fraser-McKelvie}, A., {Arag{\'o}n-Salamanca}, A., {Merrifield}, M., {et~al.}
  2018, \mnras, 481, 5580

\bibitem[{{Gao} {et~al.}(2018){Gao}, {Ho}, {Barth}, \& {Li}}]{Gao+2018}
{Gao}, H., {Ho}, L.~C., {Barth}, A.~J., \& {Li}, Z.-Y. 2018, \apj, 862, 100

\bibitem[{{Gu{\'e}rou} {et~al.}(2016){Gu{\'e}rou}, {Emsellem}, {Krajnovi{\'c}},
  {McDermid}, {Contini}, \& {Weilbacher}}]{Guerou+2016}
{Gu{\'e}rou}, A., {Emsellem}, E., {Krajnovi{\'c}}, D., {et~al.} 2016, \aap,
  591, A143

\bibitem[{{Haynes} \& {Giovanelli}(1984)}]{Haynes+Giovanelli1984}
{Haynes}, M.~P. \& {Giovanelli}, R. 1984, \aj, 89, 758

\bibitem[{{Hernquist} \& {Quinn}(1988)}]{Hernquist+1988}
{Hernquist}, L. \& {Quinn}, P.~J. 1988, \apj, 331, 682

\bibitem[{{Hernquist} \& {Quinn}(1989)}]{Hernquist+1989}
{Hernquist}, L. \& {Quinn}, P.~J. 1989, \apj, 342, 1

\bibitem[{{Hickson}(1982)}]{Hickson1982}
{Hickson}, P. 1982, \apj, 255, 382

\bibitem[{{Hickson} {et~al.}(1988){Hickson}, {Kindl}, \&
  {Huchra}}]{Hickson+1988}
{Hickson}, P., {Kindl}, E., \& {Huchra}, J.~P. 1988, \apj, 331, 64

\bibitem[{{Hickson} {et~al.}(1992){Hickson}, {Mendes de Oliveira}, {Huchra}, \&
  {Palumbo}}]{Hickson+1992}
{Hickson}, P., {Mendes de Oliveira}, C., {Huchra}, J.~P., \& {Palumbo}, G.~G.
  1992, \apj, 399, 353

\bibitem[{{Huchtmeier}(1997)}]{Huchtmeier1997}
{Huchtmeier}, W.~K. 1997, \aap, 325, 473

\bibitem[{{Johnson} {et~al.}(2007){Johnson}, {Hibbard}, {Gallagher},
  {Charlton}, {Hornschemeier}, {Jarrett}, \& {Reines}}]{Johnson+2007}
{Johnson}, K.~E., {Hibbard}, J.~E., {Gallagher}, S.~C., {et~al.} 2007, \aj,
  134, 1522

\bibitem[{{Jones} {et~al.}(2018){Jones}, {Espada}, {Verdes-Montenegro},
  {Huchtmeier}, {Lisenfeld}, {Leon}, {Sulentic}, {Sabater}, {Jones}, {Sanchez},
  \& {Garrido}}]{Jones+2018}
{Jones}, M.~G., {Espada}, D., {Verdes-Montenegro}, L., {et~al.} 2018, \aap,
  609, A17

\bibitem[{{Joye} \& {Mandel}(2003)}]{ds9}
{Joye}, W.~A. \& {Mandel}, E. 2003, in Astronomical Society of the Pacific
  Conference Series, Vol. 295, Astronomical Data Analysis Software and Systems
  XII, ed. H.~E. {Payne}, R.~I. {Jedrzejewski}, \& R.~N. {Hook}, 489

\bibitem[{{Konstantopoulos} {et~al.}(2010){Konstantopoulos}, {Gallagher},
  {Fedotov}, {Durrell}, {Heiderman}, {Elmegreen}, {Charlton}, {Hibbard},
  {Tzanavaris}, {Chandar}, {Johnson}, {Maybhate}, {Zabludoff}, {Gronwall},
  {Szathmary}, {Hornschemeier}, {English}, {Whitmore}, {Mendes de Oliveira}, \&
  {Mulchaey}}]{Konstantopoulos+2010}
{Konstantopoulos}, I.~S., {Gallagher}, S.~C., {Fedotov}, K., {et~al.} 2010,
  \apj, 723, 197

\bibitem[{{Konstantopoulos} {et~al.}(2013){Konstantopoulos}, {Maybhate},
  {Charlton}, {Fedotov}, {Durrell}, {Mulchaey}, {English}, {Desjardins},
  {Gallagher}, {Walker}, {Johnson}, {Tzanavaris}, \&
  {Gronwall}}]{Konstantopoulos+2013}
{Konstantopoulos}, I.~S., {Maybhate}, A., {Charlton}, J.~C., {et~al.} 2013,
  \apj, 770, 114

\bibitem[{{Laurikainen} {et~al.}(2010){Laurikainen}, {Salo}, {Buta}, {Knapen},
  \& {Comer{\'o}n}}]{Laurikainen+2010}
{Laurikainen}, E., {Salo}, H., {Buta}, R., {Knapen}, J.~H., \& {Comer{\'o}n},
  S. 2010, \mnras, 405, 1089

\bibitem[{{Lee-Waddell} {et~al.}(2016){Lee-Waddell}, {Spekkens}, {Chandra},
  {Patra}, {Cuillandre}, {Wang}, {Haynes}, {Cannon}, {Stierwalt}, {Sick}, \&
  {Giovanelli}}]{Lee-Waddell+2016}
{Lee-Waddell}, K., {Spekkens}, K., {Chandra}, P., {et~al.} 2016, \mnras, 460,
  2945

\bibitem[{{Lee-Waddell} {et~al.}(2012){Lee-Waddell}, {Spekkens}, {Haynes},
  {Stierwalt}, {Chengalur}, {Chandra}, \& {Giovanelli}}]{Lee-Waddell+2012}
{Lee-Waddell}, K., {Spekkens}, K., {Haynes}, M.~P., {et~al.} 2012, \mnras, 427,
  2314

\bibitem[{{Lenki{\'c}} {et~al.}(2016){Lenki{\'c}}, {Tzanavaris}, {Gallagher},
  {Desjardins}, {Walker}, {Johnson}, {Fedotov}, {Charlton}, {Hornschemeier},
  {Durrell}, \& {Gronwall}}]{Lenkic+2016}
{Lenki{\'c}}, L., {Tzanavaris}, P., {Gallagher}, S.~C., {et~al.} 2016, \mnras,
  459, 2948

\bibitem[{{Leon} {et~al.}(1998){Leon}, {Combes}, \& {Menon}}]{Leon+1998}
{Leon}, S., {Combes}, F., \& {Menon}, T.~K. 1998, \aap, 330, 37

\bibitem[{{Lisenfeld} {et~al.}(2017){Lisenfeld}, {Alatalo}, {Zucker},
  {Appleton}, {Gallagher}, {Guillard}, \& {Johnson}}]{Lisenfeld+2017}
{Lisenfeld}, U., {Alatalo}, K., {Zucker}, C., {et~al.} 2017, \aap, 607, A110

\bibitem[{{Makarov} {et~al.}(2014){Makarov}, {Prugniel}, {Terekhova},
  {Courtois}, \& {Vauglin}}]{HyperLeda}
{Makarov}, D., {Prugniel}, P., {Terekhova}, N., {Courtois}, H., \& {Vauglin},
  I. 2014, \aap, 570, A13

\bibitem[{{Malin} \& {Carter}(1983)}]{Malin+1983}
{Malin}, D.~F. \& {Carter}, D. 1983, \apj, 274, 534

\bibitem[{{Martin} {et~al.}(2005){Martin}, {Fanson}, {Schiminovich},
  {Morrissey}, {Friedman}, {Barlow}, {Conrow}, {Grange}, {Jelinsky},
  {Milliard}, {Siegmund}, {Bianchi}, {Byun}, {Donas}, {Forster}, {Heckman},
  {Lee}, {Madore}, {Malina}, {Neff}, {Rich}, {Small}, {Surber}, {Szalay},
  {Welsh}, \& {Wyder}}]{Martin+2005}
{Martin}, D.~C., {Fanson}, J., {Schiminovich}, D., {et~al.} 2005, \apjl, 619,
  L1

\bibitem[{{Mart{\'\i}nez} {et~al.}(2010){Mart{\'\i}nez}, {Del Olmo}, {Coziol},
  \& {Perea}}]{Martinez+2010}
{Mart{\'\i}nez}, M.~A., {Del Olmo}, A., {Coziol}, R., \& {Perea}, J. 2010, \aj,
  139, 1199

\bibitem[{{Martinez-Badenes} {et~al.}(2012){Martinez-Badenes}, {Lisenfeld},
  {Espada}, {Verdes-Montenegro}, {Garc{\'{\i}}a-Burillo}, {Leon}, {Sulentic},
  \& {Yun}}]{Martinez-Badenes+2012}
{Martinez-Badenes}, V., {Lisenfeld}, U., {Espada}, D., {et~al.} 2012, \aap,
  540, A96

\bibitem[{{Martini} {et~al.}(2018){Martini}, {Leroy}, {Mangum}, {Bolatto},
  {Keating}, {Sandstrom}, \& {Walter}}]{Martini+2018}
{Martini}, P., {Leroy}, A.~K., {Mangum}, J.~G., {et~al.} 2018, \apj, 856, 61

\bibitem[{{Matthee} {et~al.}(2017){Matthee}, {Schaye}, {Crain}, {Schaller},
  {Bower}, \& {Theuns}}]{Matthee+2017}
{Matthee}, J., {Schaye}, J., {Crain}, R.~A., {et~al.} 2017, \mnras, 465, 2381

\bibitem[{{McMullin} {et~al.}(2007){McMullin}, {Waters}, {Schiebel}, {Young},
  \& {Golap}}]{CASA}
{McMullin}, J.~P., {Waters}, B., {Schiebel}, D., {Young}, W., \& {Golap}, K.
  2007, in Astronomical Society of the Pacific Conference Series, Vol. 376,
  Astronomical Data Analysis Software and Systems XVI, ed. R.~A. {Shaw},
  F.~{Hill}, \& D.~J. {Bell}, 127

\bibitem[{{Mendes de Oliveira} {et~al.}(1998){Mendes de Oliveira}, {Plana},
  {Amram}, {Bolte}, \& {Boulesteix}}]{Mendes+1998}
{Mendes de Oliveira}, C., {Plana}, H., {Amram}, P., {Bolte}, M., \&
  {Boulesteix}, J. 1998, \apj, 507, 691

\bibitem[{{Morganti} \& {Oosterloo}(2018)}]{Morganti+2018}
{Morganti}, R. \& {Oosterloo}, T. 2018, arXiv e-prints, arXiv:1807.01475

\bibitem[{{Mould} {et~al.}(2000){Mould}, {Huchra}, {Freedman}, {Kennicutt},
  {Ferrarese}, {Ford}, {Gibson}, {Graham}, {Hughes}, {Illingworth}, {Kelson},
  {Macri}, {Madore}, {Sakai}, {Sebo}, {Silbermann}, \& {Stetson}}]{Mould+2000}
{Mould}, J.~R., {Huchra}, J.~P., {Freedman}, W.~L., {et~al.} 2000, \apj, 529,
  786

\bibitem[{{Oda} {et~al.}(2018){Oda}, {Ueda}, {Tanimoto}, \& {Ricci}}]{Oda+2018}
{Oda}, S., {Ueda}, Y., {Tanimoto}, A., \& {Ricci}, C. 2018, \apj, 855, 79

\bibitem[{{O'Sullivan} {et~al.}(2014{\natexlab{a}}){O'Sullivan}, {Vrtilek},
  {David}, {Giacintucci}, {Zezas}, {Ponman}, {Mamon}, {Nulsen}, \&
  {Raychaudhury}}]{OSullivan+2014b}
{O'Sullivan}, E., {Vrtilek}, J.~M., {David}, L.~P., {et~al.}
  2014{\natexlab{a}}, \apj, 793, 74

\bibitem[{{O'Sullivan} {et~al.}(2014{\natexlab{b}}){O'Sullivan}, {Zezas},
  {Vrtilek}, {Giacintucci}, {Trevisan}, {David}, {Ponman}, {Mamon}, \&
  {Raychaudhury}}]{OSullivan+2014a}
{O'Sullivan}, E., {Zezas}, A., {Vrtilek}, J.~M., {et~al.} 2014{\natexlab{b}},
  \apj, 793, 73

\bibitem[{{Paturel} {et~al.}(2000){Paturel}, {Fang}, {Petit}, {Garnier}, \&
  {Rousseau}}]{Paturel+2000}
{Paturel}, G., {Fang}, Y., {Petit}, C., {Garnier}, R., \& {Rousseau}, J. 2000,
  \aaps, 146, 19

\bibitem[{{Perley} \& {Butler}(2013)}]{Perley+Butler2013}
{Perley}, R.~A. \& {Butler}, B.~J. 2013, \apjs, 204, 19

\bibitem[{{Plauchu-Frayn} {et~al.}(2012){Plauchu-Frayn}, {Del Olmo}, {Coziol},
  \& {Torres-Papaqui}}]{Plauchu-Frayn+2012}
{Plauchu-Frayn}, I., {Del Olmo}, A., {Coziol}, R., \& {Torres-Papaqui}, J.~P.
  2012, \aap, 546, A48

\bibitem[{{Ponman} {et~al.}(1996){Ponman}, {Bourner}, {Ebeling}, \&
  {B{\"o}hringer}}]{Ponman+1996}
{Ponman}, T.~J., {Bourner}, P.~D.~J., {Ebeling}, H., \& {B{\"o}hringer}, H.
  1996, \mnras, 283, 690

\bibitem[{{Pop} {et~al.}(2018){Pop}, {Pillepich}, {Amorisco}, \&
  {Hernquist}}]{Pop+2018}
{Pop}, A.-R., {Pillepich}, A., {Amorisco}, N.~C., \& {Hernquist}, L. 2018,
  \mnras, 480, 1715

\bibitem[{{Punzo} {et~al.}(2017){Punzo}, {van der Hulst}, {Roerdink},
  {Fillion-Robin}, \& {Yu}}]{Punzo+2017}
{Punzo}, D., {van der Hulst}, J.~M., {Roerdink}, J.~B.~T.~M., {Fillion-Robin},
  J.~C., \& {Yu}, L. 2017, Astronomy and Computing, 19, 45

\bibitem[{{Punzo} {et~al.}(2016){Punzo}, {van der Hulst}, {Roerdink}, \&
  {Fillion-Robin}}]{SlicerAstro}
{Punzo}, D., {van der Hulst}, T., {Roerdink}, J., \& {Fillion-Robin}, J.-C.
  2016, {SlicerAstro: Astronomy (HI) extension for 3D Slicer}, Astrophysics
  Source Code Library

\bibitem[{{Querejeta} {et~al.}(2015){Querejeta}, {Eliche-Moral}, {Tapia},
  {Borlaff}, {Rodr{\'{\i}}guez-P{\'e}rez}, {Zamorano}, \&
  {Gallego}}]{Querejeta+2015}
{Querejeta}, M., {Eliche-Moral}, M.~C., {Tapia}, T., {et~al.} 2015, \aap, 573,
  A78

\bibitem[{{Quinn}(1984)}]{Quinn1984}
{Quinn}, P.~J. 1984, \apj, 279, 596

\bibitem[{{Ramachandran} \& {Varoquaux}(2011)}]{mayavi}
{Ramachandran}, P. \& {Varoquaux}, G. 2011, Computing in Science and
  Engineering, 13, 40

\bibitem[{{Rasmussen} {et~al.}(2008){Rasmussen}, {Ponman}, {Verdes-Montenegro},
  {Yun}, \& {Borthakur}}]{Rasmussen+2008}
{Rasmussen}, J., {Ponman}, T.~J., {Verdes-Montenegro}, L., {Yun}, M.~S., \&
  {Borthakur}, S. 2008, \mnras, 388, 1245

\bibitem[{{Ribeiro} {et~al.}(1996){Ribeiro}, {de Carvalho}, {Coziol},
  {Capelato}, \& {Zepf}}]{Ribeiro+1996}
{Ribeiro}, A.~L.~B., {de Carvalho}, R.~R., {Coziol}, R., {Capelato}, H.~V., \&
  {Zepf}, S.~E. 1996, \apjl, 463, L5

\bibitem[{{Rich} {et~al.}(2010){Rich}, {Dopita}, {Kewley}, \&
  {Rupke}}]{Rich+2010}
{Rich}, J.~A., {Dopita}, M.~A., {Kewley}, L.~J., \& {Rupke}, D.~S.~N. 2010,
  \apj, 721, 505

\bibitem[{{Rizzo} {et~al.}(2018){Rizzo}, {Fraternali}, \& {Iorio}}]{Rizzo+2018}
{Rizzo}, F., {Fraternali}, F., \& {Iorio}, G. 2018, \mnras, 476, 2137

\bibitem[{Robitaille(2019)}]{aplpy2019}
Robitaille, T. 2019, {APLpy v2.0: The Astronomical Plotting Library in Python}

\bibitem[{{Robitaille} \& {Bressert}(2012)}]{aplpy2012}
{Robitaille}, T. \& {Bressert}, E. 2012, {APLpy: Astronomical Plotting Library
  in Python}, Astrophysics Source Code Library

\bibitem[{{Rom{\'a}n} \& {Trujillo}(2017)}]{Roman+2017}
{Rom{\'a}n}, J. \& {Trujillo}, I. 2017, \mnras, 468, 4039

\bibitem[{{Rubin} {et~al.}(1991){Rubin}, {Hunter}, \& {Ford}}]{Rubin+1991}
{Rubin}, V.~C., {Hunter}, D.~A., \& {Ford}, Jr., W.~K. 1991, \apjs, 76, 153

\bibitem[{{Serra} {et~al.}(2013){Serra}, {Koribalski}, {Duc}, {Oosterloo},
  {McDermid}, {Michel-Dansac}, {Emsellem}, {Cuillandre}, {Alatalo}, {Blitz},
  {Bois}, {Bournaud}, {Bureau}, {Cappellari}, {Crocker}, {Davies}, {Davis}, {de
  Zeeuw}, {Khochfar}, {Krajnovi{\'c}}, {Kuntschner}, {Lablanche}, {Morganti},
  {Naab}, {Sarzi}, {Scott}, {Weijmans}, \& {Young}}]{Serra+2013}
{Serra}, P., {Koribalski}, B., {Duc}, P.-A., {et~al.} 2013, \mnras, 428, 370

\bibitem[{{Serra} {et~al.}(2014){Serra}, {Westmeier}, {Giese}, {Jurek},
  {Fl{\"o}er}, {Popping}, {Winkel}, {van der Hulst}, {Meyer}, {Koribalski},
  {Staveley-Smith}, \& {Courtois}}]{SoFiA}
{Serra}, P., {Westmeier}, T., {Giese}, N., {et~al.} 2014, {SoFiA: Source
  Finding Application}, Astrophysics Source Code Library

\bibitem[{{Serra} {et~al.}(2015){Serra}, {Westmeier}, {Giese}, {Jurek},
  {Fl{\"o}er}, {Popping}, {Winkel}, {van der Hulst}, {Meyer}, {Koribalski},
  {Staveley-Smith}, \& {Courtois}}]{Serra+2015}
{Serra}, P., {Westmeier}, T., {Giese}, N., {et~al.} 2015, \mnras, 448, 1922

\bibitem[{{Spekkens} \& {Karunakaran}(2018)}]{Spekkens+2018}
{Spekkens}, K. \& {Karunakaran}, A. 2018, \apj, 855, 28

\bibitem[{{Sulentic} {et~al.}(2001){Sulentic}, {Rosado}, {Dultzin-Hacyan},
  {Verdes-Montenegro}, {Trinchieri}, {Xu}, \& {Pietsch}}]{Sulentic+2001}
{Sulentic}, J.~W., {Rosado}, M., {Dultzin-Hacyan}, D., {et~al.} 2001, \aj, 122,
  2993

\bibitem[{{Tapia} {et~al.}(2017){Tapia}, {Eliche-Moral}, {Aceves},
  {Rodr{\'{\i}}guez-P{\'e}rez}, {Borlaff}, \& {Querejeta}}]{Tapia+2017}
{Tapia}, T., {Eliche-Moral}, M.~C., {Aceves}, H., {et~al.} 2017, \aap, 604,
  A105

\bibitem[{{Turner} {et~al.}(2001){Turner}, {Reeves}, {Ponman}, {Arnaud},
  {Barbera}, {Bennie}, {Boer}, {Briel}, {Butler}, {Clavel}, {Dhez}, {Cordova},
  {Dos Santos}, {Ferrando}, {Ghizzardi}, {Goodall}, {Griffiths}, {Hochedez},
  {Holland}, {Jansen}, {Kendziorra}, {Lagostina}, {Laine}, {La Palombara},
  {Lortholary}, {Mason}, {Molendi}, {Pigot}, {Priedhorsky}, {Reppin},
  {Rothenflug}, {Salvetat}, {Sauvageot}, {Schmitt}, {Sembay}, {Short},
  {Str{\"u}der}, {Trifoglio}, {Tr{\"u}mper}, {Vercellone}, {Vigroux}, {Villa},
  \& {Ward}}]{Turner+2001}
{Turner}, M.~J.~L., {Reeves}, J.~N., {Ponman}, T.~J., {et~al.} 2001, \aap, 365,
  L110

\bibitem[{{Tzanavaris} {et~al.}(2010){Tzanavaris}, {Hornschemeier},
  {Gallagher}, {Johnson}, {Gronwall}, {Immler}, {Reines}, {Hoversten}, \&
  {Charlton}}]{Tzanavaris+2010}
{Tzanavaris}, P., {Hornschemeier}, A.~E., {Gallagher}, S.~C., {et~al.} 2010,
  \apj, 716, 556

\bibitem[{{van Zee} {et~al.}(1997){van Zee}, {Maddalena}, {Haynes}, {Hogg}, \&
  {Roberts}}]{vanZee+1997}
{van Zee}, L., {Maddalena}, R.~J., {Haynes}, M.~P., {Hogg}, D.~E., \&
  {Roberts}, M.~S. 1997, \aj, 113, 1638

\bibitem[{{Verdes-Montenegro} {et~al.}(2002){Verdes-Montenegro}, {Del Olmo},
  {Iglesias-P{\'a}ramo}, {Perea}, {V{\'{\i}}lchez}, {Yun}, \&
  {Huchtmeier}}]{Verdes-Montenegro+2002}
{Verdes-Montenegro}, L., {Del Olmo}, A., {Iglesias-P{\'a}ramo}, J.~I., {et~al.}
  2002, \aap, 396, 815

\bibitem[{{Verdes-Montenegro} {et~al.}(1997){Verdes-Montenegro}, {del Olmo},
  {Perea}, {Athanassoula}, {Marquez}, \& {Augarde}}]{Verdes-Montenegro+1997}
{Verdes-Montenegro}, L., {del Olmo}, A., {Perea}, J., {et~al.} 1997, \aap, 321,
  409

\bibitem[{{Verdes-Montenegro} {et~al.}(2005{\natexlab{a}}){Verdes-Montenegro},
  {Del Olmo}, {Yun}, \& {Perea}}]{Verdes-Montenegro+2005}
{Verdes-Montenegro}, L., {Del Olmo}, A., {Yun}, M.~S., \& {Perea}, J.
  2005{\natexlab{a}}, \aap, 430, 443

\bibitem[{{Verdes-Montenegro} {et~al.}(2005{\natexlab{b}}){Verdes-Montenegro},
  {Sulentic}, {Lisenfeld}, {Leon}, {Espada}, {Garcia}, {Sabater}, \&
  {Verley}}]{Verdes-Montenegro+2005b}
{Verdes-Montenegro}, L., {Sulentic}, J., {Lisenfeld}, U., {et~al.}
  2005{\natexlab{b}}, \aap, 436, 443

\bibitem[{{Verdes-Montenegro} {et~al.}(1998){Verdes-Montenegro}, {Yun},
  {Perea}, {del Olmo}, \& {Ho}}]{Verdes-Montenegro+1998}
{Verdes-Montenegro}, L., {Yun}, M.~S., {Perea}, J., {del Olmo}, A., \& {Ho},
  P.~T.~P. 1998, \apj, 497, 89

\bibitem[{{Verdes-Montenegro} {et~al.}(2001){Verdes-Montenegro}, {Yun},
  {Williams}, {Huchtmeier}, {Del Olmo}, \& {Perea}}]{Verdes-Montenegro+2001}
{Verdes-Montenegro}, L., {Yun}, M.~S., {Williams}, B.~A., {et~al.} 2001, \aap,
  377, 812

\bibitem[{{V{\'e}ron-Cetty} \& {V{\'e}ron}(2010)}]{Veron-Cetty+Veron2010}
{V{\'e}ron-Cetty}, M.~P. \& {V{\'e}ron}, P. 2010, \aap, 518, A10

\bibitem[{{Vogt} {et~al.}(2013){Vogt}, {Dopita}, \& {Kewley}}]{Vogt+2013}
{Vogt}, F.~P.~A., {Dopita}, M.~A., \& {Kewley}, L.~J. 2013, \apj, 768, 151

\bibitem[{{Vogt} {et~al.}(2016){Vogt}, {Owen}, {Verdes-Montenegro}, \&
  {Borthakur}}]{Vogt+2016}
{Vogt}, F.~P.~A., {Owen}, C.~I., {Verdes-Montenegro}, L., \& {Borthakur}, S.
  2016, \apj, 818, 115

\bibitem[{Wilkinson {et~al.}(2016)Wilkinson, Dumontier, Aalbersberg, Appleton,
  Axton, Baak, Blomberg, Boiten, {da Silva Santos}, Bourne, Bouwman, Brookes,
  Clark, Crosas, Dillo, Dumon, Edmunds, Evelo, Finkers, Gonzalez-Beltran, Gray,
  Groth, Goble, Grethe, Heringa, {'t Hoen}, Hooft, Kuhn, Kok, Kok, Lusher,
  Martone, Mons, Packer, Persson, Rocca-Serra, Roos, van Schaik, Sansone,
  Schultes, Sengstag, Slater, Strawn, Swertz, Thompson, van~der Lei, van
  Mulligen, Velterop, Waagmeester, Wittenburg, Wolstencroft, Zhao, \&
  Mons}]{Wilkinson+2016}
Wilkinson, M.~D., Dumontier, M., Aalbersberg, I.~J., {et~al.} 2016, Scientific
  Data, 3, 160018

\bibitem[{{Williams} \& {Rood}(1987)}]{Williams+Rood1987}
{Williams}, B.~A. \& {Rood}, H.~J. 1987, \apjs, 63, 265

\bibitem[{{Zucker} {et~al.}(2016){Zucker}, {Walker}, {Johnson}, {Gallagher},
  {Alatalo}, \& {Tzanavaris}}]{Zucker+2016}
{Zucker}, C., {Walker}, L.~M., {Johnson}, K., {et~al.} 2016, \apj, 821, 113

\end{thebibliography}

\appendix

\section{Full workflow and reproducibility}
\label{sec:workflow}

Aside from the scientific aims of this work we have also aimed to follow the scientific method as fully as possible by making the reduction and analysis completely reproducible. Our intention is to provide any other astronomer with access to our data and methodology such that they are able to understand what was done to the data (at any level they wish), to verify our analysis, or to reuse any of our outputs. We have not been able to fully meet these goals for every aspect of our analysis, but have done so for the majority. The full findings of this experiment will be detailed in another publication, but in this section we provide a brief description of our methodology and experiences with regard to reproducibility. Our full workflow is available at \url{github.com/AMIGA-IAA/hcg-16}, while the final plots can be recreated and modified at \url{mybinder.org/v2/gh/AMIGA-IAA/hcg-16/master} using the default version of the final data product (as in this paper) without needing to execute the whole workflow.

\subsection{Need and motivation}

Reproducibility is a key pillar of the scientific method, but in the current age of digital research it is frequently not possible to reproduce or reuse previous results. Although we could point to a number of studies indicating that science (of all sub-disciplines) is currently in a reproducibility crisis, indicating the necessity of methodological improvements, it is perhaps more informative to highlight this point with some concrete examples from our own project.
\begin{itemize}
    \item Figure \ref{fig:group_spec} shows that the spectral profile and total flux measurements do not quite agree with those from HIPASS. Although we have identified the baseline subtraction around this source as a potential problem, without a substantial investment of time and an expert from within the HIPASS team it would not be possible to fully understand this difference. While the HIPASS reduced data are publicly available and the general methodology has been published, it is difficult to know exactly what has happened to a specific part of the data. This meant our investigation of this issue was only possible from one side, which was ultimately not completely successful.
    \item The VLA data used in this paper have been used in previous publications \citep{Verdes-Montenegro+2001,Borthakur+2010}. We had access to one of the previous cubes (reduced using \texttt{AIPS}), which an author still had on their hard drive. Initially we were making comparisons between that cube and the one from this work to verify our results and investigate the effect of using multi-scale CLEAN. However, these comparisons were eventually abandoned because, even with the file history in the \texttt{fits} file, it was difficult to know exactly what had been done previously. Furthermore, if we disagreed with a step from the previous analysis it would have been a considerable amount of additional work (if even possible) to reproduce the previous reduction and modify that step.
\end{itemize}
Our goal is to take a step forward in this work to show how such barriers to reproducibility can be minimised in future. 

In recent years this difficulty with reproducibility in modern science has received considerable work and attention (including by members of the research team at the Instituto de Astrof\'{i}sica de Andaluc\'{i}a). One such recent work, \citet{Wilkinson+2016}, describes a set of generic principles regarding how scientific data should be stored and documented in order to make them as reusable as possible, in particular, for automated processing. These are the FAIR principles (Findable, Accessible, Interoperable, and Reusable), \url{www.go-fair.org}. Ourselves, \citet{Wilkinson+2016} and others argue that these principles should not only be applied to data, but also to algorithms and tools in order to enable full transparency and reproducibility in practice. We have used these principles to help guide our approach, which we discuss below.

\subsection{Discussion of our approach}

Before embarking on creating a workflow framework it is necessary to decide at what level a project should be reproducible. This ranges from providing just the raw and reduced data, to allowing every value in the final paper to be re-generated in a completely automated way, such that if any error were to be identified in the reduction or analysis (or that of previous works on which we depend) then all dependent values throughout the paper would be automatically changed once it was corrected. Unfortunately the latter is not yet fully achievable (more discussion below), but we also wish to go beyond providing just the data by allowing other astronomers to fully understand our methodology and to modify or reuse it. Therefore, we have elected to make every digital object created in this work a reproducible product of the workflow. In this case this corresponds to every figure and table which displays results of this work. This includes Table \ref{tab:HIprops} as the properties it contains are all non-trivial results of this work, however, Tables \ref{tab:optprops} \& \ref{tab:gas_time} are not included as they are either values taken from the literature or the result of trivial calculations with literature values. Unfortunately, these aims have not been fully realised and there are a few results which are not included in our workflow, which we discuss below.

To fully describe a workflow it is not sufficient to provide only the data (raw or reduced) and the reduction scripts as different versions of the same software might not produce identical results, even given the same input, and installing unfamiliar software can often be a barrier. Therefore, it is also necessary to regularise the software environment. To do this we made extensive use of \texttt{docker} (\url{www.docker.com}) containers and \texttt{Conda} (\url{docs.conda.io}) environments. Each piece of software used, in this case \texttt{CASA} and \texttt{SoFiA}, was placed inside a \texttt{docker} container so that exactly the same software environment can be used on any machine. A \texttt{Conda} environment was created to regularise the \texttt{Python} environment which was used for generating the final plots. All the steps of the reduction, analysis, and plotting were chained together in a single workflow using the \texttt{CGAT-core} (\url{github.com/cgat-developers/cgat-core}) workflow management software. This allows the final workflow to be executed using a single script which downloads the raw data and scripts, builds the \texttt{Conda} environments, pulls the \texttt{docker} containers, runs the data reduction and analysis steps, and finally generates the plots displayed in the paper. Furthermore, any one of these steps can be executed individually (provided the prerequisite steps have already been executed) and the reduction and analysis scripts themselves are stored separately to the software environment so that they can be modified as any end user wishes and the workflow re-run. For example, if a particular user wanted the data to be CLEANed to a lower threshold than was used, this would require only the threshold value in the imaging script to be changed and the workflow executed again.\footnote{Here we should note that the separation of features (\S \ref{sec:HIsep}) was an unavoidably manual process (part of the reason why \texttt{SlicerAstro} is not included in our workflow), thus the regions which we defined for each galaxy and feature will be unchanged by any modification to the scripts, however, the resulting spectra and moment maps within those regions might change slightly as a result of different choices in the data reduction steps.}

As mentioned above, our original goal has not been fully achieved and some figures are missing from the workflow. Figures \ref{fig:pv_plot} \& \ref{fig:deep_opt} are not included and the velocity labels in the first moments of the galaxies (Figures \ref{fig:HCG16a_moms}--\ref{fig:HCG16d_moms}, \ref{fig:NGC848_moms}, and \ref{fig:PGC8210_moms}) had to be added manually, so will not be updated if the iso-velocity contours change. The X3D interactive plot was also not included in the workflow as at present this still requires manual alterations to circumvent bugs in the software. 

Figure \ref{fig:pv_plot} was not included as this was the only figure for which the tools in \texttt{SlicerAstro} were required (unless we were to completely reinvent the ``wheel''). While \texttt{SlicerAstro} was a key tool for separating emission from features and galaxies, this process ultimately results in a set of masks which define the regions assumed to contain each source. As these masks are created in a manual way, the result can only be preserved rather than the process made completely reproducible. Hence, the final masks were imported into \texttt{CASA} and used to generate the moments, which meant that creating a \texttt{docker} container for \texttt{SlicerAstro} was not necessary, with the exception of this figure which uses its segmented PV slice feature. Although it is possible to script processes in \texttt{SlicerAstro} the parent \texttt{Slicer} software is not designed in a way that makes this straightforward for general users, for example it is not possible to record steps performed in the GUI as a script. Therefore we decided to omit this plot from the final workflow due to the time investment it would take to include it. Figure \ref{fig:deep_opt} was not included because it relies on an \texttt{IDL} pipeline to mask the bright emission and enhance the LSB features in the DECaLS images. This pipeline is in the process of being converted to \texttt{Python}, but as this is a work in progress it has not been included.

One of the key reasons why we did not attempt to make every value in the paper (including those that only appear in the text) fully reproducible is because at present this is not feasible. It is normal for astronomers to upload large data tables or large survey datasets to services such as CDS (Centre de Donn\'{e}es astronomiques de Strasbourg) where they can be queried and accessed in an automated way, however, this frequently does not cover all the required information. To take an explicit example consider our Table \ref{tab:gas_time}. This table lists the SFRs and gas masses of the four core galaxies in HCG 16 along with the trivial result of the calculation of gas consumption timescale for each. The SFR data were taken from \cite{Lenkic+2016} and the molecular mass data from \citet{Martinez-Badenes+2012}, while the \hi \ mass values are from this work. In order to make this table completely reproducible (e.g. if a hypothetical mistake in one of those papers was to be found and corrected) all these data would need to be stored in a service like CDS and in addition we would also need an automated way to read the distance used (or the Hubble constant if distance is inferred directly from velocity). Our intention here is not to criticise these works, indeed their data tables are (mostly) available in CDS, and we are also guilty of shortcomings in this respect. For example, someone wishing to automatically access the distance which we used for HCG 16 would face the same problem, while the B-band luminosities we list in Table \ref{tab:optprops} come from a private database compiled by the AMIGA team. Another issue is when the information provided by another paper is an equation, for example we use a scaling relation from \citet{Jones+2018}, but equations can also have mistakes and typographical errors, and there is no automated method to connect them between different works.

To overcome these shortcomings would require a standardised electronic method of reporting all this information so that all external information used in a paper can be fully clarified within a workflow. For data, services like CDS and the International Virtual Observatory Alliance are suitable solutions, but all data tables need to be included, even those which seem insignificant. However, for values such as the Hubble constant that a paper uses, the distances to objects (for papers about a small sample of objects), or any other values which appear only in the text, there is currently no good solution of which we are aware. The same is true for equations. Ideally such information would be included in some form of metadata that accompany a paper or its data. 

At present making projects reproducible requires an enormous additional effort and it is unsurprising that it is so often neglected, however, once tools and standardised approaches have been established the burden placed on individual scientists will be greatly decreased. There are many other tools and services available to help this process and we plan to continue to explore and test them in future works.

\subsection{Workflow overview and final products}

The workflow (\url{github.com/AMIGA-IAA/hcg-16}) begins by constructing the necessary software environments as discussed above and then runs the following steps:
\begin{enumerate}
    \item Download data: The raw VLA data, the CLEANing mask,  the masks that we made using \texttt{SlicerAstro} to separate each galaxy and feature, and the GBT and HIPASS cube are all downloaded from the EUDAT B2SHARE service (\url{b2share.eudat.eu}) which provides a cloud repository for data. The DOI for this dataset is 10.23728/b2share.f8fcd84bcd454bdc8ea0ec2d69bdfe9a.
    
    The reason the VLA data were re-hosted here is both because it was necessary to include the masks in addition to the raw data and because of the nature of the VLA archive it is not currently possible to access its data in an entirely automated manner.
    \item Flagging and calibration: The exact steps (performed in \texttt{CASA}) to flag and calibrate the raw data are run.
    \item Imaging: The imaging steps are run to generate the \hi \ cubes with \texttt{CASA}. A simple threshold moment 0 map is also created and the masks for each separate feature and galaxy are used to generate mini-cubes and moments.
    \item Masking: \texttt{SoFiA} is used to make two masks of the entire group (one with a lower threshold intended to included all real emission and one with a higher threshold to make the velocity field). The first mask is later used to generate the integrated spectrum. This step also produces a moment 0 and moment 1 map. \texttt{SoFiA} is also used to make a mask of HCG 16 in the HIPASS cube.
\end{enumerate}

The above steps generate all the final cubes and maps needed to produce the figures in the paper. For each figure in the paper we have provided a \texttt{Python} script in the form of a \texttt{Jupyter} (\url{jupyter.org}) notebook. These notebooks begin by listing the input data used in making each plot and explain which steps of the workflow those data dependent on and direct the user to the relevant scripts, should they wish to modify them and re-run the workflow. As we do not anticipate the majority of readers will want to modify our methodology, but may be interested in modifying or reusing our plots, these notebook and the final data products have also been provided (separate from the workflow) and may be accessed through the notebook server \url{mybinder.org/v2/gh/AMIGA-IAA/hcg-16/master}.

At present the workflow can be downloaded and run by any user locally (or the notebook can be accessed in the cloud). We have also tested running the workflow in the cloud using a virtual machine hosted by the EGI Federated Cloud, which we accessed through the EOSC (European Open Science Cloud) catalogue, and containers on the \texttt{SciServer} service (www.sciserver.org). While there are definite advantages to working in these environments, for example on \texttt{SciServer} the software environment is already fixed for the end user which immediately simplifies the process of reproducibility, thus far our workflow has not been fully integrated with either of these services, so is only available for local execution. In the near future, with the data volumes that upcoming surveys will produce, there will cease to be a choice between working locally and in the cloud, and we will continue to work on cloud-based implementations of reproducible workflows.

\section{Integrated flux and profile discrepancies}
\label{sec:flux_disc}

Figure \ref{fig:group_spec} shows the comparison of the VLA spectrum measured in this work and the spectrum we extracted from the relevant HIPASS cube. In addition to the slight excess of flux in the VLA spectrum (which is formally not possible) the profile shapes of the \hi \ emission are quite different, suggesting more investigation is needed. 

While regions that are brighter in HIPASS could be explained by the presence of diffuse gas that is resolved out by the VLA, the reverse does not apply. The HIPASS profile ends at $\sim$4100 \kms, whereas the VLA spectrum continues until $\sim$4300 \kms. Investigating this further we found faint negative regions in the HIPASS data, possibly indicating problems with baseline subtraction around this extended source. We also found that the disagreement between the two datasets is a function of position, with them agreeing well on a diagonal line stretching SE to NW, but diverging (in opposite senses in opposite directions) away from this line. As this is not a radial dependence it is unlikely to result from the VLA primary beam correction. We also found a similar trend when comparing HIPASS with the previous reduction of the same VLA data \citep{Borthakur+2010}, indicating the problem is also not due to the new reduction in \texttt{CASA} as oppose to \texttt{AIPS}, or MS-CLEAN versus standard CLEAN. Unfortunately we were unable to identify the exact cause of this disagreement.

Here we also note that the flux scale in our cube is slightly higher than the previous reduction published in \citet{Borthakur+2010}. That paper only shows the VLA spectrum weighted to correspond to the GBT observation, finding that there is 5\% less flux in the VLA spectrum than in the GBT spectrum, whereas we find 12\% more. Part of this difference can be explained by the masking of the VLA data. The \texttt{SoFiA}-generated mask is designed to smooth over several scales in order to recover as much flux as possible and is thus more extended than more traditional masks. Applying our mask to the cube of \citet{Borthakur+2010} results in a profile that falls between our data and the GBT data. Therefore, the mask is only a partial explanation for the difference in flux between our reduction of the VLA data and that of \citet{Borthakur+2010}. That work quotes the flux of the flux calibrator (3C48) as 16.5 Jy, while we use the \texttt{CASA}-based model \citep{Perley+Butler2013} which has a flux between 16.43 and 16.49 Jy across the bandwidth. This is a negligible difference that, if anything, would reduce rather than increase the flux we measure. 

It might be suggested that using MS-CLEAN instead of standard CLEAN could have resulted in us recovering more flux. However, this is a misinterpretation of what MS-CLEAN does (discussed in Appendix \ref{sec:MSC_flux}). In fact, the integrated flux in our MS-CLEAN image is lower than an equivalent reduction using standard CLEAN.

It is also worth noting that the flagging performed in this work and in \citet{Borthakur+2010} is markedly different, but from the final image cube that we have available it is impossible to say exactly how. The header implies that almost none of the C array data were flagged previously and only about 7\% of the D array data. We flagged a total of 9\% of the C array data and 37\% of the D array data, either due to RFI, shadowing, or otherwise inconsistent baselines. This minimalist versus zealous flagging approach may contribute to the differences we find.

In conclusion we have not been able to fully resolve the small discrepancies between these datasets and we argue that without a complete workflow, as we have sought to create for this project, or completely repeating previous data reductions (which is both beyond the scope of this work and not guaranteed to be successful) it is not possible to reliably identify the exact cause.

\section{Flux recovery with MS-CLEAN}
\label{sec:MSC_flux}

The MS-CLEAN algorithm is sometimes regarded as being able to recover more flux because the CLEAN model (for MS-CLEAN) for an extended source will typically have more flux in its components than a standard CLEAN model based on the same data. In fact we find that our MS-CLEAN cube contains \textit{less} flux than when we imaged using standard CLEAN, even though the MS-CLEAN model components contain more flux. This is because it is more able to CLEAN diffuse emission and can therefore remove the positive platform (and surrounding negative bowl) on which the entire group sits in a standard CLEAN image. Although in the case of HCG 16 this feature is quite subtle, it nevertheless appears in almost every channel where these is emission from the core group, and thus contributes significantly to the total flux of a standard CLEAN image. Therefore, it is more accurate to say that the MS-CLEAN algorithm gives a more robust measure of the total \hi \ flux in the group and a more realistic representation of the morphology of the \hi, but it is misleading to say that it recovers more flux.

\section{Channel maps}

The online version of the paper is accompanied by channel maps showing the contours of \hi \ emission in individual spectral channels in the region of the group core and around NGC 848 and the SE tail.

\newpage
\begin{landscape}
\begin{figure}
    \centering
    \includegraphics[width=0.33\columnwidth]{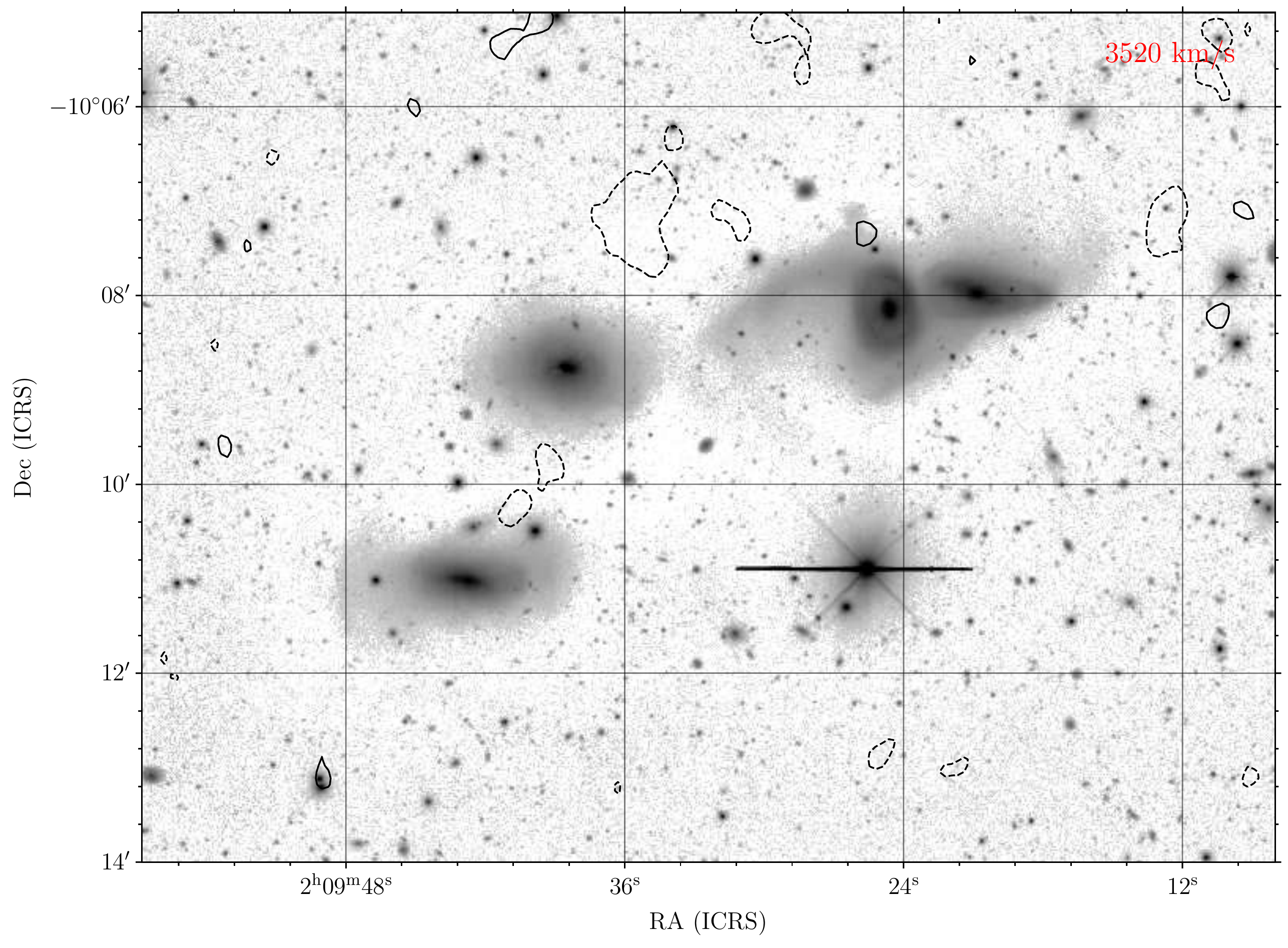}
    \includegraphics[width=0.33\columnwidth]{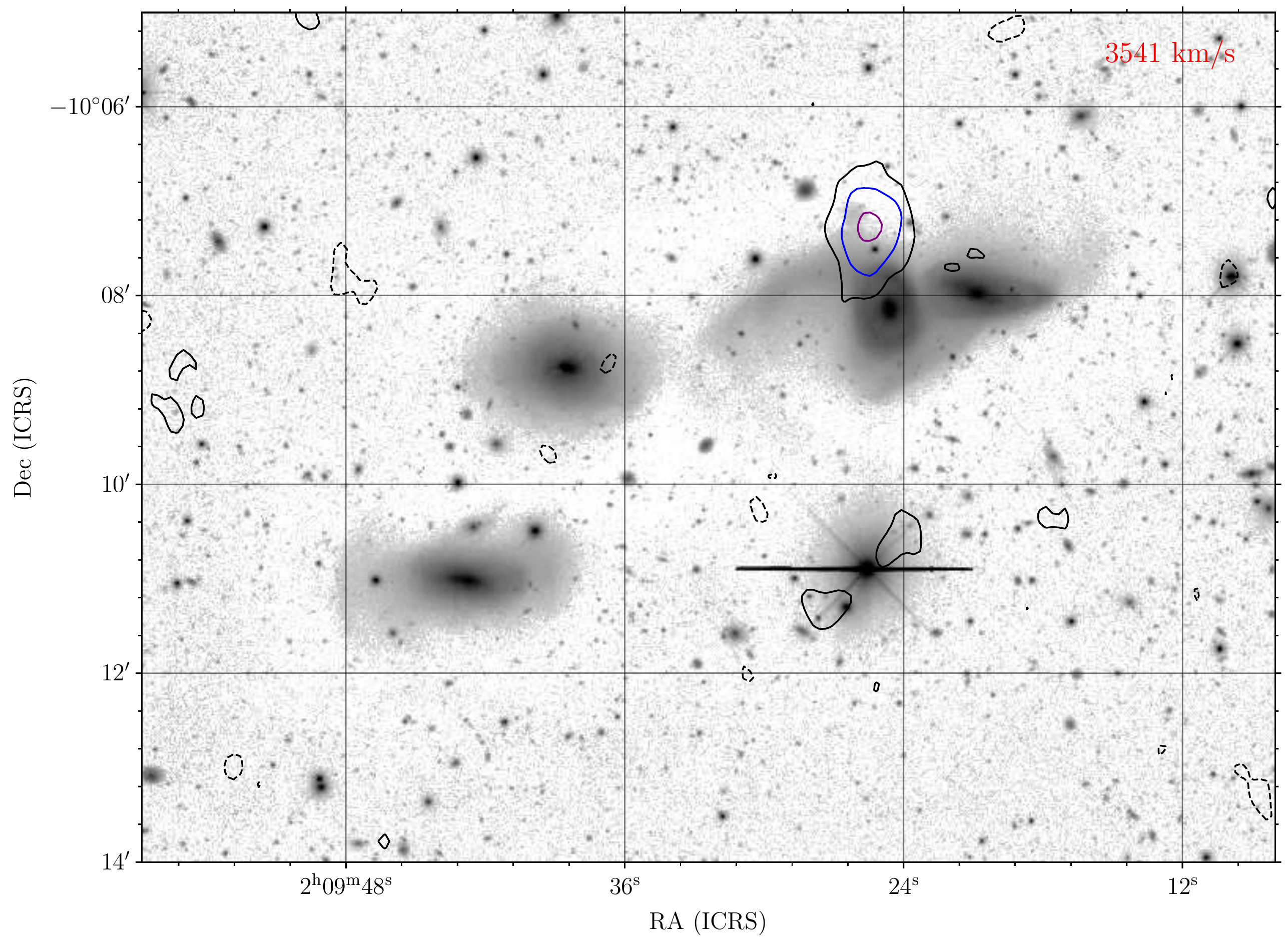}
    \includegraphics[width=0.33\columnwidth]{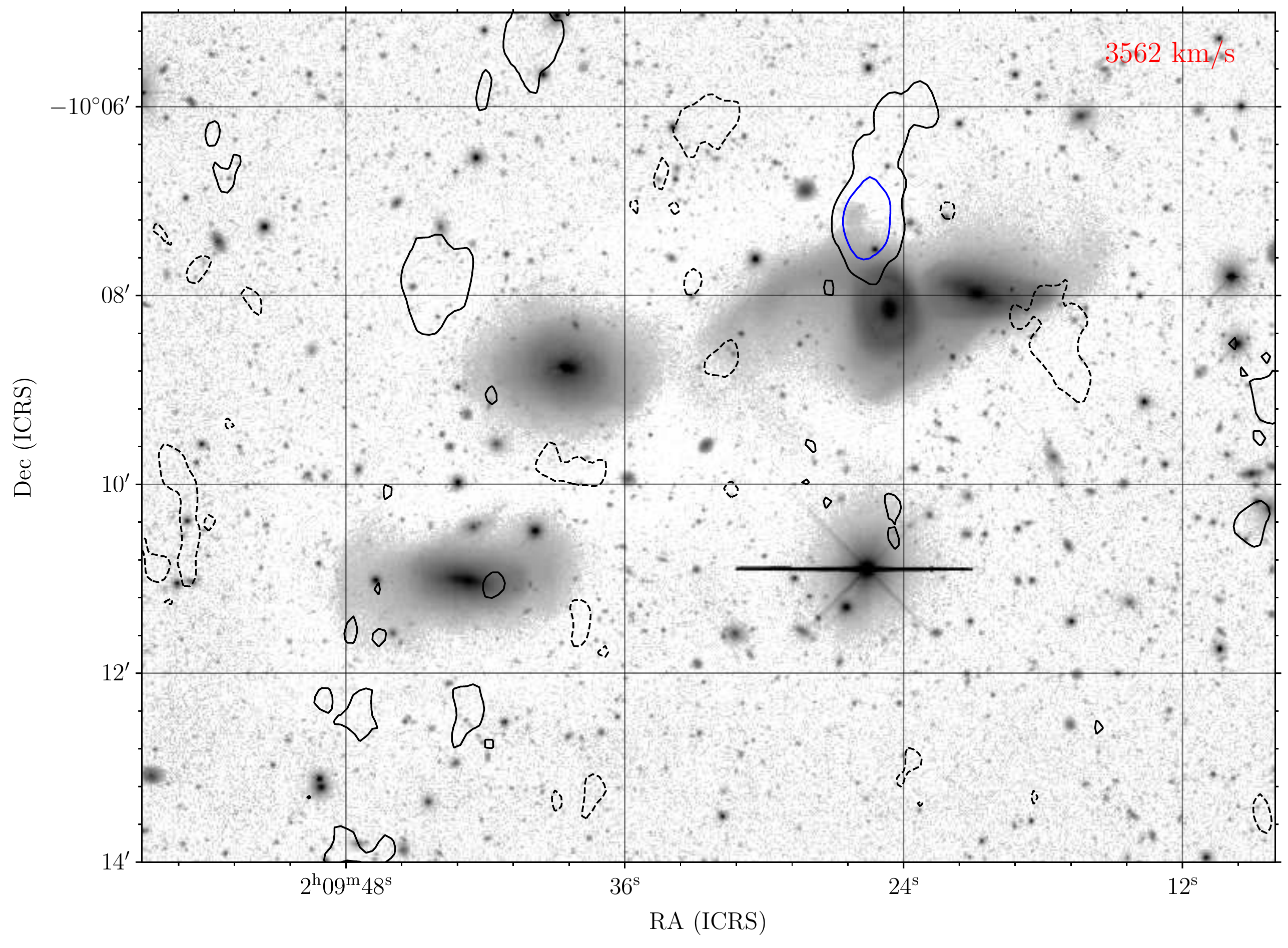}
    \includegraphics[width=0.33\columnwidth]{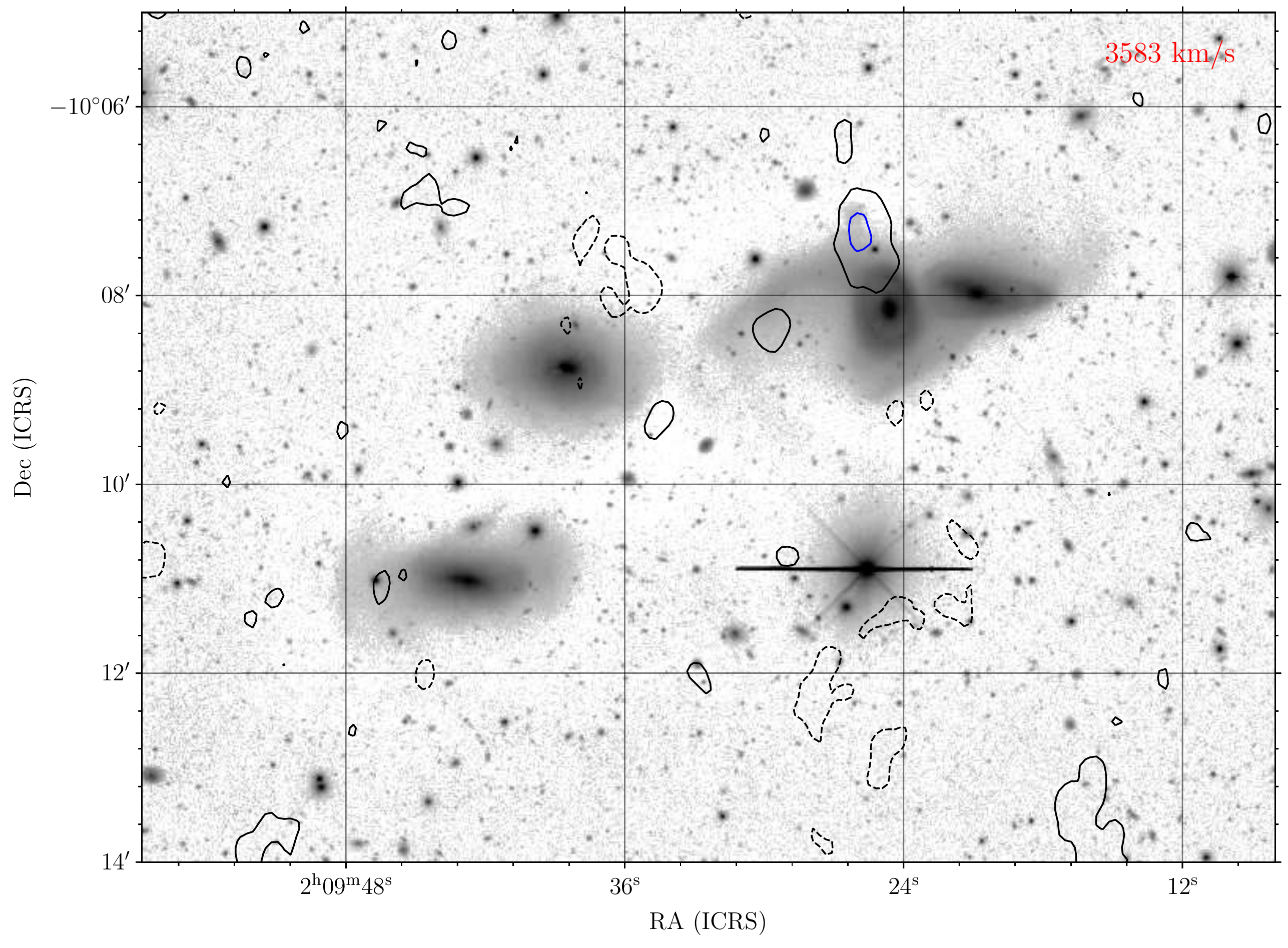}
    \includegraphics[width=0.33\columnwidth]{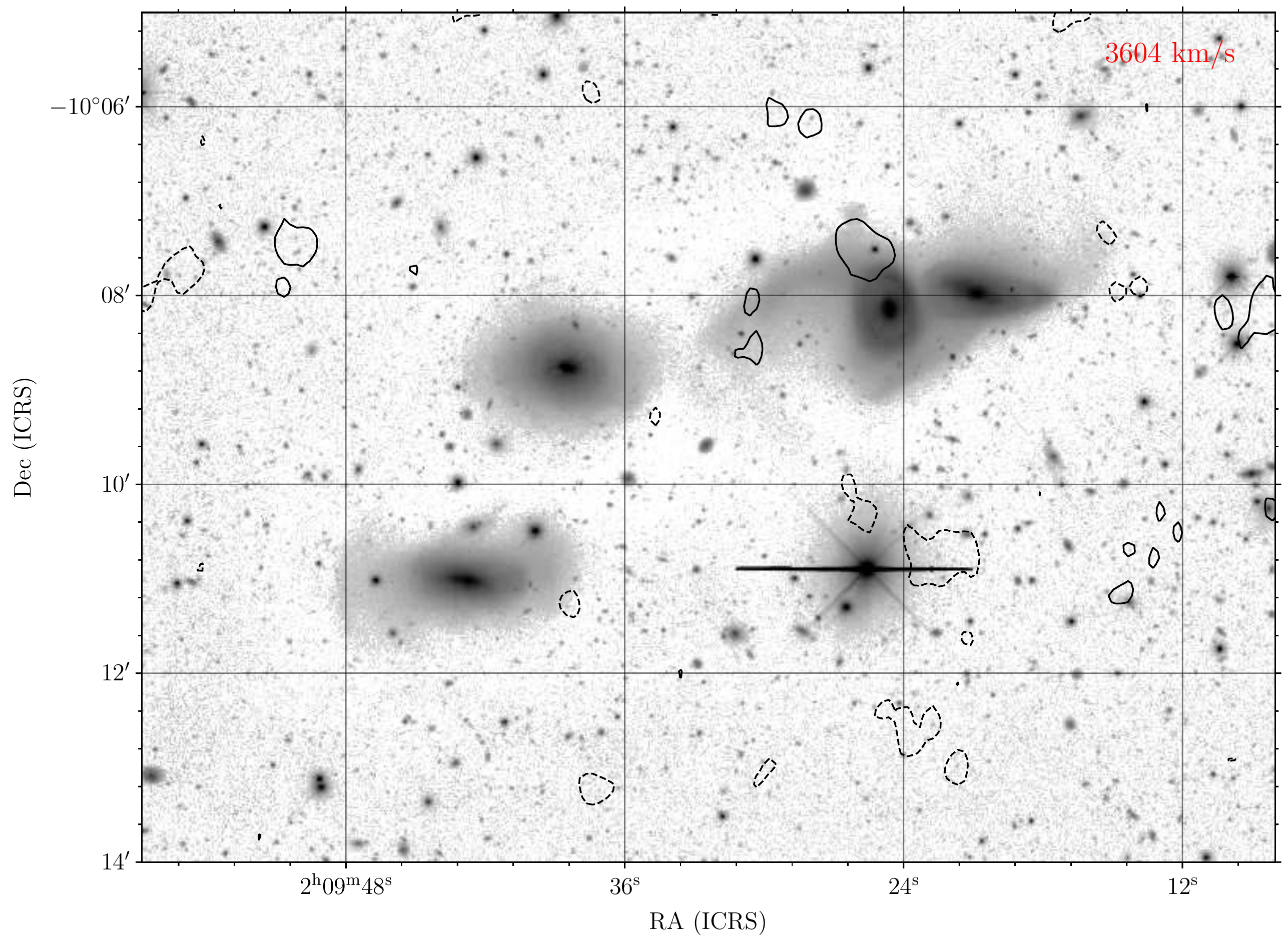}
    \includegraphics[width=0.33\columnwidth]{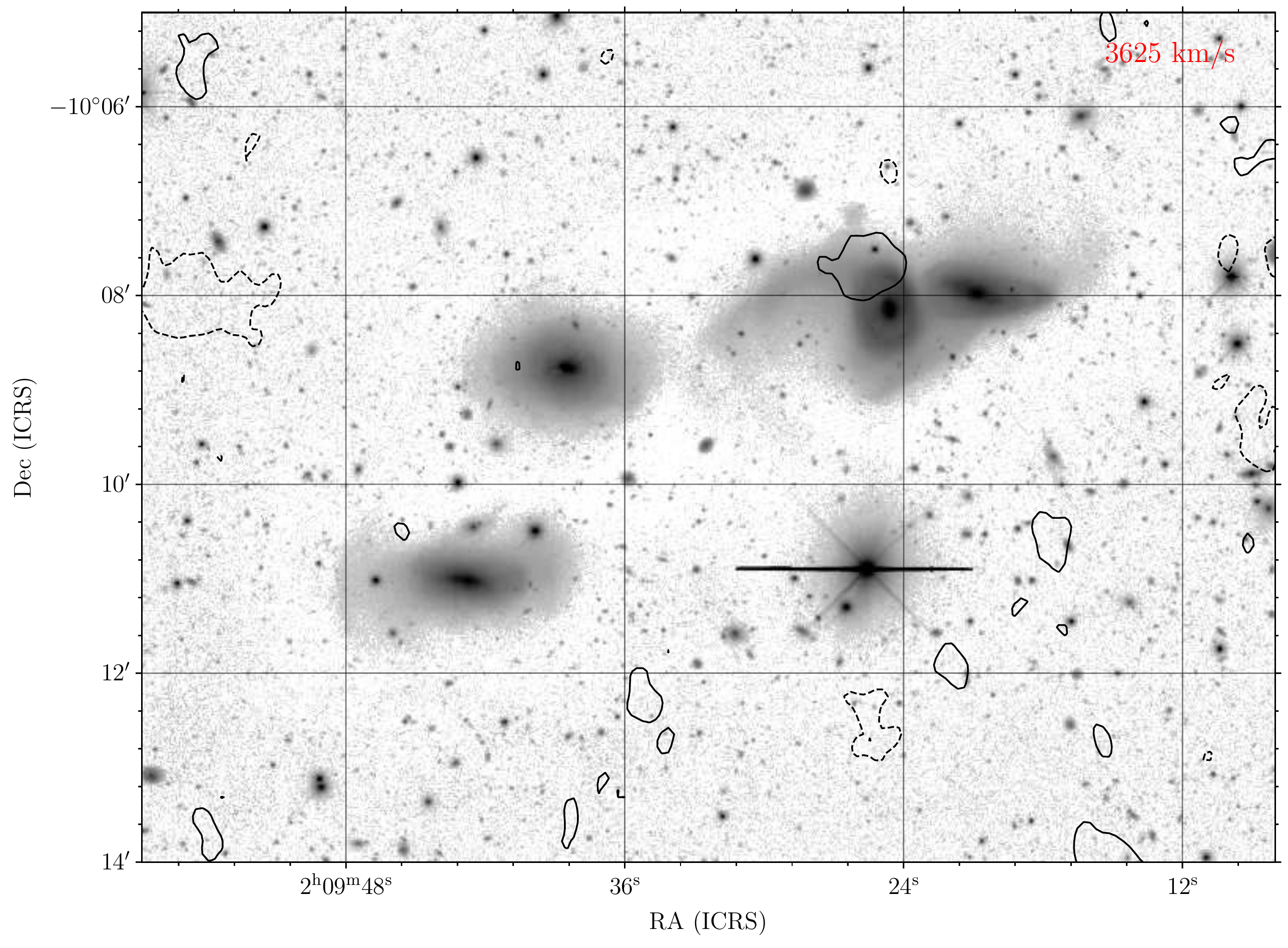}
    \includegraphics[width=0.33\columnwidth]{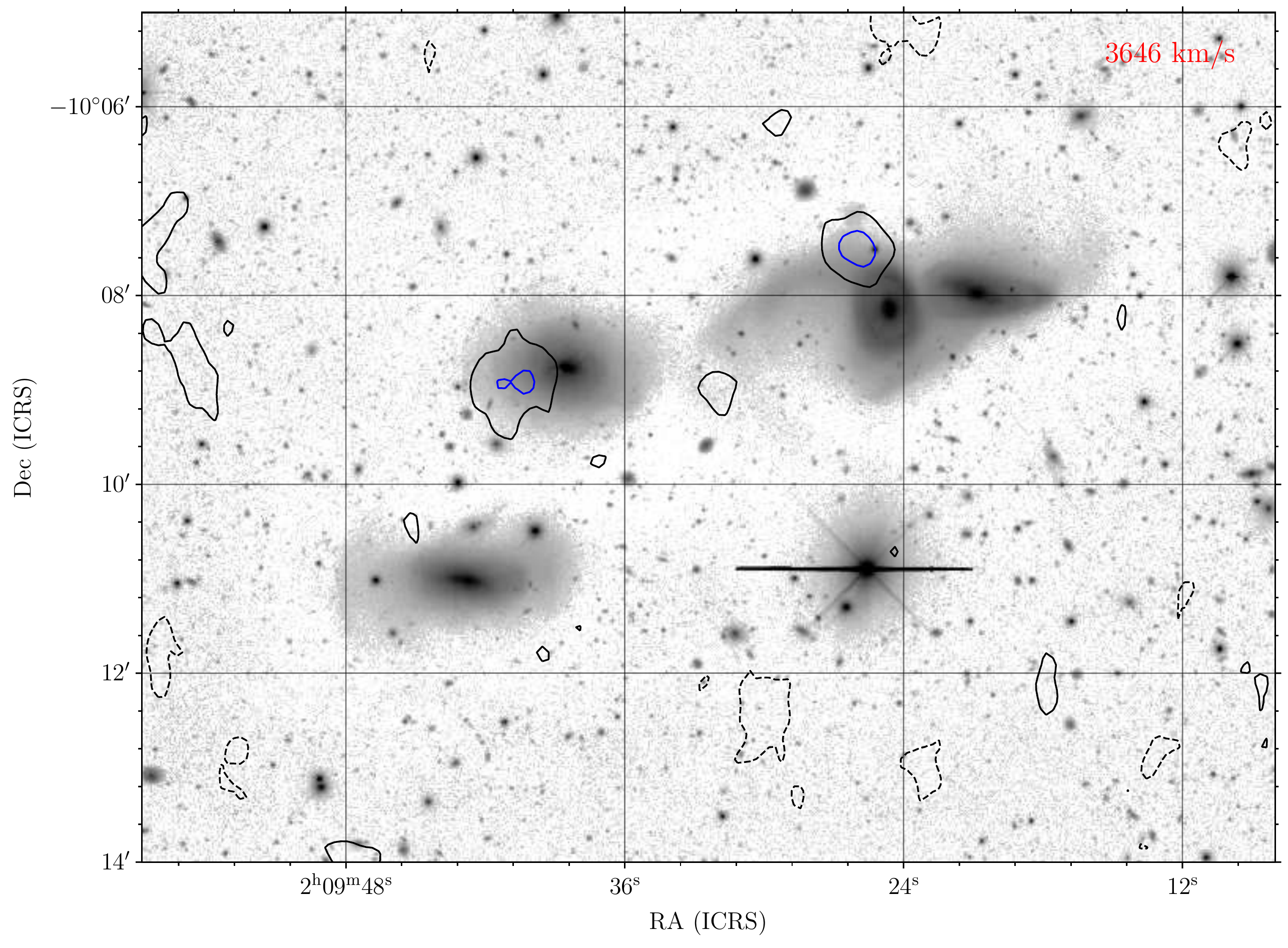}
    \includegraphics[width=0.33\columnwidth]{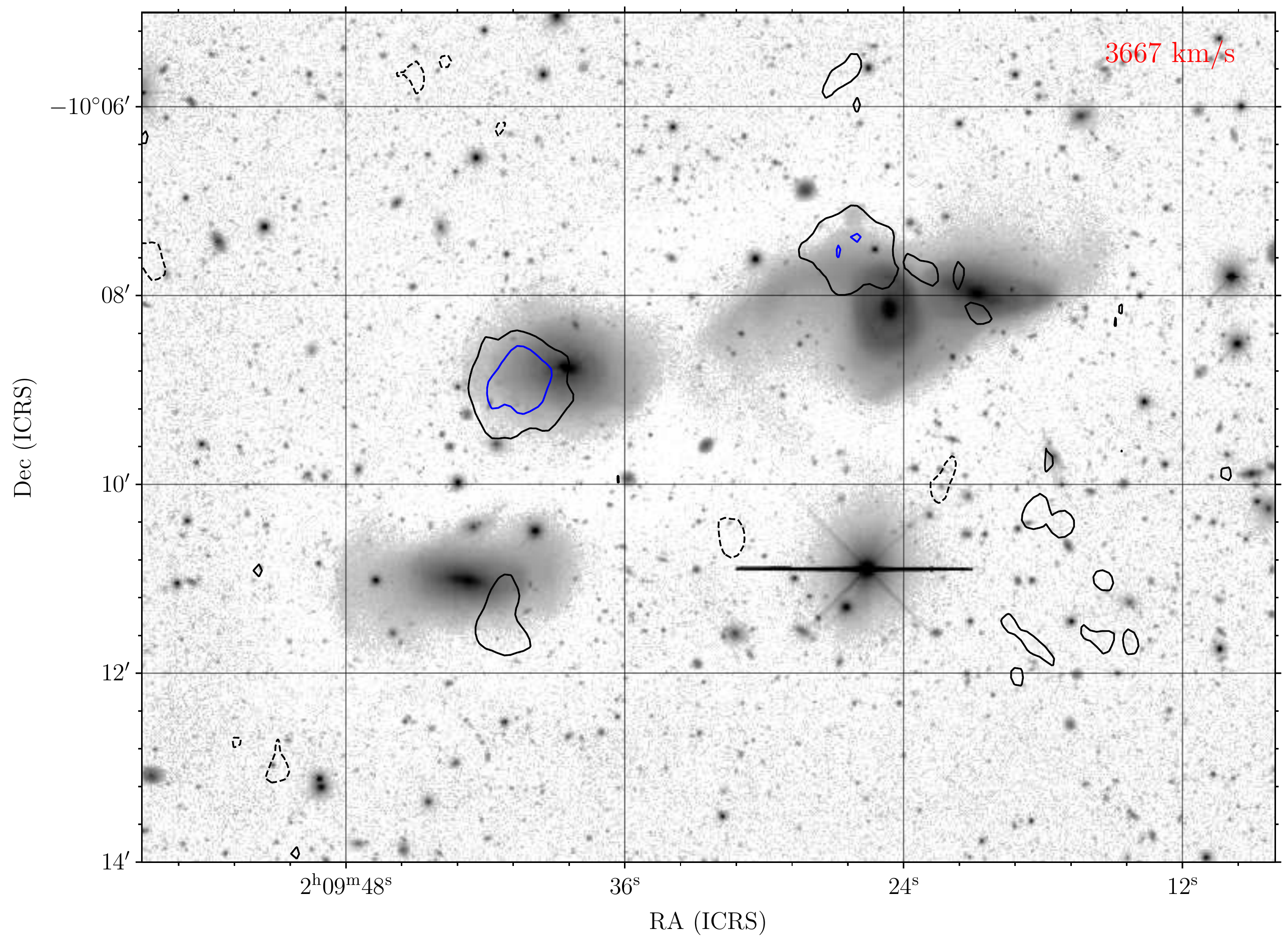}
    \includegraphics[width=0.33\columnwidth]{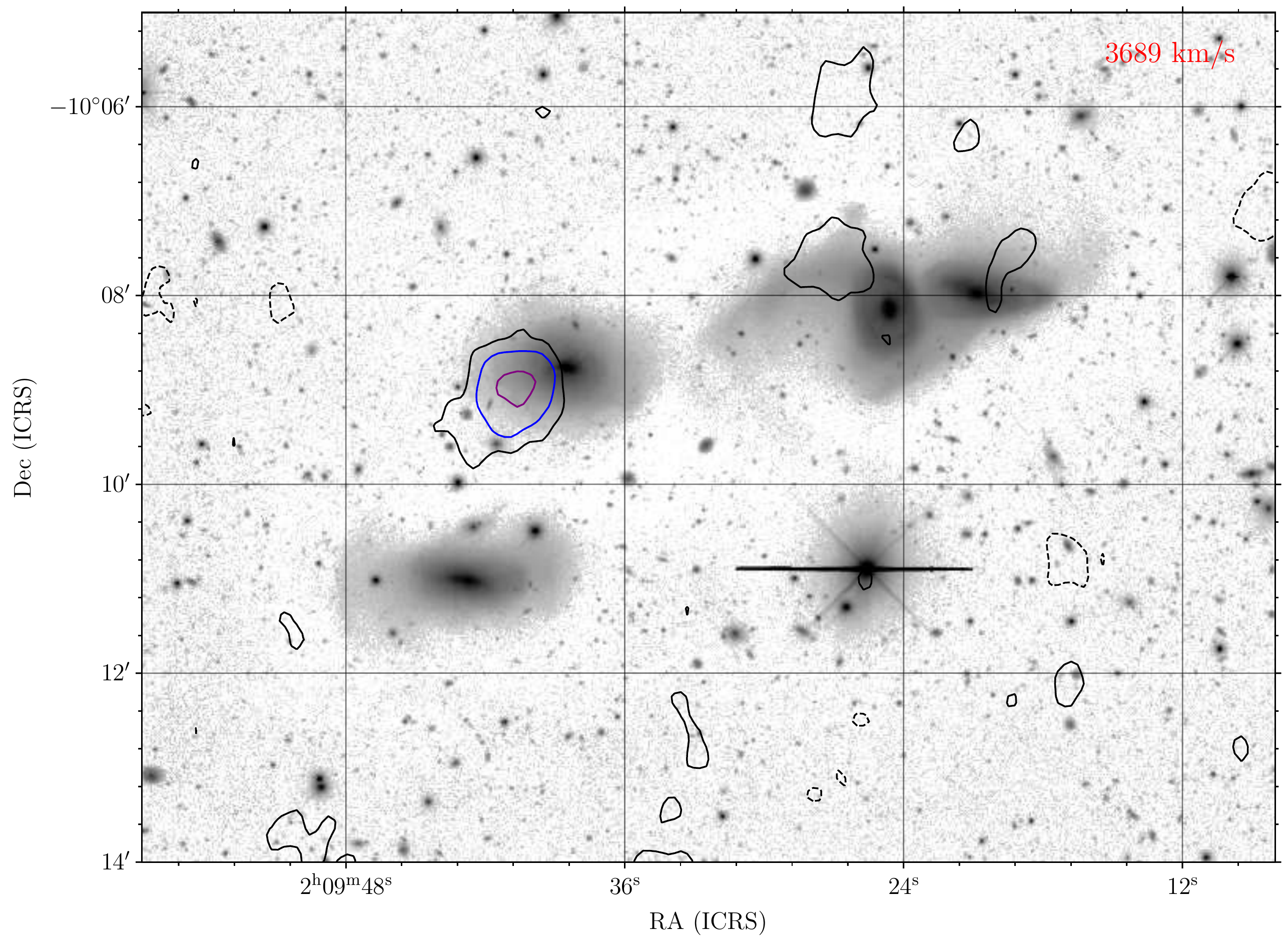}
    \caption{Channel maps showing the \hi \ emission in the core group overlaid on the DECaLS $r$-band image. Contour levels: -1,1,2,4,8,16 $\times 0.9$ mJy per beam, $\times 1.8 \times 10^{19} \; \mathrm{cm^{-2}}$, or $\times 0.15 \; \mathrm{M_{\odot}\,pc^{-2}}$. The contours are coloured (in order of increasing flux): black (dashed), black, blue, purple, red, and orange.}
    \label{fig:chn_maps_core}
\end{figure}
\end{landscape}

\newpage
\begin{landscape}
\begin{figure}
\centering
    \includegraphics[width=0.33\columnwidth]{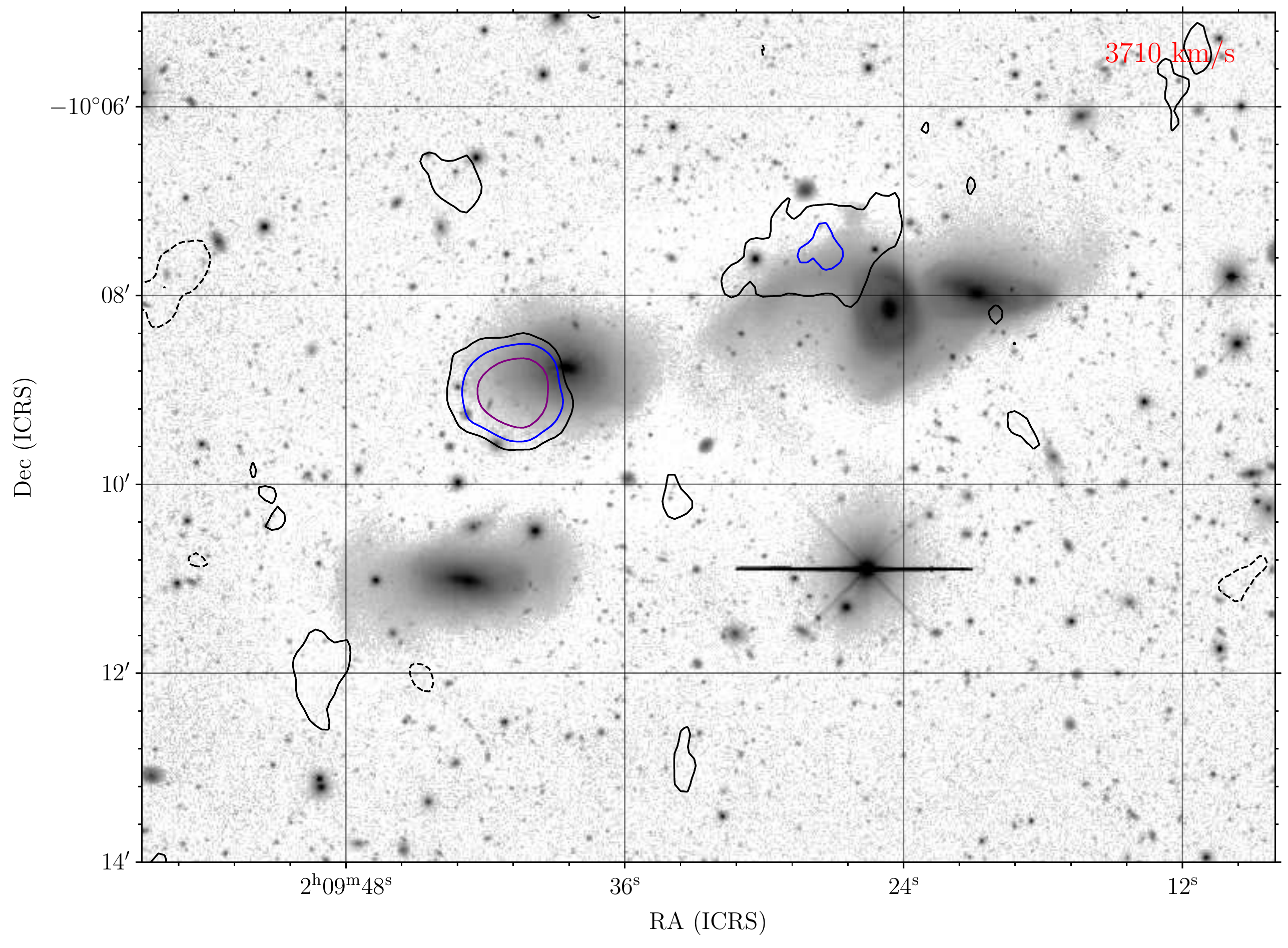}
    \includegraphics[width=0.33\columnwidth]{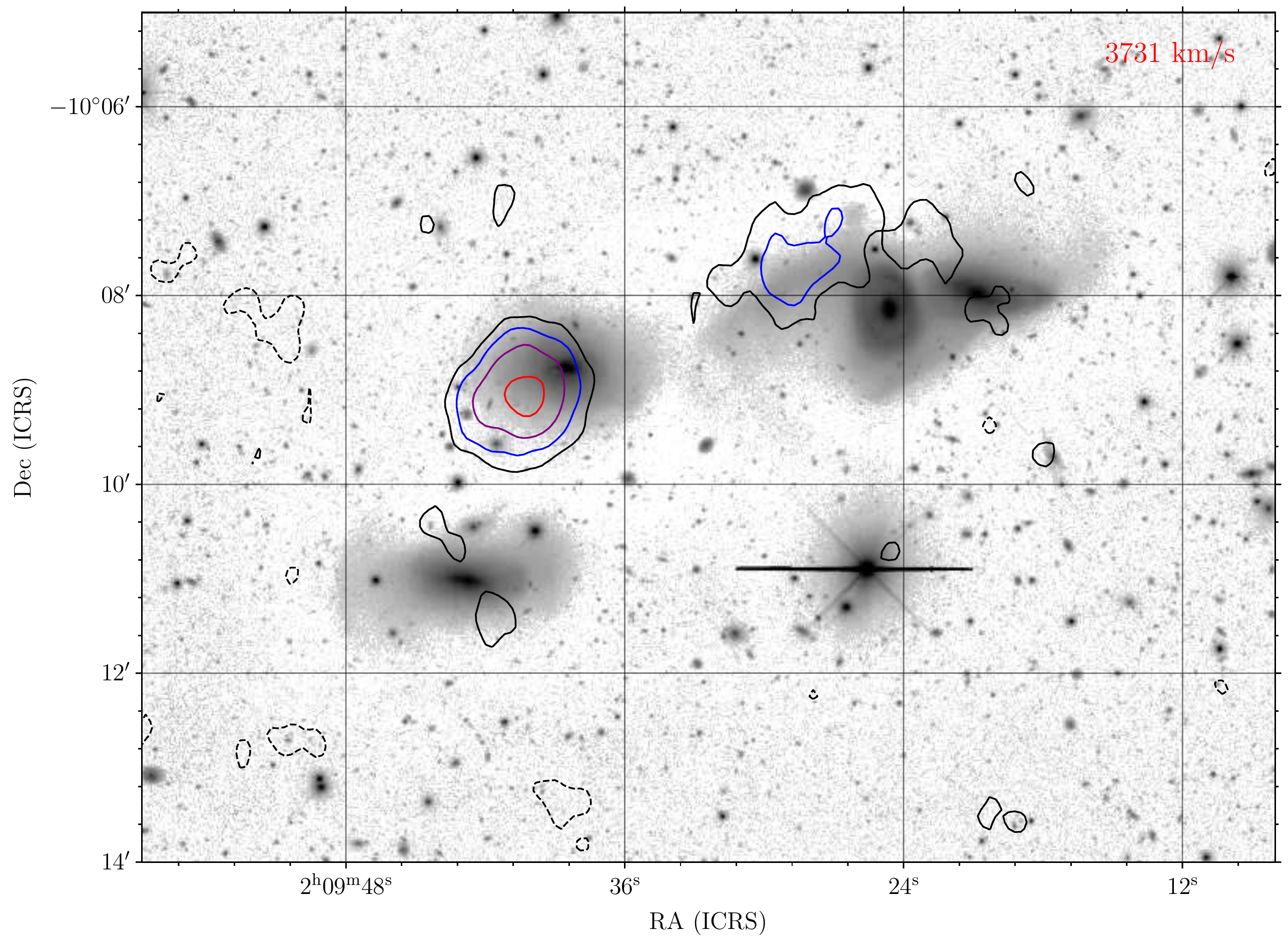}
    \includegraphics[width=0.33\columnwidth]{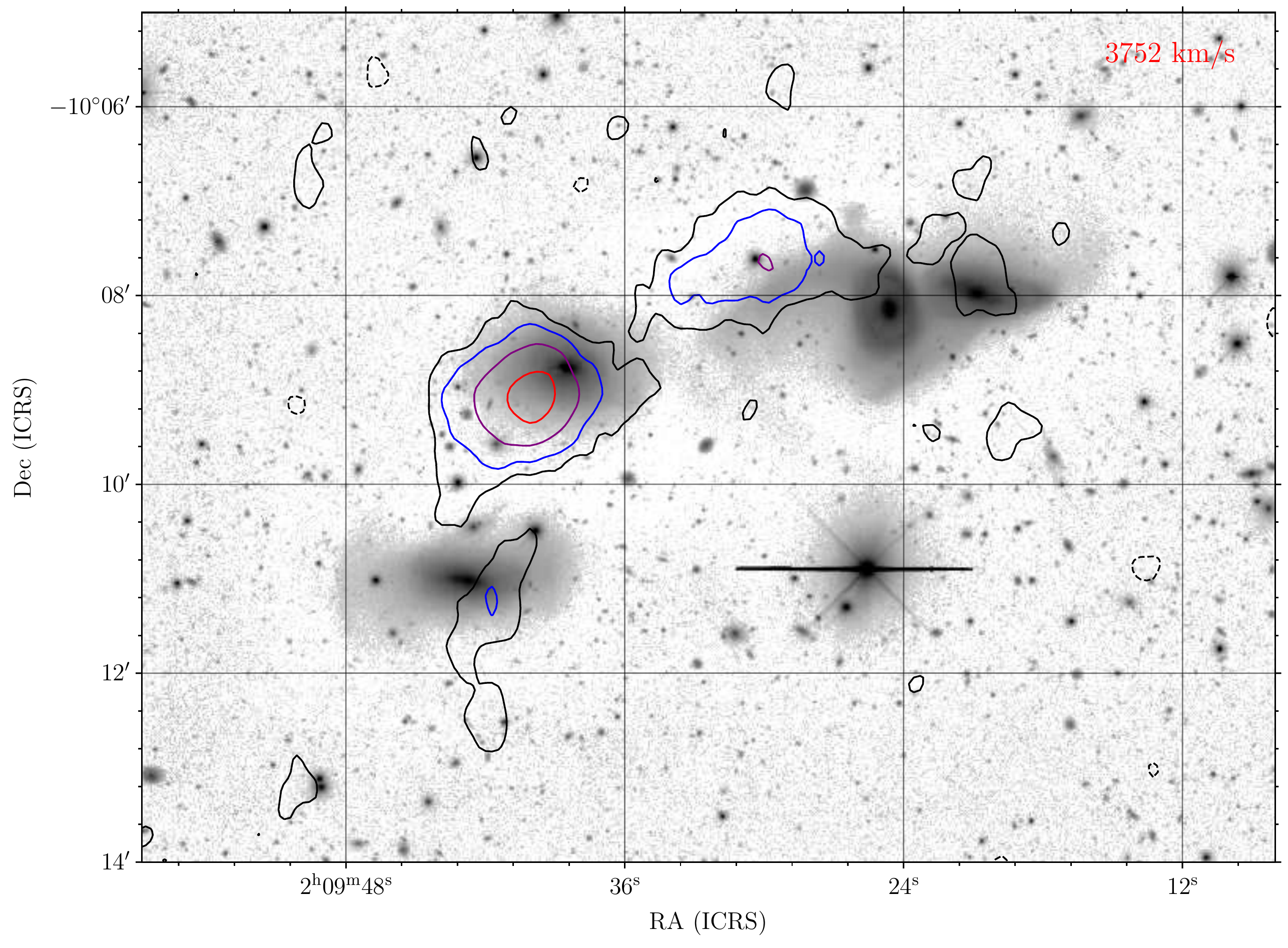}
    \includegraphics[width=0.33\columnwidth]{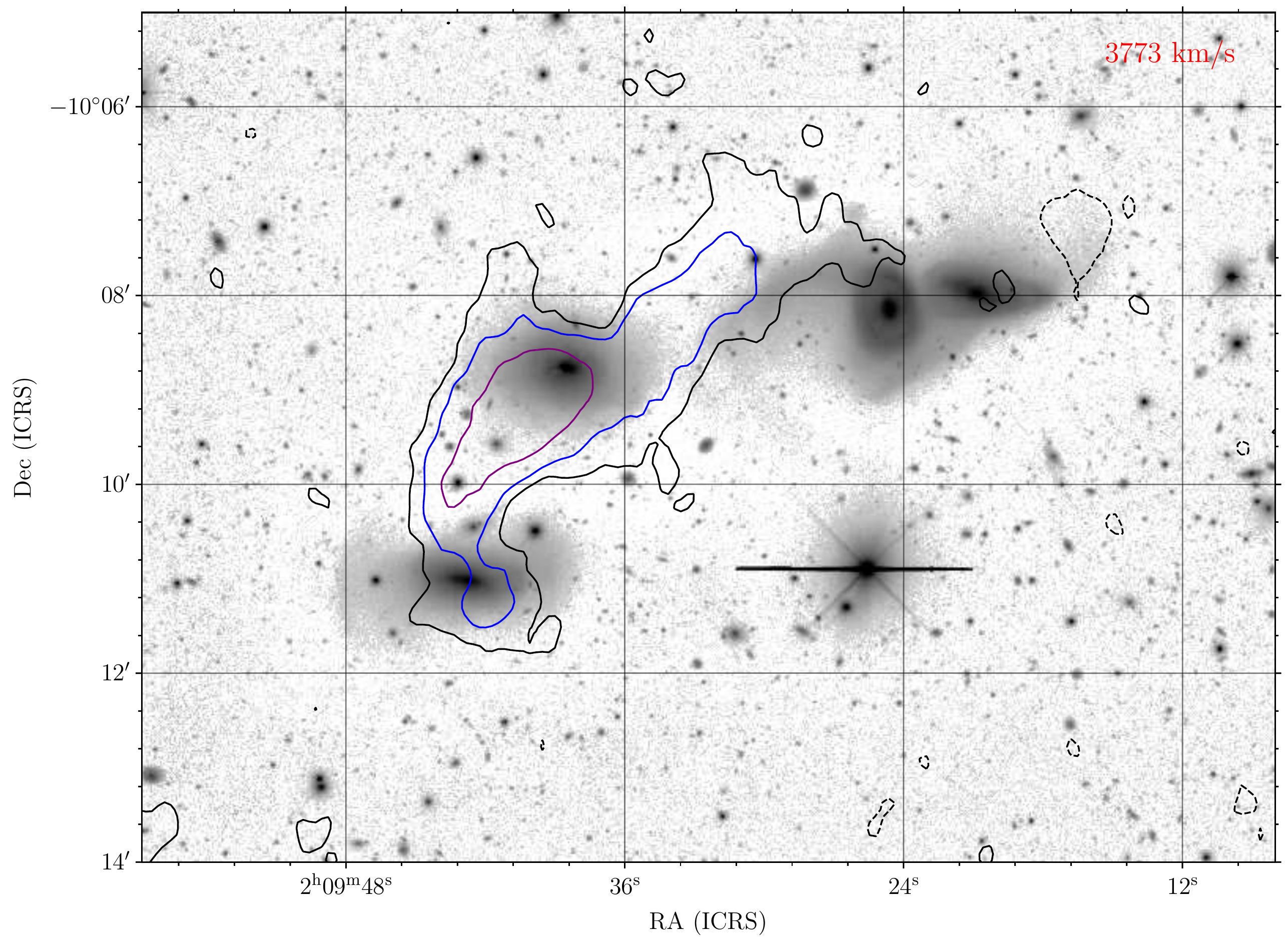}
    \includegraphics[width=0.33\columnwidth]{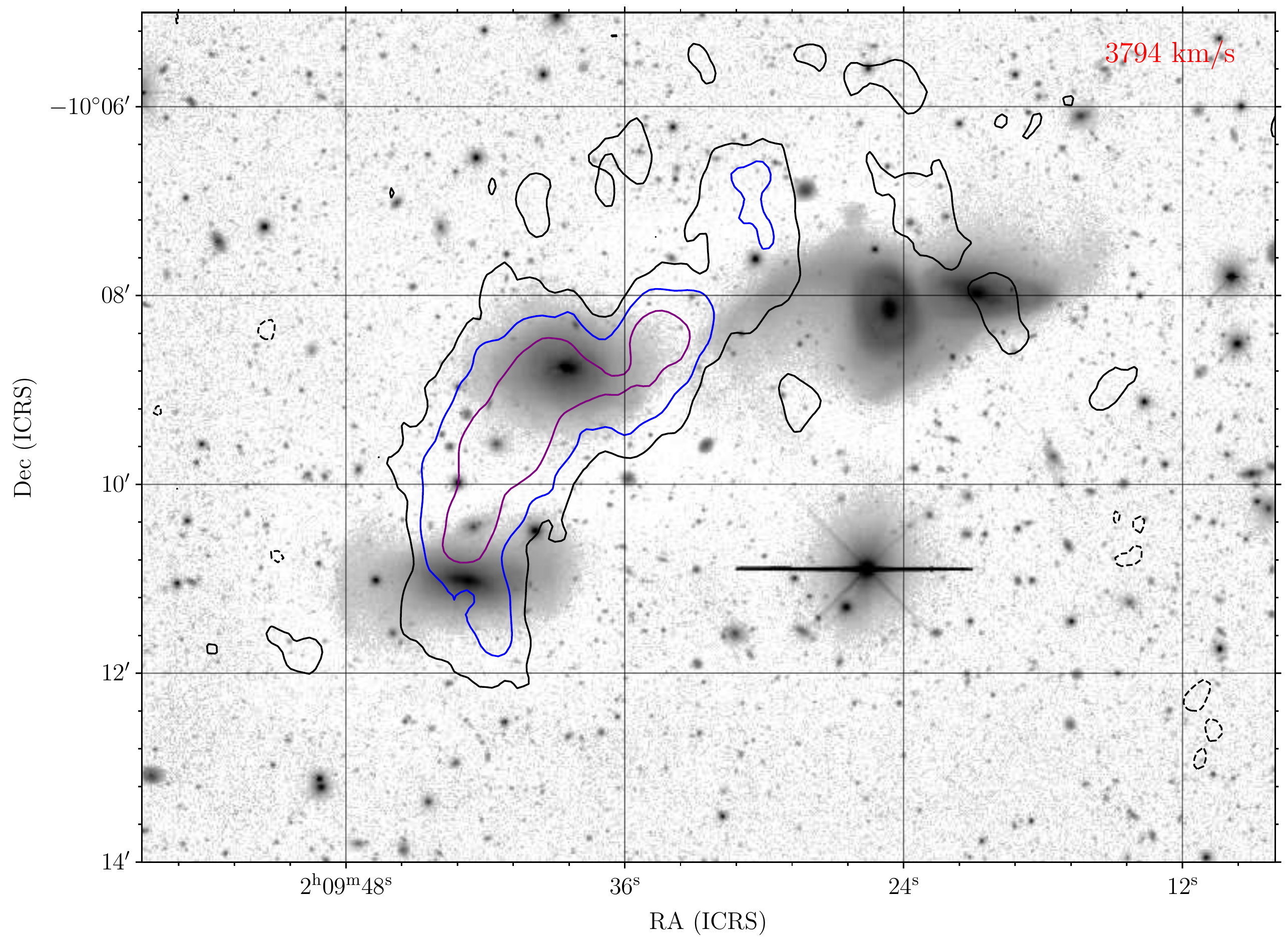}
    \includegraphics[width=0.33\columnwidth]{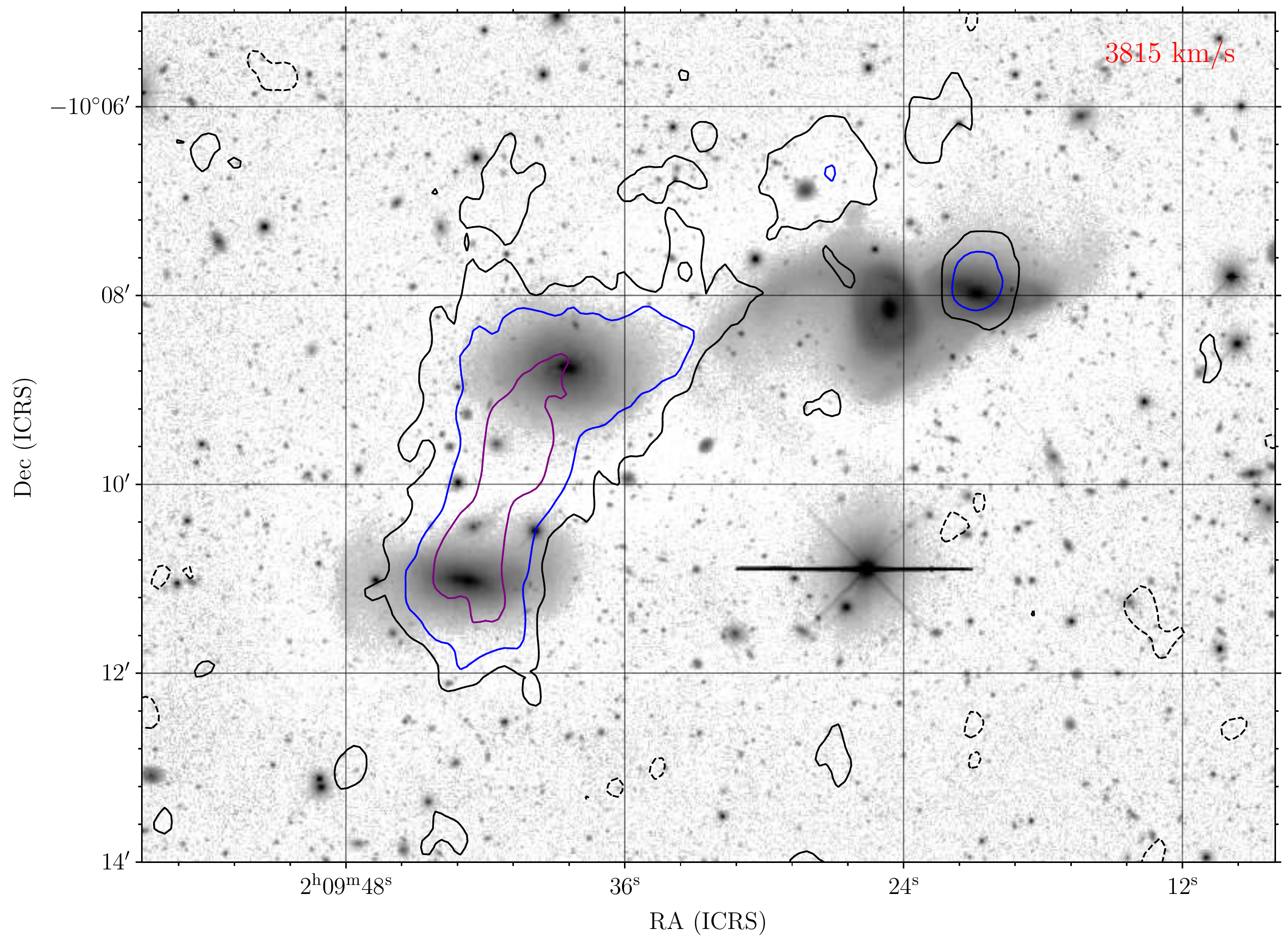}
    \includegraphics[width=0.33\columnwidth]{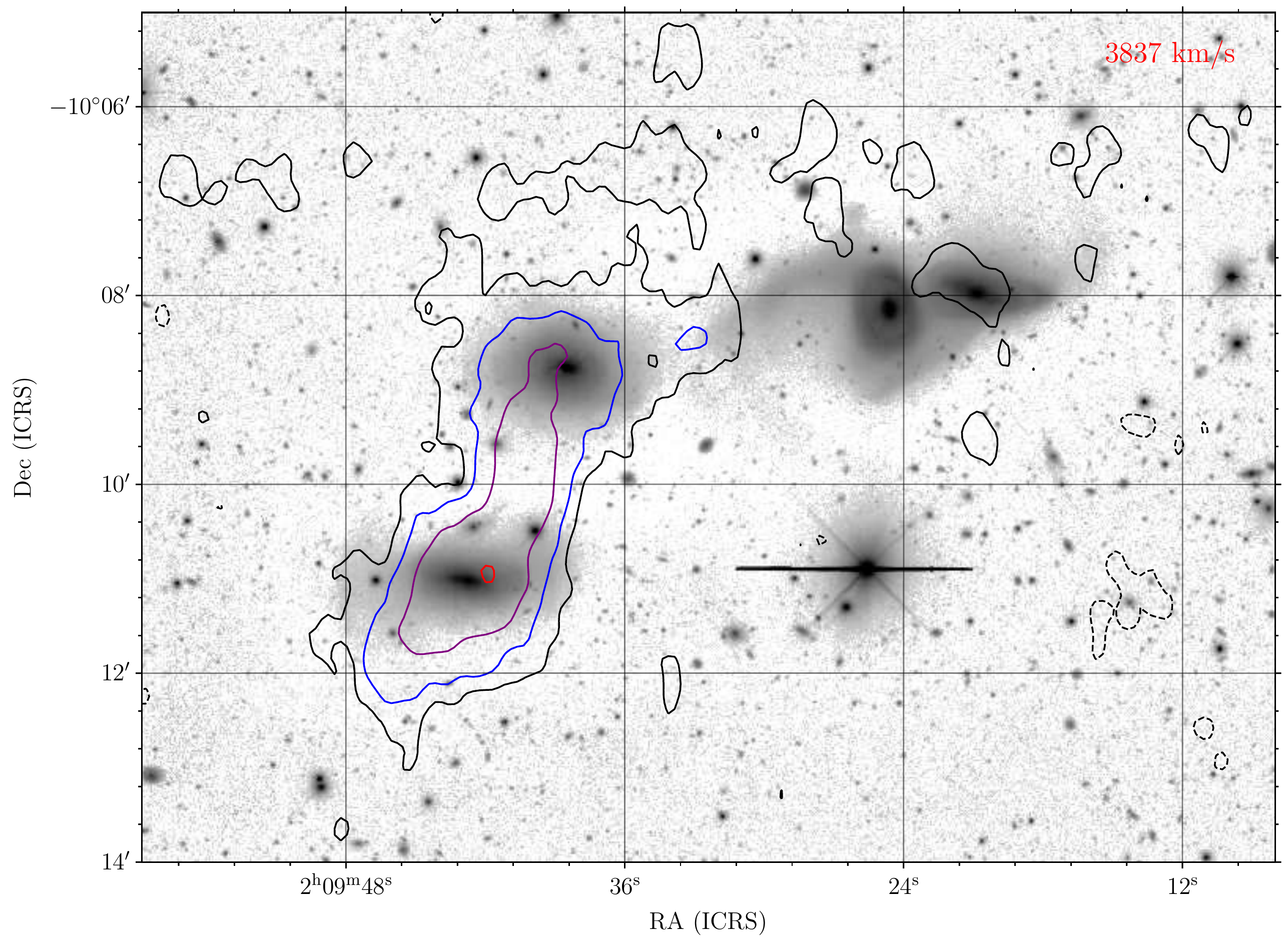}
    \includegraphics[width=0.33\columnwidth]{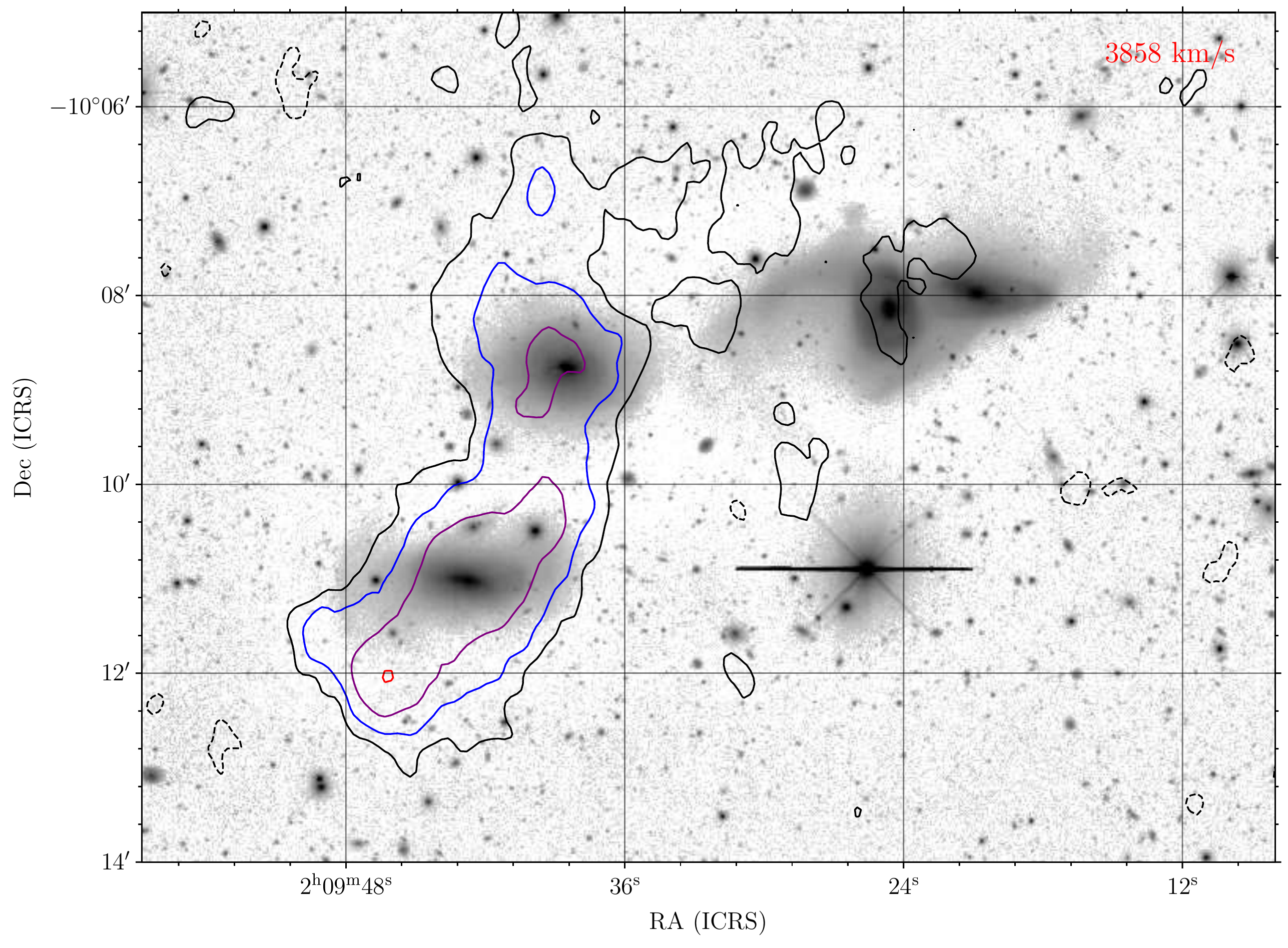}
    \includegraphics[width=0.33\columnwidth]{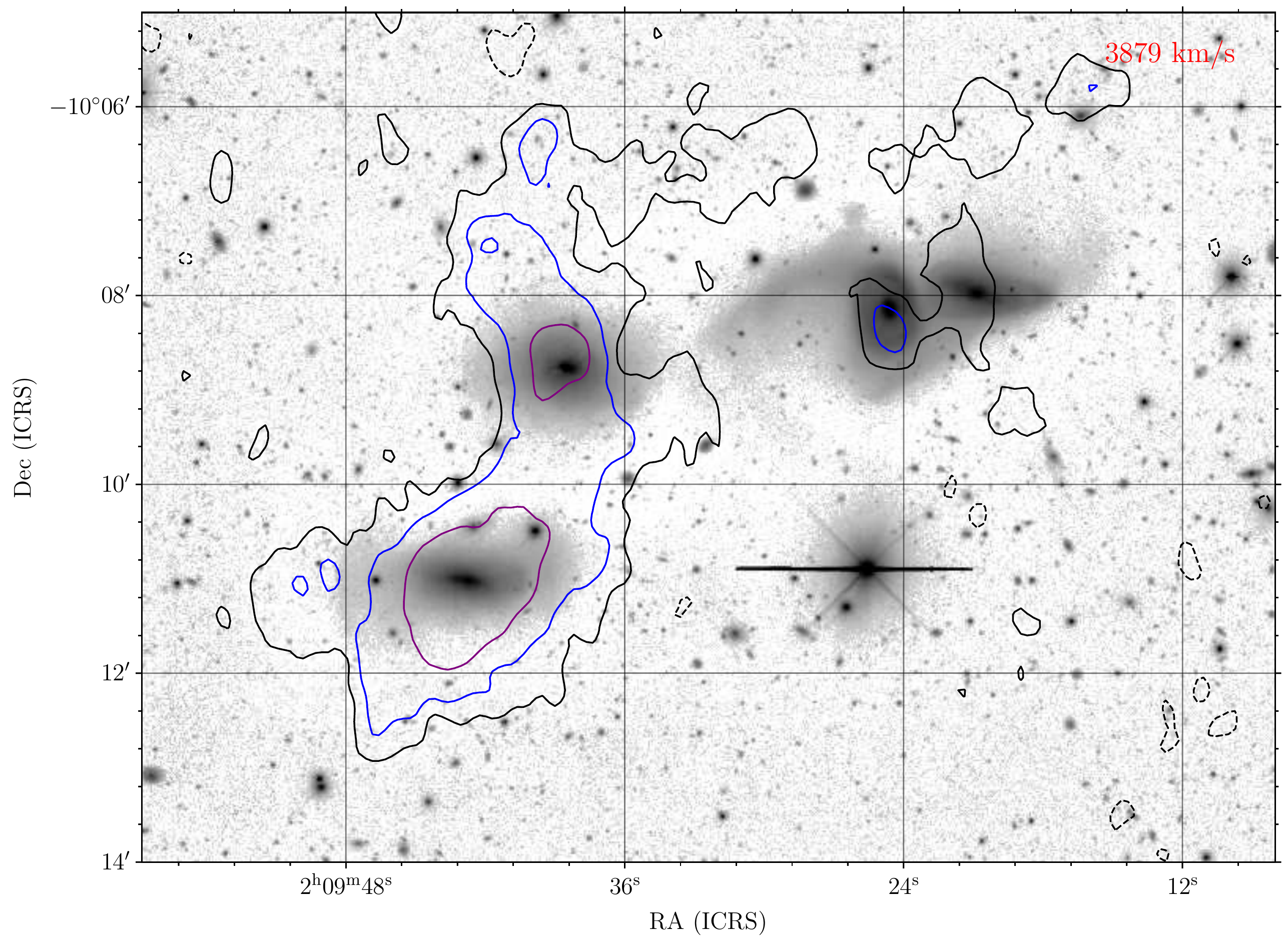}
\end{figure}
\end{landscape}

\newpage
\begin{landscape}
\begin{figure}
\centering
    \includegraphics[width=0.33\columnwidth]{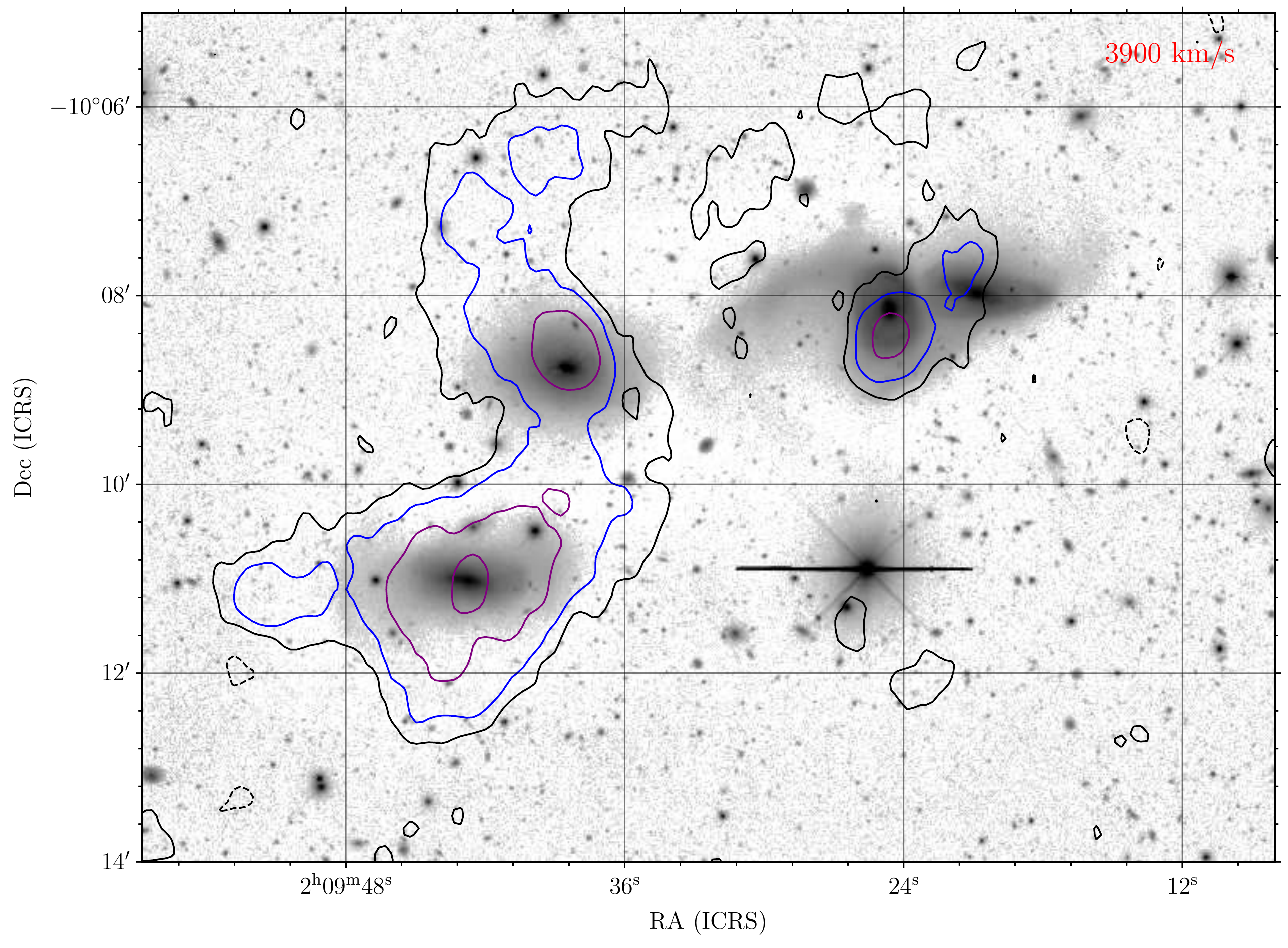}
    \includegraphics[width=0.33\columnwidth]{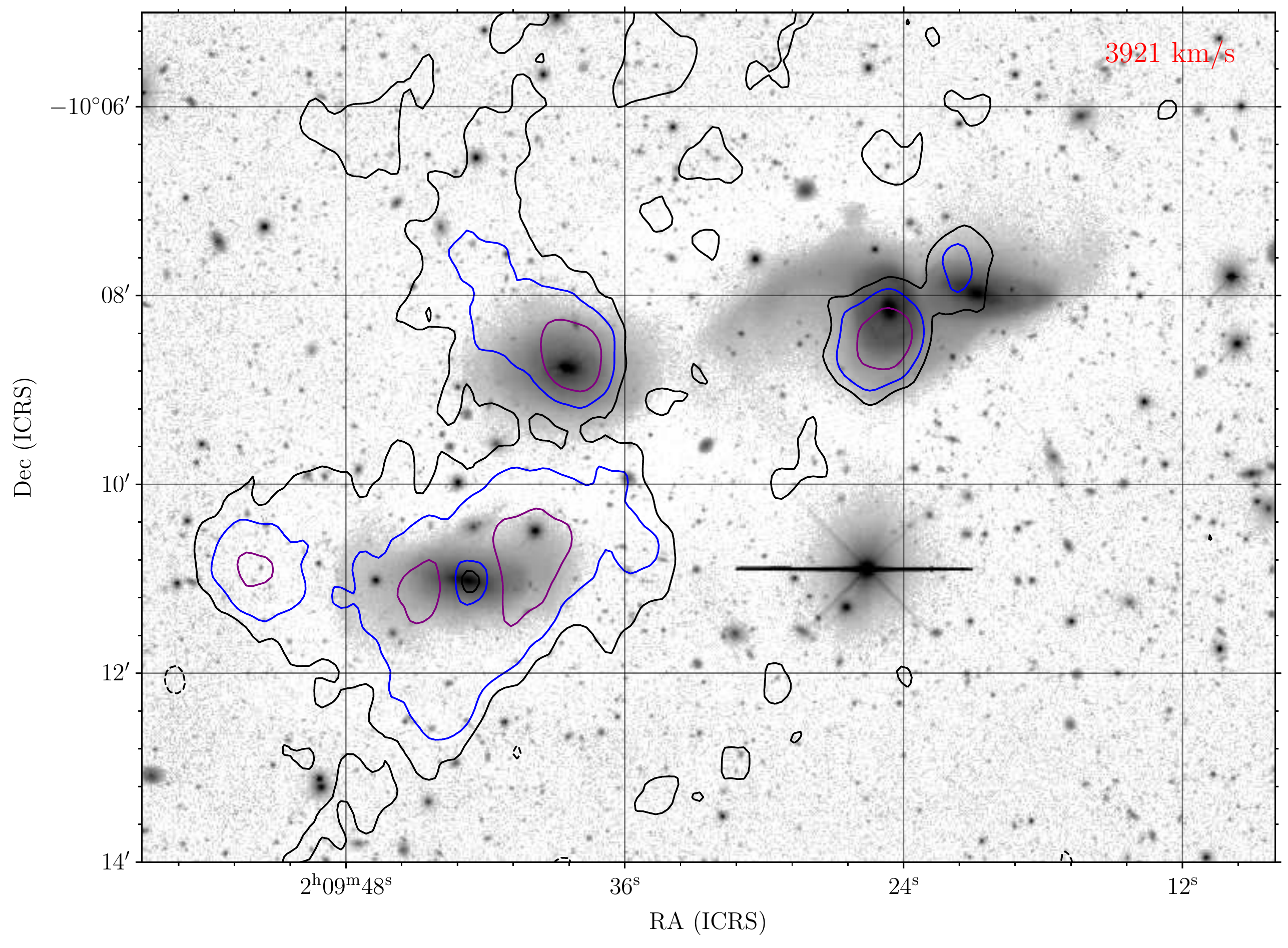}
    \includegraphics[width=0.33\columnwidth]{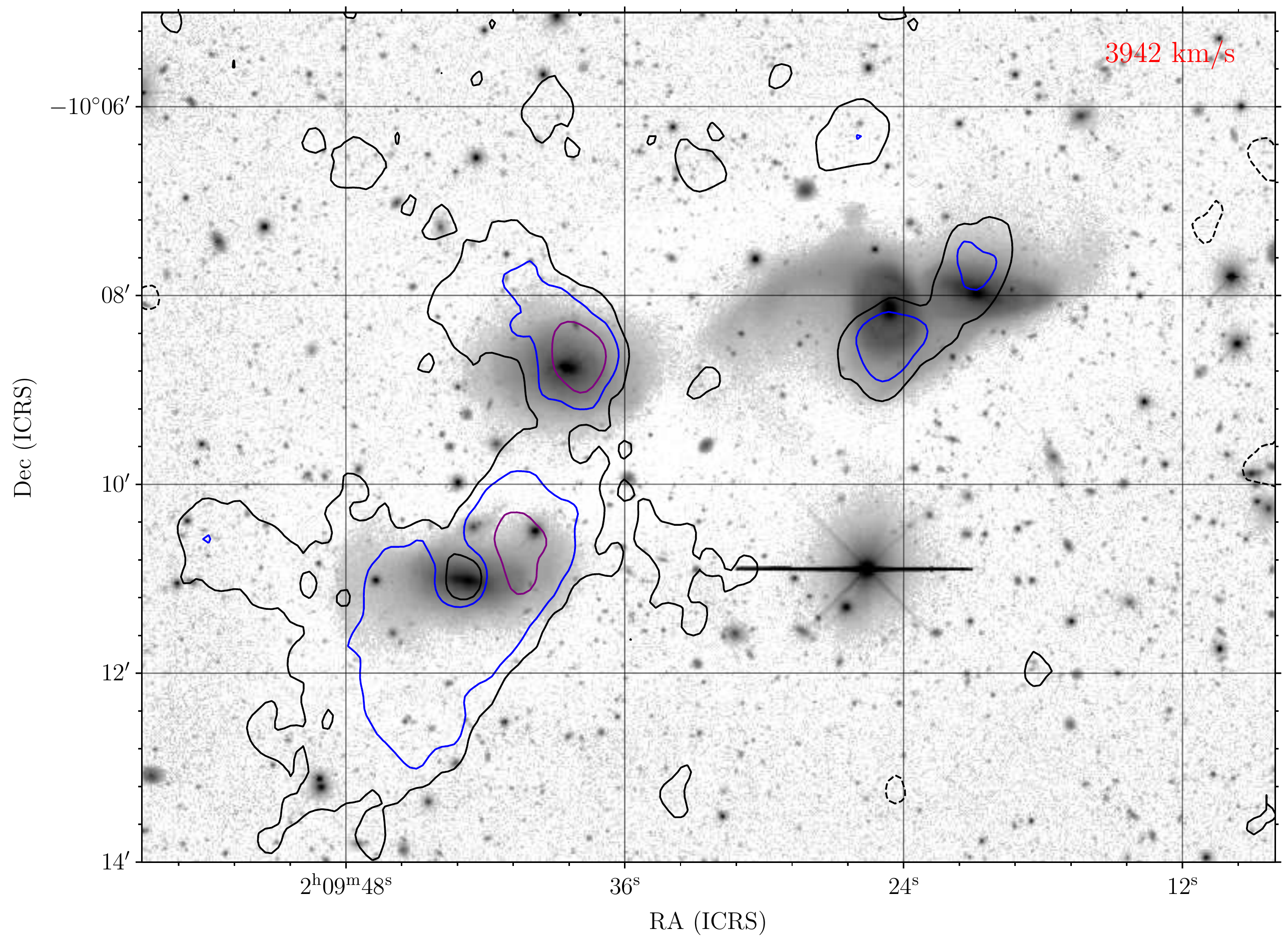}
    \includegraphics[width=0.33\columnwidth]{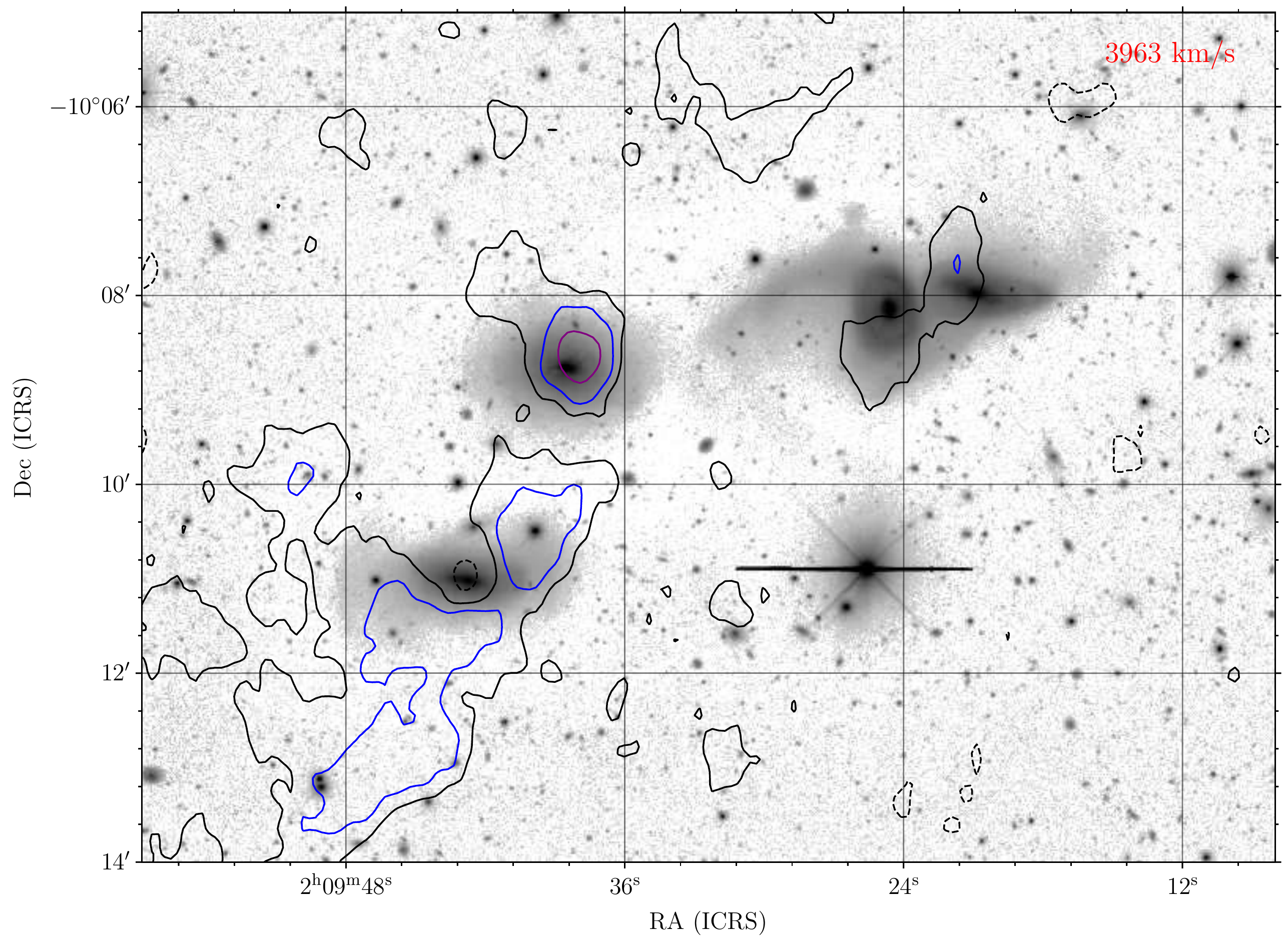}
    \includegraphics[width=0.33\columnwidth]{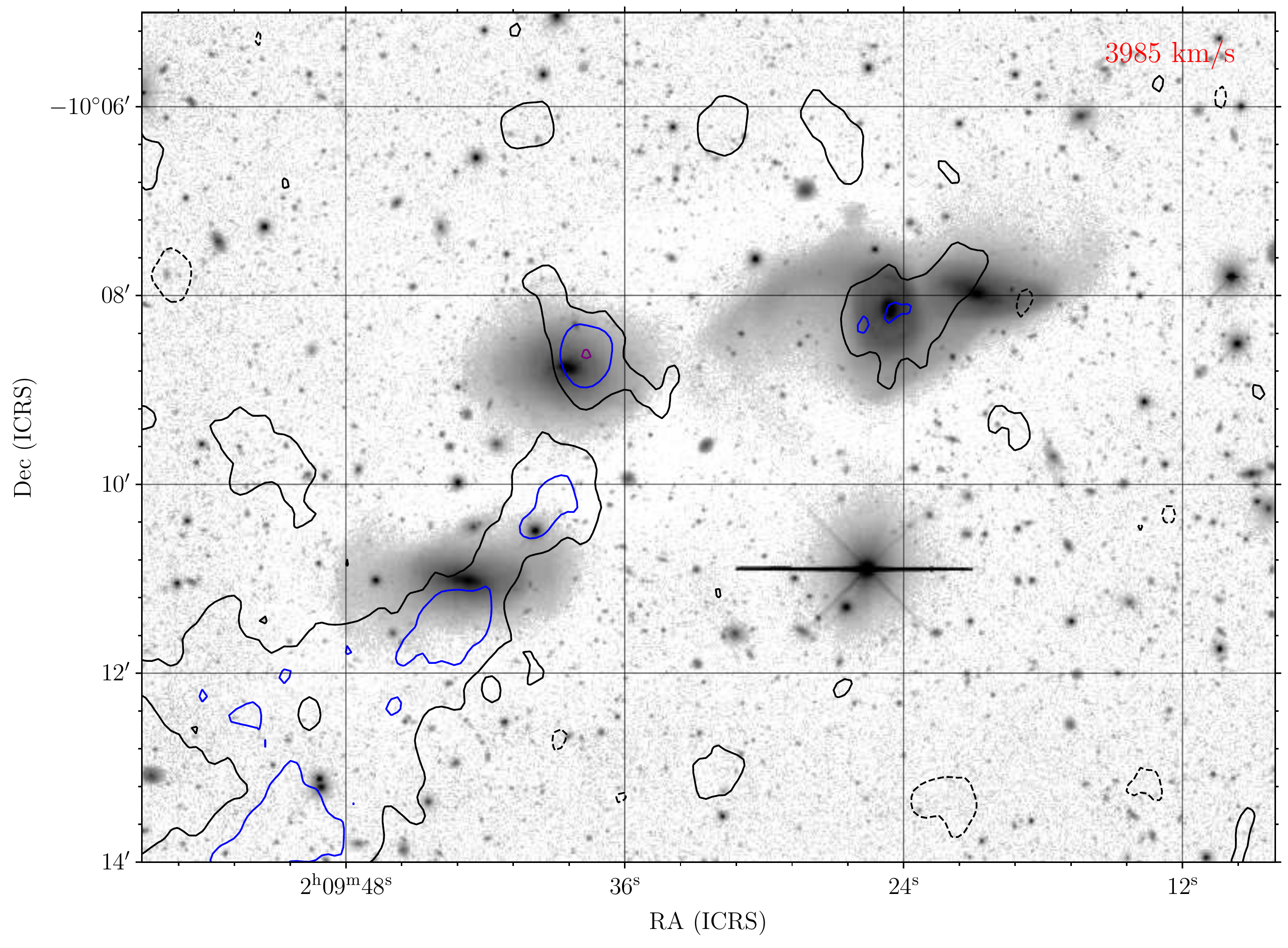}
    \includegraphics[width=0.33\columnwidth]{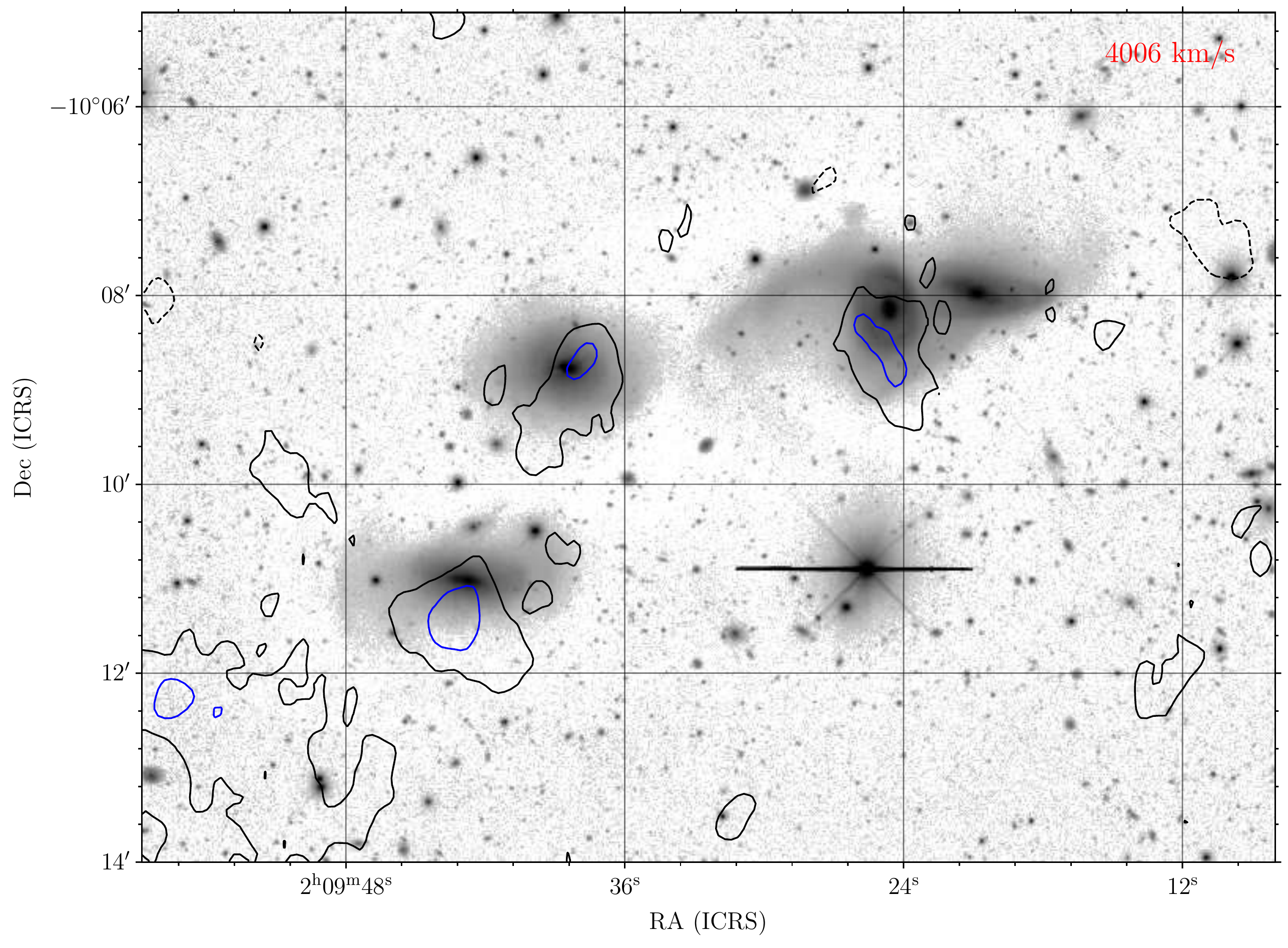}
    \includegraphics[width=0.33\columnwidth]{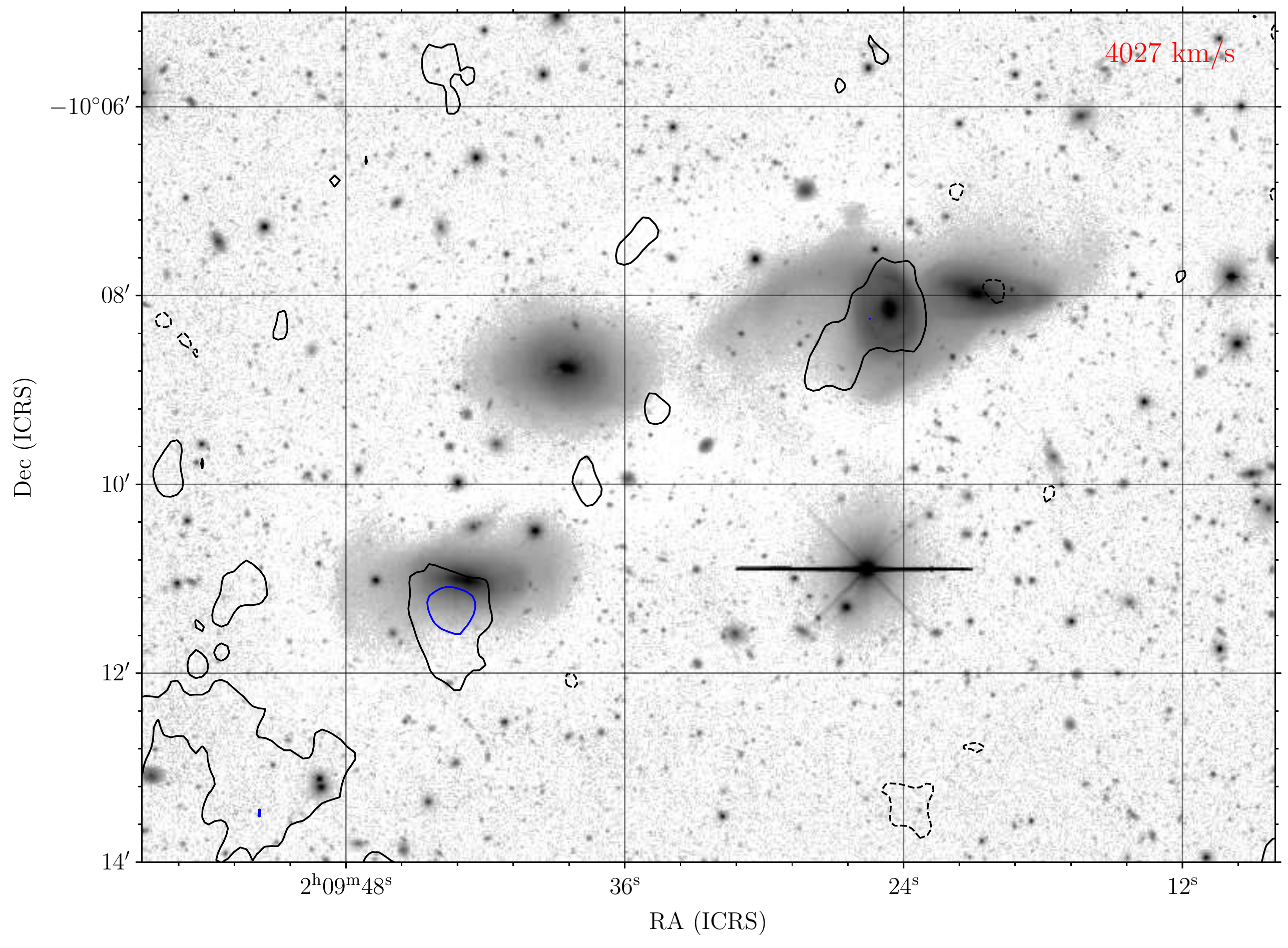}
    \includegraphics[width=0.33\columnwidth]{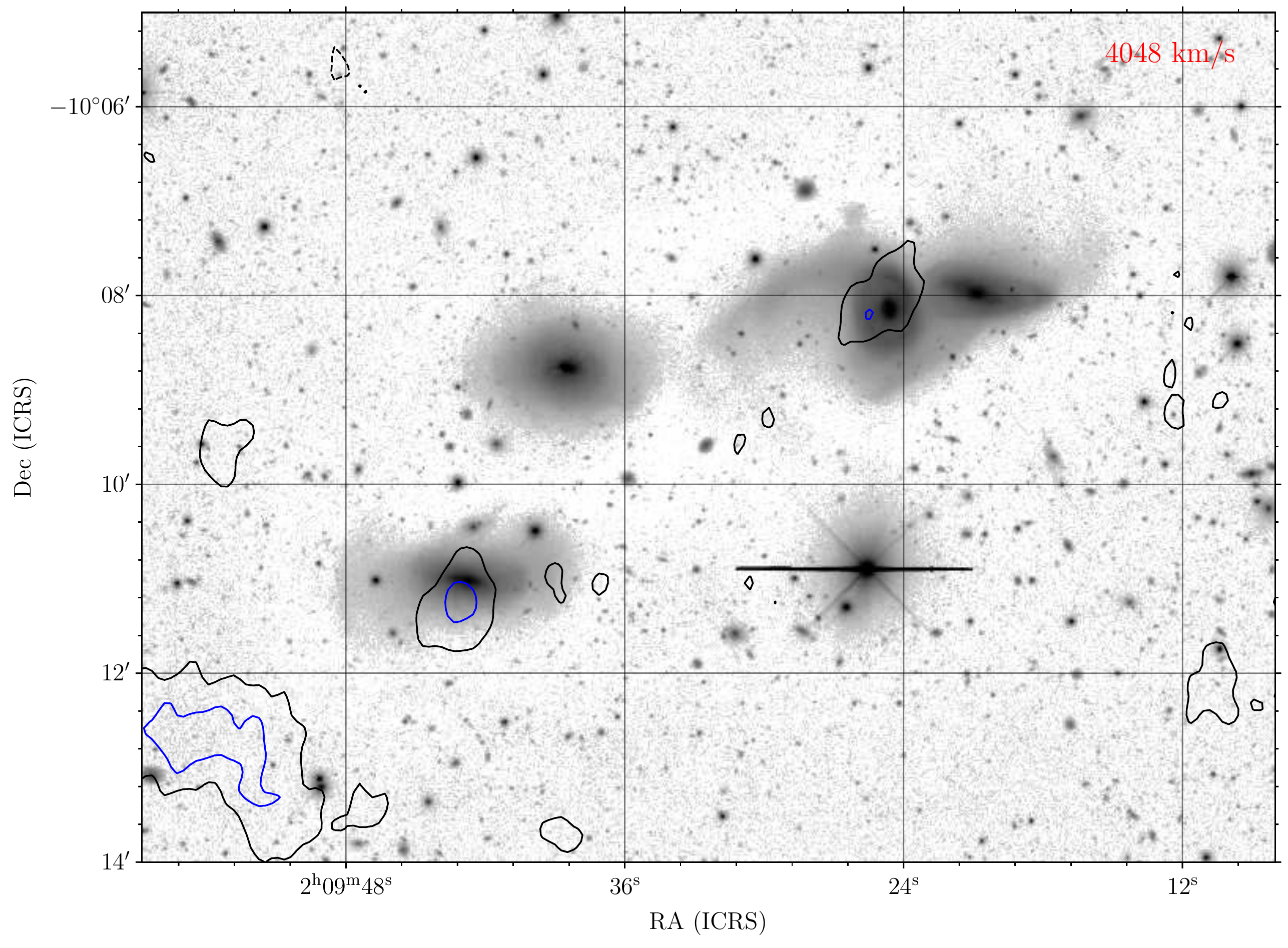}
    \includegraphics[width=0.33\columnwidth]{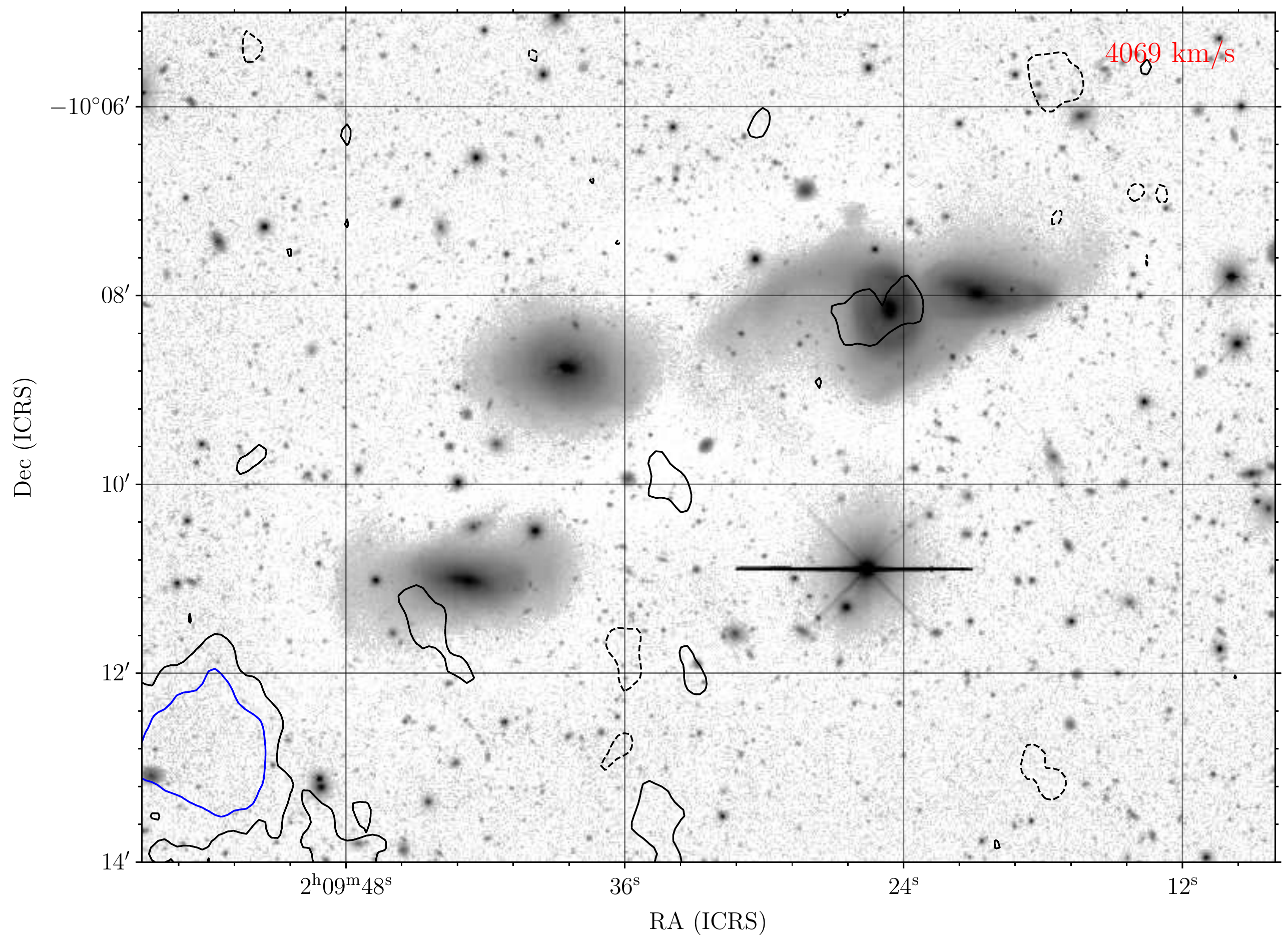}
\end{figure}
\end{landscape}

\newpage
\begin{landscape}
\begin{figure}
\centering
    \includegraphics[width=0.33\columnwidth]{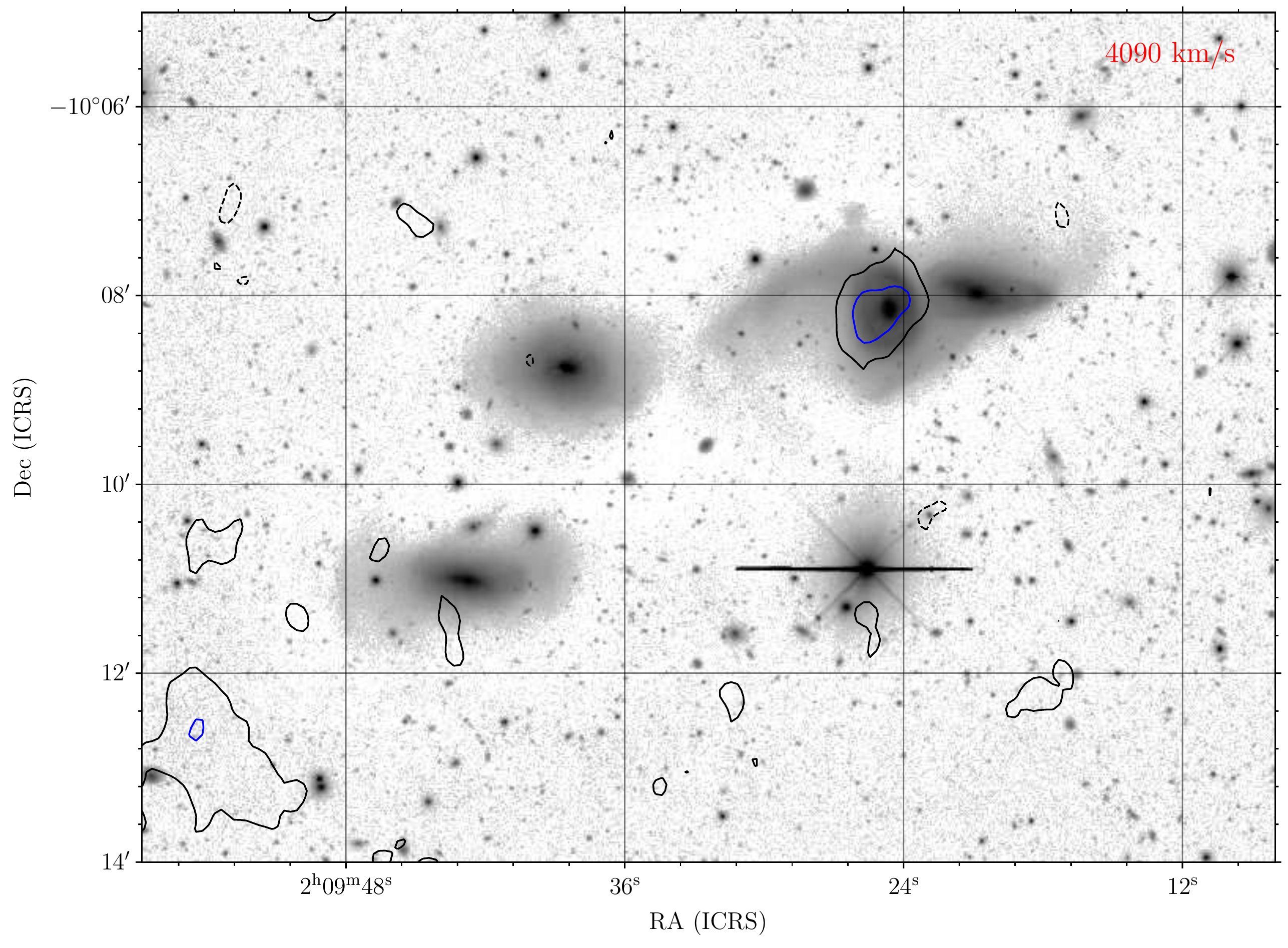}
    \includegraphics[width=0.33\columnwidth]{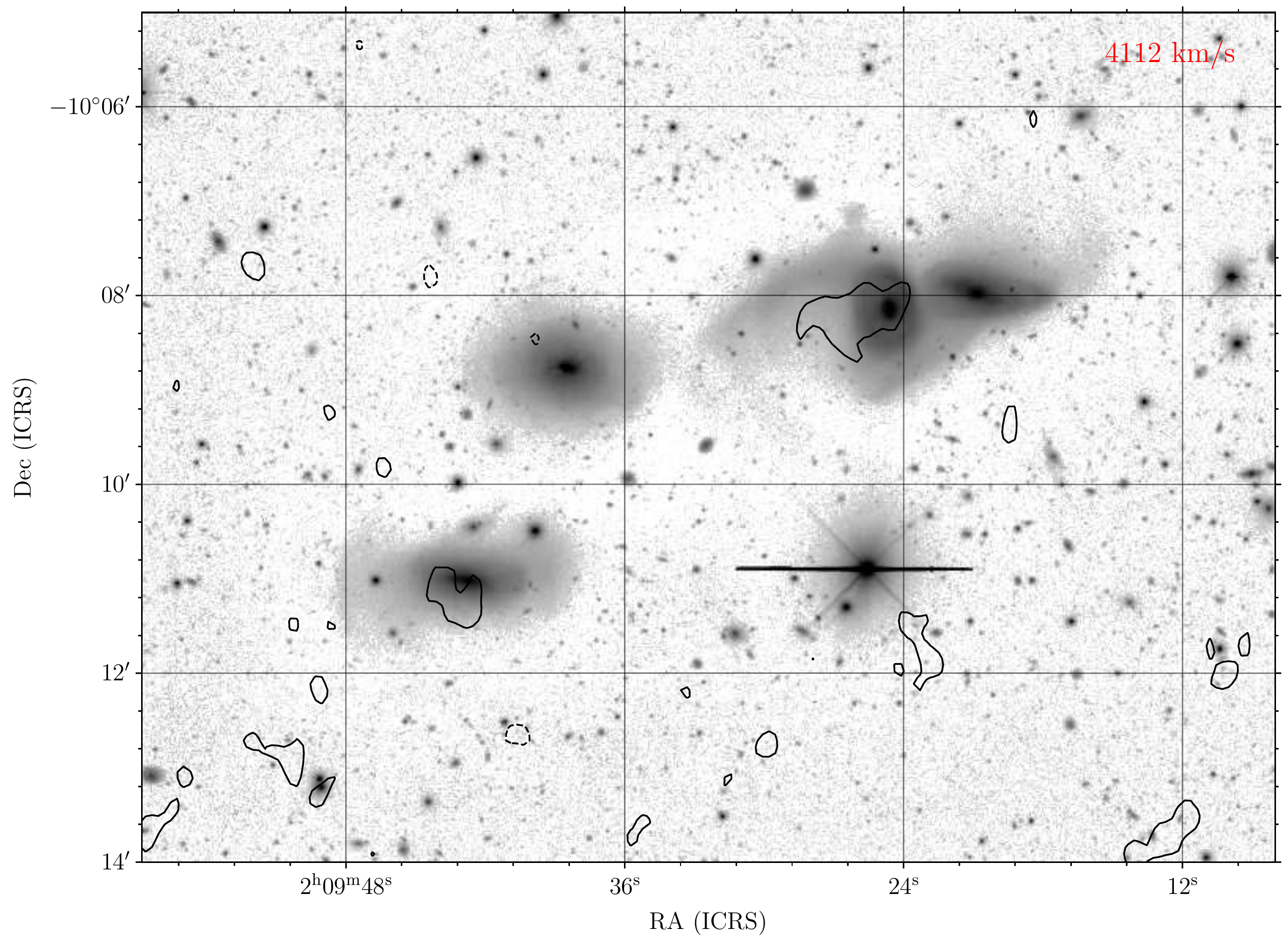}
    \includegraphics[width=0.33\columnwidth]{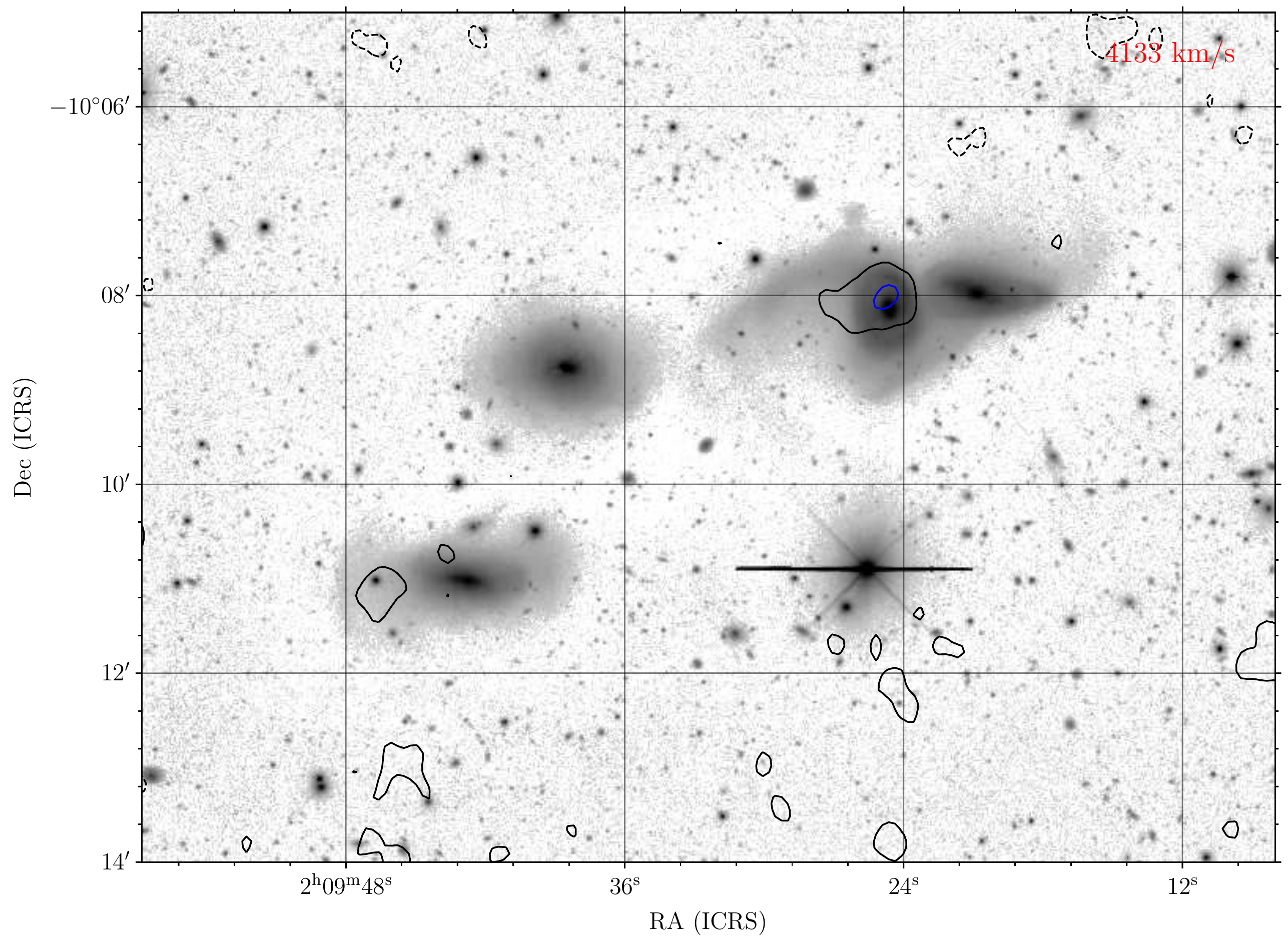}
    \includegraphics[width=0.33\columnwidth]{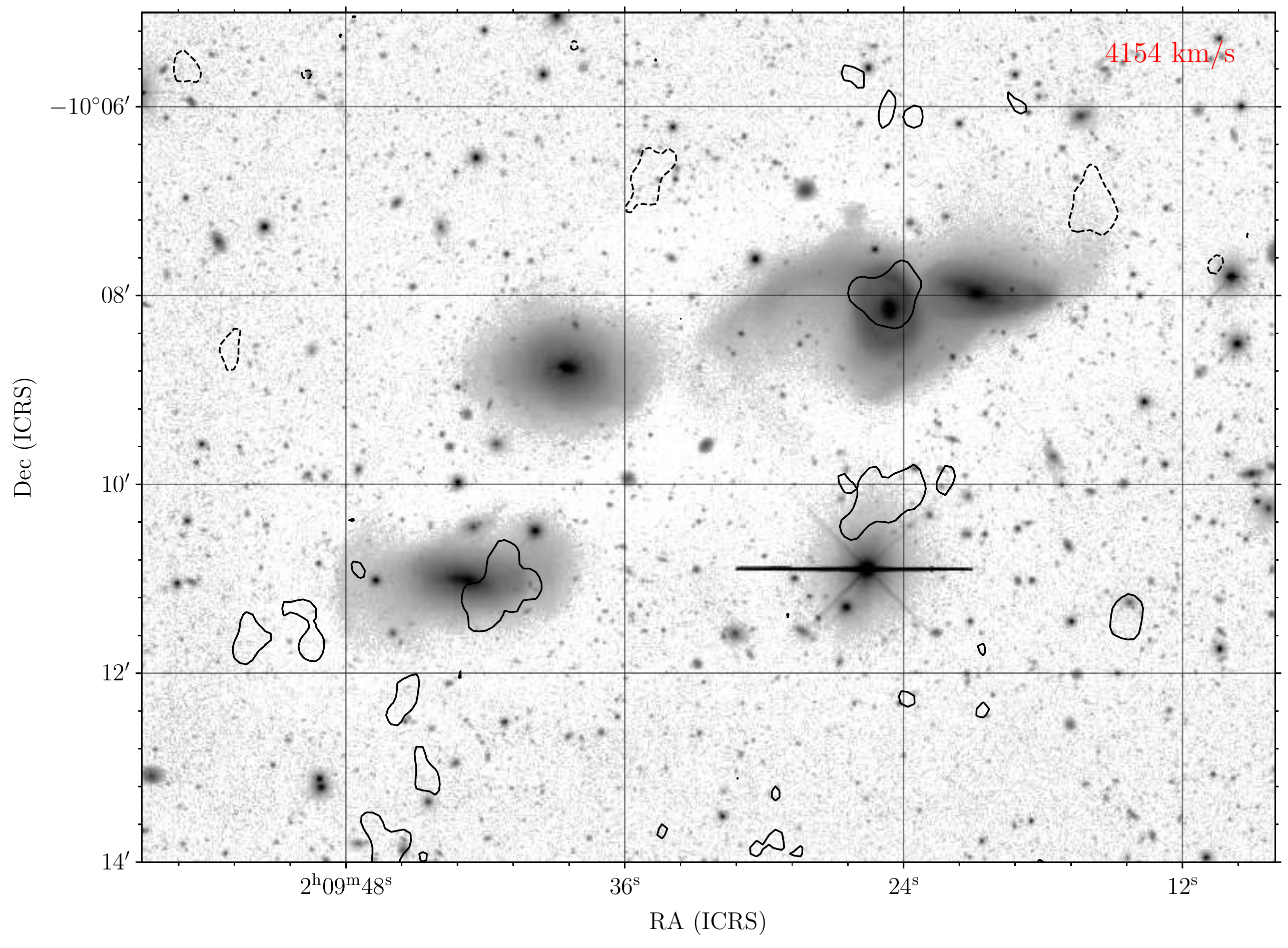}
    \includegraphics[width=0.33\columnwidth]{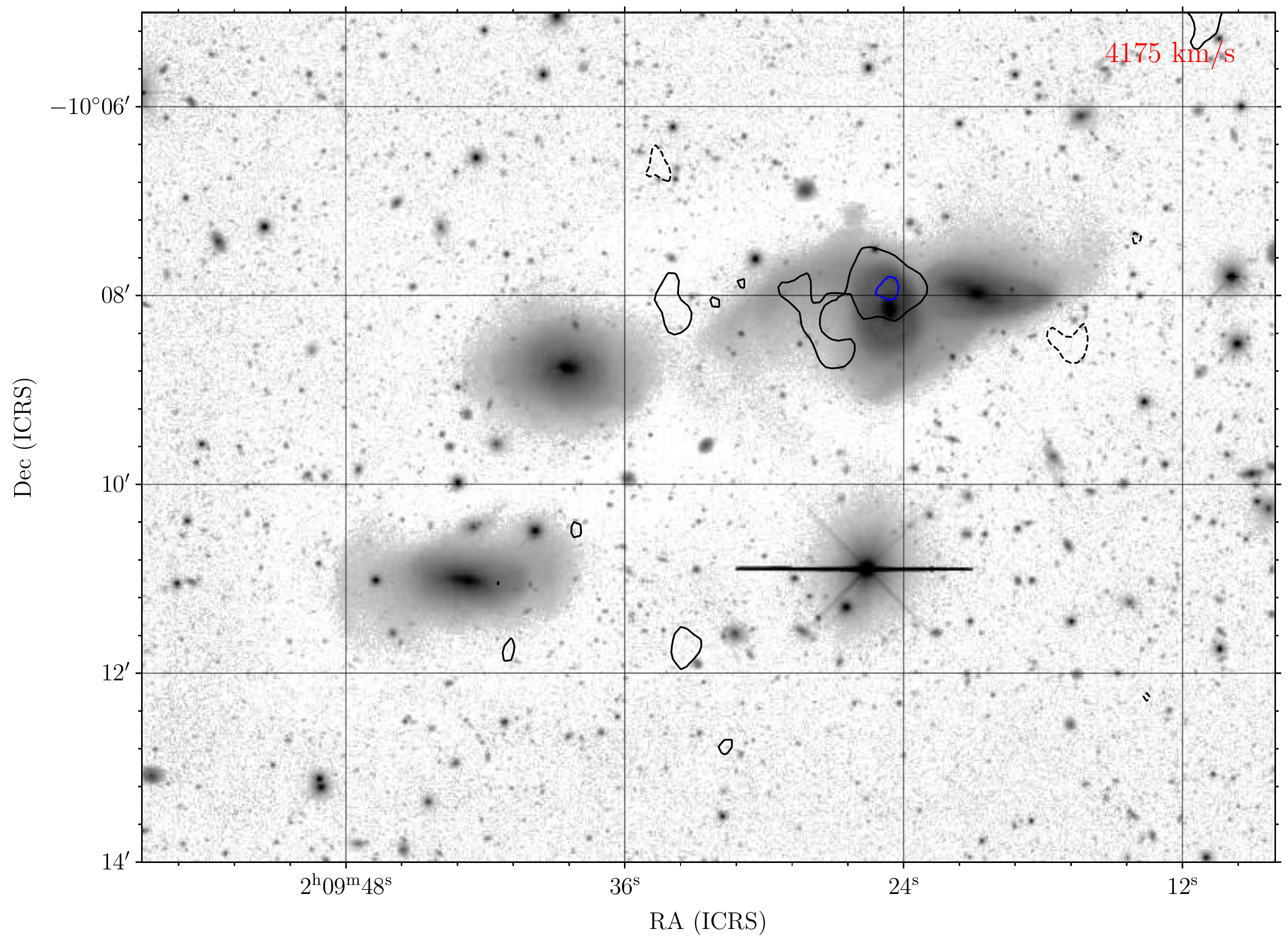}
    \includegraphics[width=0.33\columnwidth]{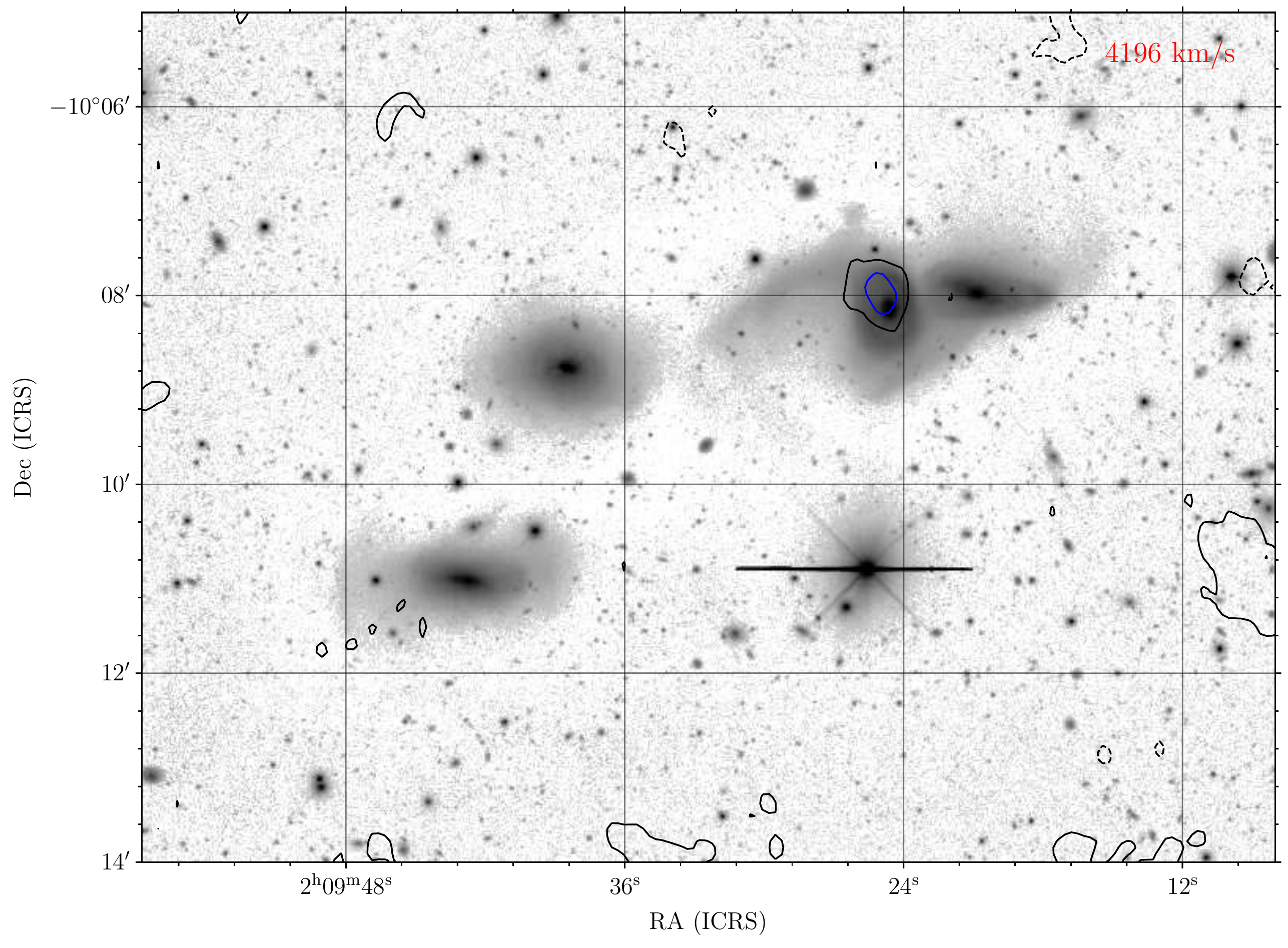}
    \includegraphics[width=0.33\columnwidth]{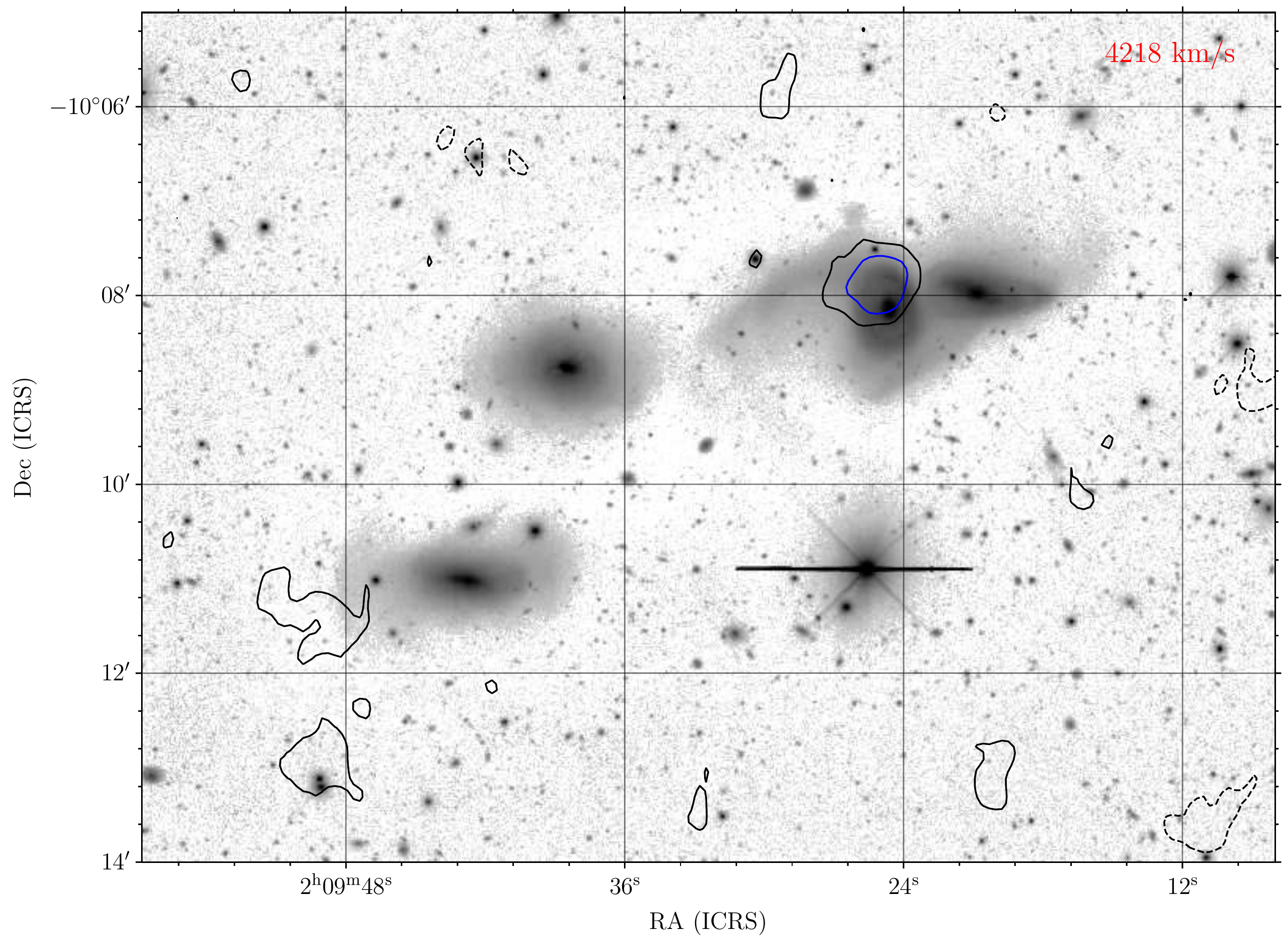}
    \includegraphics[width=0.33\columnwidth]{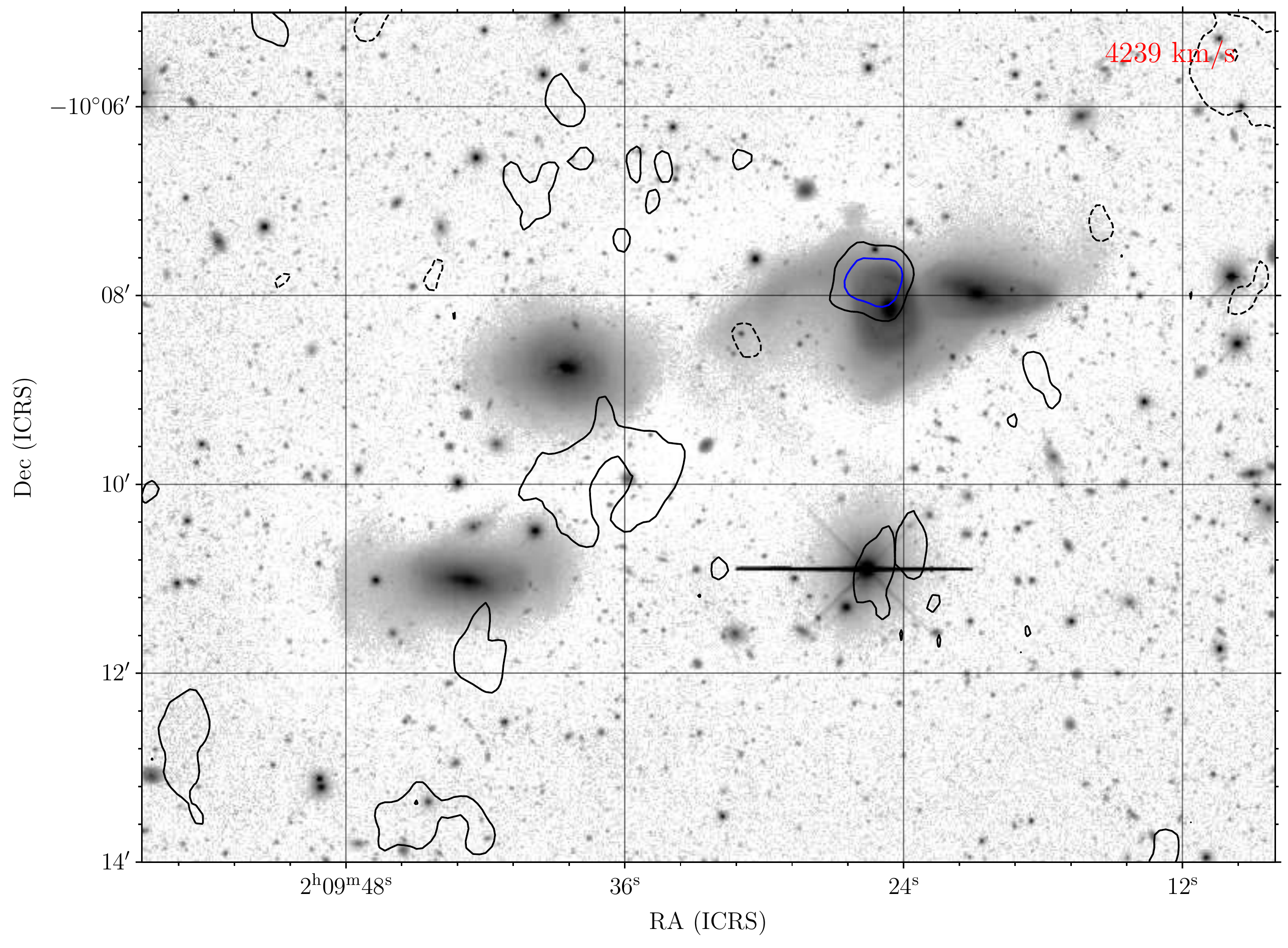}
    \includegraphics[width=0.33\columnwidth]{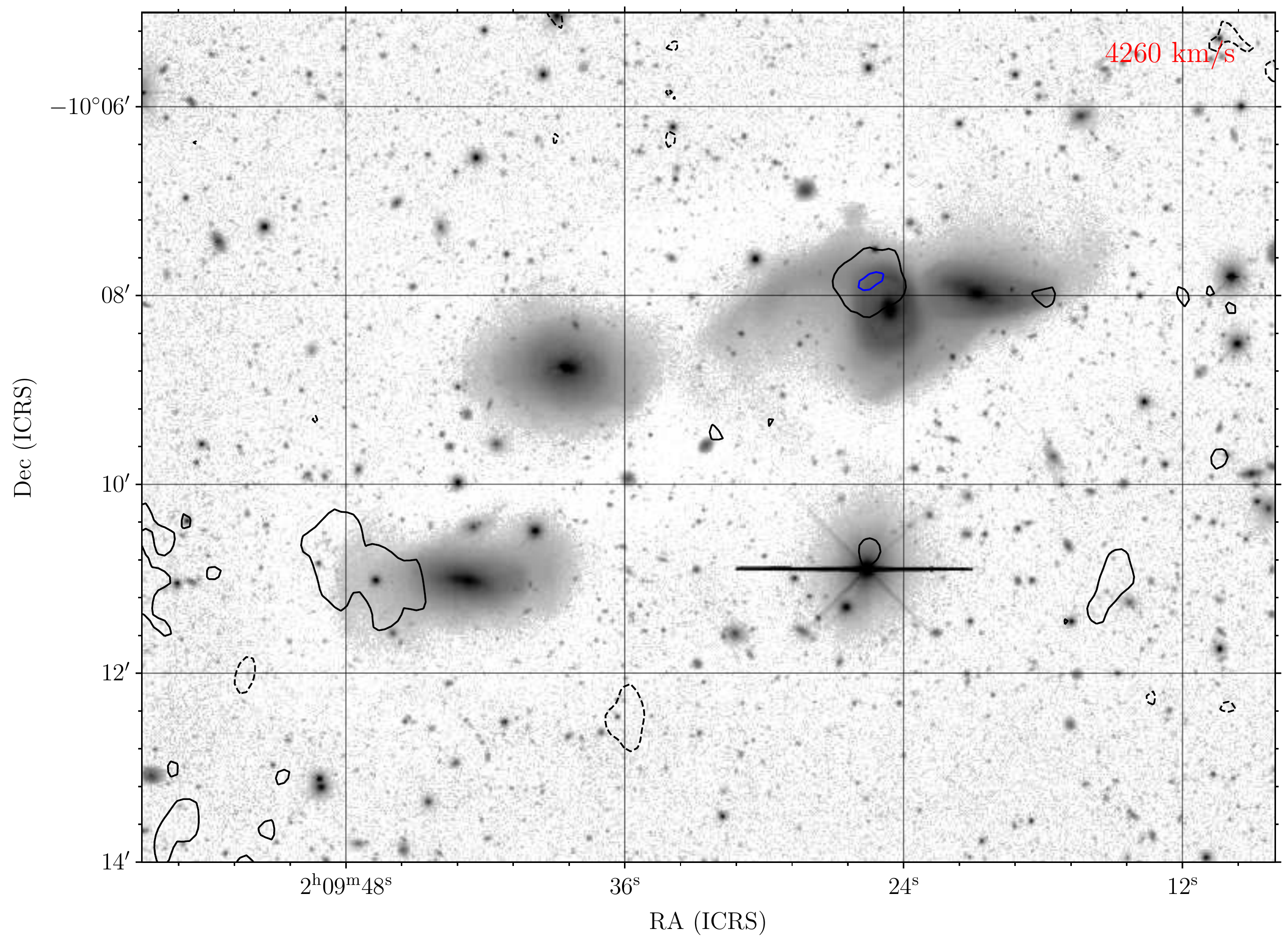}
\end{figure}
\end{landscape}

\newpage
\begin{figure*}
    \centering
    \includegraphics[width=\columnwidth]{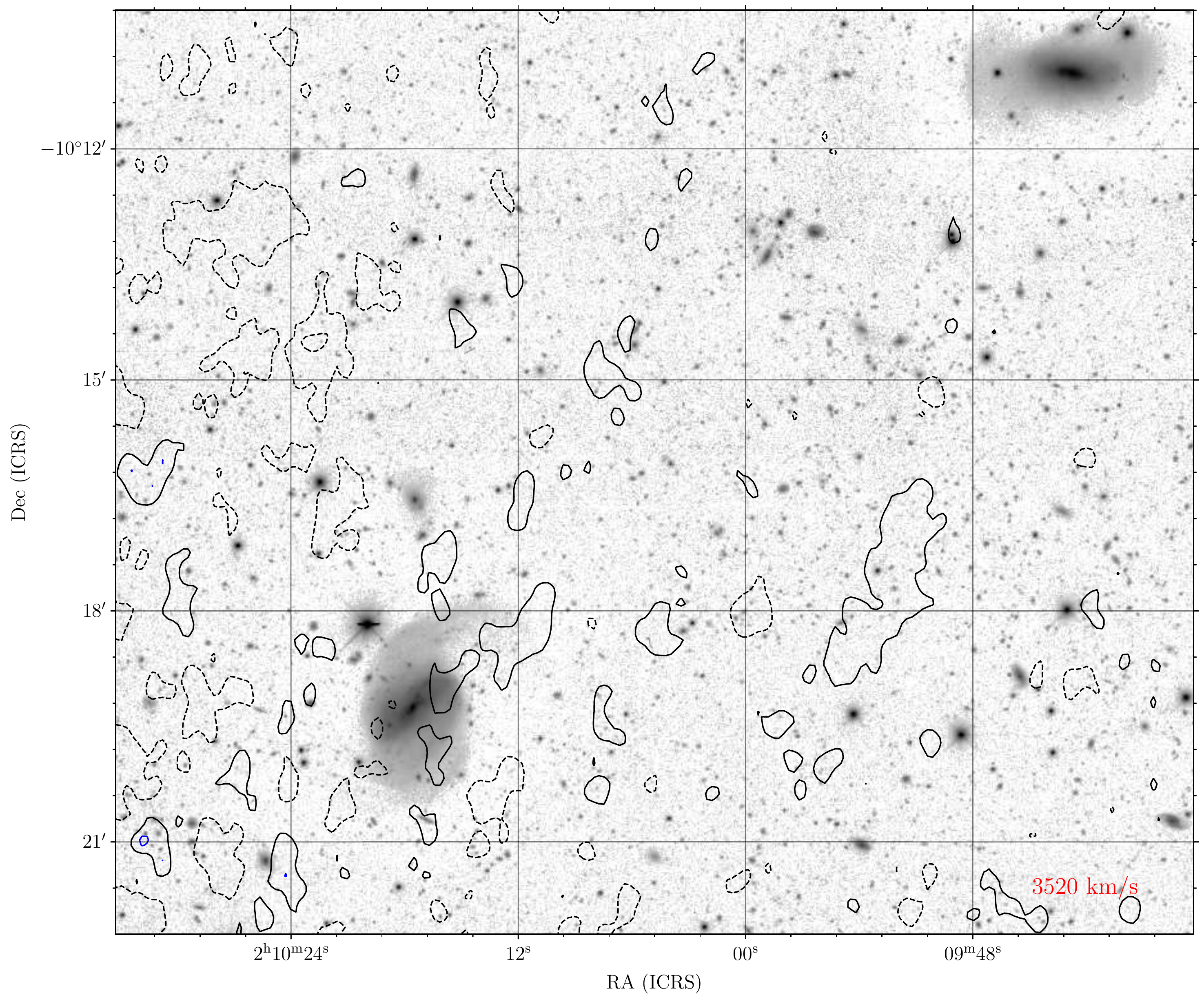}
    \includegraphics[width=\columnwidth]{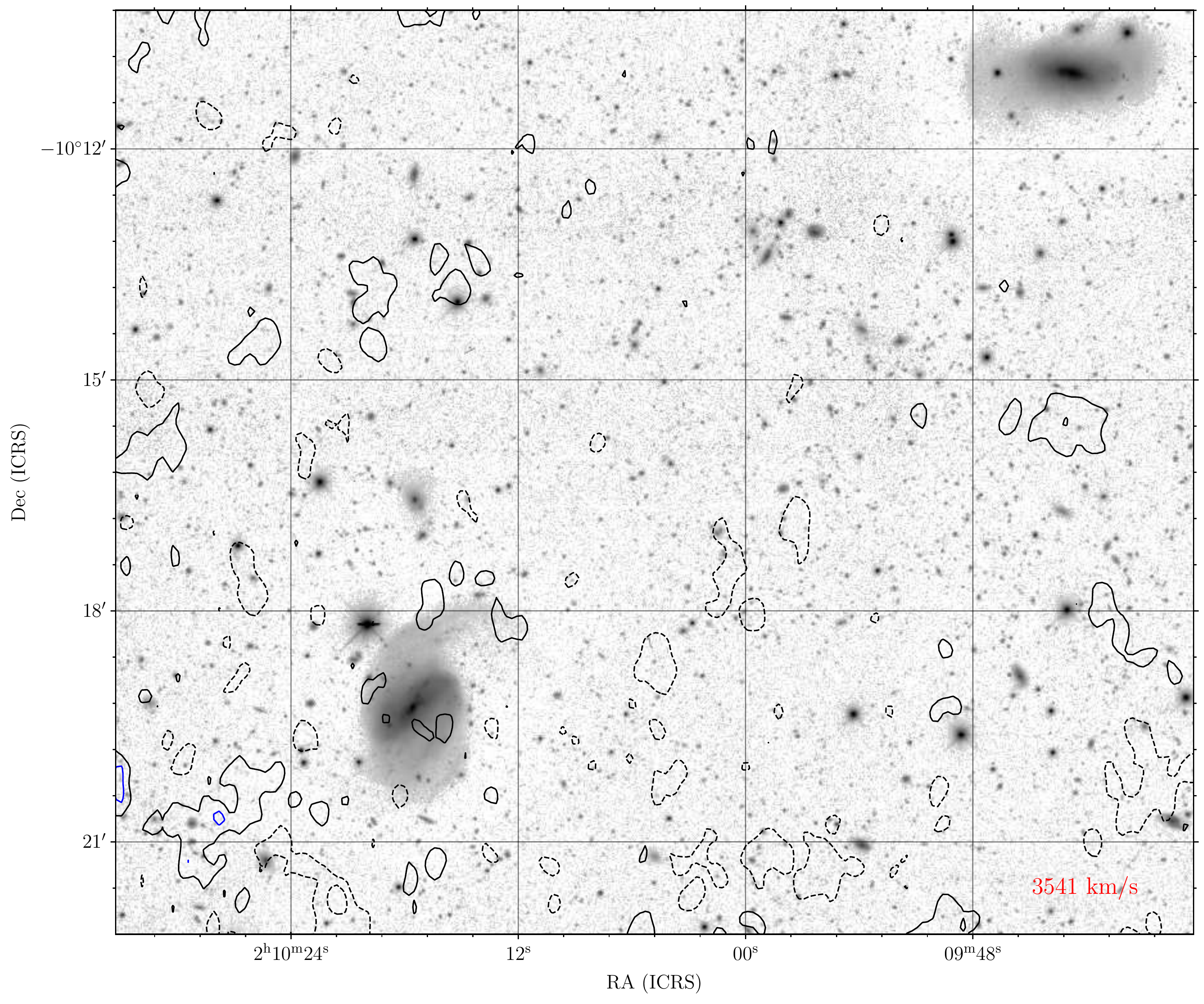}
    \includegraphics[width=\columnwidth]{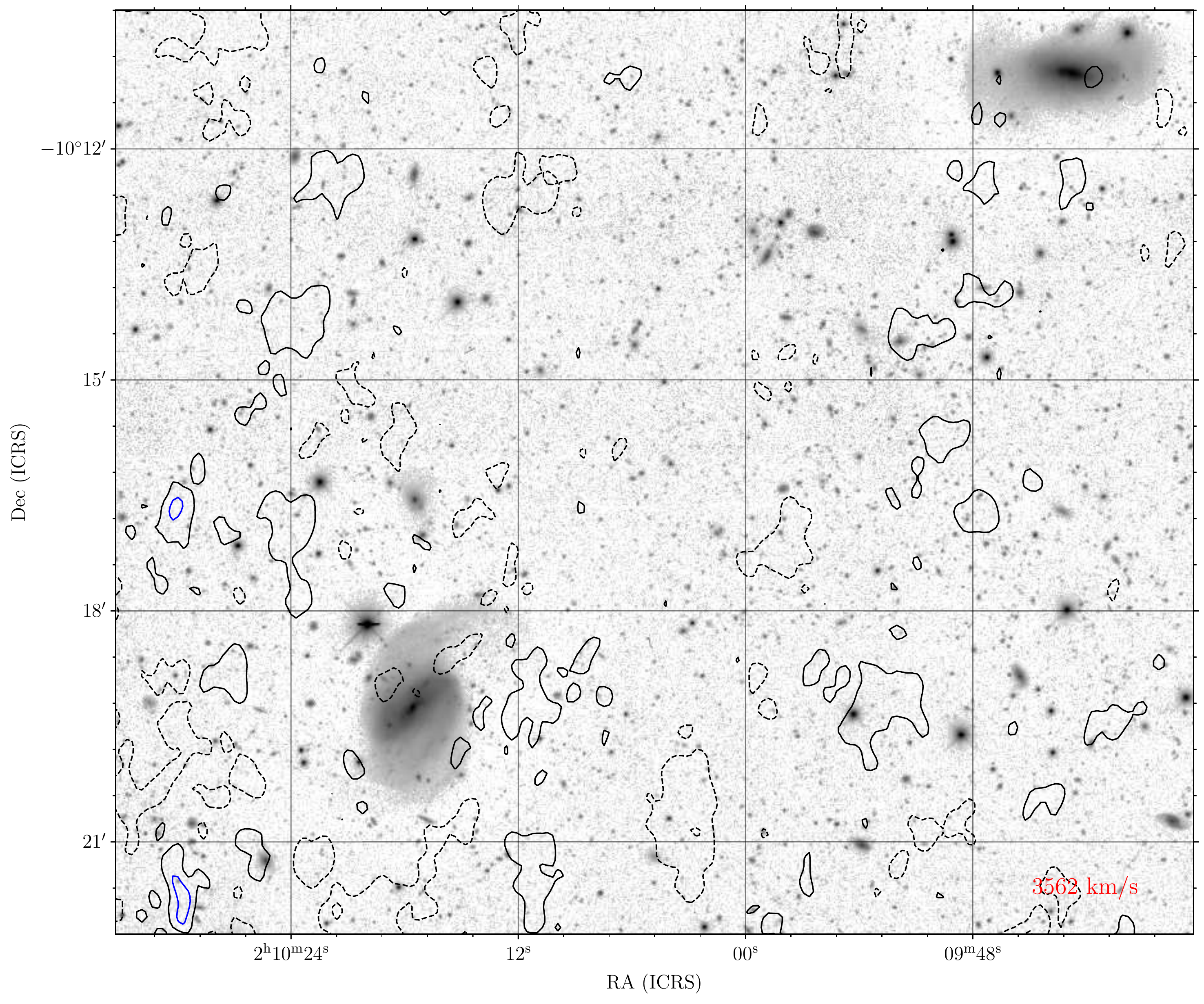}
    \includegraphics[width=\columnwidth]{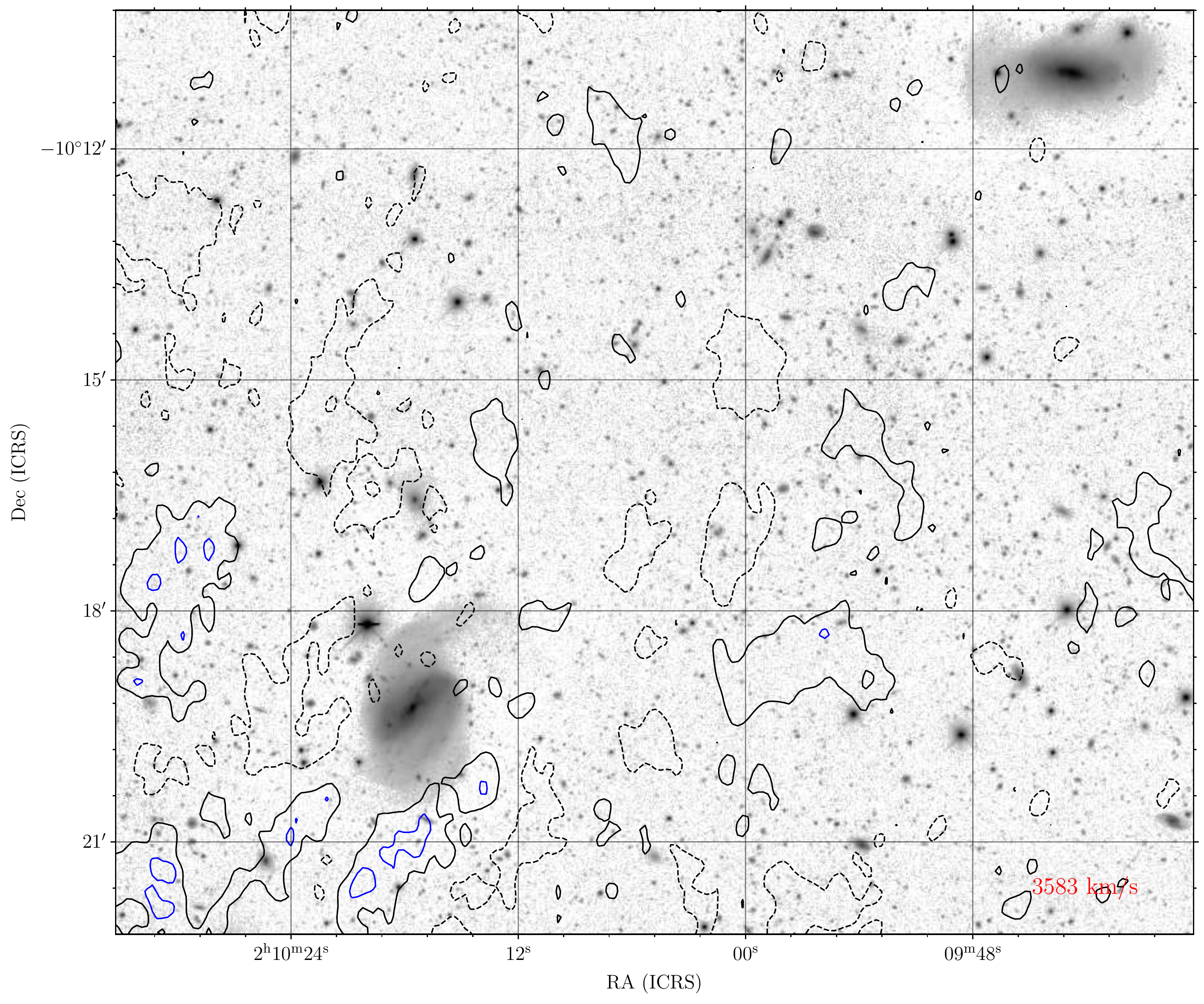}
    \includegraphics[width=\columnwidth]{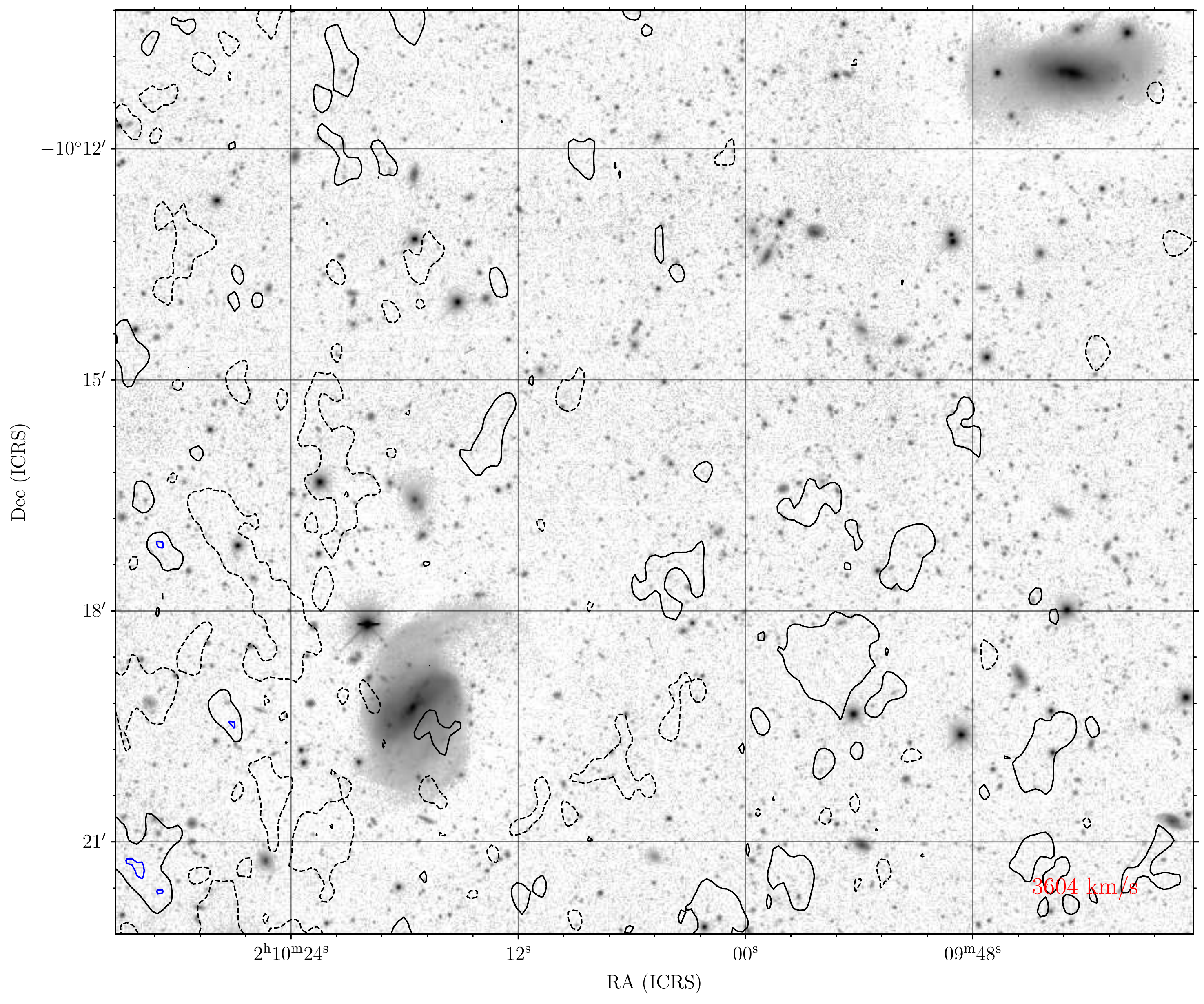}
    \includegraphics[width=\columnwidth]{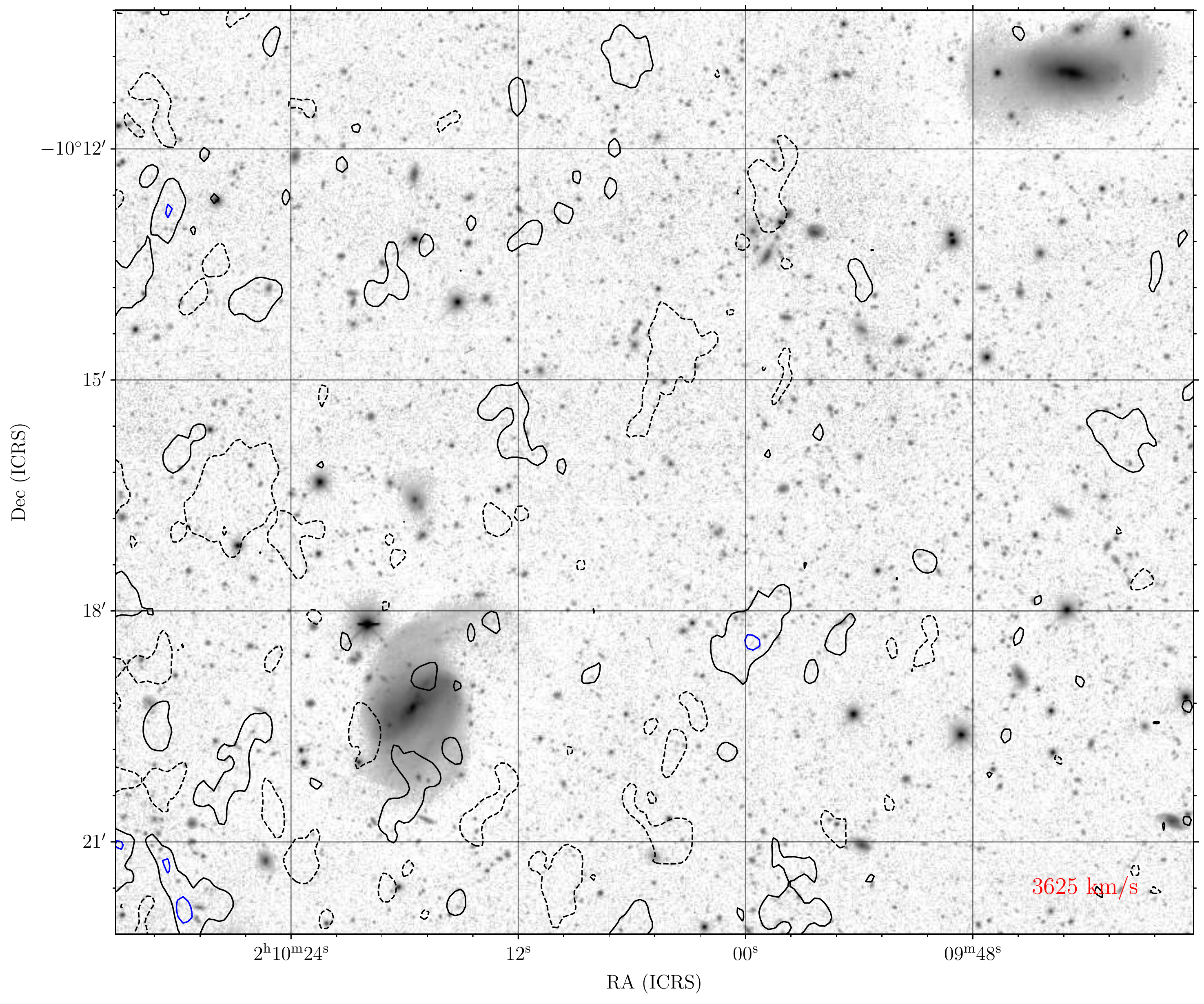}
    \caption{Channel maps showing the \hi \ emission in the SE tail and NGC 848 overlaid on the DECaLS $r$-band image. Contour levels: -1,1,2,4,8,16 $\times 0.9$ mJy per beam, $\times 1.8 \times 10^{19} \; \mathrm{cm^{-2}}$, or $\times 0.15 \; \mathrm{M_{\odot}\,pc^{-2}}$. The contours are coloured (in order of increasing flux): black (dashed), black, blue, purple, red, and orange. The noise level rises towards the South East as this it is near the edge of the VLA primary beam.}
    \label{fig:chn_maps_tail}
\end{figure*}

\newpage
\begin{figure*}
    \centering
    \includegraphics[width=\columnwidth]{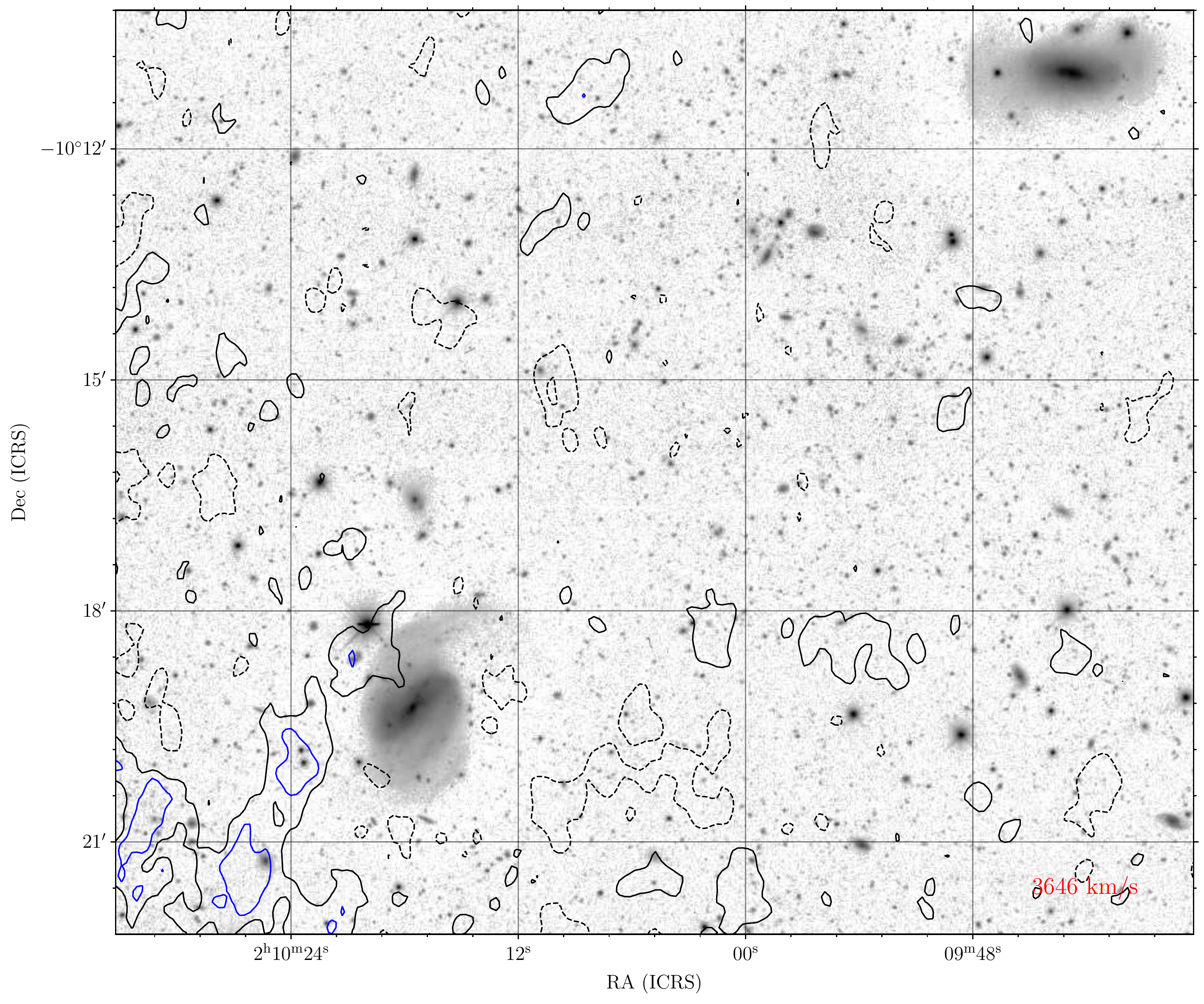}
    \includegraphics[width=\columnwidth]{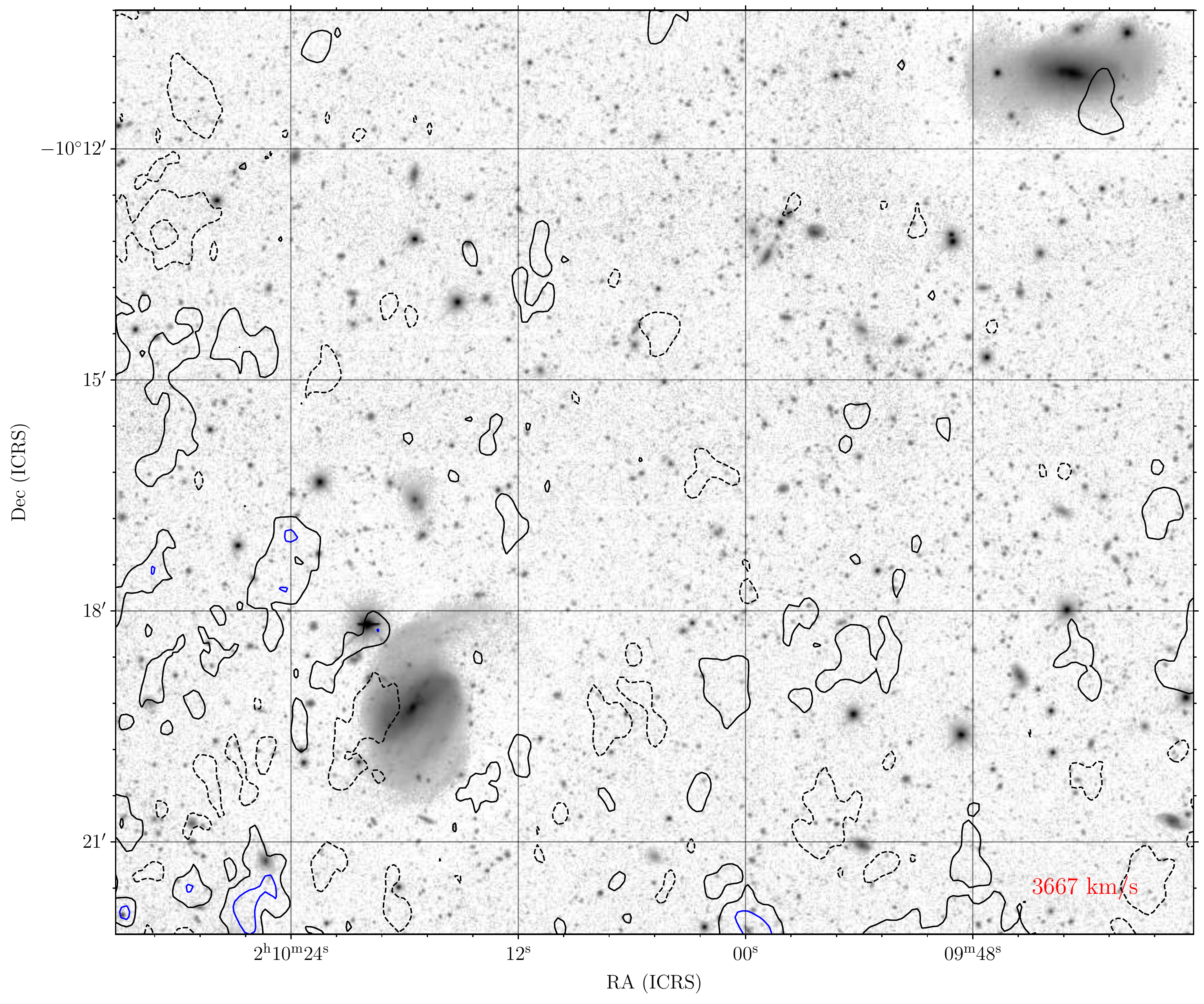}
    \includegraphics[width=\columnwidth]{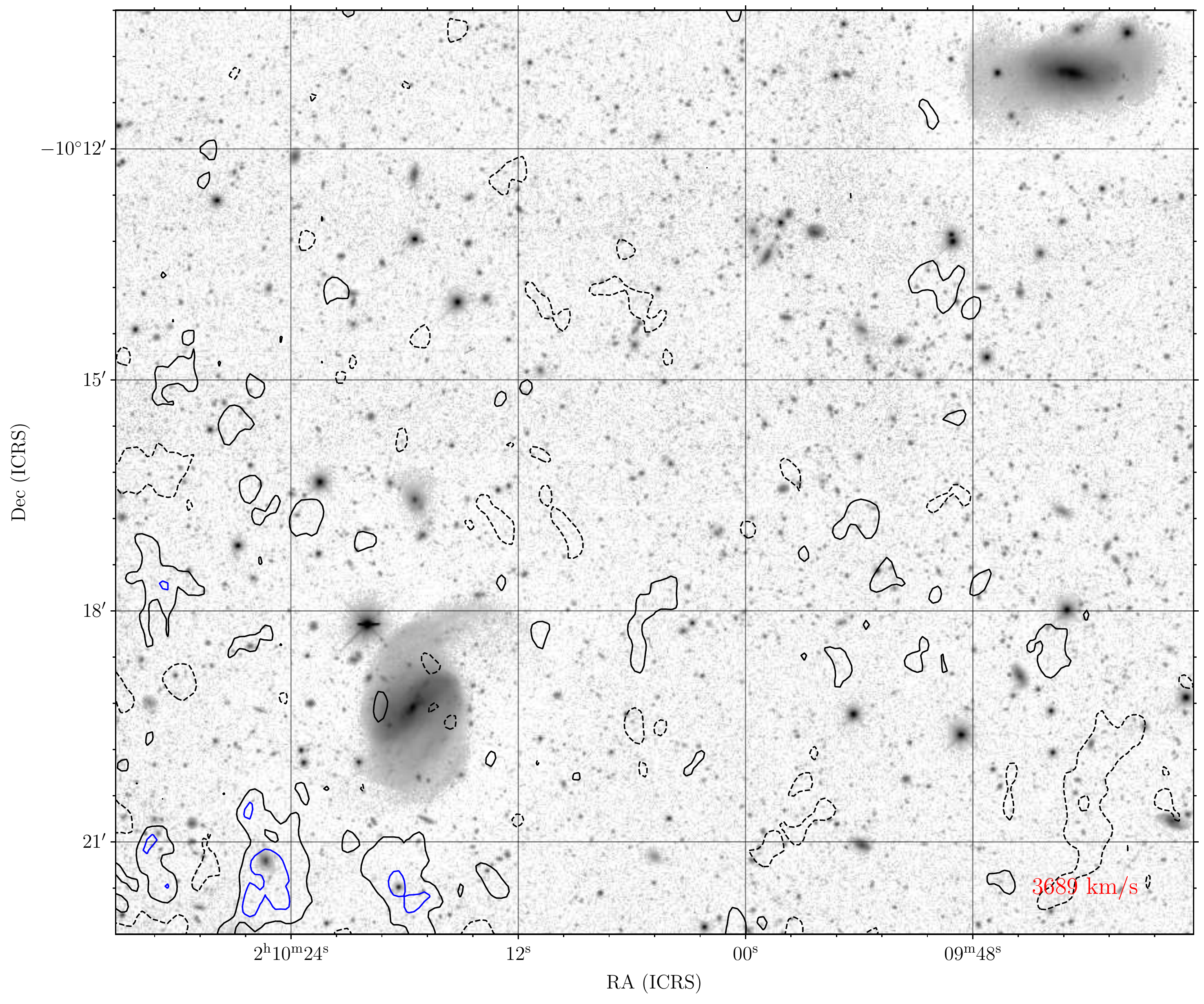}
    \includegraphics[width=\columnwidth]{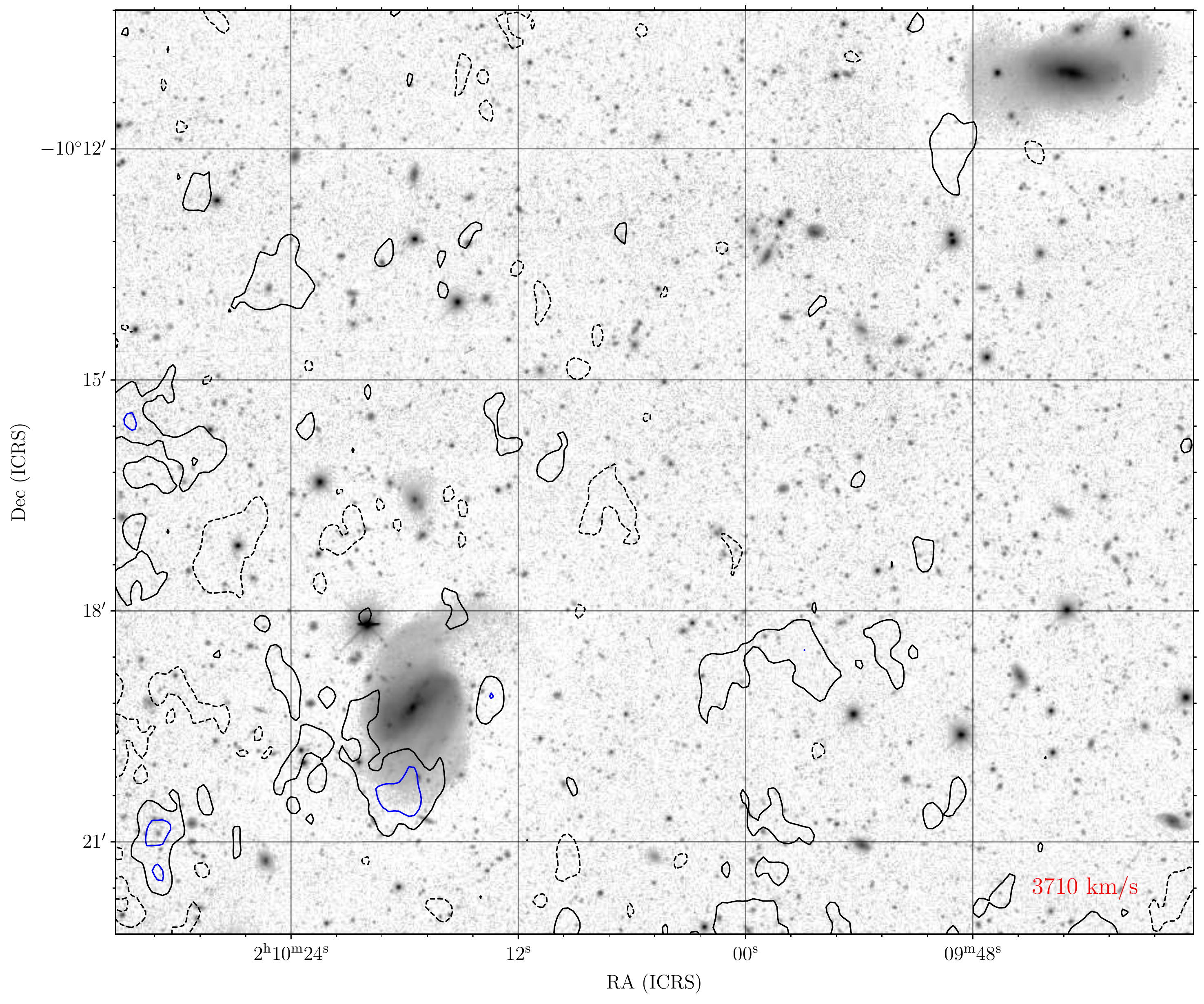}
    \includegraphics[width=\columnwidth]{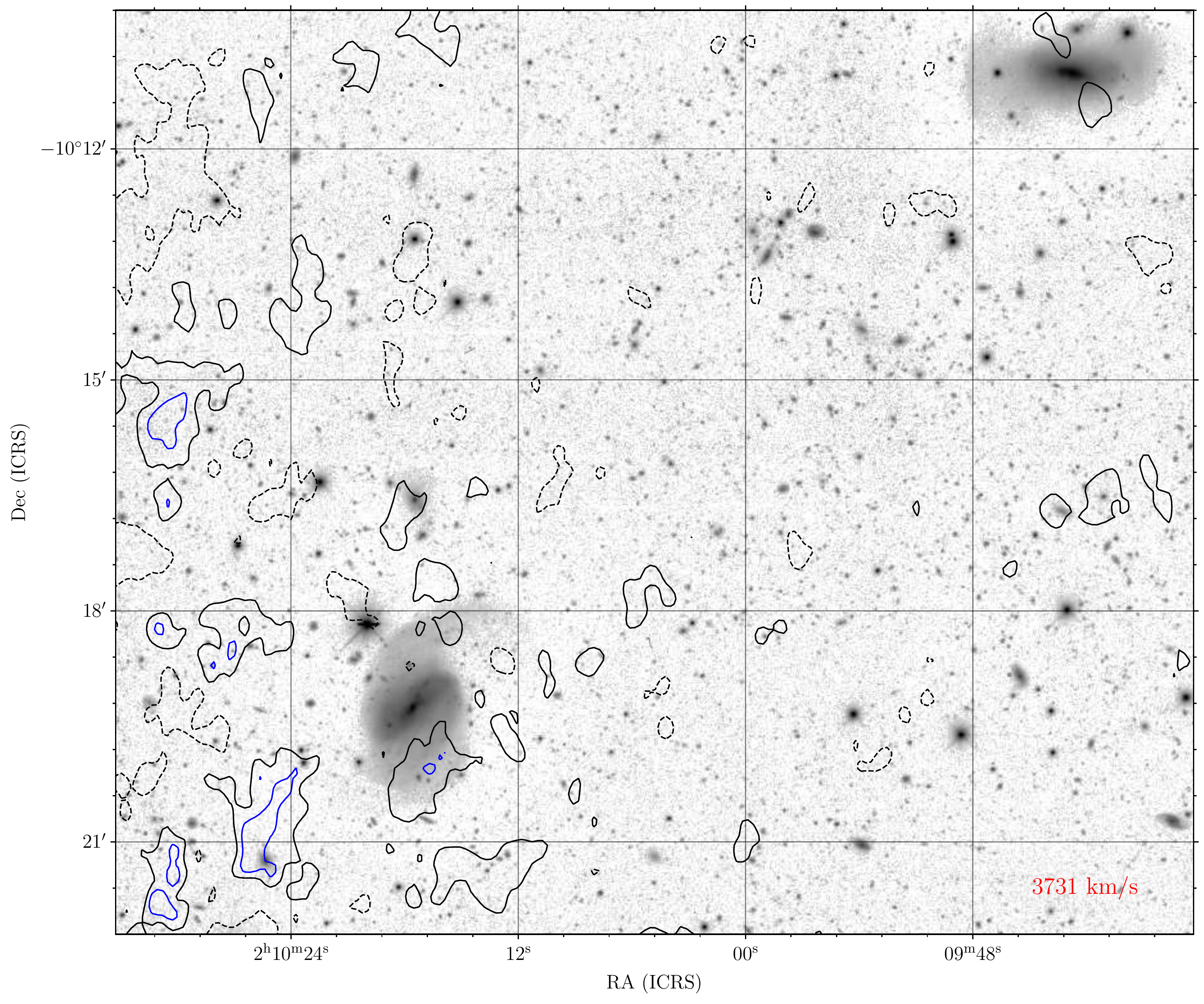}
    \includegraphics[width=\columnwidth]{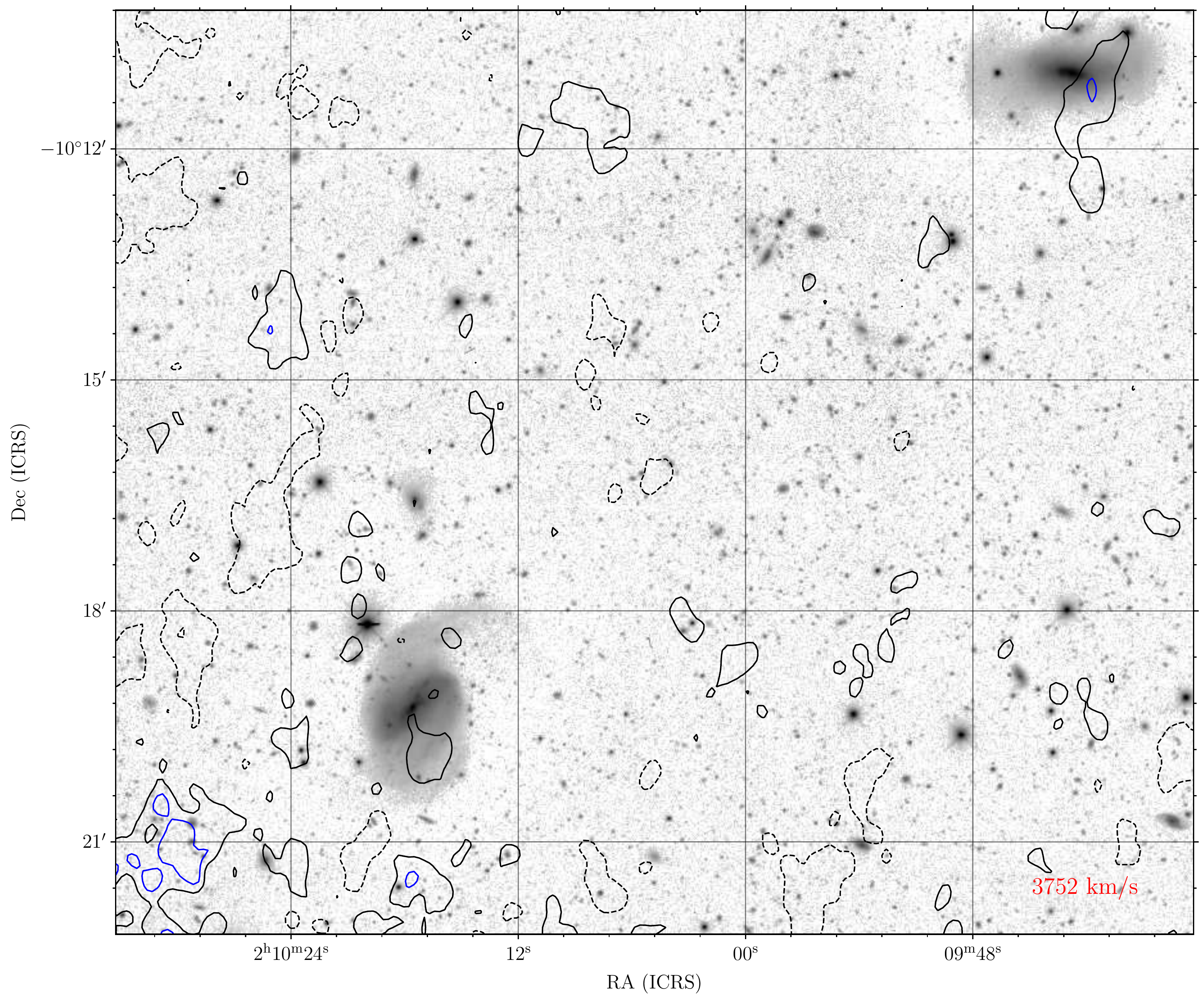}
\end{figure*}

\newpage
\begin{figure*}
    \centering
    \includegraphics[width=\columnwidth]{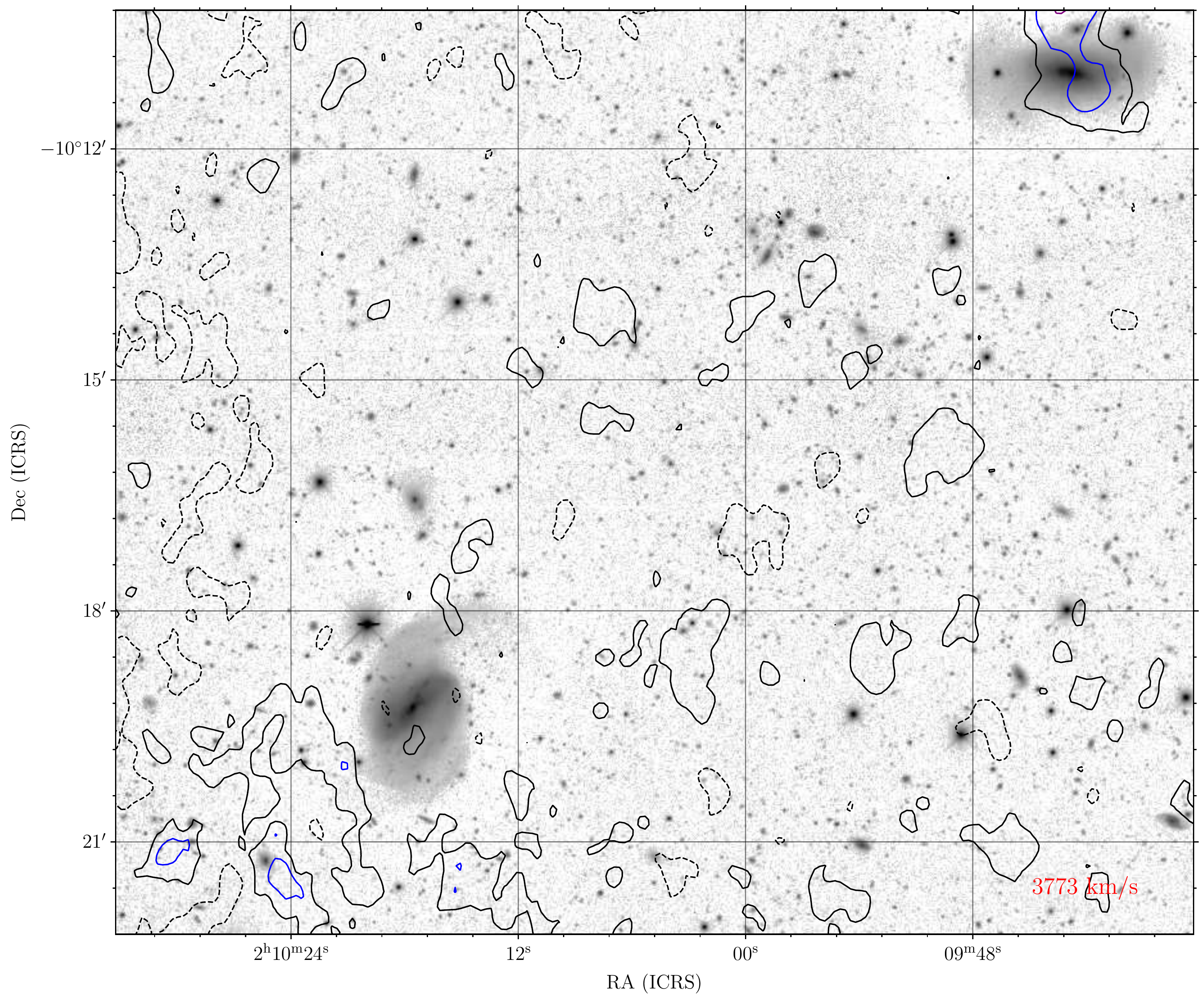}
    \includegraphics[width=\columnwidth]{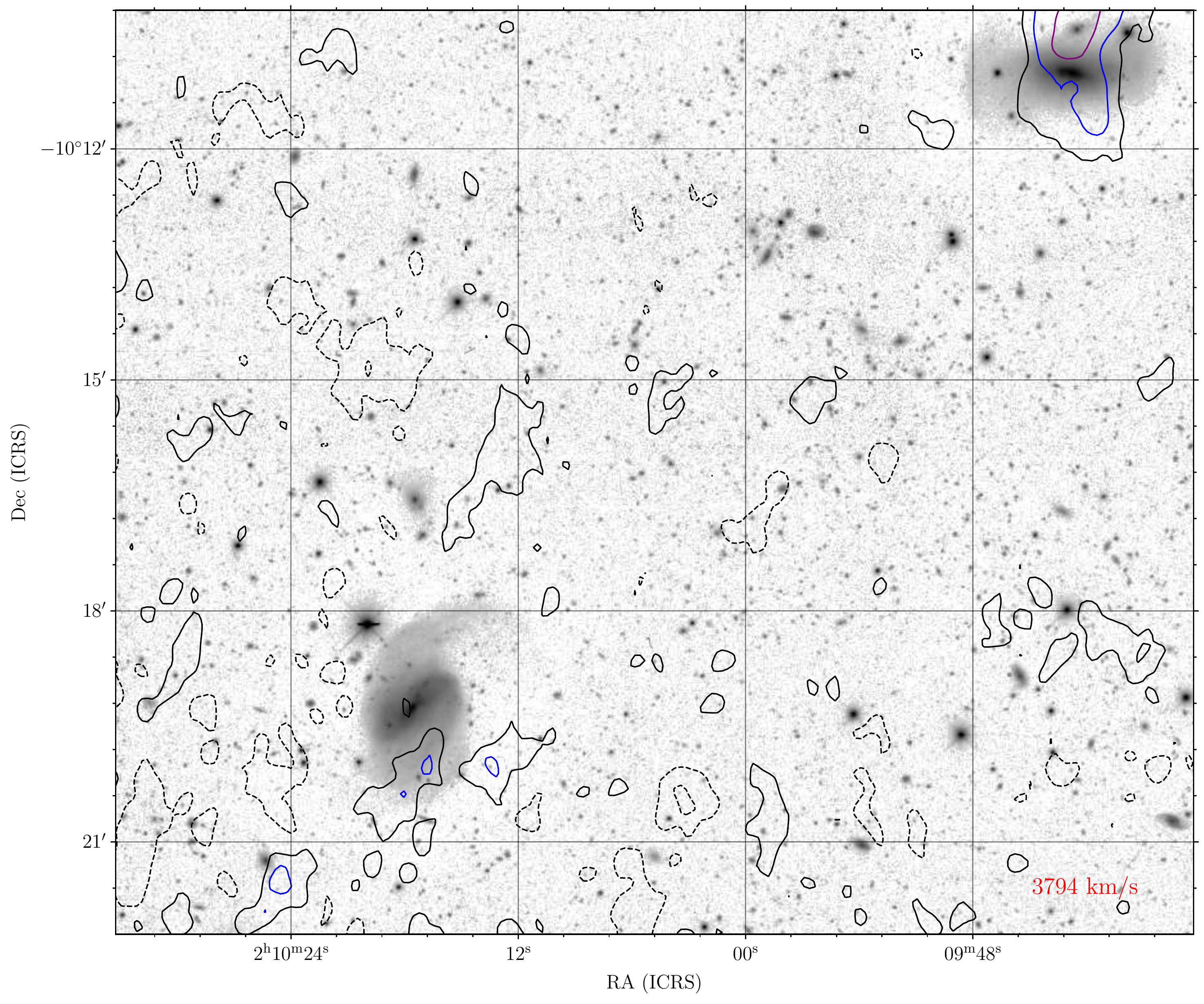}
    \includegraphics[width=\columnwidth]{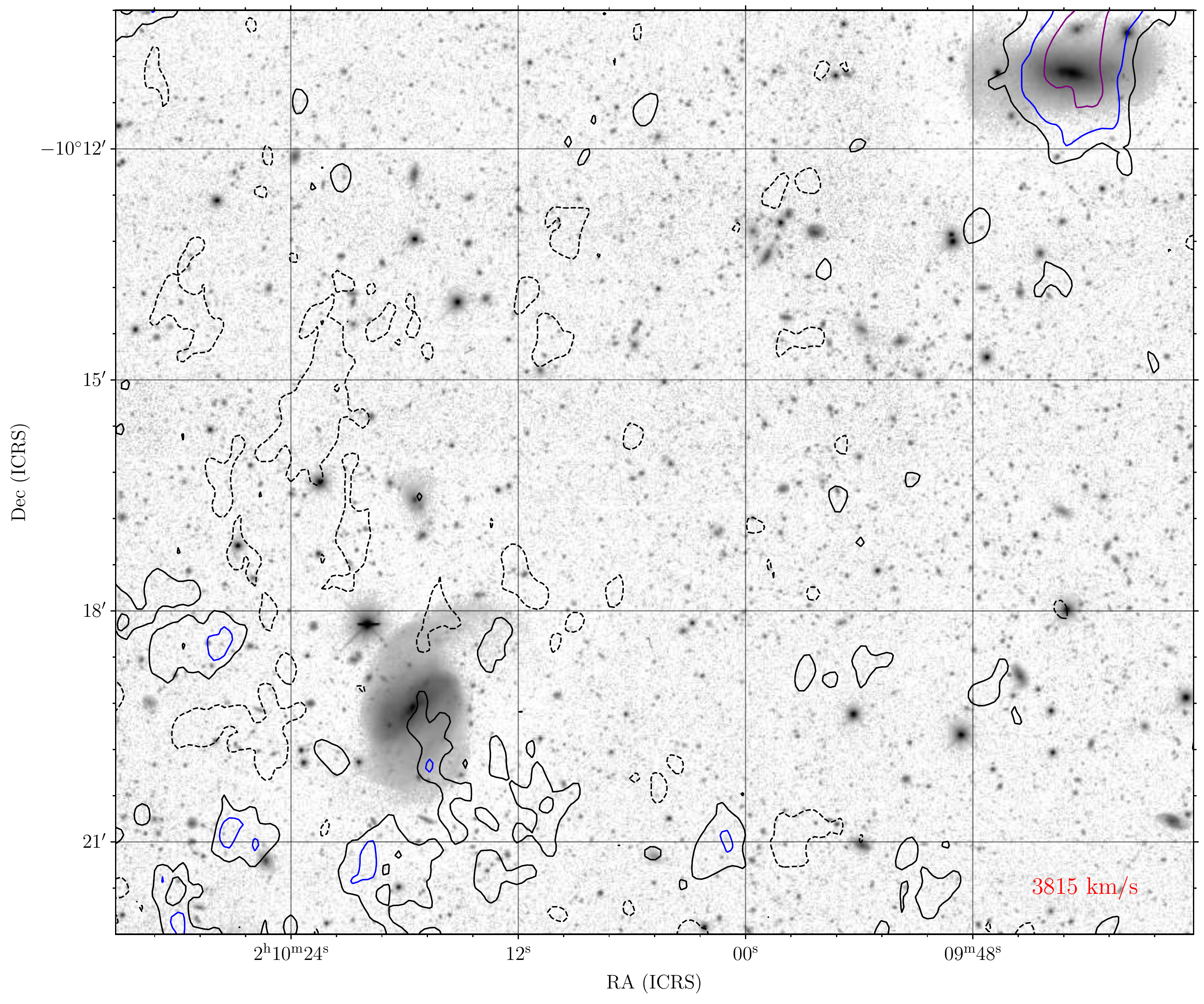}
    \includegraphics[width=\columnwidth]{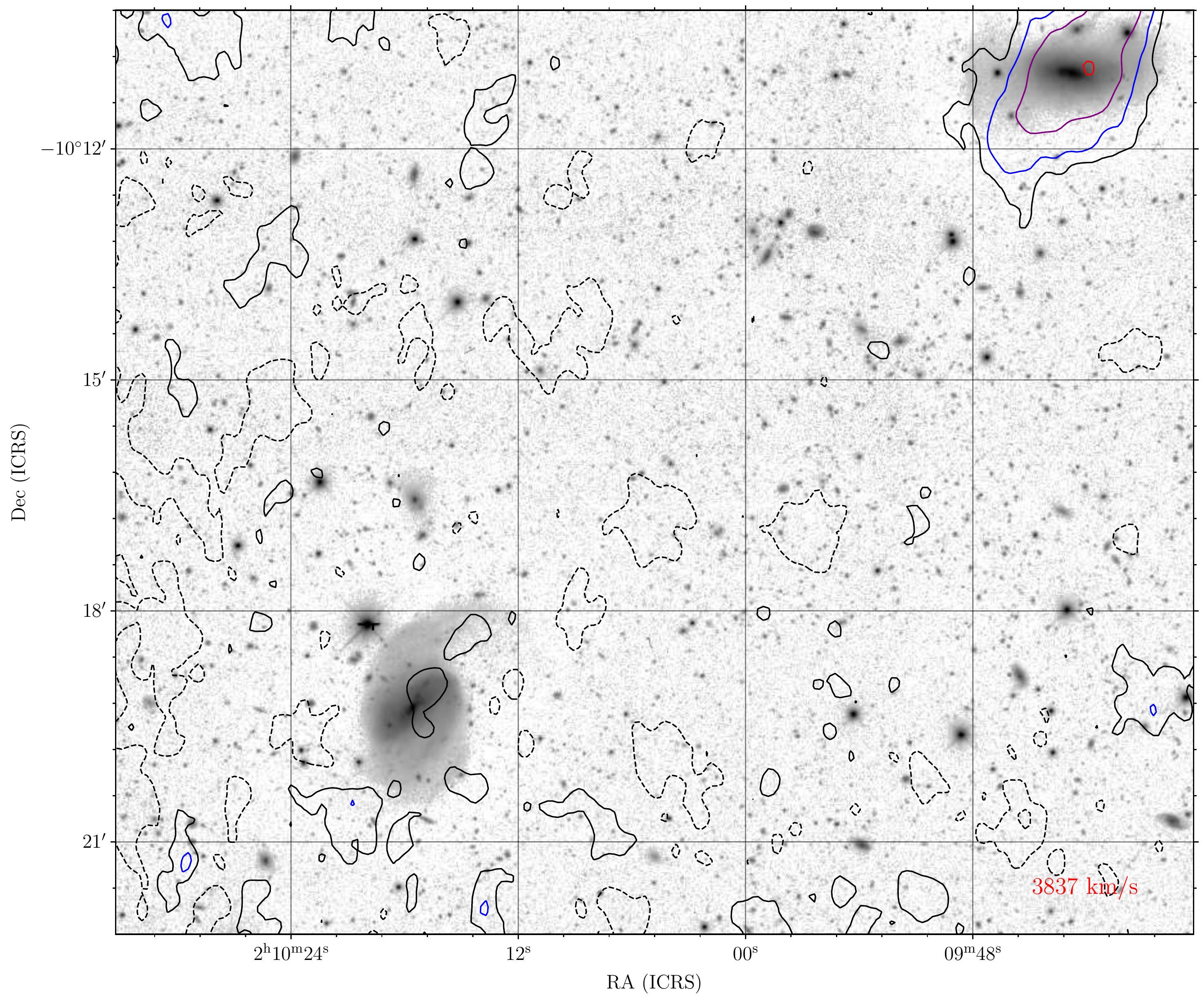}
    \includegraphics[width=\columnwidth]{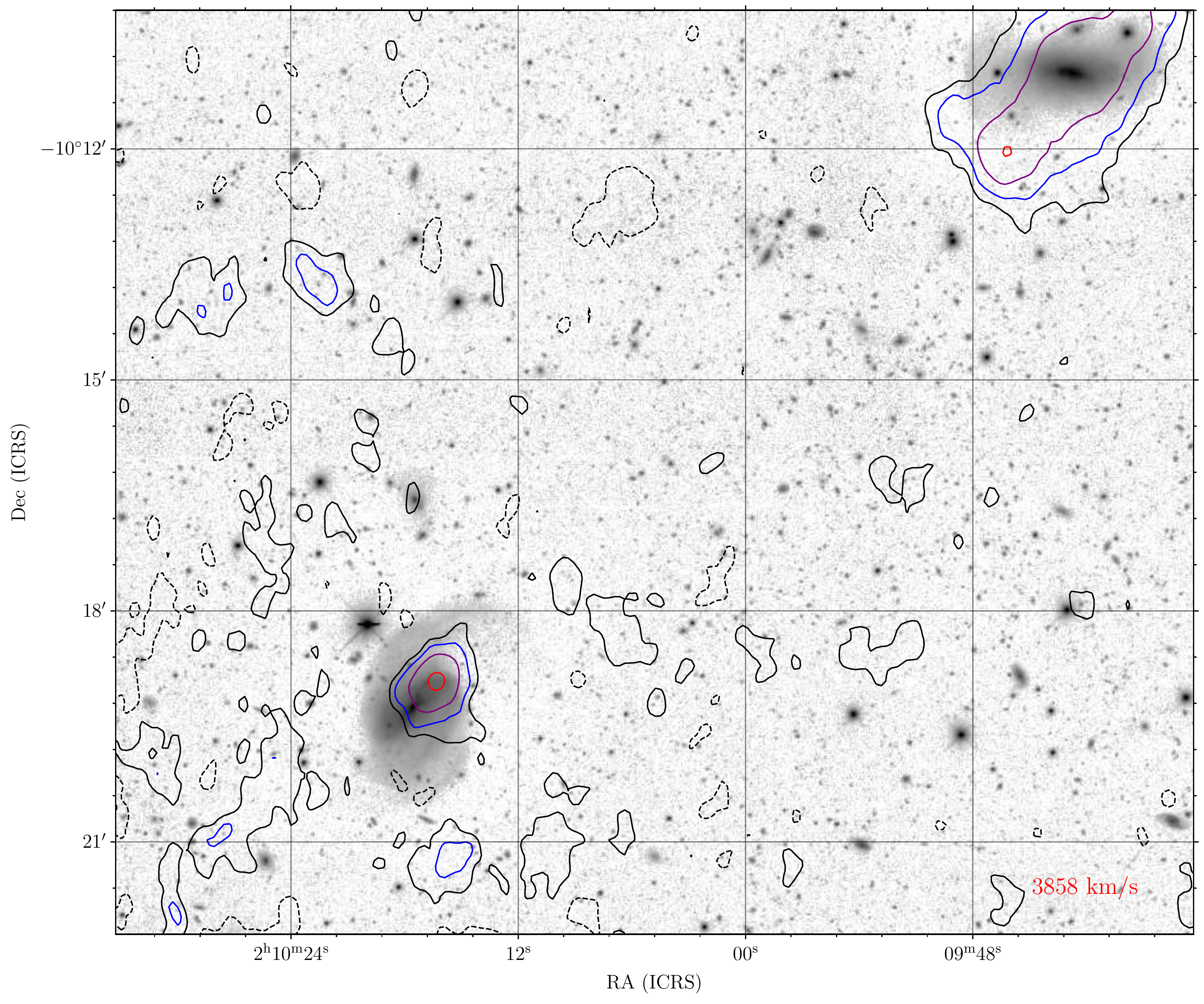}
    \includegraphics[width=\columnwidth]{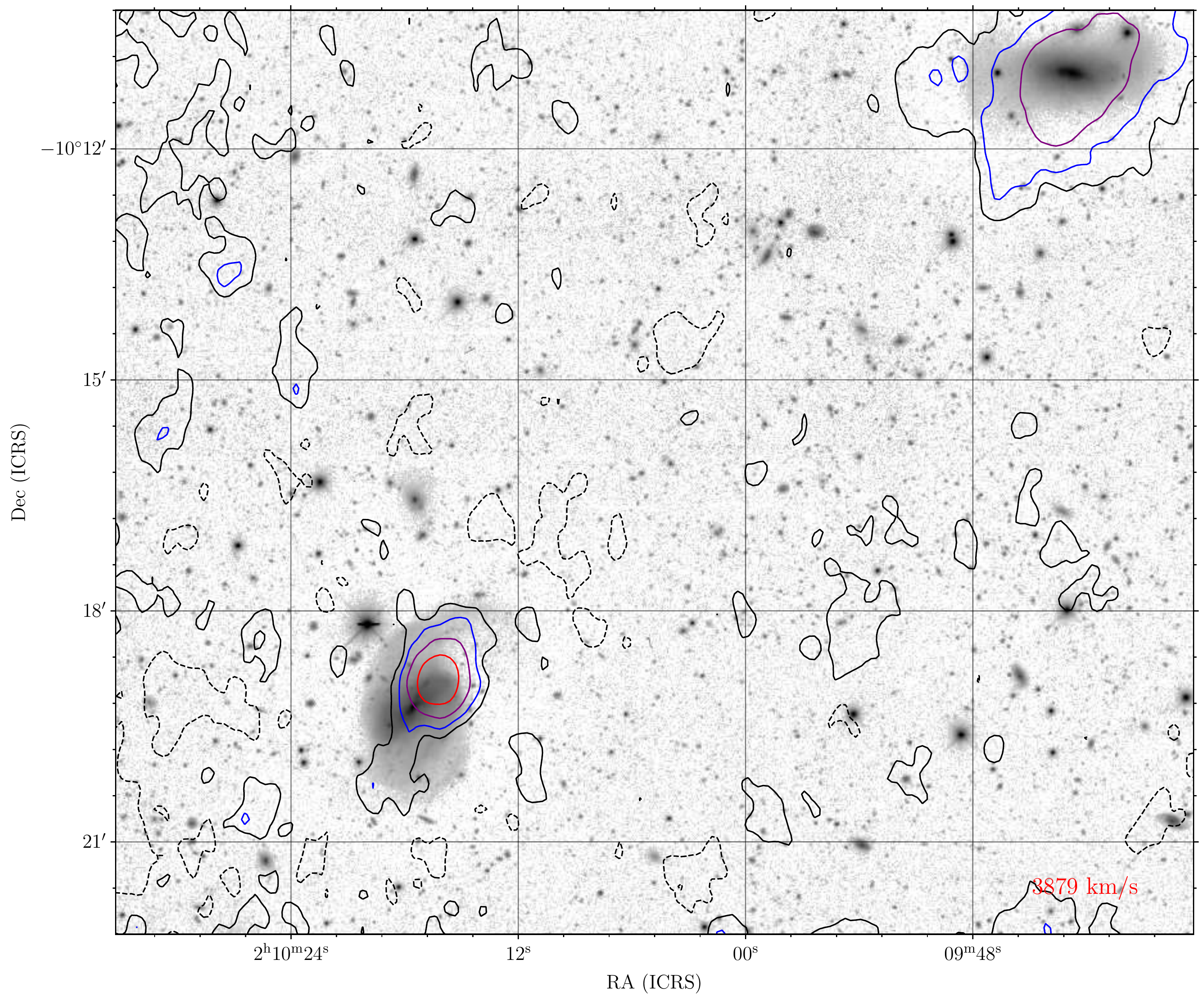}
\end{figure*}

\newpage
\begin{figure*}
    \centering
    \includegraphics[width=\columnwidth]{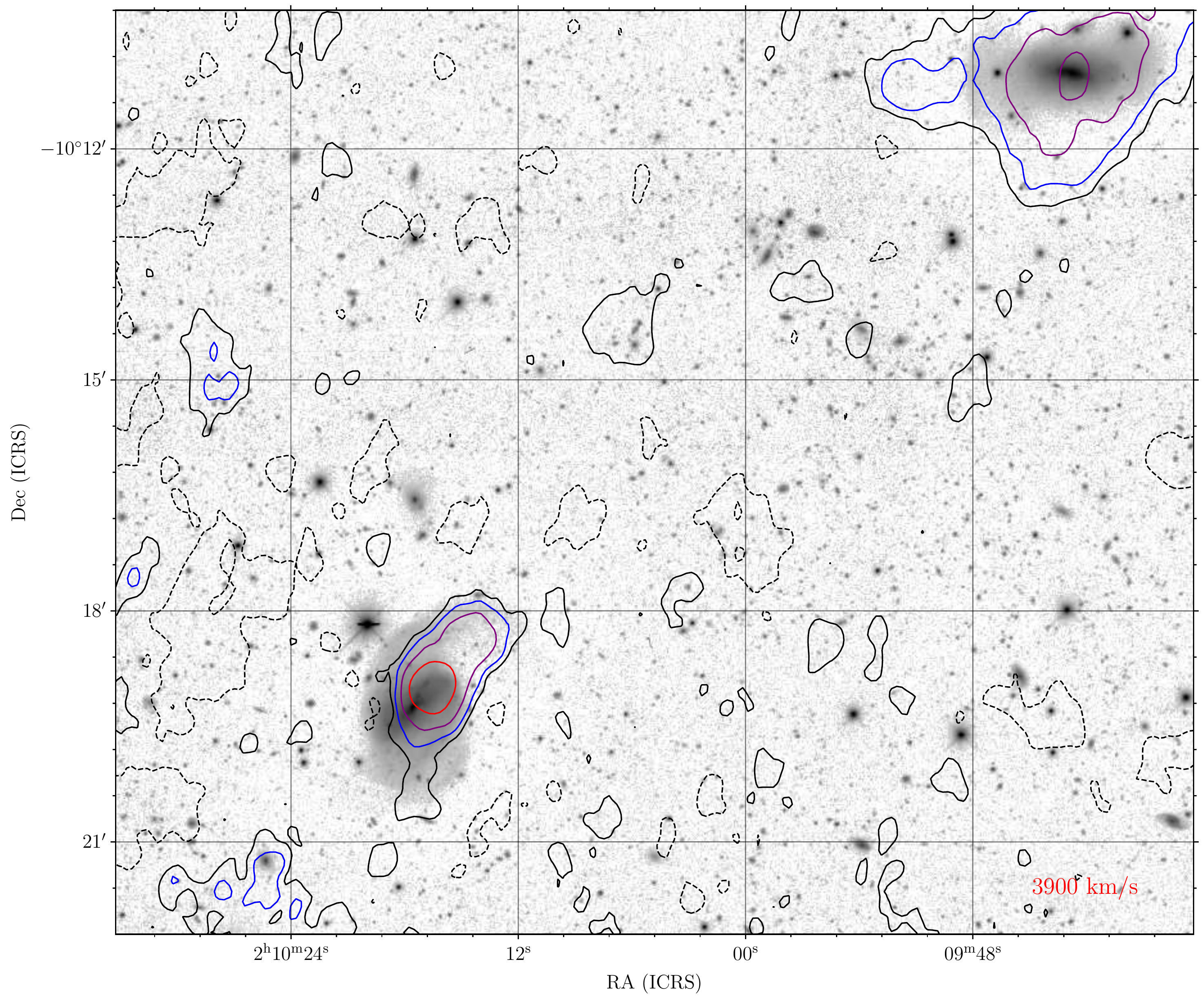}
    \includegraphics[width=\columnwidth]{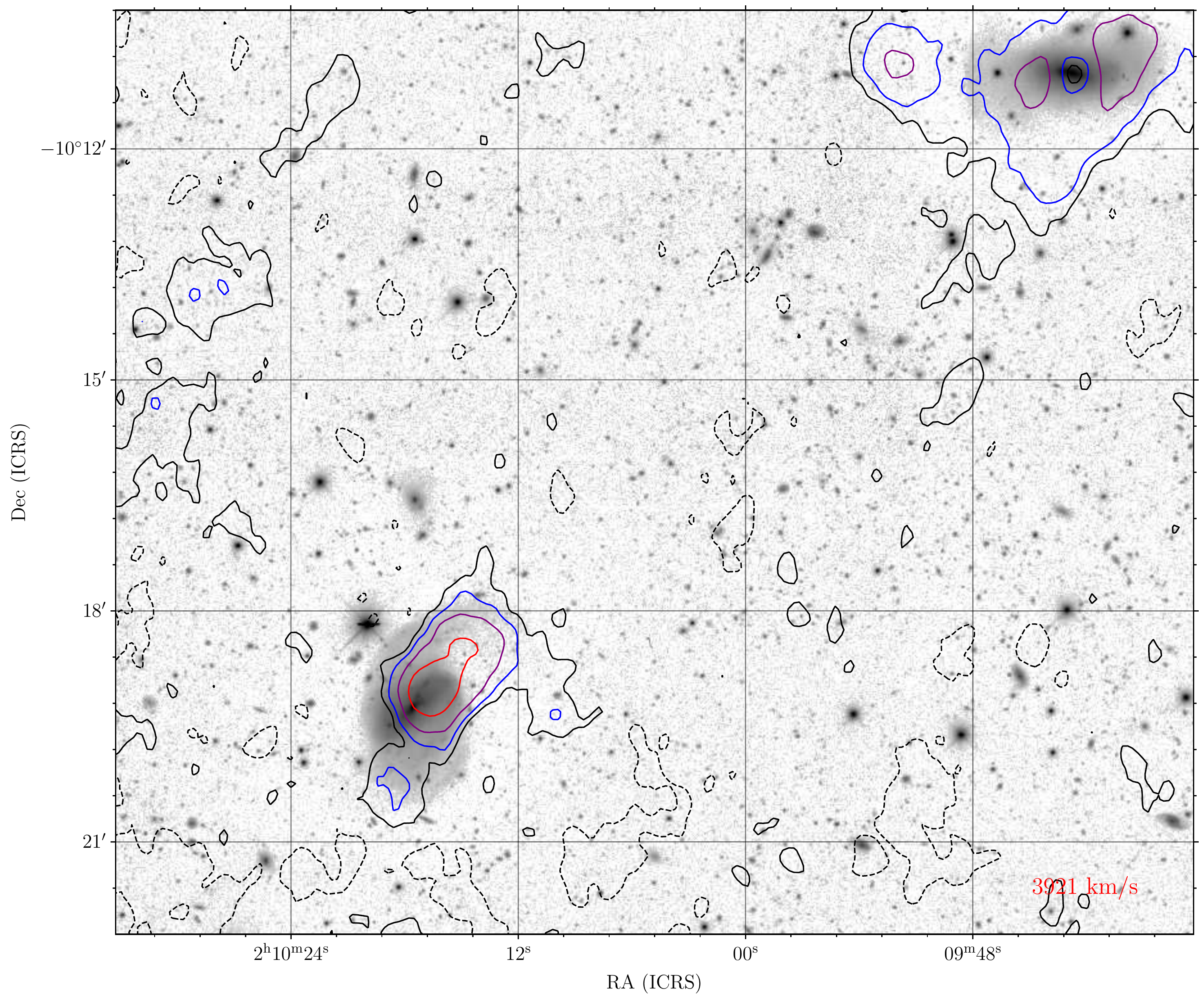}
    \includegraphics[width=\columnwidth]{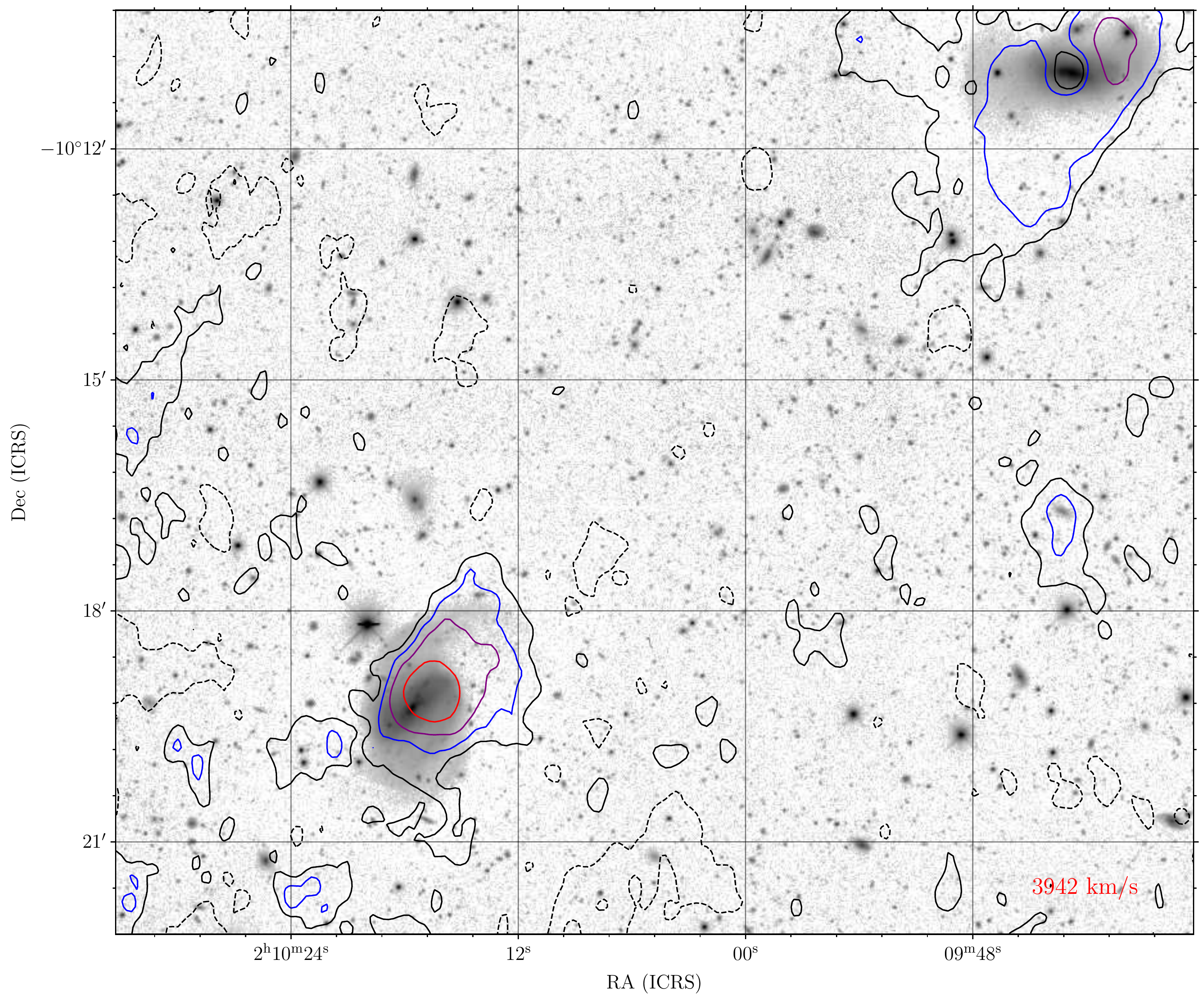}
    \includegraphics[width=\columnwidth]{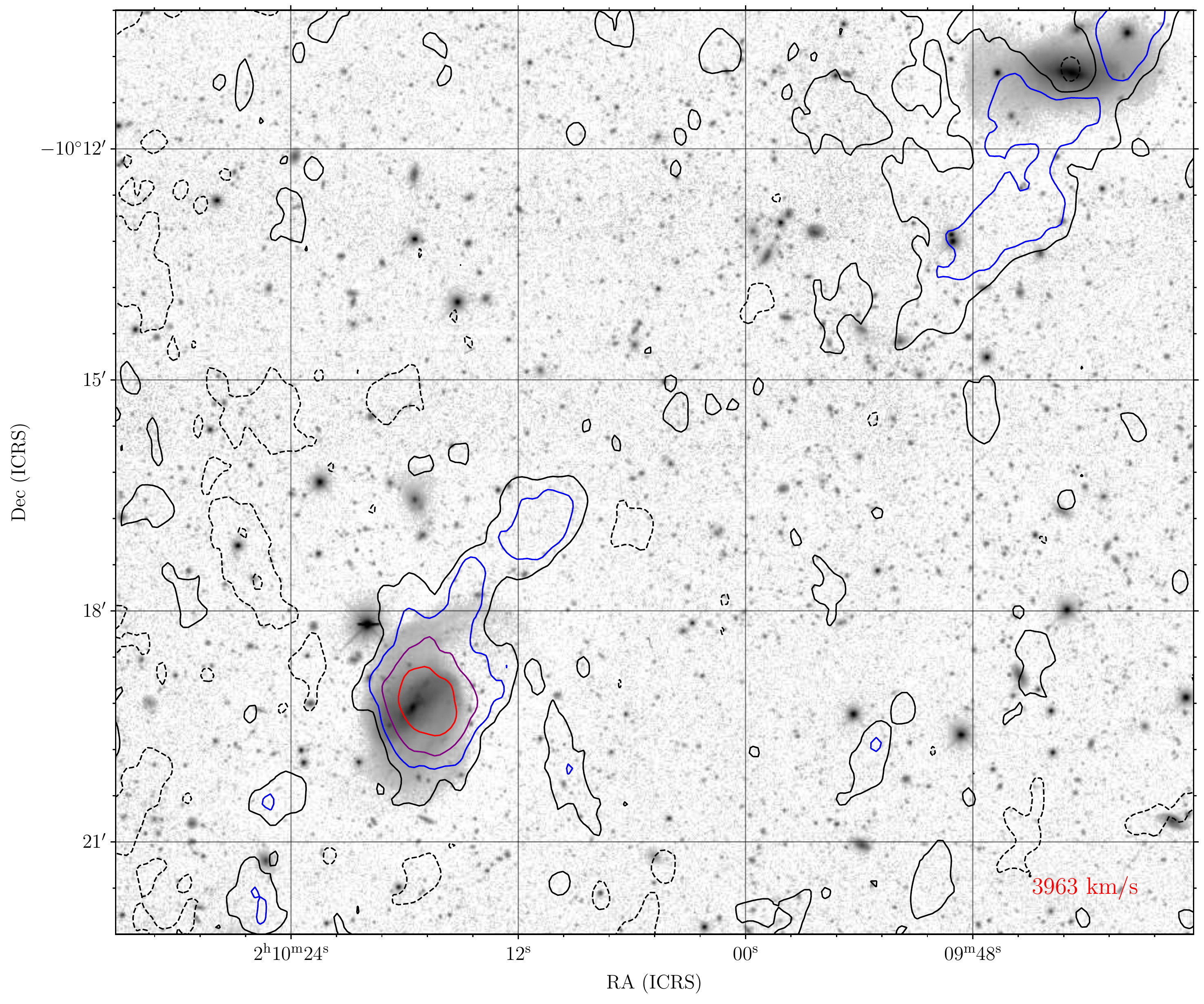}
    \includegraphics[width=\columnwidth]{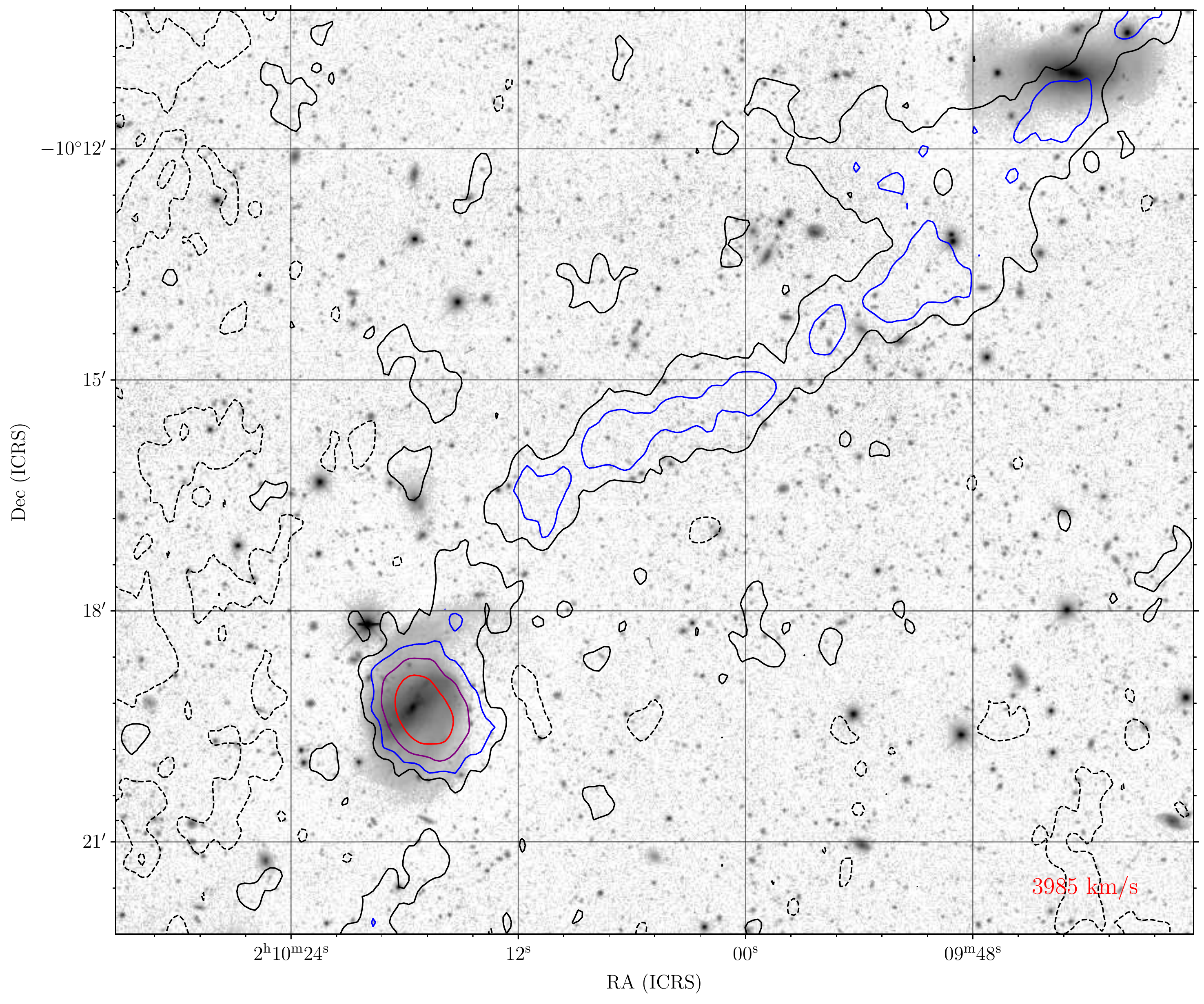}
    \includegraphics[width=\columnwidth]{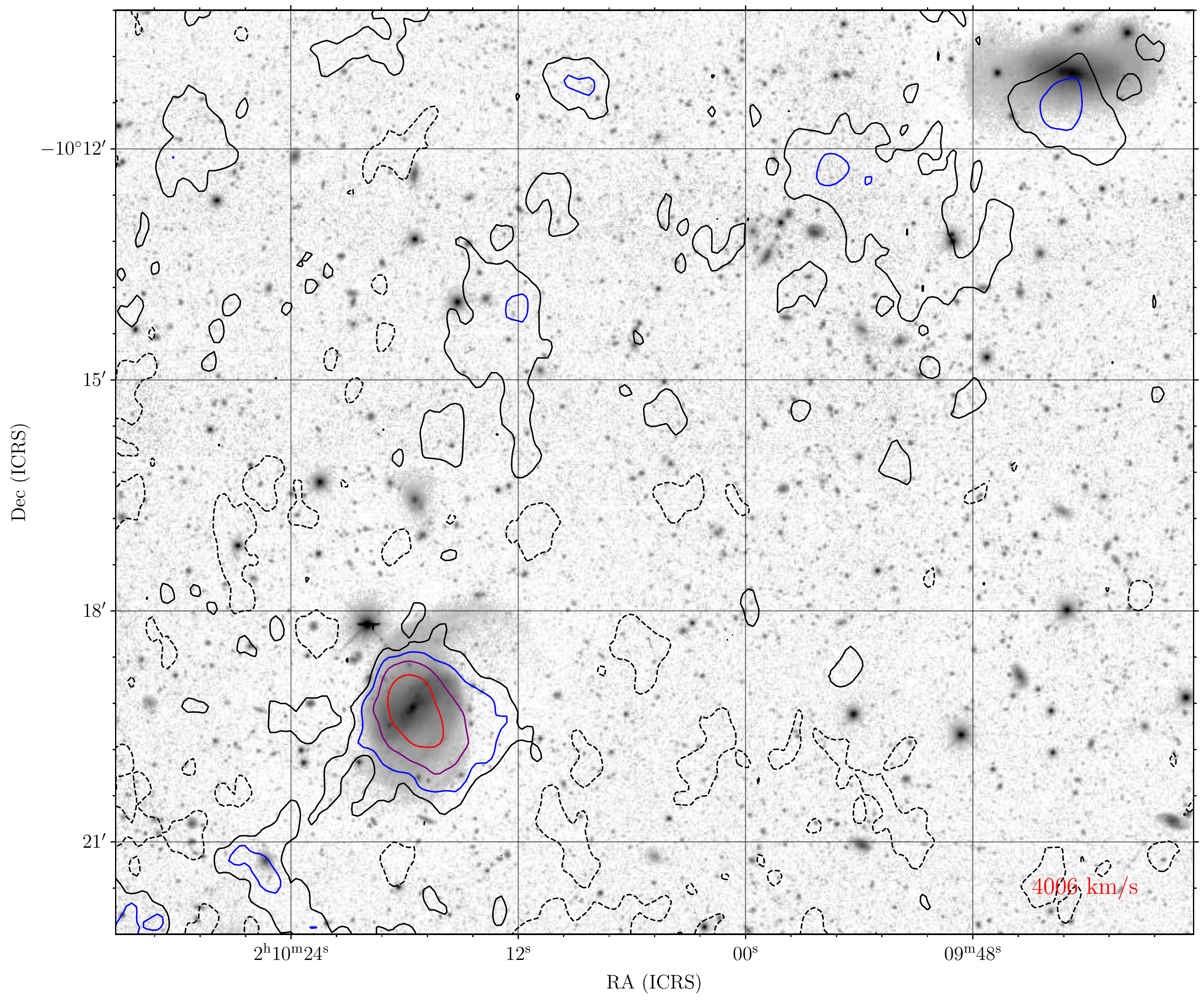}
\end{figure*}

\newpage
\begin{figure*}
    \centering
    \includegraphics[width=\columnwidth]{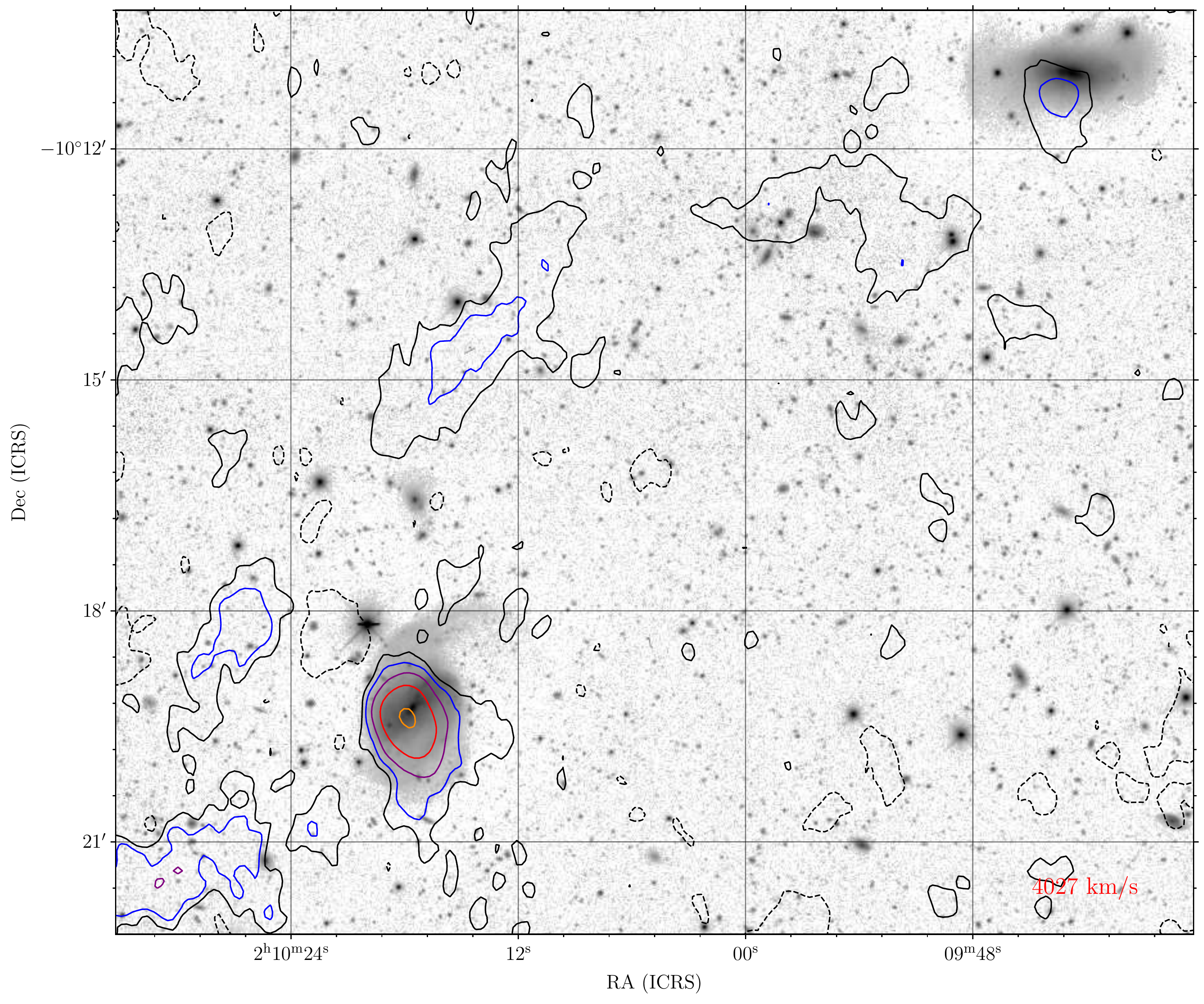}
    \includegraphics[width=\columnwidth]{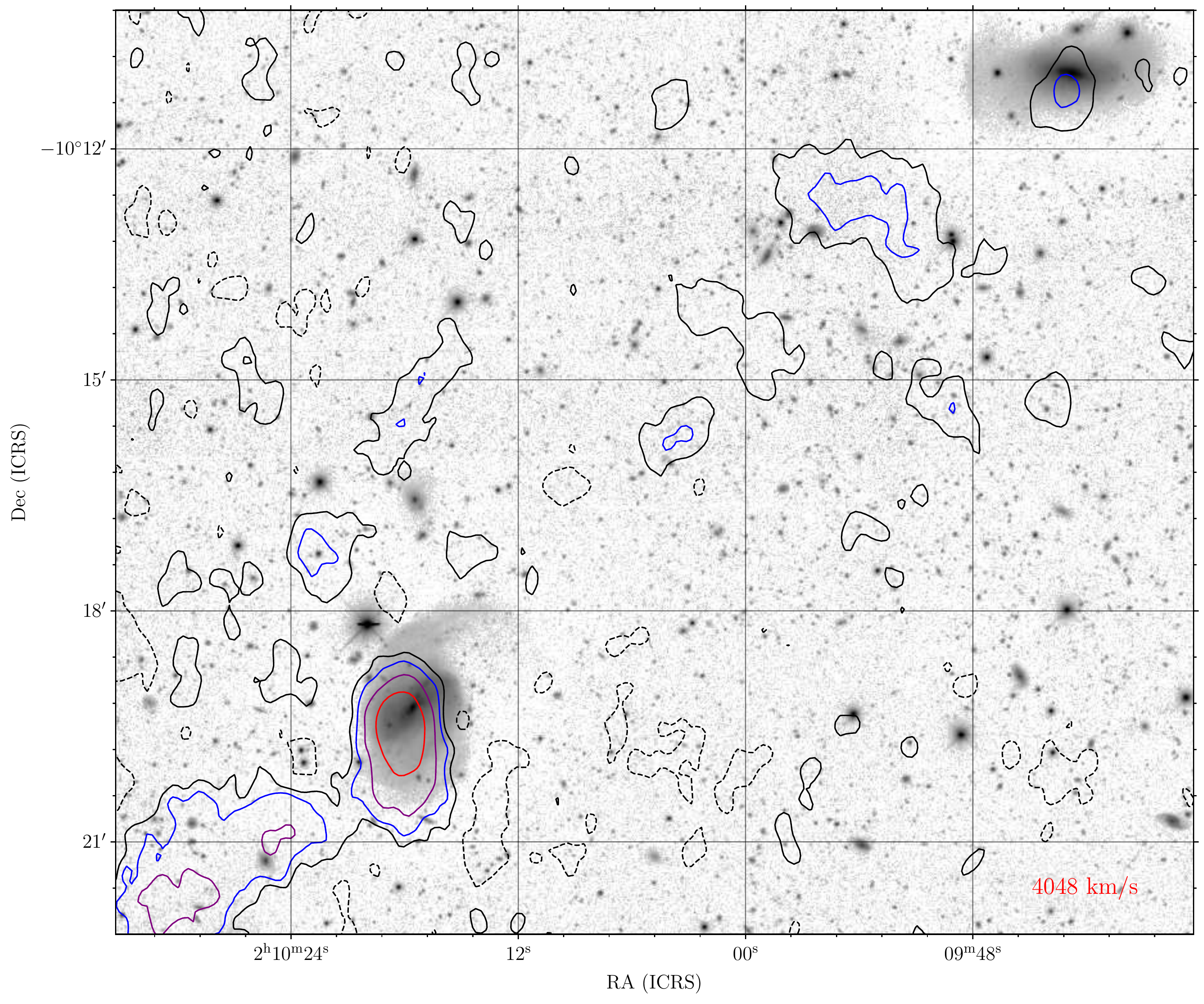}
    \includegraphics[width=\columnwidth]{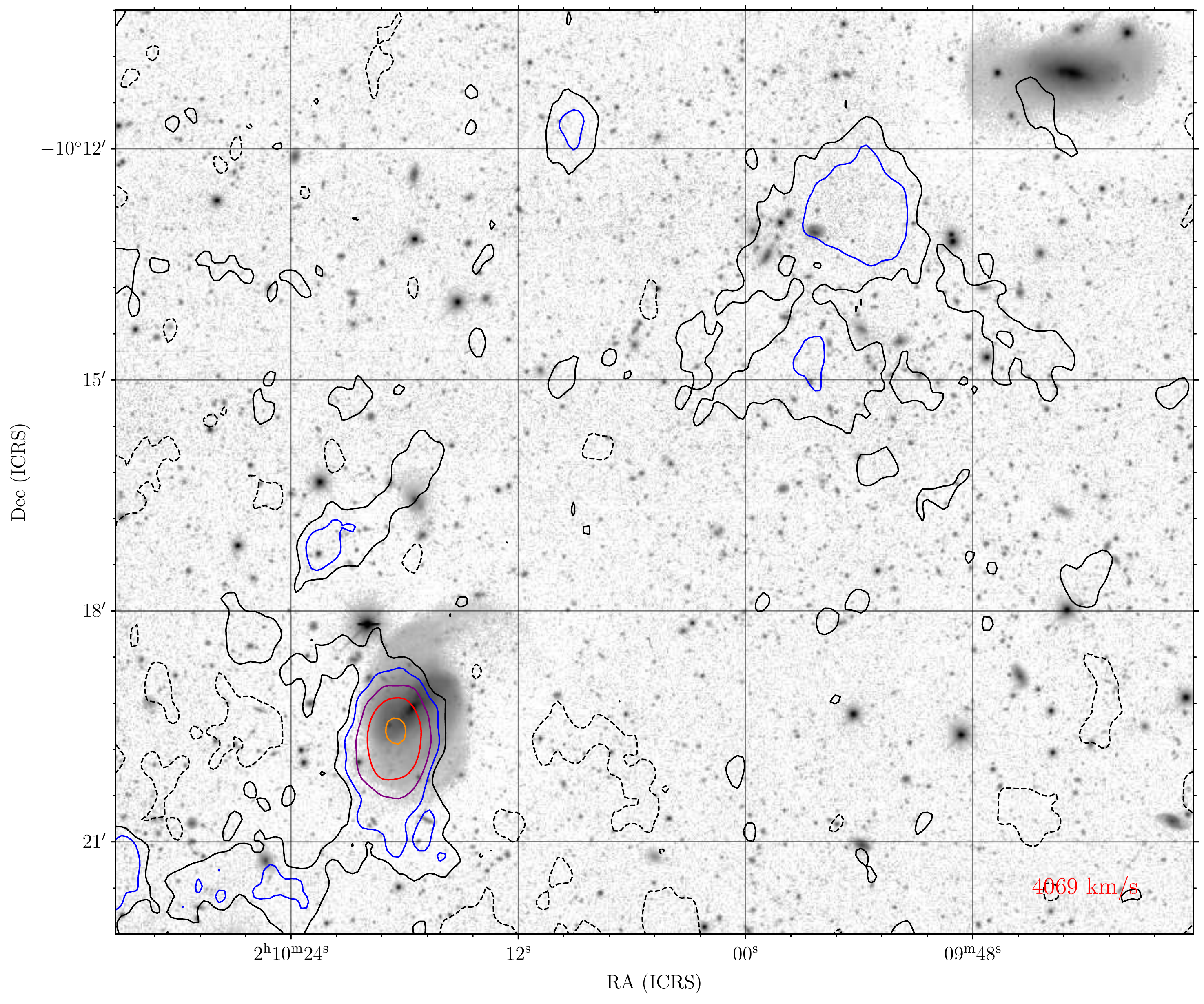}
    \includegraphics[width=\columnwidth]{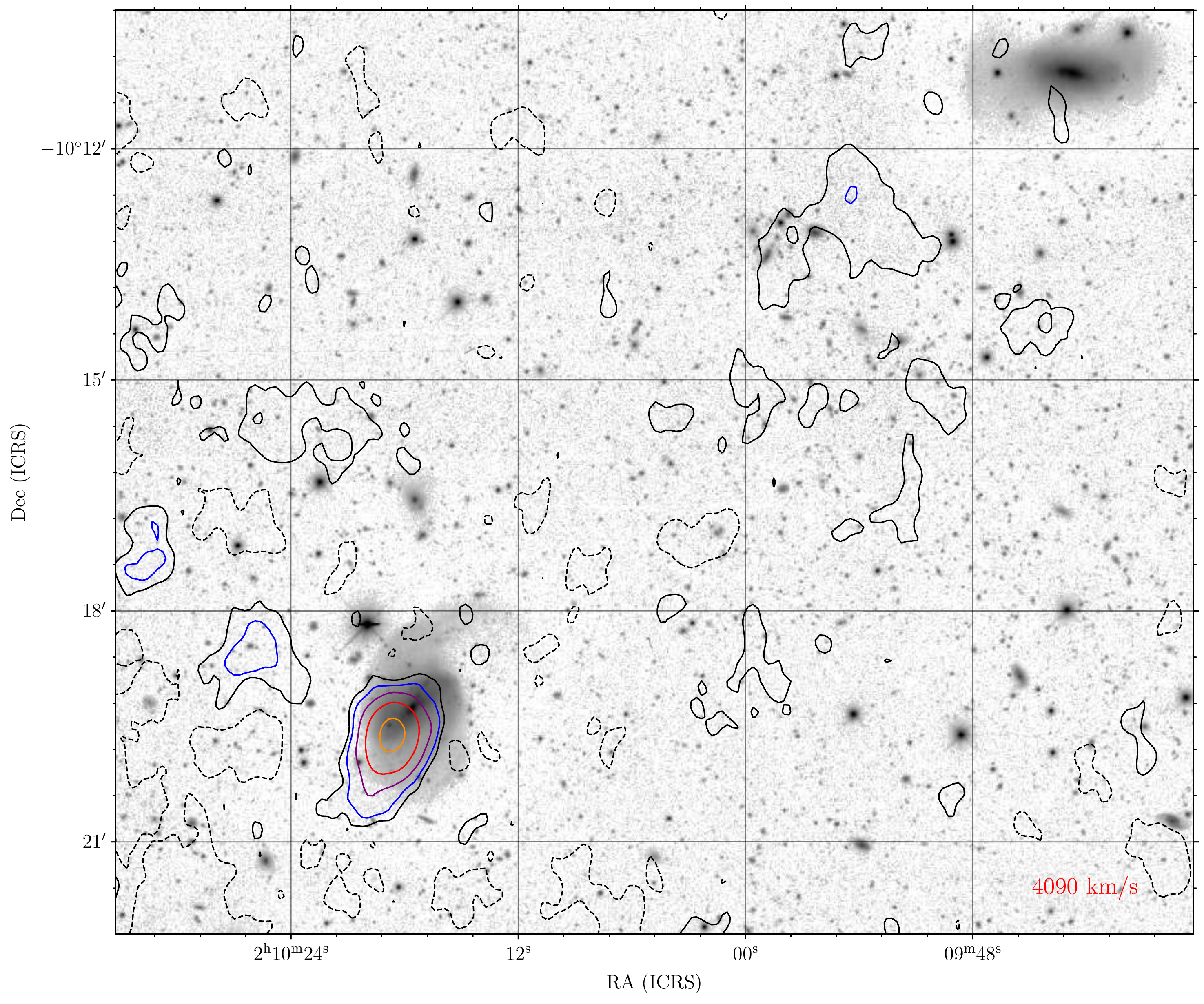}
    \includegraphics[width=\columnwidth]{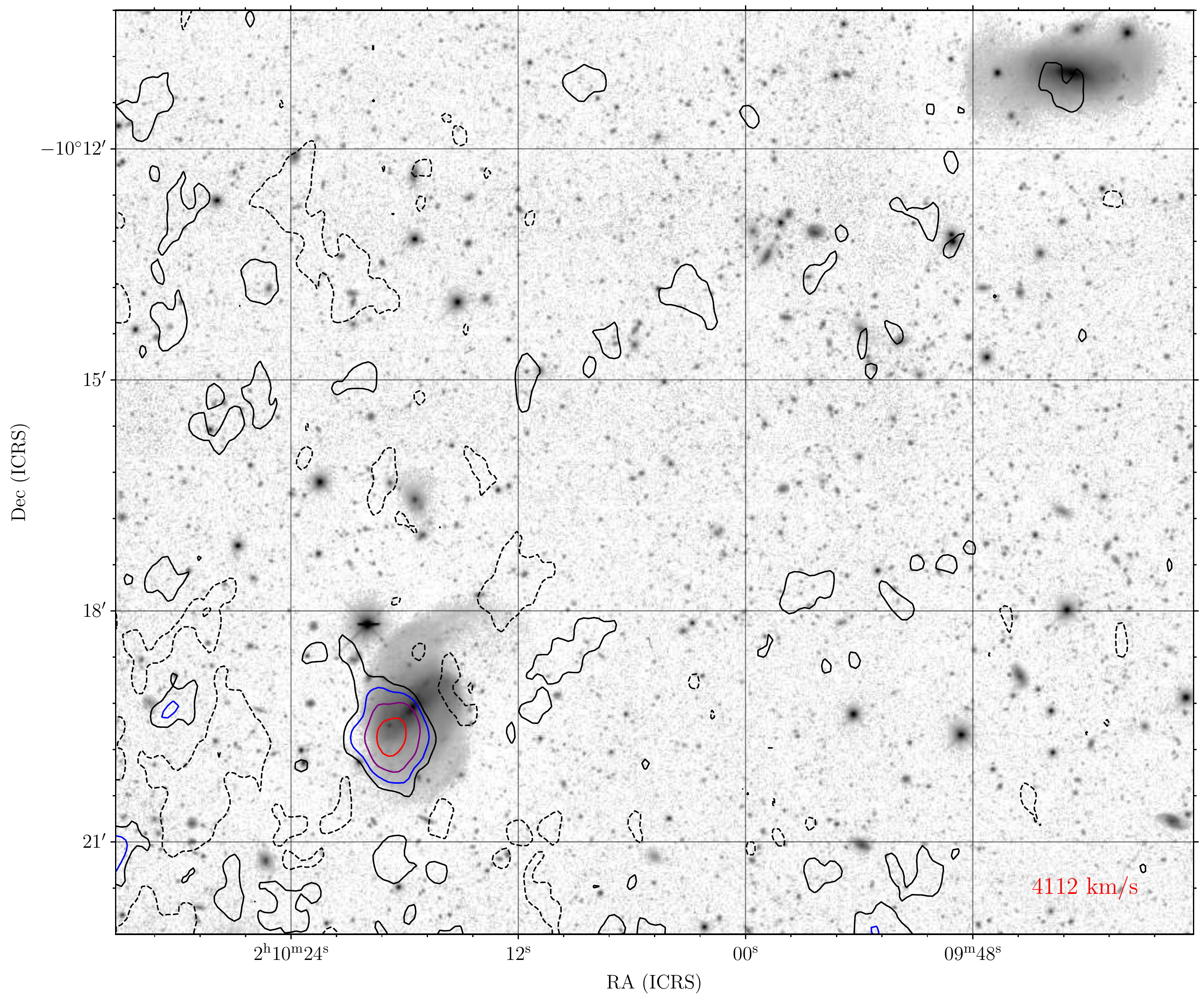}
    \includegraphics[width=\columnwidth]{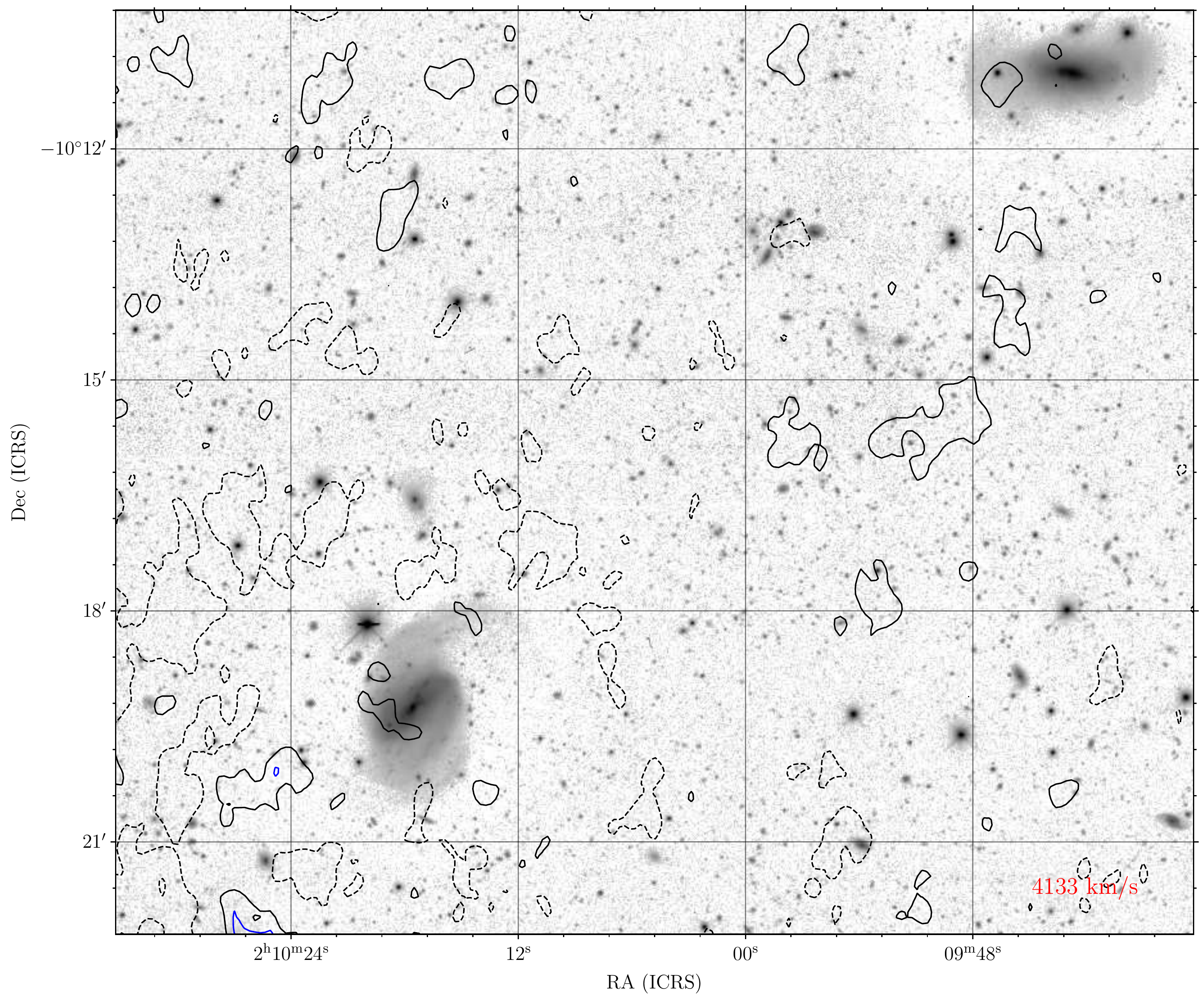}
\end{figure*}

\newpage
\begin{figure*}
    \centering
    \includegraphics[width=\columnwidth]{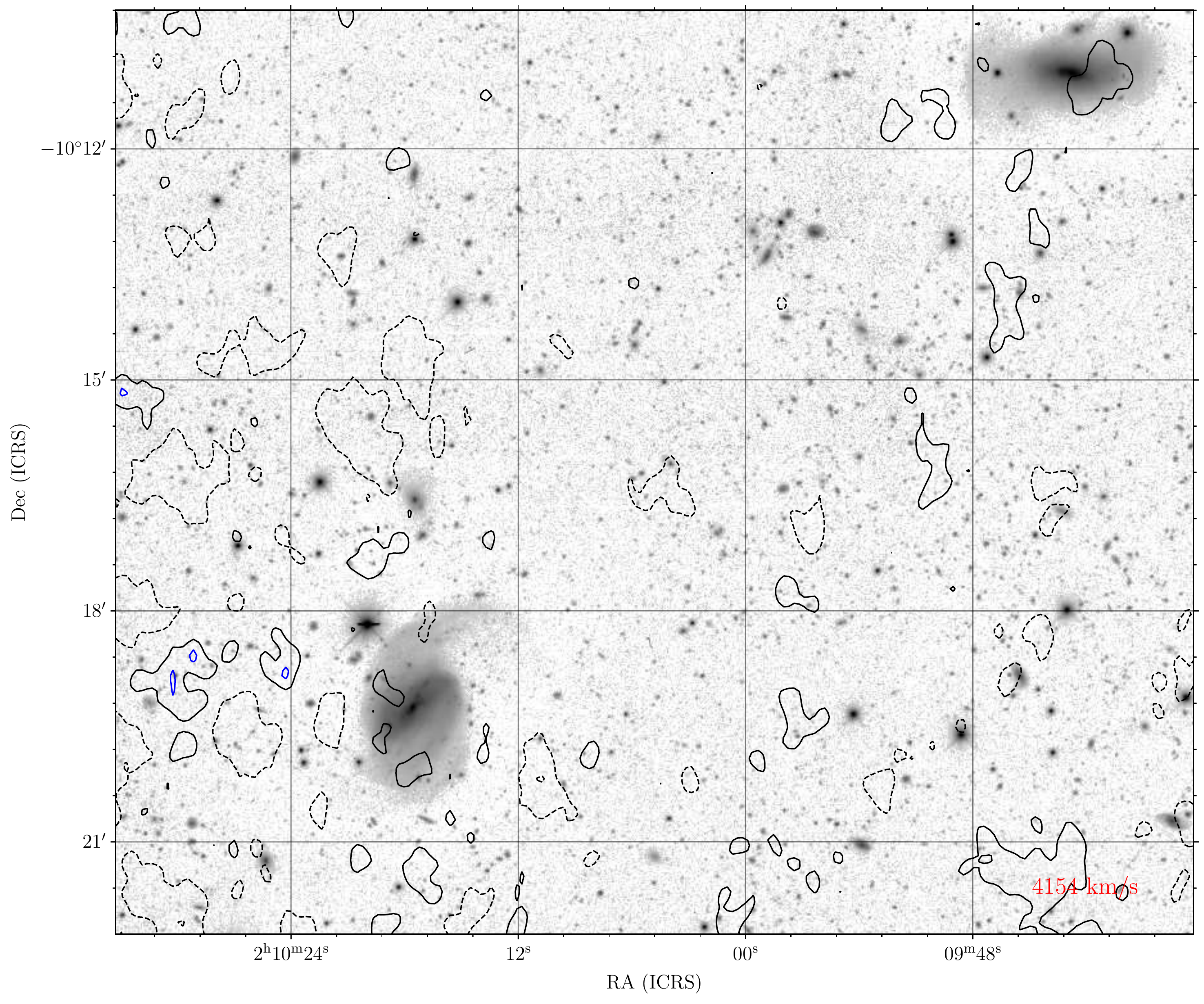}
    \includegraphics[width=\columnwidth]{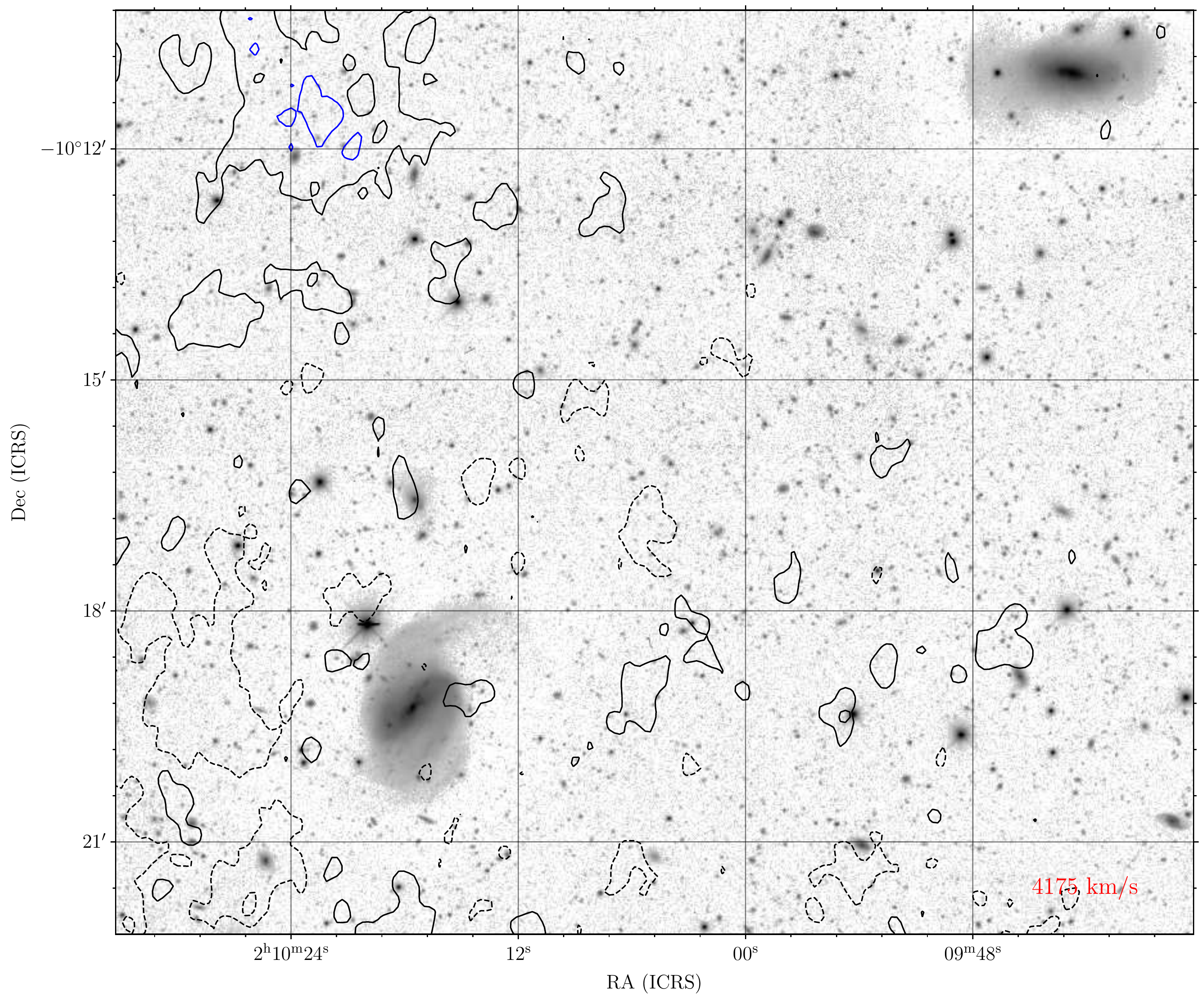}
    \includegraphics[width=\columnwidth]{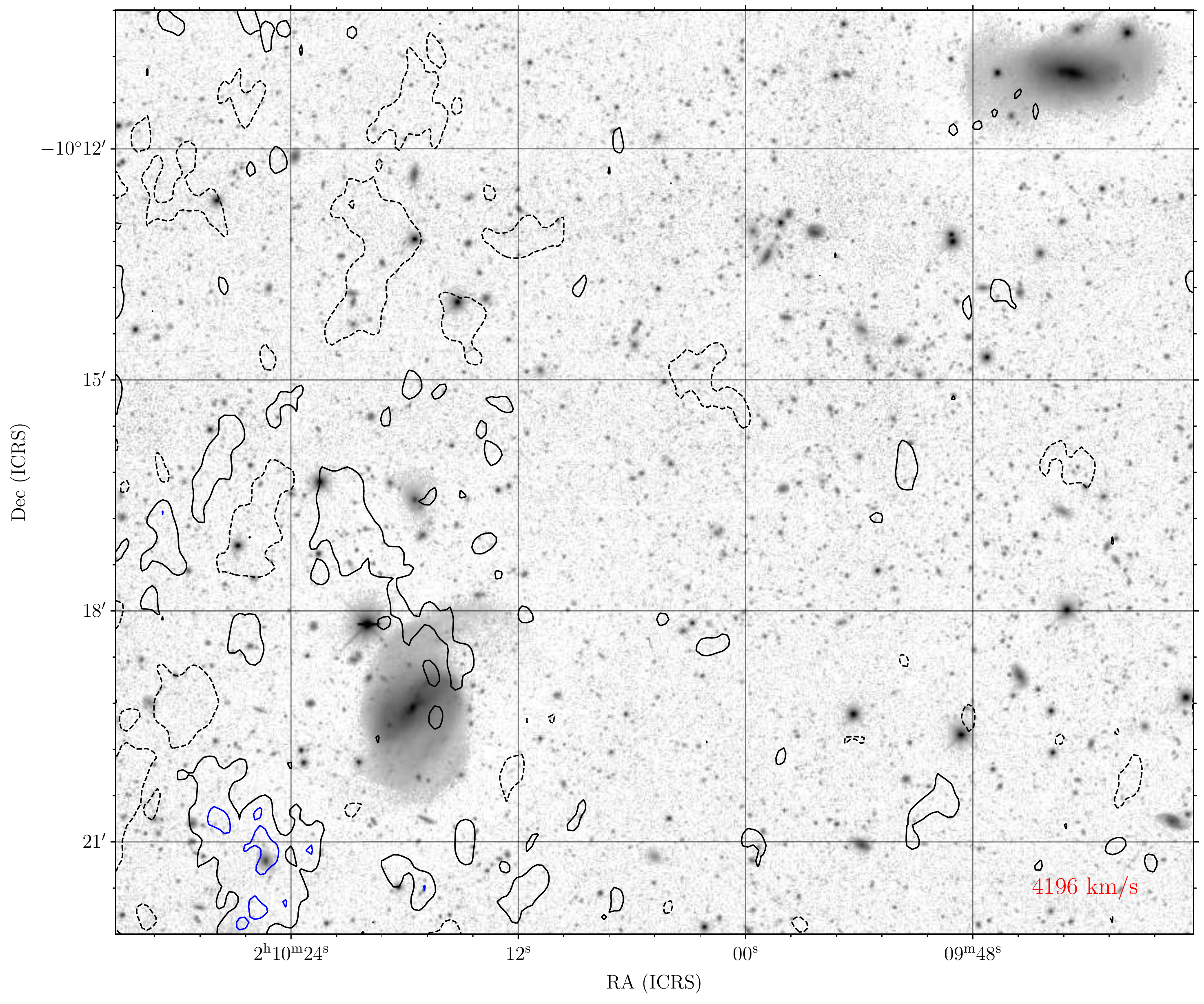}
    \includegraphics[width=\columnwidth]{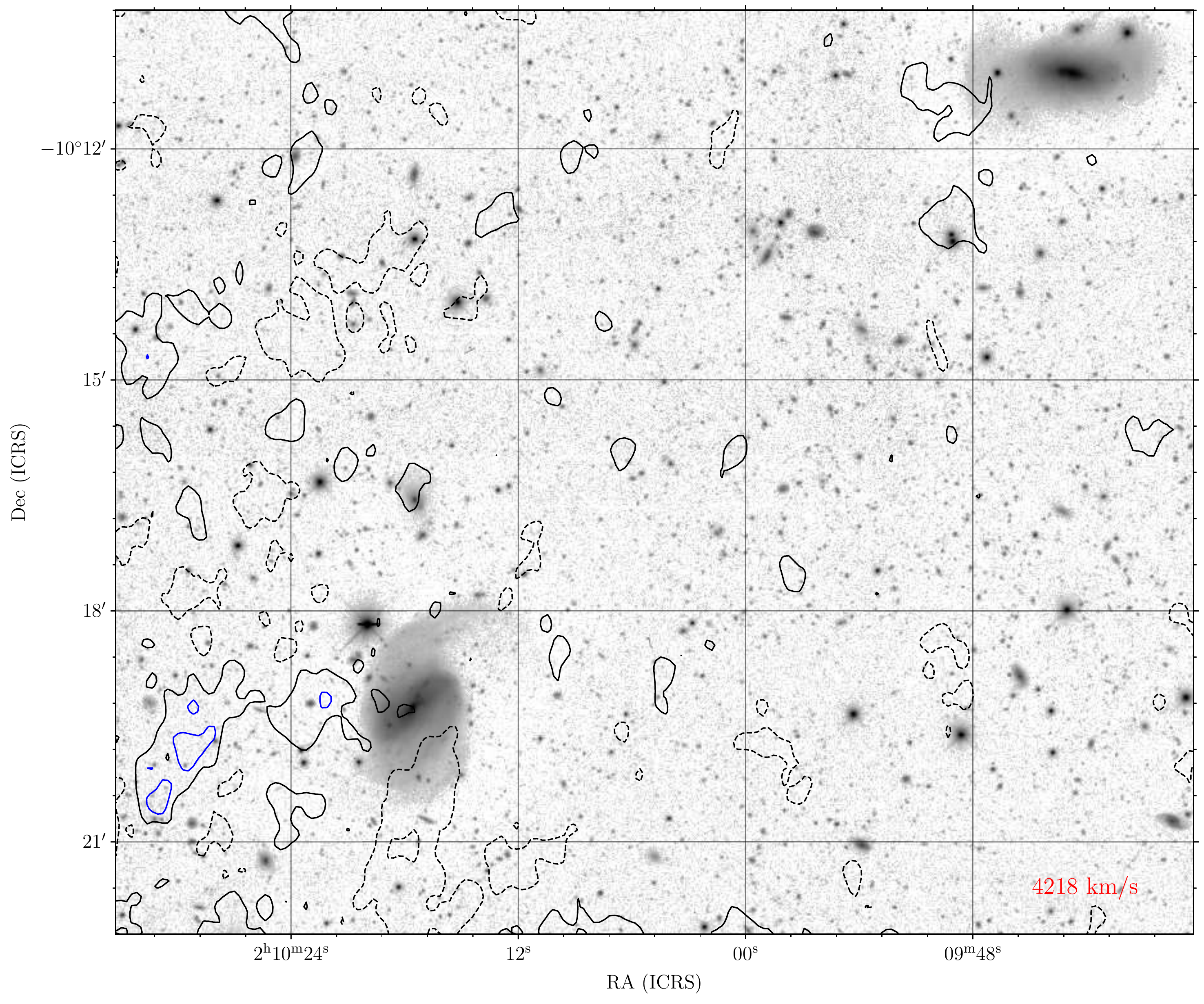}
    \includegraphics[width=\columnwidth]{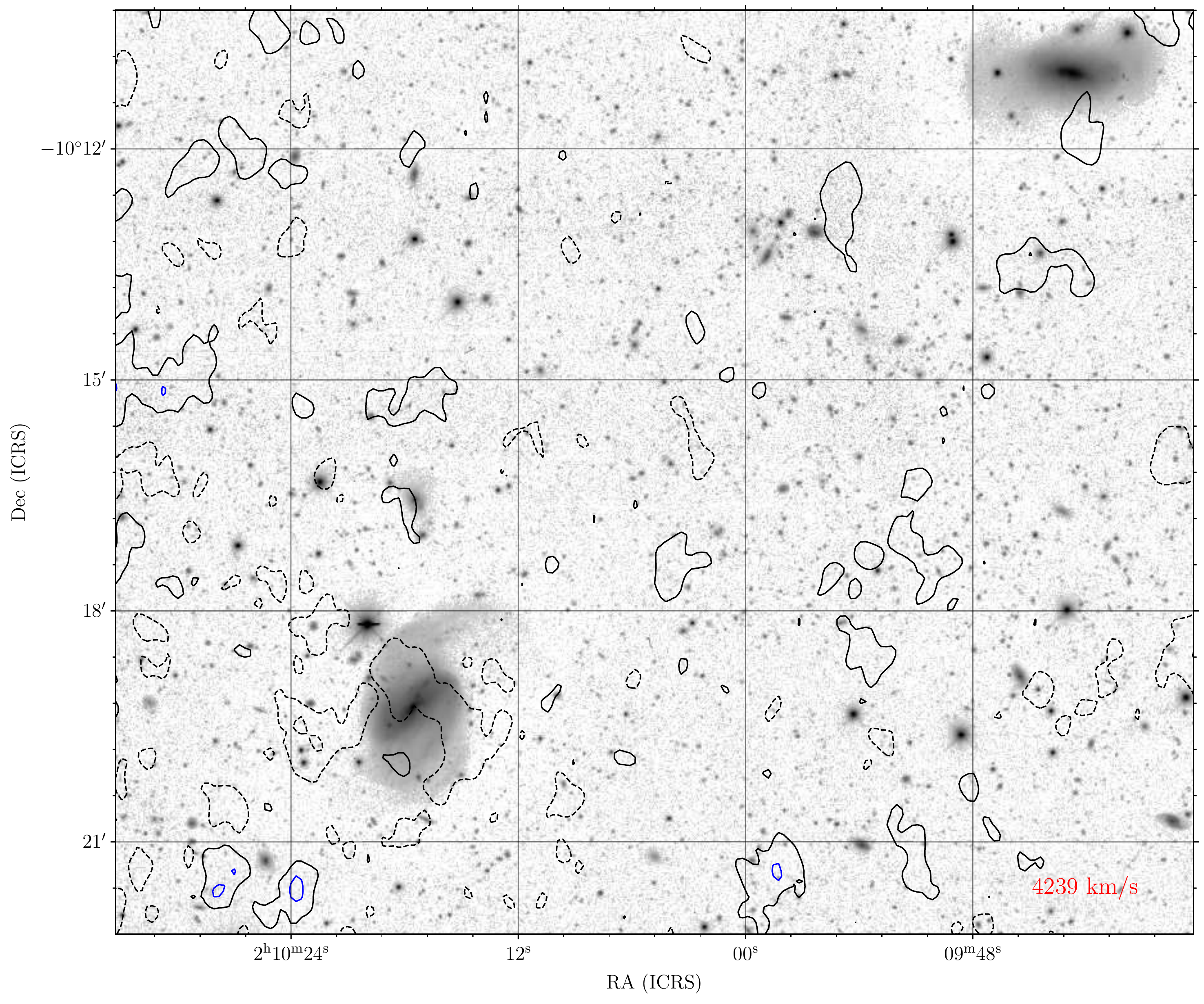}
    \includegraphics[width=\columnwidth]{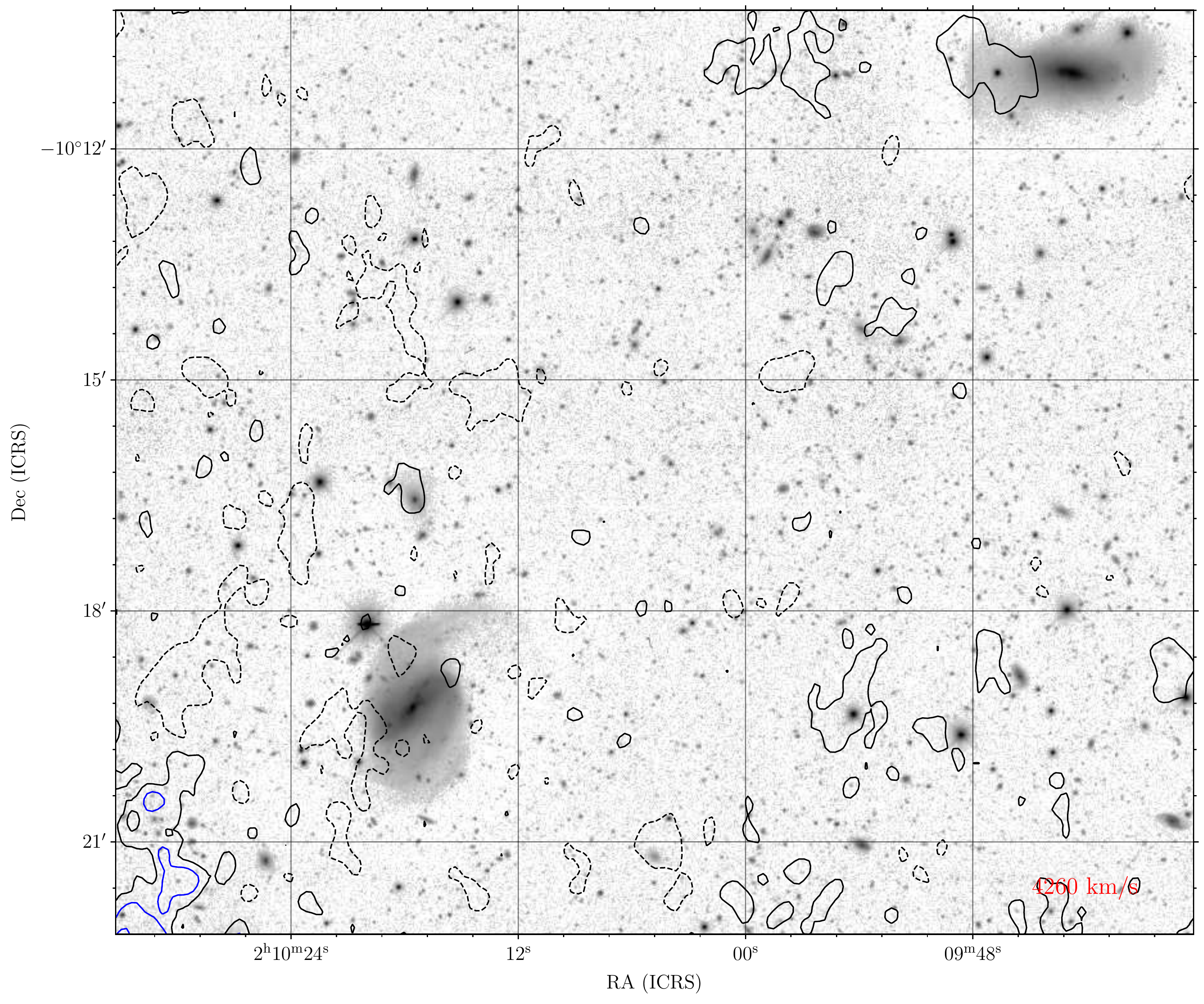}
\end{figure*}

\end{document}